\pdfoutput=1
\documentclass[a4paper,fleqn,usenatbib]{mnras}

\usepackage[T1]{fontenc}
\usepackage{ae,aecompl}
\usepackage{times}
\usepackage{pdflscape}

\usepackage{subfloat}
\usepackage{graphicx}	
\usepackage{amsmath}	
\usepackage{amssymb}	

\title[SDSS-IV MaNGA: Spatially resolved star formation histories in galaxies]{SDSS-IV MaNGA: Spatially resolved star formation histories in galaxies as a function of galaxy mass and type}

\author[Goddard et al.]{D.~Goddard$^{1}$\thanks{E-mail: daniel.goddard@port.ac.uk},
D.~Thomas$^{1}$,
C.~Maraston$^{1}$,
K.~Westfall$^{1}$,
J.~Etherington$^{1}$,
R.~Riffel$^{2, 3}$, 
\newauthor{N.~D.~Mallmann$^{2,3}$, Z.~Zheng$^{4}$, M.~Argudo-Fern{\'a}ndez$^{5}$, J.~Lian$^{1}$, M.~Bershady$^{6}$,}
\newauthor{K.~Bundy$^{7}$, N.~Drory$^{8}$, D.~Law$^{9}$, R.~Yan$^{10}$, D.~Wake$^{6,11}$, A.~Weijmans$^{12}$, D.~Bizyaev$^{13}$,}
\newauthor{J.~Brownstein$^{14}$, R. R.~Lane$^{15}$, R.~Maiolino$^{16,17}$, K.~Masters$^{1}$, M.~Merrifield$^{18}$,}
\newauthor{C.~Nitschelm$^{19}$, K.~Pan$^{13}$, A.~Roman-Lopes$^{20}$, T.~Storchi-Bergmann$^{2,3}$, D. P.~Schneider$^{21,22}$}\\
Affiliations are listed at the end of the paper}

\date{Accepted XXX. Received YYY; in original form ZZZ}

\pubyear{2016}

\begin{document}
\bibliographystyle{mnras}
\label{firstpage}
\pagerange{\pageref{firstpage}--\pageref{lastpage}}
\maketitle


\begin{abstract}
We study the internal gradients of stellar population properties within $1.5\;R_{\rm e}$ for a representative sample of 721 galaxies with stellar masses ranging between $10^{9}\;M_{\odot}$ to $10^{11.5}\;M_{\odot}$ from the SDSS-IV MaNGA IFU survey. Through the use of our full spectral fitting code FIREFLY, we derive light and mass-weighted stellar population properties and their radial gradients, as well as full star formation and metal enrichment histories. We also quanfify the impact that different stellar population models and full spectral fitting routines have on the derived stellar population properties, and the radial gradient measurements. In our analysis, we find that age gradients tend to be shallow for both early-type and late-type galaxies. {\em Mass-weighted} age gradients of early-types are positive ($\sim 0.09\; {\rm dex}/R_{\rm e}$) pointing to "outside-in" progression of star formation, while late-type galaxies have negative {\em light-weighted} age gradients ($\sim -0.11\; {\rm dex}/R_{\rm e}$), suggesting an "inside-out" formation of discs. We detect negative metallicity gradients in both early and late-type galaxies, but these are significantly steeper in late-types, suggesting that radial dependence of chemical enrichment processes and the effect of gas inflow and metal transport are far more pronounced in discs. Metallicity gradients of both morphological classes correlate with galaxy mass, with negative metallicity gradients becoming steeper with increasing galaxy mass. The correlation with mass is stronger for late-type galaxies, with a slope of $d(\nabla [Z/H])/d(\log M)\sim -0.2\pm 0.05\;$, compared to $d(\nabla [Z/H])/d(\log M)\sim -0.05\pm 0.05\;$ for early-types. This result suggests that the merger history plays a relatively small role in shaping metallicity gradients of galaxies.
\end{abstract}

\begin{keywords}
galaxies: formation -- galaxies: evolution -- galaxies: elliptical and lenticular, cD -- galaxies: spiral -- galaxies: stellar content -- galaxies: star formation
\end{keywords}


\section{Introduction}
\label{sec:introduction}
The formation and evolution of galaxies is one of the key problems in modern astrophysics governed by complex physics of star formation and suppression, as well as the interplay between dark and baryonic matter. It is well established that the formation epochs of the stellar populations in galaxies follow a pattern such that the mean stellar age increases and the formation timescale decreases with increasing galaxy mass, which is often referred to as "downsizing" \citep[e.g.][]{cowie1996,heavens2004,nelan2005,Thomas2005,renzini2006,thomas2010}. The true physical causes for this remain yet to be understood, and the key processes must involve a complex interplay between the gas accretion, satellite accretion, star formation, chemical enrichment, and the suppression of star formation through heating and/or galactic winds.\\
\\
The analysis of stellar population properties provide an important tool for constraining the relative importance and underlying physics of these processes. A key to differentiating between various processes lies in the analysis of spatially resolved stellar population properties, as processes such as feedback from central supermassive black holes or the accretion of gas or satellite galaxies, for instance, will lead to different effects at different radii. As one example, chemical enrichment models based on a simple monolithic collapse scenario with radially dependent triggers of galactic winds yield strong negative gradients in metallicity \citep{carlberg1984,thomas1999,pipino2006,pipino2010}. In such models, galaxies are predicted to undergo "outside-in" formation where star formation ceases earlier in the outermost regions due to the earlier onset of galactic winds at large radii with shallower gravitational potential wells \citep{pipino2006}. Galaxy merging, however, will dilute stellar population gradients \citep{white1980,ogando2005}. Recent cosmological models indeed show that age and metallicity gradients in early-type galaxies are predicted to be shallow due to merging \citep{hirschmann2015}. We also know from such models that accreted stellar material is expected to lie in the outskirts \citep{lackner2012}.\\
\\
The spatial distributions of stellar populations in galaxies has been studied for several decades, starting with early work by \citet{Faber1977} on radial gradients of colours and absorption line-strength indices. More detailed analyses in the 1990s from long-slit spectroscopy, mostly focusing on early-type galaxies, established the general presence of negative metallicity gradients in their stellar populations \citep{peletier1990,gorgas1990,franx1990,bender1992,davies1993,carollo1993,kobayashi1999,jorgensen1999}, suggesting an "outside-in" process of metal enrichment. More detailed follow up analyses, based on long-slit spectroscopy and a few on colours, placed the first constraints on the gradients in other stellar population parameters such as age and chemical element ratios \citep{mehlert2003,tamura2004,depropris2005,forbes2005,mendez2005,proctor2005,wu2005,moorthy2006,baes2007,jablonka2007,reda2007,sanchez2007,spolaor2008,rawle2008,clemens2009,coccato2010,kuntschner2010,rawle2010,spolaor2010,tortora2010,prochaska2011,bedregal2011,koleva2011,harrison2011,loubser2011,morelli2012,greene2012,labarbera2012,greene2013,sanchez2014,morelli2015,roig2015,gonz2015}.
The general consensus from these studies is that early-type galaxies and bulges of spiral galaxies exhibit significant negative gradients in metallicity, but no or very mild positive gradients in age and $\alpha$/Fe element ratio. The absence of a significant age gradient in early-type galaxies is further confirmed by the redshift evolution of colour gradients \citep{saglia2000,tamura2000}. Several of the studies above detect a mild positive age gradient, which suggests an "outside-in" progression of star formation in early-type galaxies and bulges where galaxy centres are more metal-rich, and star formation continues in the central regions, with enriched material, after it had stopped in the outskirts \citep{bedregal2011}. Relatively few studies focused on larger samples of late-type galaxies. Recent work by \citet{sanchez2014} and \citet{gonz2015} find generally negative light-weighted age and metallicity gradients in late-type galaxies.\\
\\
Hence, the next crucial step is now to establish the dependence of stellar population gradients on basic galaxy population properties such as galaxy mass and galaxy type, as well as the environment in which the galaxy resides. These questions have been addressed in previous work, but no consensus on the existence of such fundamental dependencies has been reached so far. \citet{forbes2005} and \citet{roig2015} find an indication for steepening of the metallicity gradient with galaxy mass for early-type galaxies, whereas \citet{sanchez2007} see a dependency with isophote shape rather than galaxy mass or velocity dispersion, and most of the other studies quoted above have not detected significant correlations with any of the galaxy parameters explored. \\
\\
The advent of large-scale Integral-Field-Unit (IFU) surveys of the local galaxy population are providing the next step forward in these studies. IFU spectroscopy, pioneered by the SAURON and ATLAS-3D projects \citep{bacon2001,davies2001,cappellari2011}, allows spatially resolved studies of stellar populations at unprecedented detail. The new generation of such surveys, CALIFA \citep{sanchez2012}, SAMI \citep{croom2012}, and MaNGA \citep{bundy2015} are providing IFU data for large galaxy samples for the first time, allowing for statistically meaningful studies of spatially resolved stellar population properties in the multivariate space of galaxy parameters to be conducted. Studies from the CALIFA survey, for example, have shown a dependence of stellar population properties with central velocity dispersion \citep{sanchez2014} and concluded that galaxy morphology is more important than mass in shaping stellar population gradients \citep{gonz2015} using a sample of 300 galaxies (73 early-types and 227 late-types). In this paper, we report on the first year data collected by the MaNGA survey and analyse radial gradients in stellar population properties using a statistically superior galaxy sample of 721 galaxies. We also reconstruct full star formation histories as a function of galaxy mass and type for both early and late-type galaxies. The MaNGA galaxy sample is also large enough to conduct an unbiased investigation into the dependence of stellar population gradients on galaxy environment, and the results of this study are presented in an accompanying paper \citep[hereafter Paper 2]{goddard2016b}.
\\
\\
This paper is organised in the following manner: Section 2 explains details of the MaNGA survey and reduction/analysis of the data and the methods used to determine galaxy morphology. Section 3 describes the numerical tools used for full spectral fitting and for obtaining radial gradients. In Section 4, we present the results of our study, then briefly provide a discussion in Section 5 and finally describe our conclusions in Section 6. Throughout this paper, the redshifts and stellar masses quoted are taken from the Nasa Sloan Atlas catalogue (NSA, \cite{blantoncat}). The Effective radius ($R_{\rm e}$) of each galaxy is measured from Sloan Digital Sky Survey photometry by performing a Se\'rsic fit in the r-band. When quoting luminosities, masses and distances we make use of a $\Lambda$CDM cosmology with $\Omega_{m}$ = 0.3 and H$_{0}$ = 67 km$^{-1}$ s$^{-1}$ Mpc$^{-1}$ \citep{planck2015}.

\begin{figure*}
\includegraphics[width=1.0 \textwidth]{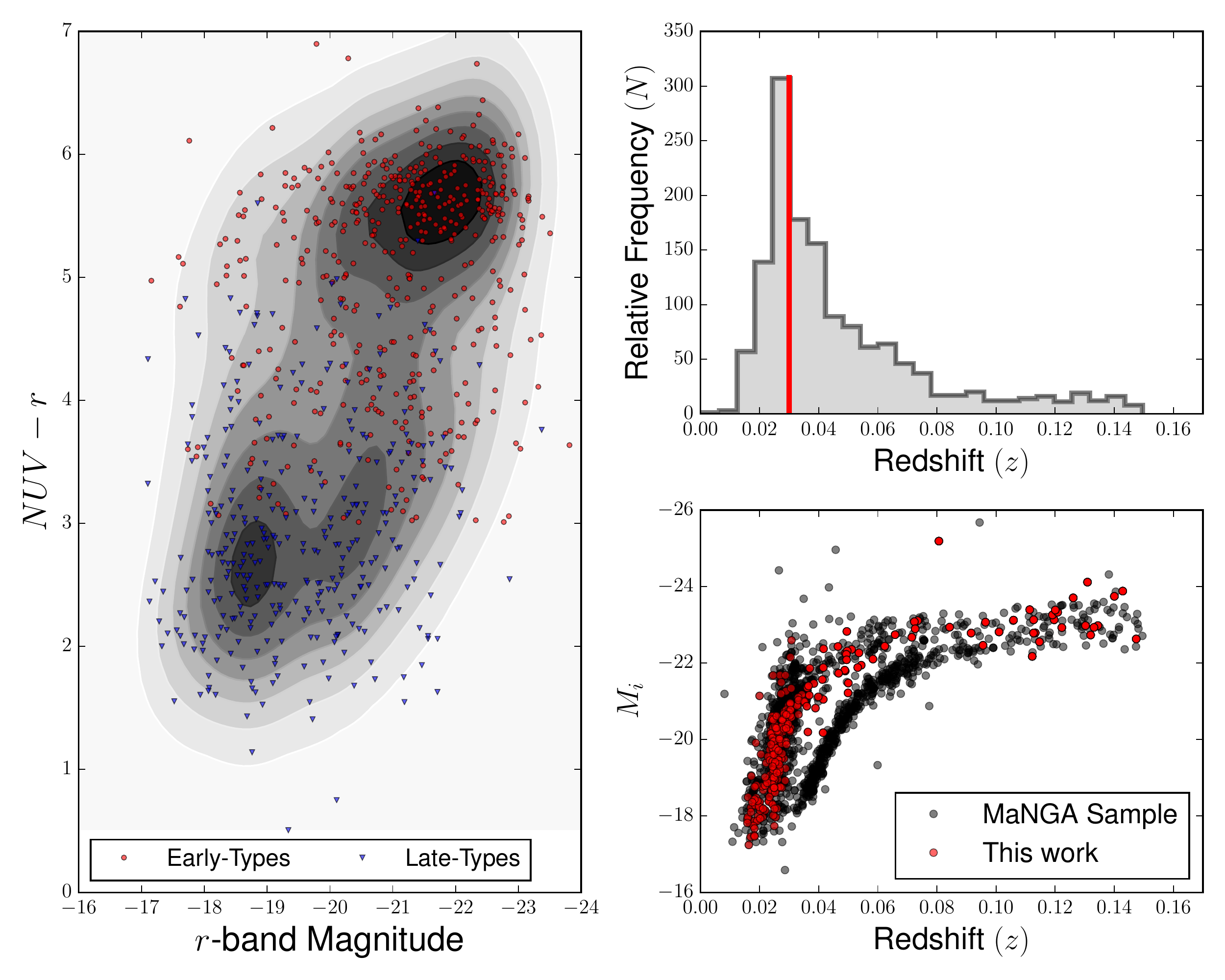}
\caption{Characterisation of the MaNGA galaxy sample used in this work. The left panel shows a colour-magnitude diagram for the 806 MaNGA galaxies used in this work. The contours represent the parent MaNGA galaxy sample and the individual points correspond to MaNGA galaxies used in this work. The upper right hand panel shows the redshift distribution of the sample, where the red line signifies the median redshift value of 0.03. The bottom right hand panel shows the $i$-band magnitude as a function of redshift for the parent MaNGA sample and for the galaxies used in this work.}
\label{fig:manga_sample}
\end{figure*}

\section{Observations and Data}
\begin{figure*} 
\includegraphics[width=0.49\textwidth]{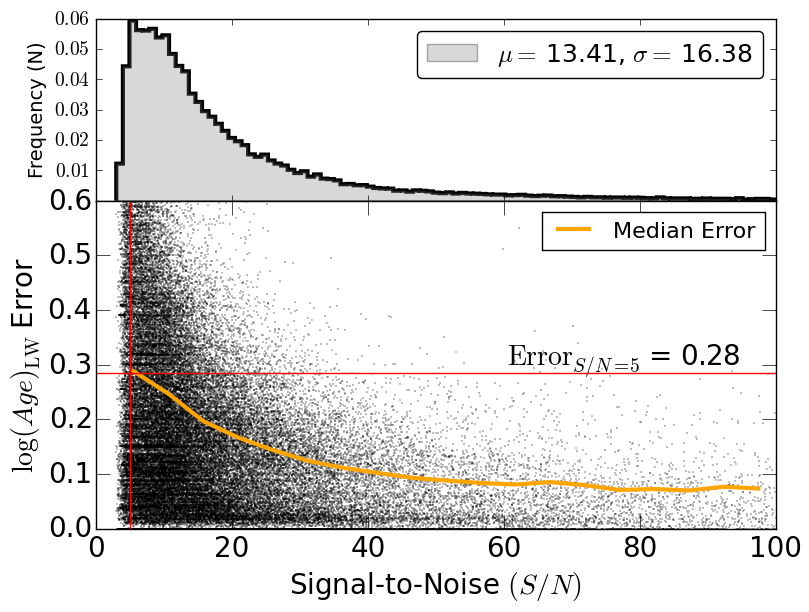}
\includegraphics[width=0.49\textwidth]{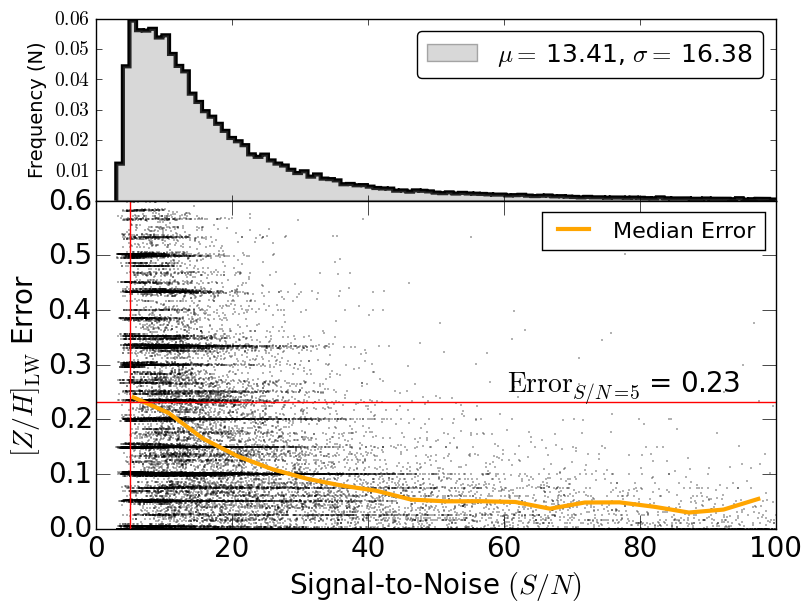}\\[12pt]
\caption{Error on luminosity-weighted age (left-hand panel) and metallicity (right-hand panel) as a function of $S/N$ as derived by FIREFLY for 60 galaxies in this work. Individual points correspond to individual Voronoi cells. The orange line in both plots represents the median error as a function of S/N and the red lines show where the running median corresponds to $S/N=5$. The average error on $\log(Age)$ is 0.28 dex and the error on $[Z/{\rm H}]$ is 0.23 dex at this $S/N$ threshold, respectively. The $S/N$ distributions can be seen at the top of each panel.}
\label{fig:ff_signal-to-noise}
\end{figure*}
The MaNGA survey \citep{bundy2015} is part of the fourth generation of the Sloan Digital Sky Survey \citep[SDSS]{york2000} and aims to obtain spatially resolved information of 10,000 nearly galaxies (median redshift $z\sim 0.03$) by 2020. The MaNGA galaxy sample is divided into a `Primary' and `Secondary' sample, following a 2:1 split. The `Primary' sample observes a galaxies optical extension out to $1.5\; R_{\rm e}$, whereas the `Secondary' sample observes galaxies out to $2.5\; R_{\rm e}$. MaNGA uses five different types of integral field unit (IFU), with sizes that range from 19 fibres ($12.5''$ diameter) to 127 fibres ($32.5''$ diameter), to optimise these observations. Fibre bundle size and galaxy redshift are selected such that the fibre bundle provides the desired radial coverage (see Wake et al. in prep for further details on sample selection and bundle size optimisation and \citet{law2015} for observing strategy). Each fibre has a diameter of 2 arcsecs and is fed into the highly sensitive BOSS spectrograph \citep{smee2013, drory2015}, which is attached to the Sloan 2.5 m telescope \citep{gunn2006, yan2015}. The BOSS spectrograph provides extensive coverage in two different wavelength channels ($3600\AA$ to $10300\AA$) with spectral resolution $R \sim 2000$ ($R \sim 1600$ at $4000\AA$ \& $R \sim 2300$ at $8500\AA$). Typical integration times are $2-3$ hours, consisting of 15 minute individual exposures dithered by roughly a fibre radius along the vertices of an equilateral triangle to ensure uniform coverage across each IFU. \\
\\
In this work, we selected an initial sample of 806 Primary sample galaxies from the MaNGA data release MPL4 \citep[equivalent to the public data release SDSS DR13]{sdssdr13}, that were observed during the first year of operation. The distribution of redshifts for the galaxies in this work and the corresponding colour-magnitude diagram can be seen in Figure~\ref{fig:manga_sample}. Galaxies with M$_{i}$ brighter than $-21.5$ mag have lower physical resolution due to the increasing redshift of the sample. The observational data was reduced using the MaNGA data-reduction-pipeline (DRP, \citet{lawdrp}) and then analysed using the MaNGA data analysis pipeline (DAP, Westfall et al, in prep). We concisely highlight some of the important steps below.

\subsection{Data Reduction Pipeline (DRP)}
Firstly, using a row-by-row algorithm the individual fibre flux and inverse variance are extracted and then wavelength calibrated using a sequence of Neon-Mercury-Cadmium arc lines. Flat-field corrections are computed using internal quartz calibration lamps. For sky subtraction, a cubic basis spline model is constructed using the background flux seen by 92 individual fibres that are placed on blank regions of the sky. The model is then subtracted from the resulting composite spectrum and shifted to the native wavelength solution of each fibre. Flux calibration is performed using the MaNGA 7 fibre mini-bundles. This procedure differs from that applied to the Prototype-MaNGA (P-MaNGA) work \citep{wilkinson2015,li2015,belfiore2015}, where the flux calibration was performed by fitting \citet{kurucz1979} model stellar spectra to the spectra of calibration standard stars covered with single fibres at each of the three dither positions. Flux calibration vectors differed by up to 10$\%$ from exposure to exposure; however, using the 7 fibre IFU mini-bundles results in a $\sim 1\%$ photometric uncertainty \citep{yan2015}. Combining the flux-calibrated spectra from the blue and the red cameras across the dichroic break is done using an inverse weighted basis spline function. Astrometric solutions are derived for each individual fibre spectrum that incorporate information about the IFU bundle metrology (location of fibre, dithering and atmospheric chromatic differential refraction, among other effects). The individual fibre spectra from all exposures for a given galaxy are then combined into a single data cube using the astrometric solution and a nearest neighbour sampling algorithm. The spatial pixel (spaxel) size of the final data cube is 0.5 arcsec.

\subsection{Data Analysis Pipeline (DAP)}
The MaNGA data analysis pipeline (DAP) is the main survey-level software package that analyses the DRP-reduced datacubes to provide properties such as stellar kinematics and emission/absorption-line fluxes. The development of the DAP is ongoing and a detailed discussion of the DAP, its algorithms, products, and robustness will be provided in a forthcoming paper. Here, we have used the current version of the DAP (version 1.1.1) to provide us with a number of inputs required for running our full spectral fitting code, FIREFLY \citep[see Section~\ref{sec:fitting_code}]{wilkinson2015, wilkinson2016}. First, we combine spectra in the datacube to reach a minimum signal-to-noise ratio ($S/N$) $=5$ in the $r$-band ($5600.1 - 6750.0\AA$) using the Voronoi binning algorithm of \cite{cappellari2004}. Given that the datacubes are generated by redistributing flux from fibres with a Gaussian Full width at half maximum (FWHM) of $\sim$$2\farcs5$ into $0\farcs5$ spaxels, there is significant covariance between adjacent spaxels \citep[for a more detailed discussion of this, see][]{law2016}. This fact is critical to our binning algorithm and covariance is taken into account for the calculation of the resulting S/N ratio of the binned spectrum. The DAP applies a simple calibration of the S/N that approximately accounts for this covariance (see also \cite{califa2013}):
\begin{equation}
\text{S/N}_{\text{Adjusted}} = \text{S/N}_{\text{NC}} / (1+1.62 \cdot\log(N_{b}))
\end{equation}
where $\text{S/N}_{\text{Adjusted}}$ is the S/N corrected for covariance, $\text{S/N}_{\text{NC}}$ is the nominally calculated S/N and $N_{b}$ is the number of spaxels in the binned spectrum. It is important to note that $S/N = 5$ is the minimum threshold for the Voronoi cell binning, and thus many cells exceed this value. This is highlighted by the histograms at the top of Figure~\ref{fig:ff_signal-to-noise}.
\\
\\
The use of our full spectral fitting code (see Section~\ref{sec:fitting_code}) requires two additional inputs provided by the DAP; measurements of the stellar velocity dispersion ($\sigma$) and fits to the strong nebular emission lines.  The determination of the stellar velocity dispersion is done using the Penalised Pixel-Fitting (pPXF) method of \cite{emsellem}. During the fit, all emission lines are masked and the stellar continuum is fit using a subset of the MILES stellar library \citep{miles2006} as the spectral templates. Rather than using the full library, the DAP splits the library into a 12x12x12 grid in $\log(t_{\mathrm{eff}})$, $[Fe/H]$ and $\log(\it{g})$ space, and selects a single stellar template from each grid cell.  This selection yields 219 templates.  The spectral resolution of the modified MILES library is correctly matched to the MaNGA data using the DRP-provided resolution measurements in the same way as is done with FIREFLY. In detail, the application of pPXF to the binned spectra adopts a 6th order multiplicative polynomial and describes the line-of-sight velocity distribution (LOSVD) as a Gaussian parameterised by the velocity, $V$, and velocity dispersion, $\sigma$. The comparison of the velocity dispersion measurements with those from the DiskMass survey \citep{martin2013} for a small subset of galaxies revealed a bias in the measurement close to the spectral resolution limit, which, however, is not relevant in the context of the present work.\\
\\
Finally, the MaNGA DAP fits individual Gaussians to the strong nebular emission lines after subtracting the best-fit stellar-continuum model from pPXF.  The best-fitting parameters for all the fitted lines [\ion{O}{II}], [\ion{O}{III}], [\ion{O}{I}], H$\alpha$,  H$\beta$, [\ion{N}{II}], and [\ion{S}{II}] are used to construct a model, emission-line only spectrum for each binned spectrum.  These models are subtracted from the binned spectra to produce emission-free spectra for analysis using FIREFLY.  Although additional weak emission lines may be present in the binned spectra, such as H$\delta$, they are not sufficiently strong to affect the best-fit parameters provided by FIREFLY (see Appendix of \citet{wilkinson2016}).

\subsection{Morphological Classification of Galaxies}
\begin{figure}
\begin{center}
\includegraphics[height=0.30\textwidth,width=0.40\textwidth]{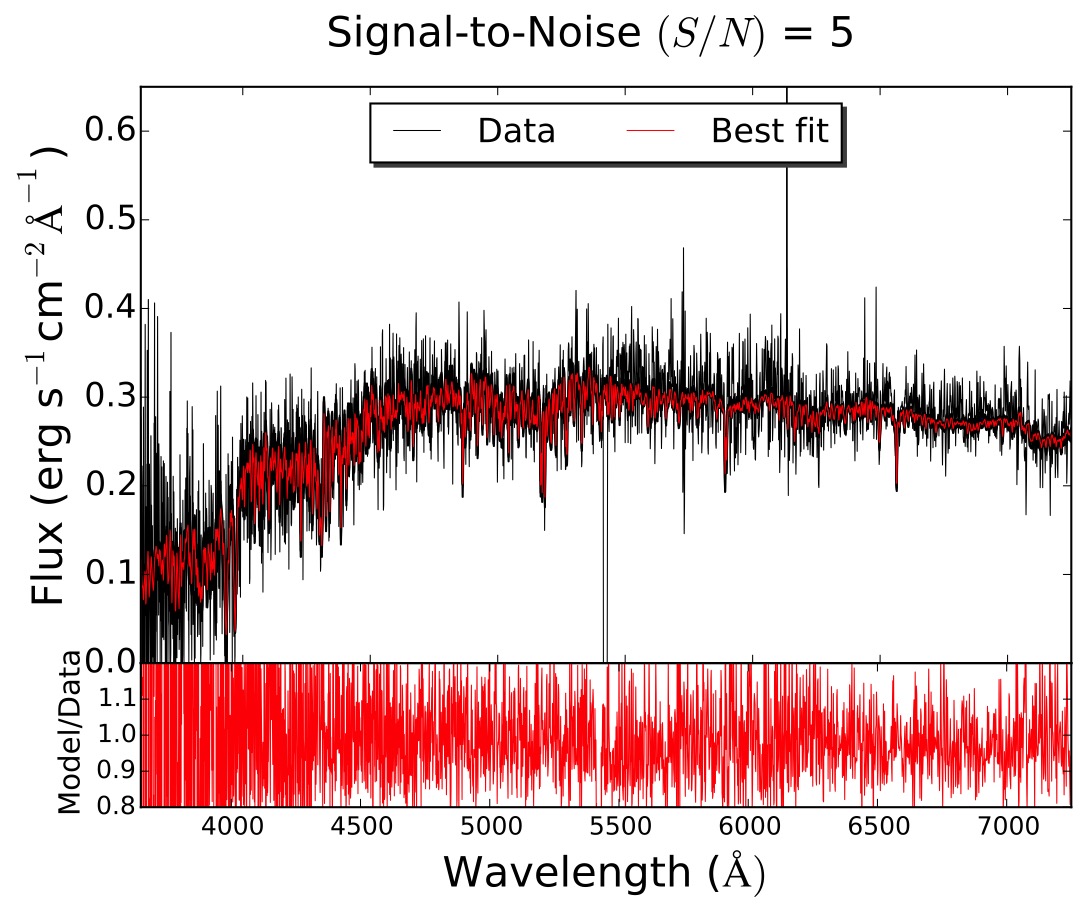} 
\includegraphics[height=0.30\textwidth,width=0.40\textwidth]{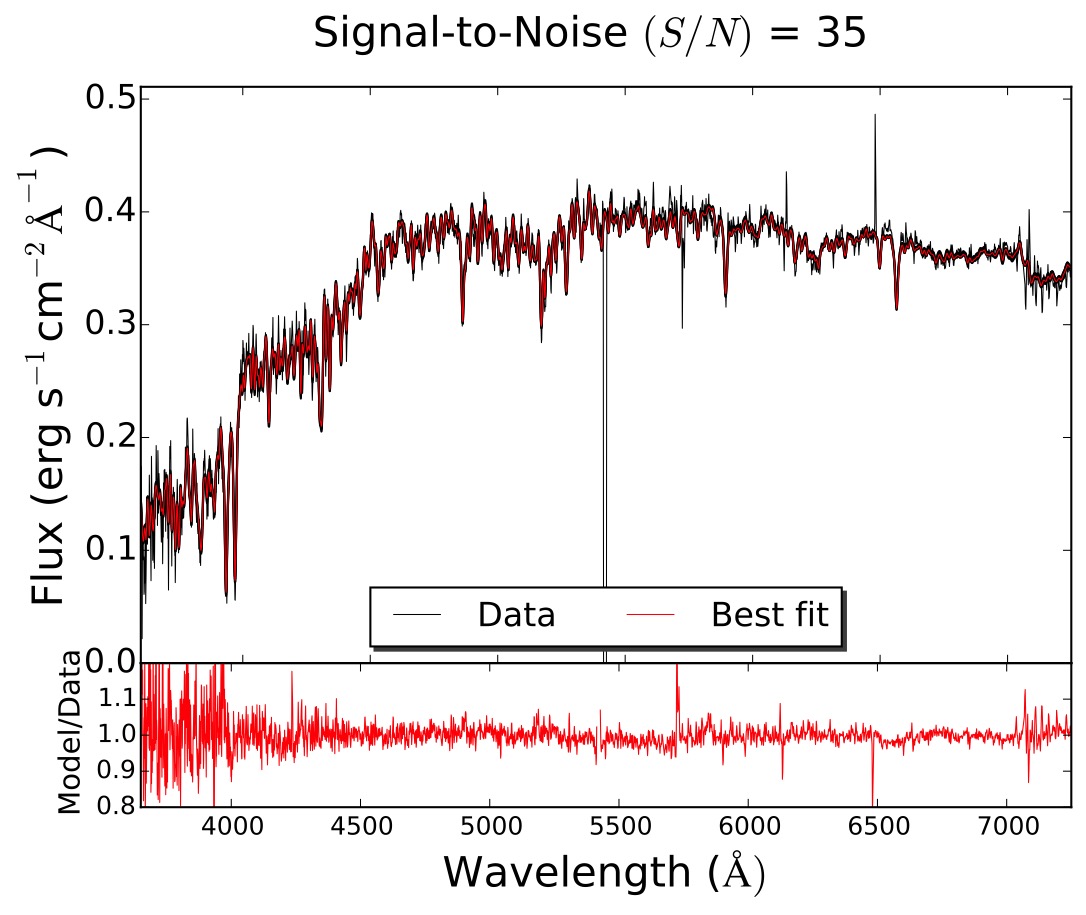}
\includegraphics[height=0.30\textwidth,width=0.40\textwidth]{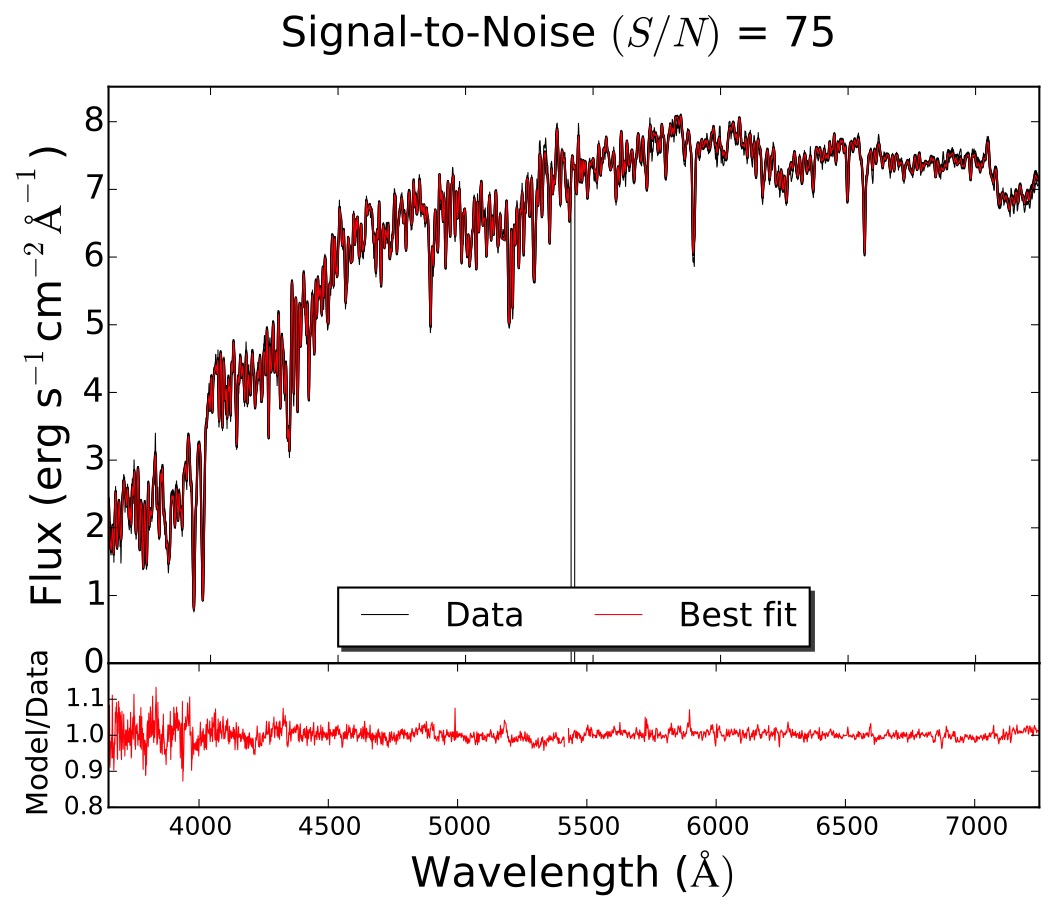}
\includegraphics[height=0.30\textwidth,width=0.40\textwidth]{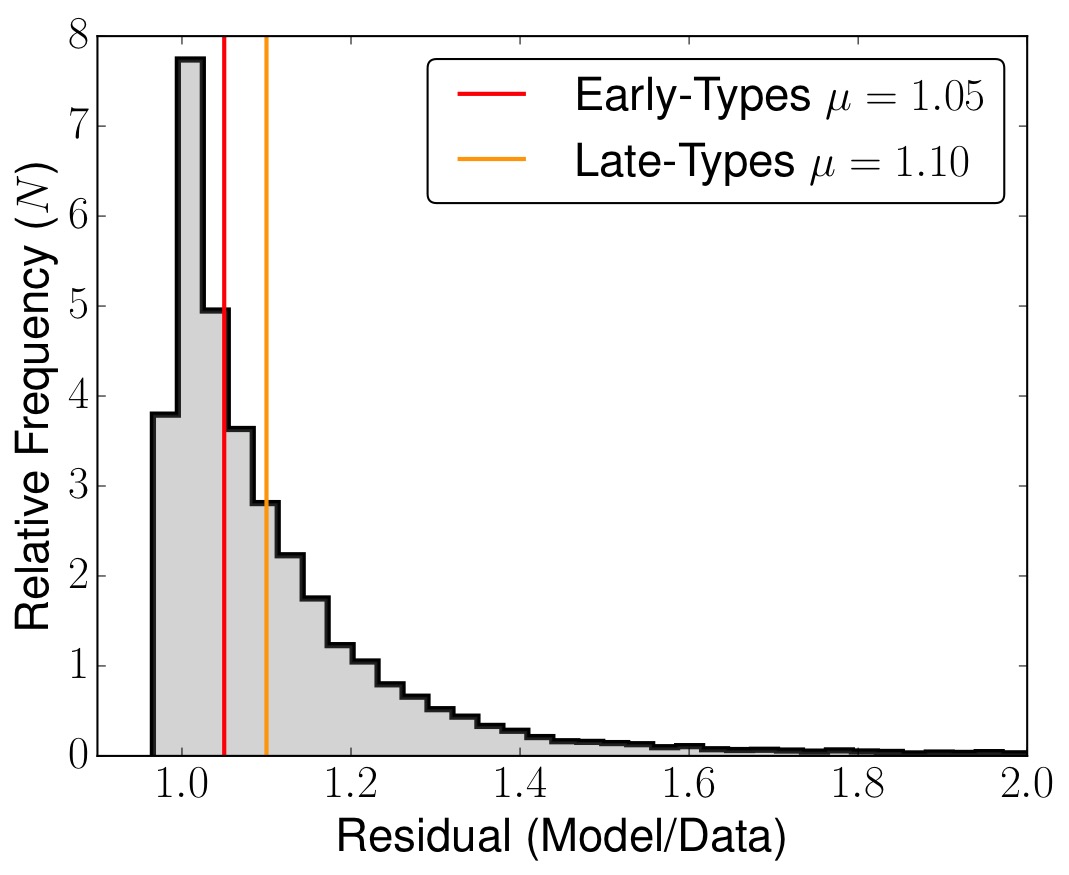}
\end{center}
\caption{From top to bottom, we show example spectral fits at three different $S/N$ ratios obtained using FIREFLY. In each plot, the black spectrum is the example MaNGA data and the red spectrum is the FIREFLY best fit. The sub-panel under each of the spectra shows the residual (Model/Data) of the fit to the data. The very bottom panel shows the distribution of residuals for all the spectral fits in this work and the red and orange line represent the median residual for early- and late-type galaxies, respectively.}
\label{fig:ff_spectral}
\end{figure}
\begin{figure*}
\includegraphics[width=0.33\textwidth]{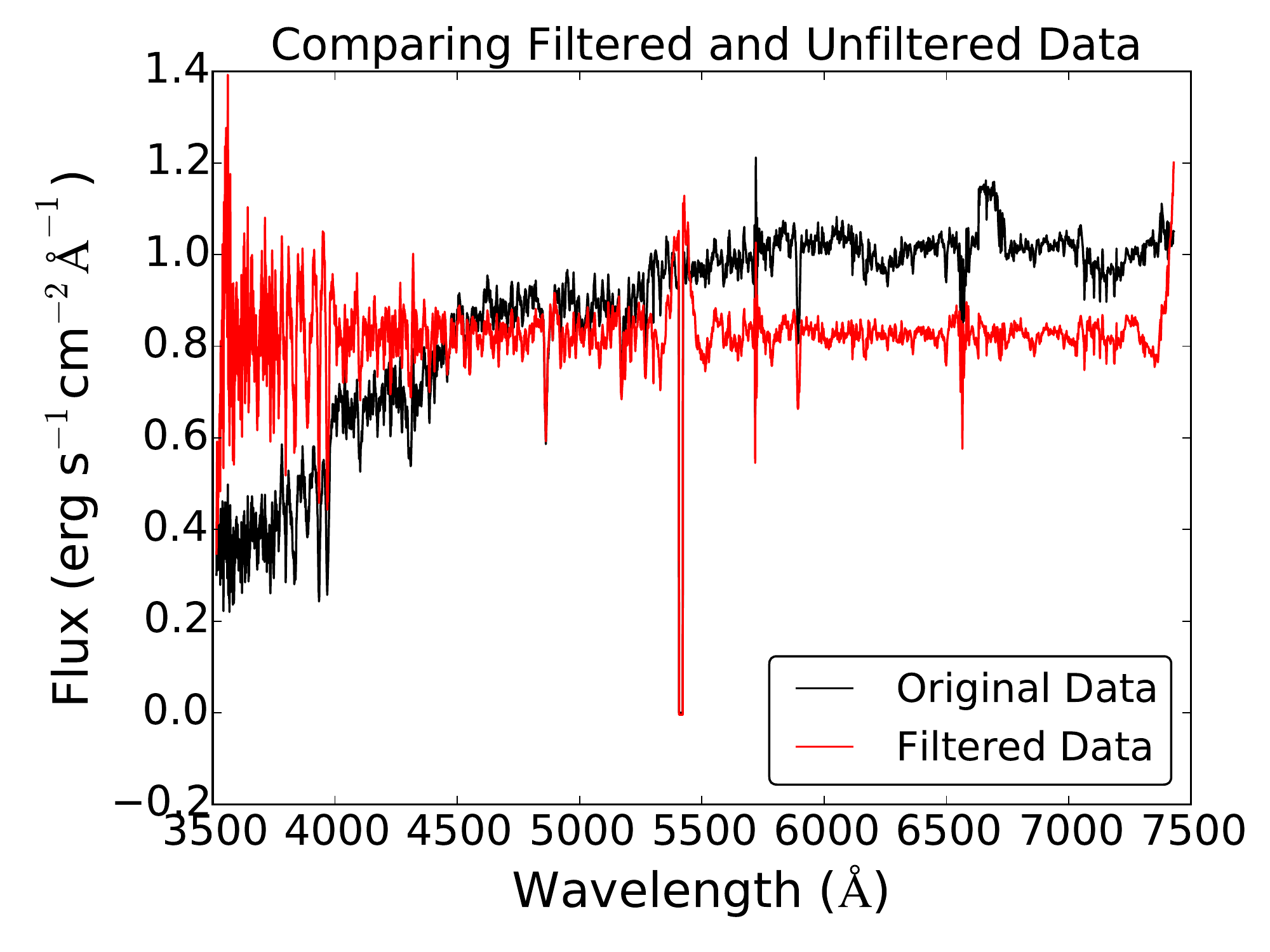}
\includegraphics[width=0.33\textwidth]{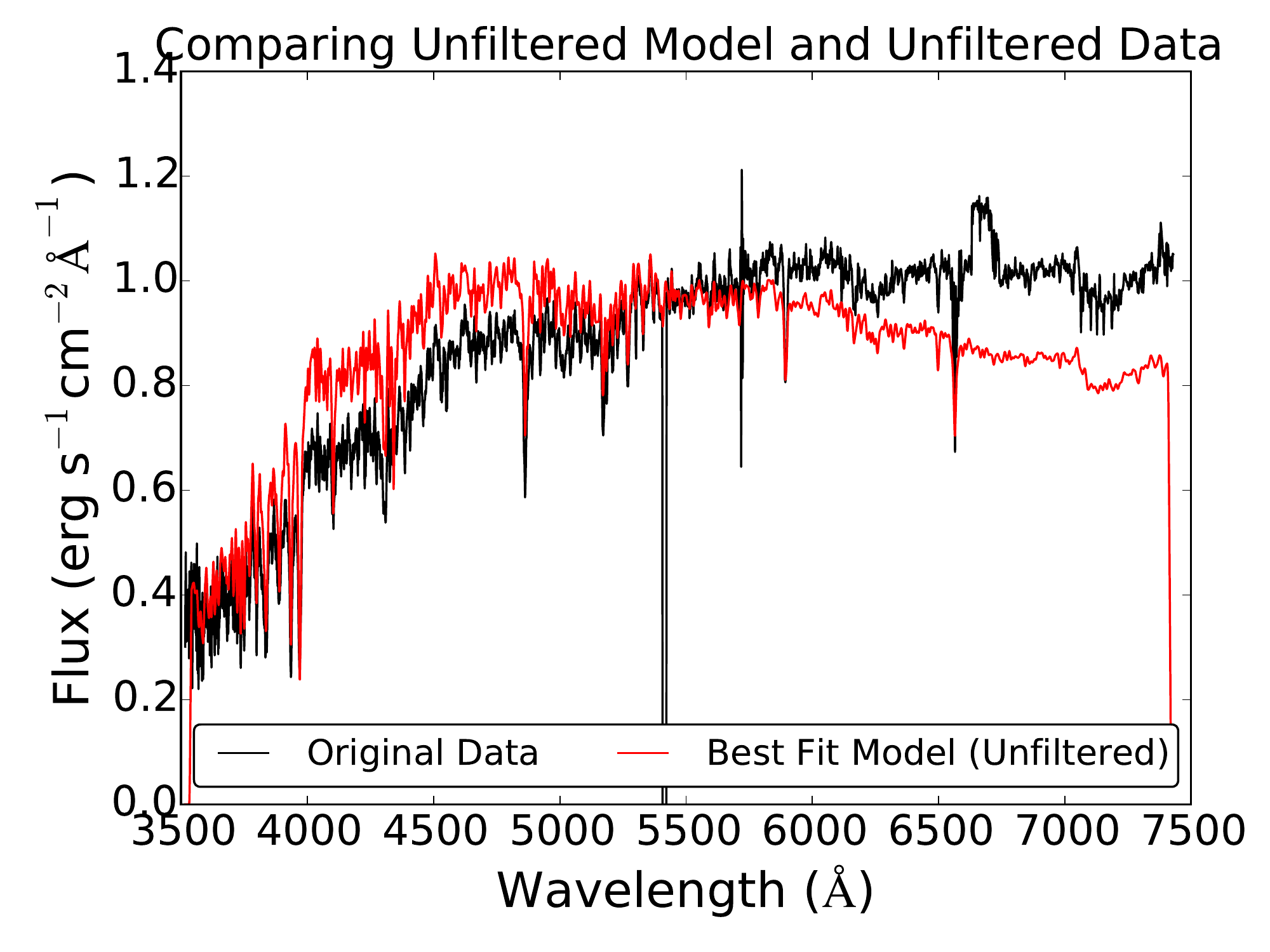}
\includegraphics[width=0.33\textwidth]{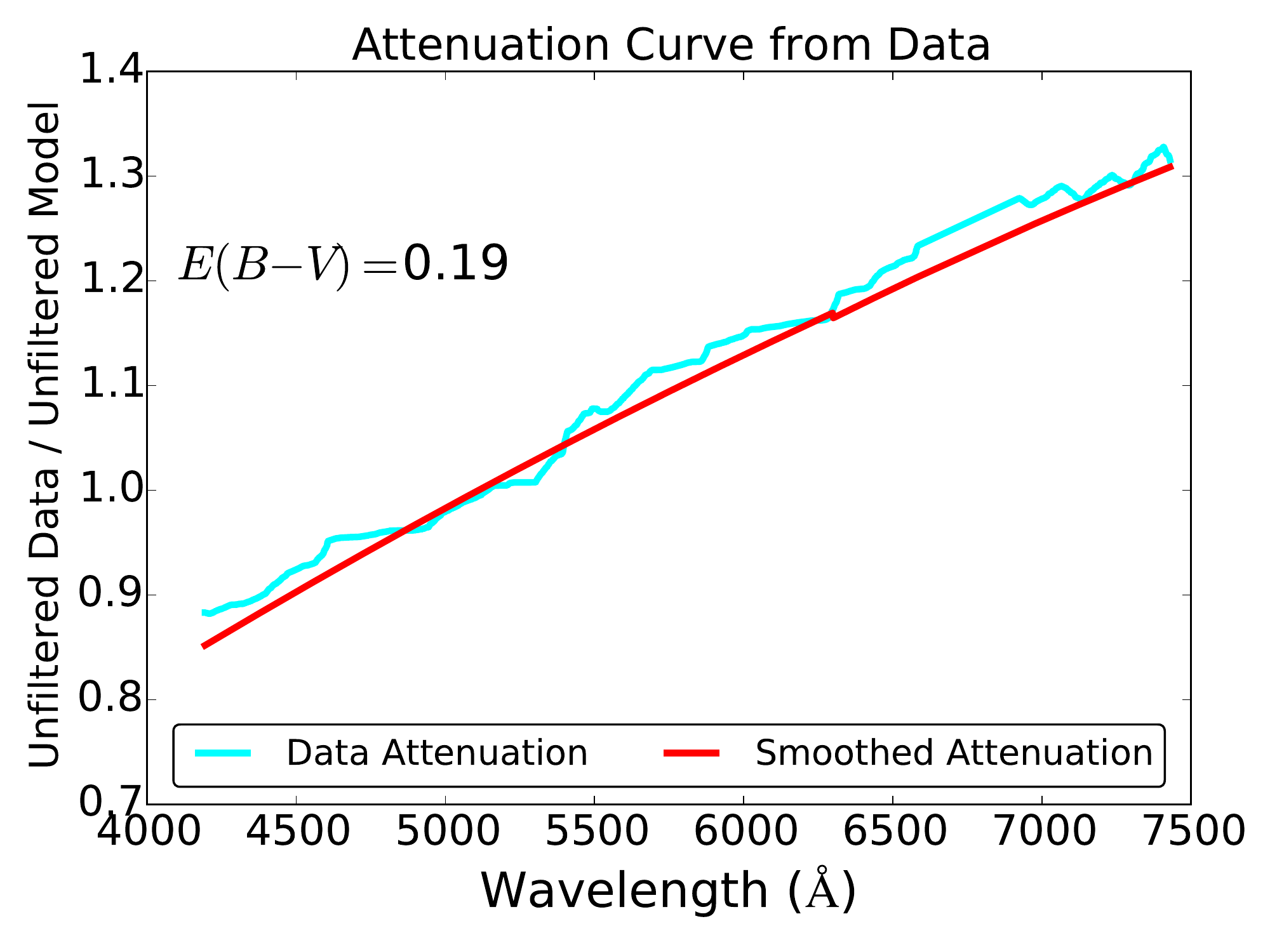}
\caption{Figure outlining the process FIREFLY uses to measure dust extinction. The left panel shows a comparison between the initial input spectrum (black) and spectrum produced after the High Pass Filter has been applied (red) to remove the large scale continuum shape. The middle panel shows the best fit unfiltered model (red) and initial input spectrum data (black). The right hand panel shows the residual between the best fit unfiltered model and the initial input data (cyan), and the smoothed attenuation curve derived from FIREFLY (red), which gives an $E(B-V) = 0.19$. The derived attenuation curve is then applied to all base models and these are then used to fit the original data, as described in Section \ref{sec:spectral_fit}.}
\label{fig:dust_plots}
\end{figure*}
By construction, our initial MaNGA galaxy sample of 806 galaxies includes a variety of morphological types. To classify galaxies by morphology, we used Galaxy Zoo \citep{lintott2011, willett2013}. The Galaxy Zoo catalogue provides statistics from the general public, who vote which galaxy morphology they believe best describes the SDSS galaxy. As the MaNGA galaxy sample is drawn from the SDSS, we can simply cross match with the Galaxy Zoo catalogue in order to establish the morphological classification. In this work we split the galaxies into two subsets, namely `Early-type' galaxies (Elliptical/Lenticular) and `Late-type' galaxies (Spiral/Irregular). Galaxies with an 80$\%$ majority vote for a specific morphological type were selected for this analysis. Galaxies which did not fulfil this criterion were visually inspected and classified by the authors ($\sim$15\% of the sample). We choose to only distinguish between two morphologies in an attempt to keep a large enough sample to carry out a meaningful statistical analysis. More detailed morphological investigations (breaking down late-types into more distinct Hubble types) will require larger data sets than the one used in this study, and are therefore the subject of future MaNGA work.

\subsection{Final Sample}
\begin{figure}
\includegraphics[width=0.49\textwidth]{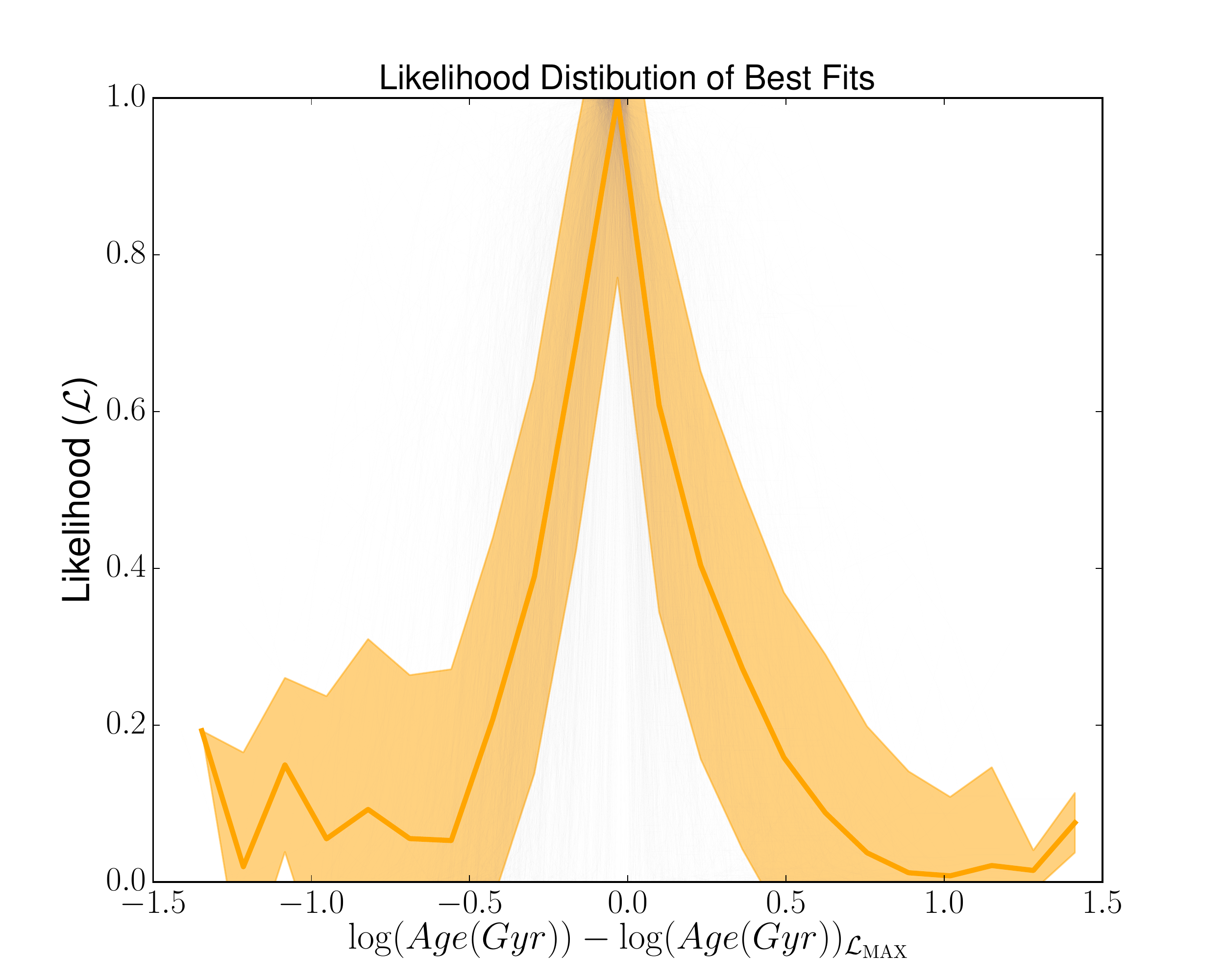}
\caption{Example likelihood distributions of light-weighted age, constructed from the array of best fits produced by FIREFLY for a sample of 60 galaxies. The grey lines correspond to individual spectra, the orange line shows the median value and the orange shaded region shows the standard deviation. FIREFLY uses the median and 1-$\sigma$ interval of these distributions to deduce the average stellar population property and its associated error as described in Section~\ref{sec:spectral_fit}.}
\label{fig:likelihood}
\end{figure}
In Paper 2, we study the effect of galaxy environment on the derived stellar population gradients, using the same galaxy sample as in this work. Due to the complex geometry of the SDSS survey area (which consists of an array of parabolic strips), some MaNGA galaxies that reside close to the footprint edge had to be excluded from both analyses because an accurate measure of environment was not possible. Furthermore, a number of galaxies that were in the final morphologically classified sample had to also be neglected due to having unreliable velocity dispersion estimates from the DAP, which led to poor spectral fits. This led to the exclusion of 85 galaxies (33 early-type galaxies and 52 late-type galaxies spanning a range of environments and masses), from our original sample leaving 505 early-type galaxies and 216 late-type galaxies (70\% and 30\% of the sample, respectively). 

\section{Stellar Population Gradients}
\label{sec:fitting_code}
We apply the full spectral fitting code FIREFLY \citep[{\bf F}itting {\bf I}te{\bf R}ativ{\bf E}ly {\bf F}or {\bf L}ikelihood anal{\bf Y}sis]{wilkinson2015,wilkinson2016} and the models of \cite{maraston2011} to the MaNGA DAP Voronoi binned spectra ($S/N = 5$) to determine stellar population ages and metallicities. We then calculate the radial gradients of these properties and investigate the dependence of these gradients on stellar mass and galaxy type. As mentioned previously, in Paper 2 the dependence of the gradients on galaxy environment is investigated. A parallel paper by \citet{zheng2016} also looks at environmental dependence of stellar population gradients but uses different fitting codes, stellar population models and environmental measure than what is used here and in Paper 2.

\subsection{Stellar Population Properties}
To obtain stellar population properties, such as age and metallicity, we use a newly developed full spectral fitting code FIREFLY \citep{wilkinson2015,wilkinson2016}. FIREFLY was developed in order to map out inherent spectral degeneracies, work well at low $S/N$ ratios and include a robust method for determining reddening. The fitting approach also employs minimal priors. In particular, no prior is set on the range of the stellar population parameters, the star formation history or the reddening. This means that the only prior will be the adopted stellar population model grid. This approach has been shown to be an accurate way to recover the stellar population properties of mock galaxies in \citet{wilkinson2016}. In the following sections, we briefly describe the FIREFLY approach to stellar population analysis.

\subsubsection{Full Spectral Fitting}
\label{sec:spectral_fit}
\begin{figure}
\includegraphics[width=0.48\textwidth]{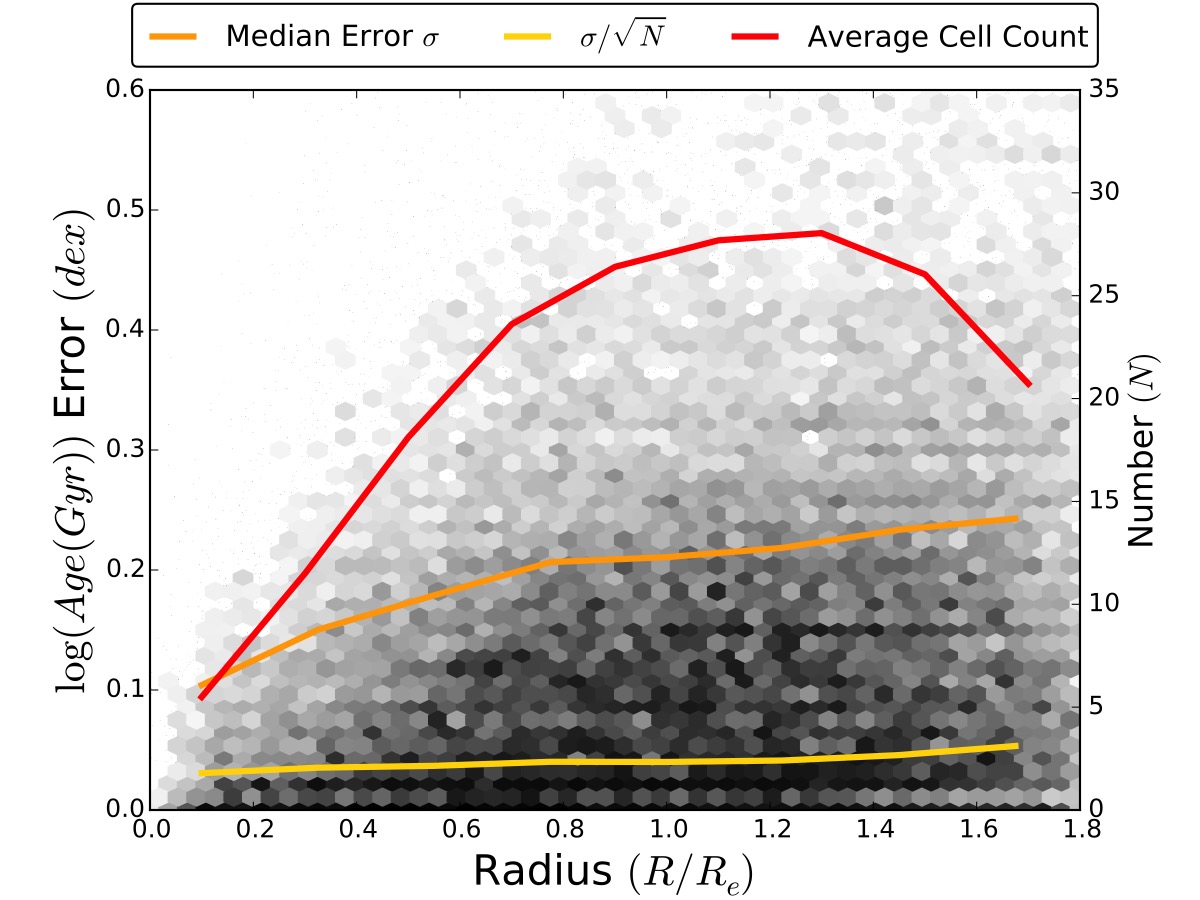}
\includegraphics[width=0.48\textwidth]{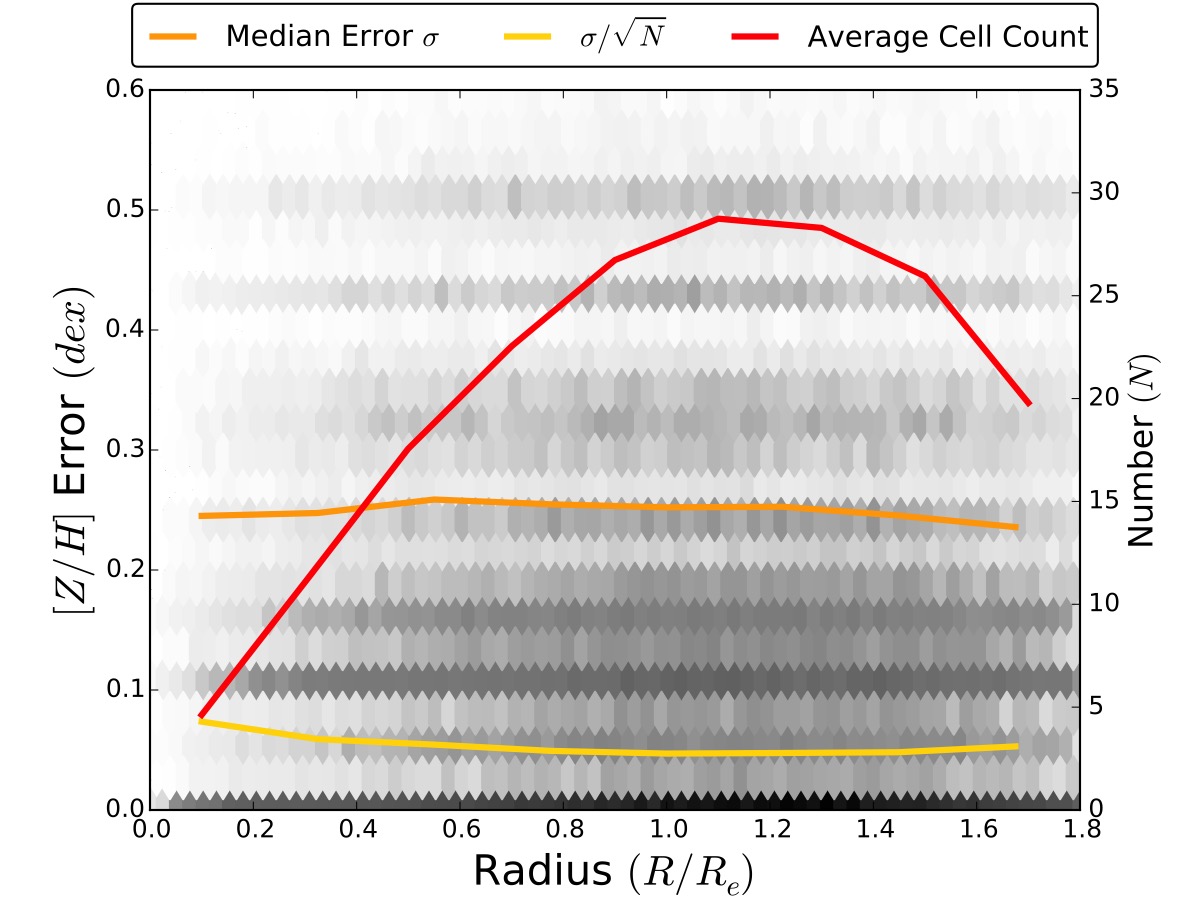}
\caption{Precision in the determination of age (top panel) and metallicity (bottom panel) as a function of radial bin. The grey scale indicates the number density of Voronoi cells.}
\label{fig:dx_r}
\end{figure}
FIREFLY is a $\chi^2$ minimisation fitting code\footnote{Calculated as $\chi^2= \sum_{\lambda} \frac{(O(\lambda)-M(\lambda))^2}{E(\lambda)^2}$, where $O(\lambda)$ is the observed SED, $M(\lambda)$ is the model spectrum and $E(\lambda)$ is the error.} that for a given input Spectral Energy Distribution (SED), fits Simple Stellar Populations (SSPs), checking iteratively whether their $\chi^2$ values are improved by adding different SSPs components, with various luminosity contributions. At each iteration, a selection of the best set of fits from both the one and the multiple-SSP fits are saved for the next cycle. This process is repeated until adding further SSPs no longer improves the $\chi^2$. SSP components that already exist in the previous iteration of the fit can still be added to create new fits, thus allowing final solutions with arbitrary proportions of the initial grid of SSPs.
\\
\\
To allow for the number of SSPs used in each linear combination to converge, each iteration must improve the previous fit by a statistically significant amount, which is governed by the Bayesian Information Criterion (BIC, \cite{2007MNRAS.377L..74L}),
\begin{equation}
BIC=\chi^2 + k\ln(n),
\end{equation}
where $k$ is the number of fitting parameters (number of SSPs used) and $n$ is the number of flux points. This BIC adds a penalty term to the $\chi^2$ value that scales linearly to the number of SSPs used. Once the fitting procedure has converged, an array of typically $\sim 100-1000$ model fits is returned, with each of these model fits being constructed from an arbitrarily weighted, linear combination of SSPs. The luminosity of each fit is given by:
\begin{equation}
L(\lambda) = \sum_i^{\mathrm{N_{\mathrm{SSP}}}} \omega_{i} L_{\mathrm{N_{\mathrm{SSP}}}}(\lambda),
\end{equation}
where $\omega_{i}$ is the weight of the SSP, $\mathrm{N_{\mathrm{SSP}}}$ is the number of SSPs used and $L_{\mathrm{N_{\mathrm{SSP}}}}(\lambda)$ is the luminosity of each SSP.
\\
\\
The final array of fits obtained by FIREFLY provide light and mass-weighted ages $\log(Age(Gyr))$, metallicities $[Z/H]$\footnote{Defined as the abundance of heavy elements relative to Hydrogen \begin{equation} [Z/H] = \log \left(Z/Z_{\odot}\right) - \log \left(H/H_{\odot}\right)  \end{equation}}, stellar mass densities $\Sigma_{*}$, reddening E($B-V$), and corresponding $\chi^2$ values. Light-weighted properties involve weighting each stellar population component by their geometrically averaged total luminosity across the fitted wavelength range, whereas mass-weighted properties involve weighting by the stellar mass contribution over the same wavelength range. These two weightings can be complementary as light-weighted properties will be affected by recently formed young stellar populations, whereas mass-weighted properties tell us about the cumulative evolution of the galaxy. The likelihoods of these fits are then calculated by using the chi-squared cumulative probability distribution:
\begin{equation}
\mathcal{P}(X = \chi_{0}^2)=\int_{\chi_{0}^2}^{\inf} \frac{1}{2^{k/2} \Gamma(\frac{k}{2})} X^{(k/2)-1} \exp^{-X/2}  \mathrm{d}X,
\end{equation}
where $\Gamma$ is the gamma function, $\chi_{0}^2$ is the critical value of chi-squared, $k$ is the degrees of freedom which can be expressed as $k=N-\nu-1$, where $N$ is the number of data points and $\nu$ number of SSPs used. Gaussian profiles are then fit to each of the marginal distributions to estimate each properties average value and the associated error. Figure~\ref{fig:likelihood} shows how to visualise these distributions.

\subsubsection{Dust Attenuation}
Prior to fitting the model templates to the data, FIREFLY takes into account galactic and interstellar reddening of the spectra. Foreground Milky Way reddening is accounted for by using the foreground dust maps of \cite{schlegel1998} and the extinction curve from \cite{fitzpatrick1999}. The intrinsic dust attenuation of each galaxy is then determined following an original procedure, fully described in \citet{wilkinson2015,wilkinson2016}, which we summarise here. 
The central point to FIREFLYs determination of dust attenuation is that rather than assuming a pre-determined attenuation curve, the code derives an attenuation law from the data itself. This follows the notion that the continuum shape is distorted from its intrinsic shape by dust attenuation\footnote{In addition to dust attenuation, flux mis-calibration will also alter the continuum shape. However, this effect will be automatically accounted for in the dust attenuation procedure described here.}. The attenuation is calculated in the following way. 
\\
\\
Initially, all SEDs (both base stellar population models and data) are normalised to their total light and then preprocessed using a `High Pass Filter (HPF)'. The HPF uses an analytic function across all wavelengths to rectify the continuum before deriving the stellar population parameters. This is done using a window function applied to the Fourier transform of the spectra:
\begin{equation}
F_{\lambda}=  f_{\lambda} \otimes W_{\lambda},
\end{equation}
where $F_{\lambda}$ is the output flux, $f_{\lambda}$ is the input flux and $W_{\lambda} = \mathcal{F}^{-1} W_{k}$ describes which modes are removed by the Fourier filter. The window function $W_{k}$ is given by:
\begin{displaymath}
   W_{k} = \left\{
     \begin{array}{lr}
       0 & : k  \leq k_{crit}\\
       1 & : \text{Otherwise,}
     \end{array}
   \right.
\end{displaymath} 
where we set $k_{crit}$ = 40 in this paper. This means that modes in the spectrum with fewer than 40 oscillations are removed, and higher frequencies pass through the filter unchanged. $k_{crit}$ = 40 was tested in \cite{wilkinson2015} on mock galaxies and P-MaNGA data and was shown to be an appropriate value for studies such as this one. \\
\\
The rectified models are then fit to the rectified observed spectrum (using the method described in Section~\ref{sec:spectral_fit}) in order to derive the intrinsic stellar population parameters of the best fit. The final best fit model in its original form, i.e. non-rectified, is then compared to the original unfiltered data. Any shape mismatch between the two is attributed to dust attenuation. The offset between the HPF filtered data and the original unfiltered data is calculated as a function of wavelength and then smoothed, allowing for an attenuation curve to be deduced (see Figure~\ref{fig:dust_plots} for visualisation of the process). The derived attenuation curve is then applied to the base stellar population models, which are then used to fit the original unfiltered input data as explained in Section~\ref{sec:spectral_fit}. Example $E(B-V)$ maps and radial profiles calculated using this method are shown in Figures~\ref{fig:maps} and \ref{fig:example_gradients}, respectively.

\subsubsection{Stellar Population Models}
\label{sec:models}
For the full spectral fitting in this work, we use the stellar population models of \cite{maraston2011} (M11), which utilise the MILES stellar library \citep{miles2006} and assume a Kroupa stellar initial mass function [IMF, \cite{kroupa2011}]. The model library spans ages from 6.5 Myr to 15 Gyr and metallicities $[Z/{\rm H}]=-2.3, -1.3 ,-0.3, 0.0, 0.3$. It has a spectral resolution of 2.5$\AA$ FWHM \citep{beifiori2011,falcon-barroso2011,prugniel2011} and a wavelength coverage of $3500-7428\AA$; this is smaller than the wavelength range of the MaNGA data $3600-10300\AA$, and we restrict our full spectral fit to the wavelength range of the models. The broadening of absorption features due to velocity dispersion and the difference in resolution between the data and the models must be taken in to account before fitting to the data. For this, we combine the spectral resolution provided by the DRP and the velocity dispersion provided by the DAP to adjust the resolution of the models to match the data, and take into account the broadening of absorption features.

\subsubsection{Quality of Spectral Fits}
FIREFLY can recover stellar population properties with acceptable accuracy down to a $S/N$ ratio as low as 5. We chose this $S/N$ threshold for our spatial binning scheme as the best compromise between spatial sampling and accuracy in spectral fitting. Figure~\ref{fig:ff_signal-to-noise} shows the distribution with running median of the errors on luminosity-weighted age (left-hand panel) and metallicity (right-hand panel) as a function of $S/N$ ratio. The surveys target accuracy in age and metallicity determination at all radii sampled is 0.1 dex \citep{yan2016}. The figure shows that this accuracy is reached with $S/N\sim 20$ \citep[see also][]{johansson2012,wilkinson2015}. The median error on age and metallicity at our minimum threshold of $S/N=5$ is 0.28 dex and 0.23 dex, respectively, with a large scatter up to 0.5 dex. Higher $S/N$ and accuracy in the determination of stellar population properties will then be obtained by combining Voronoi cells in radial bins. Examples of sample spectral fits for three different $S/N$ ratios are shown in the top three panels of Figure~\ref{fig:ff_spectral}, as well as the distribution of residuals for all the the galaxies fitted in this work (bottom panel).

\subsubsection{Example Maps}
\begin{landscape}
\begin{figure}
\includegraphics[height=0.17\textwidth,width=0.22\textwidth, angle=0]{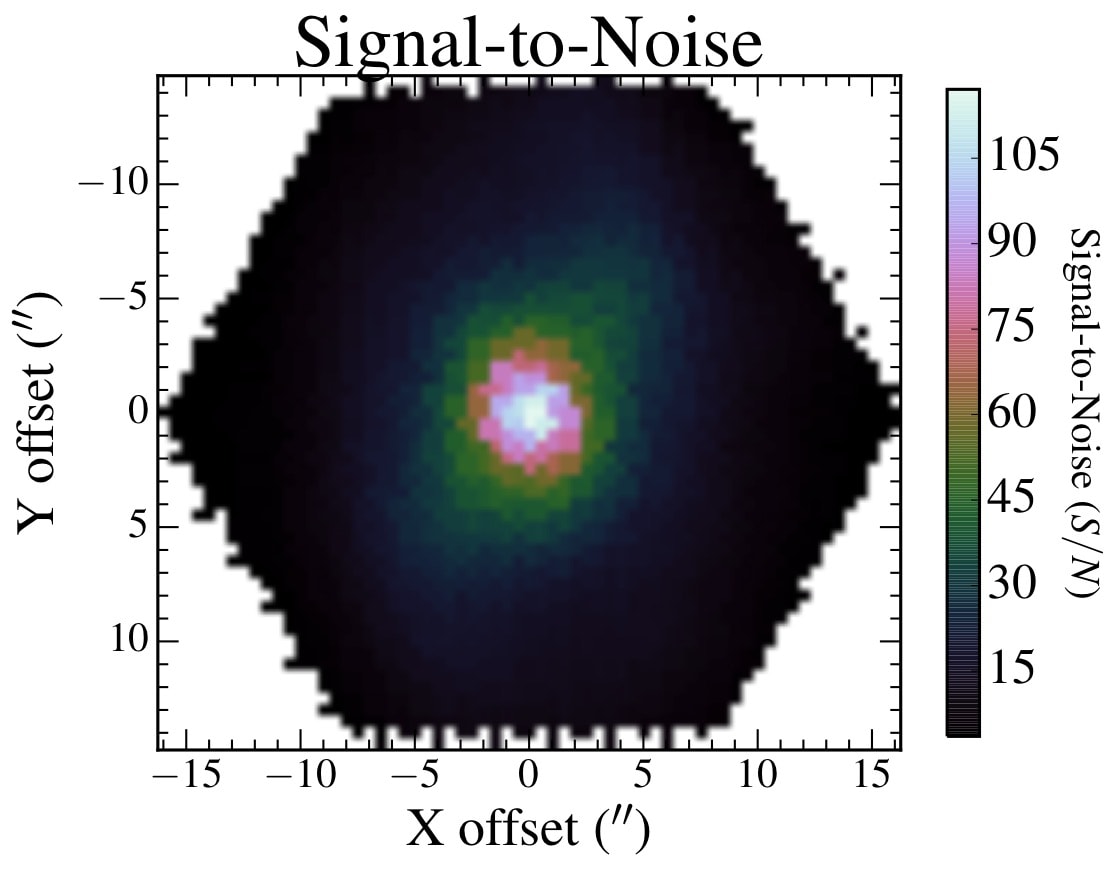}
\includegraphics[height=0.17\textwidth,width=0.22\textwidth, angle=0]{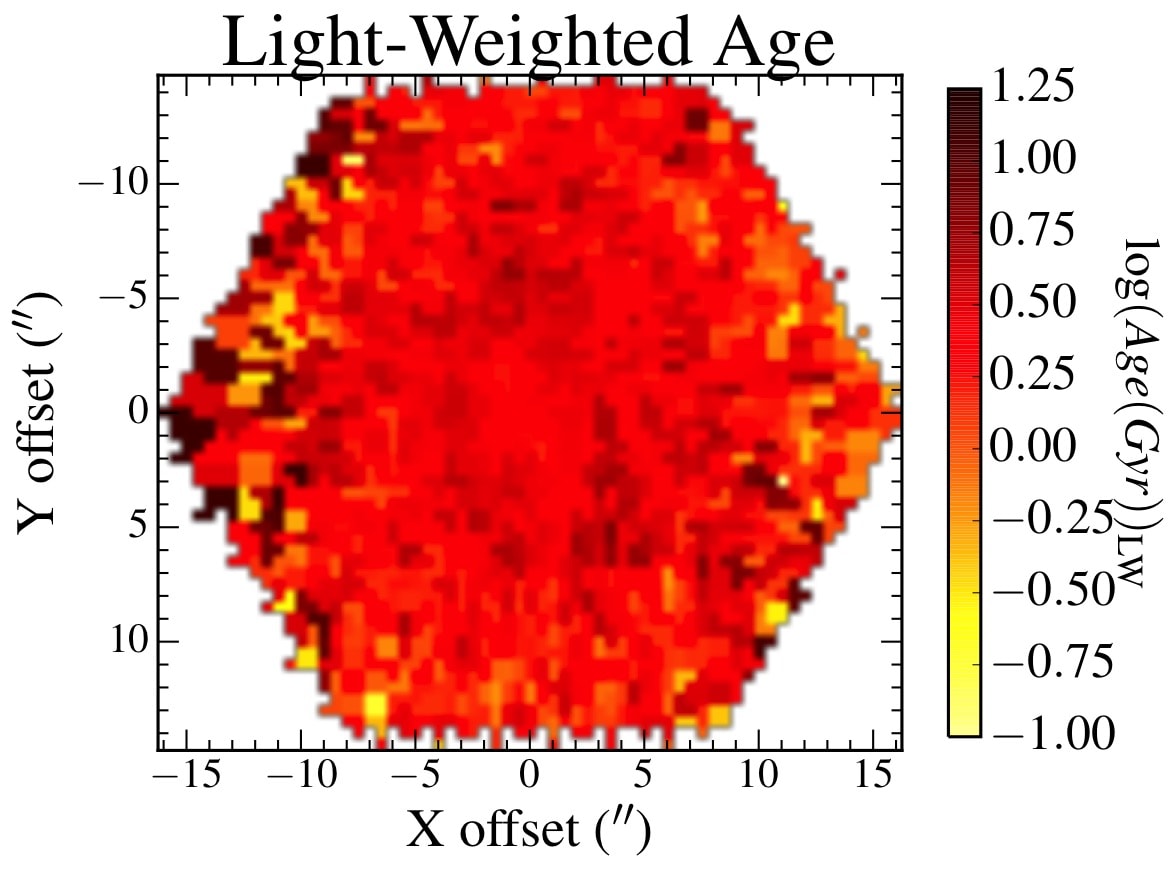}
\includegraphics[height=0.17\textwidth,width=0.223\textwidth, angle=0]{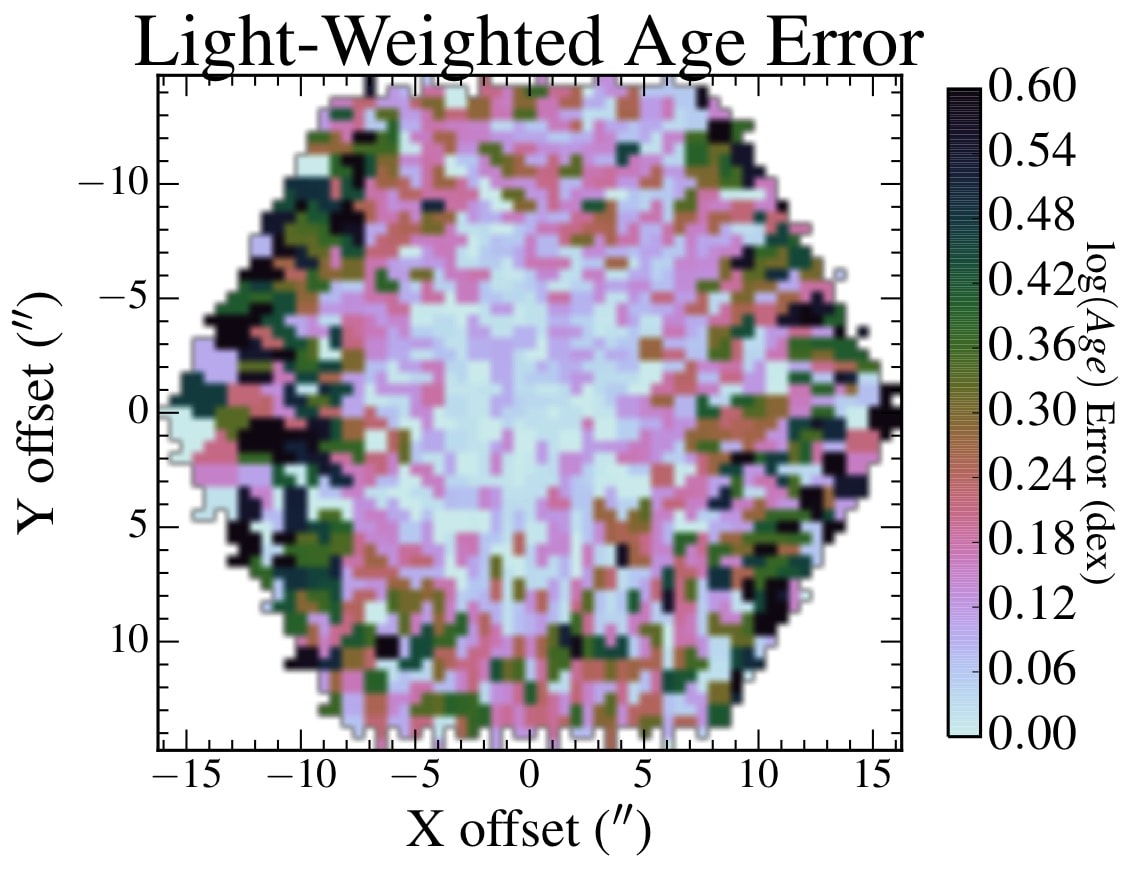}
\includegraphics[height=0.17\textwidth,width=0.223\textwidth, angle=0]{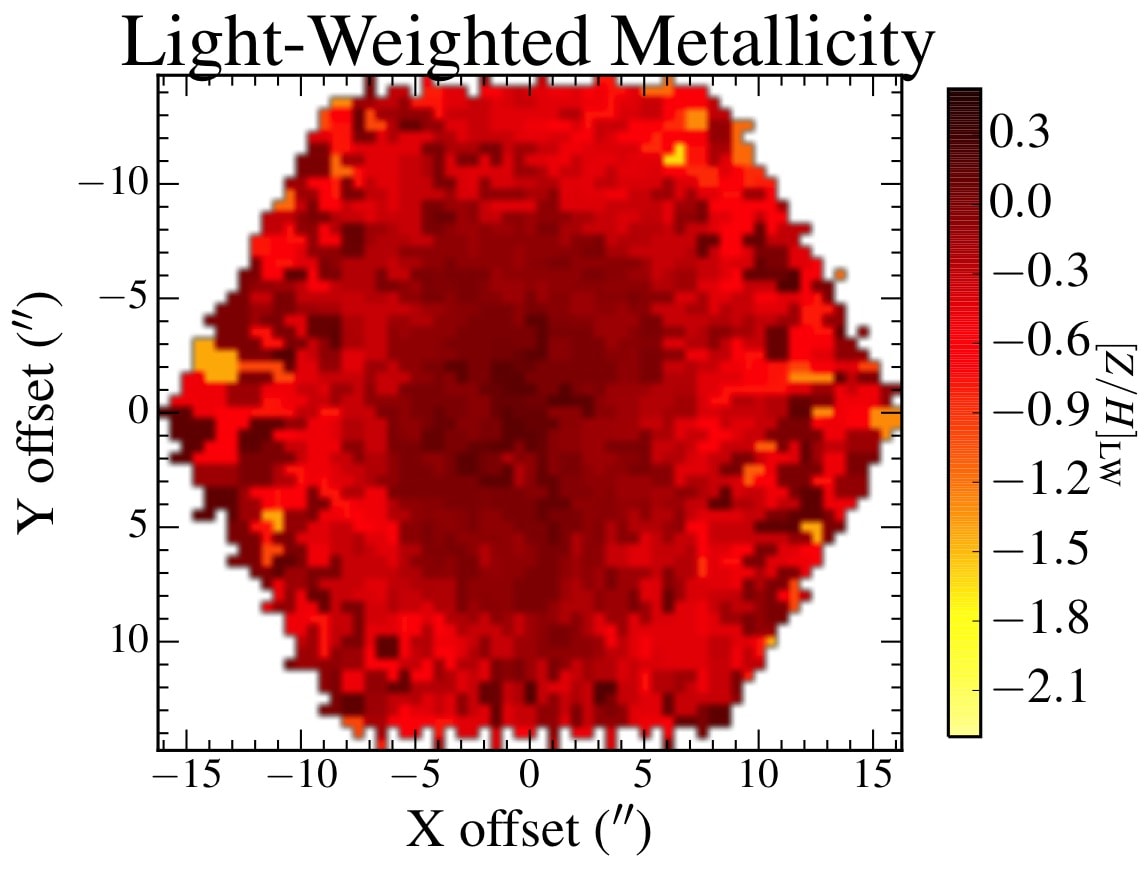}
\includegraphics[height=0.17\textwidth,width=0.223\textwidth, angle=0]{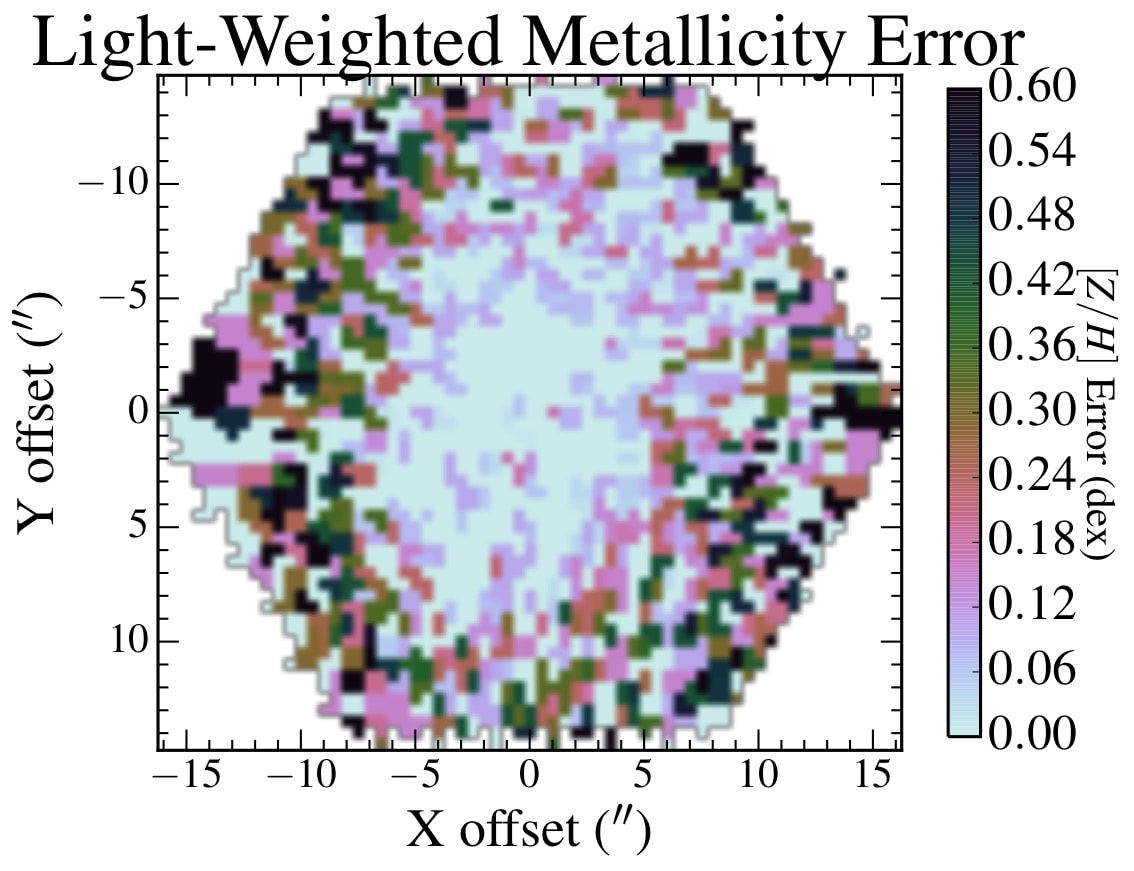}
\includegraphics[height=0.17\textwidth,width=0.223\textwidth, angle=0]{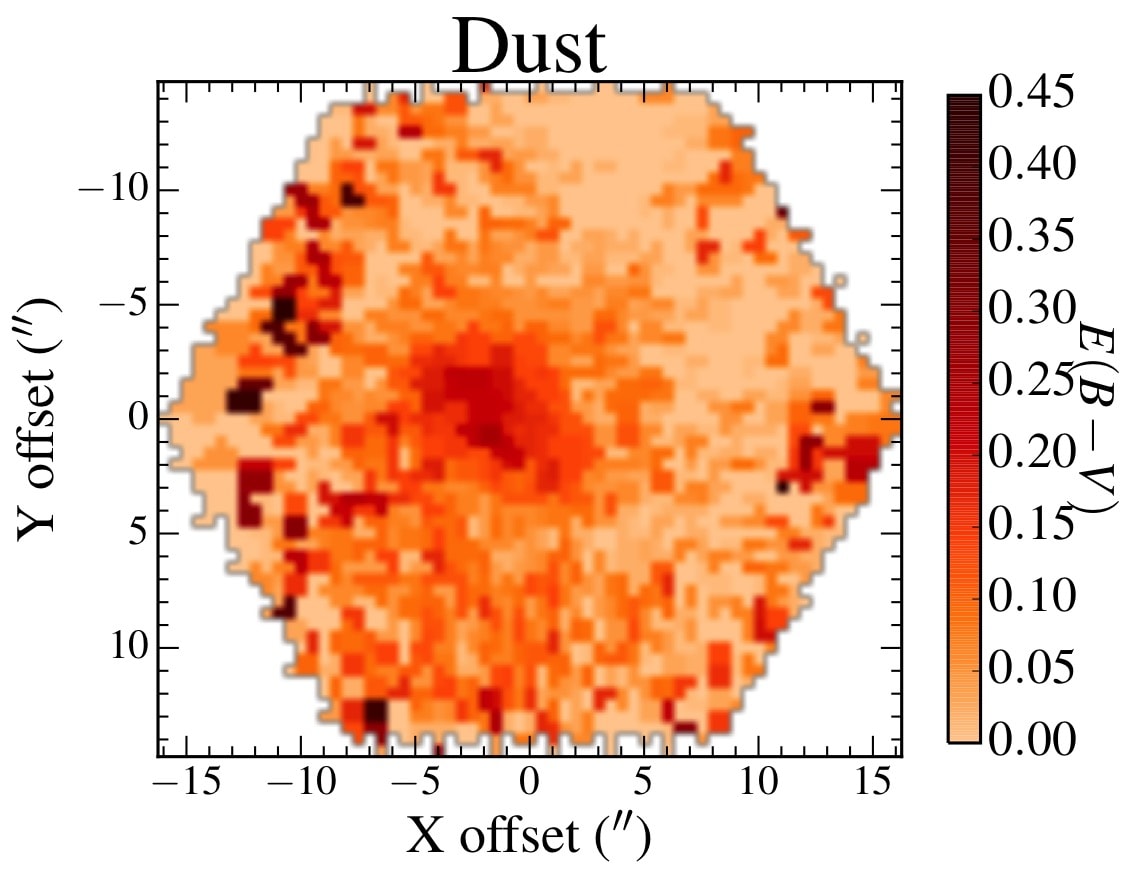}
\includegraphics[height=0.17\textwidth,width=0.223\textwidth, angle=0]{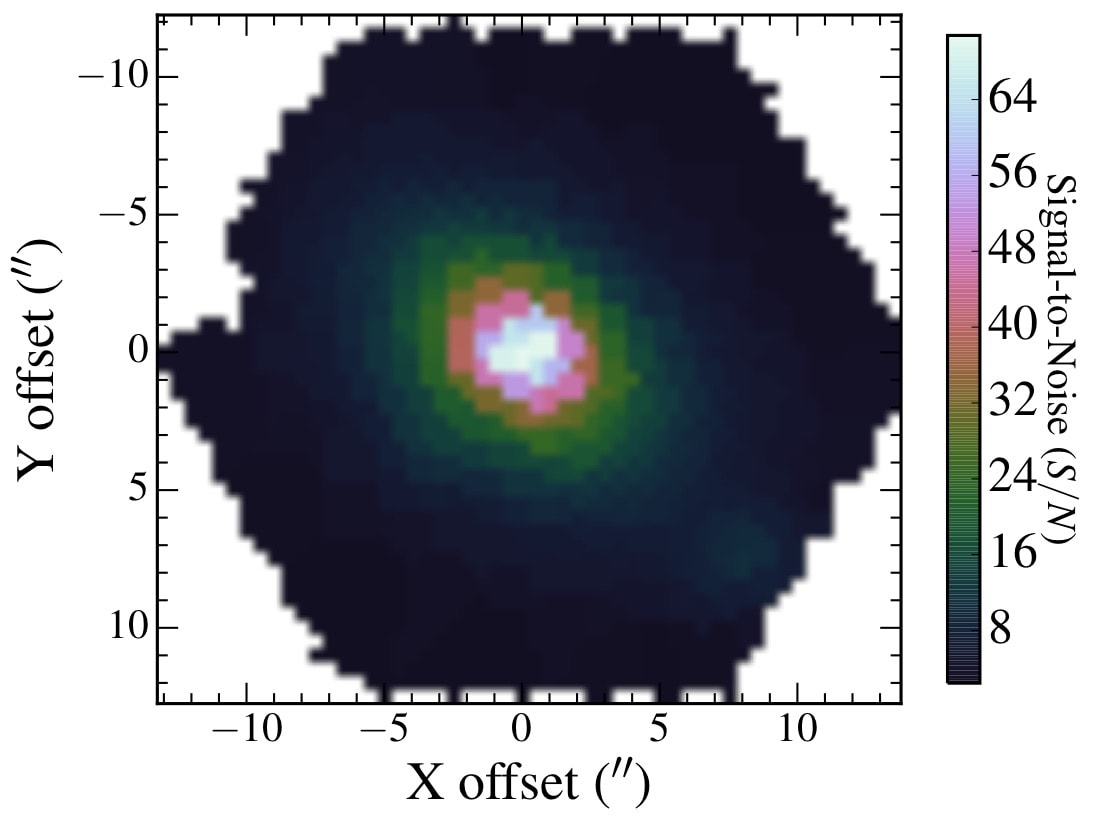}
\includegraphics[height=0.17\textwidth,width=0.223\textwidth, angle=0]{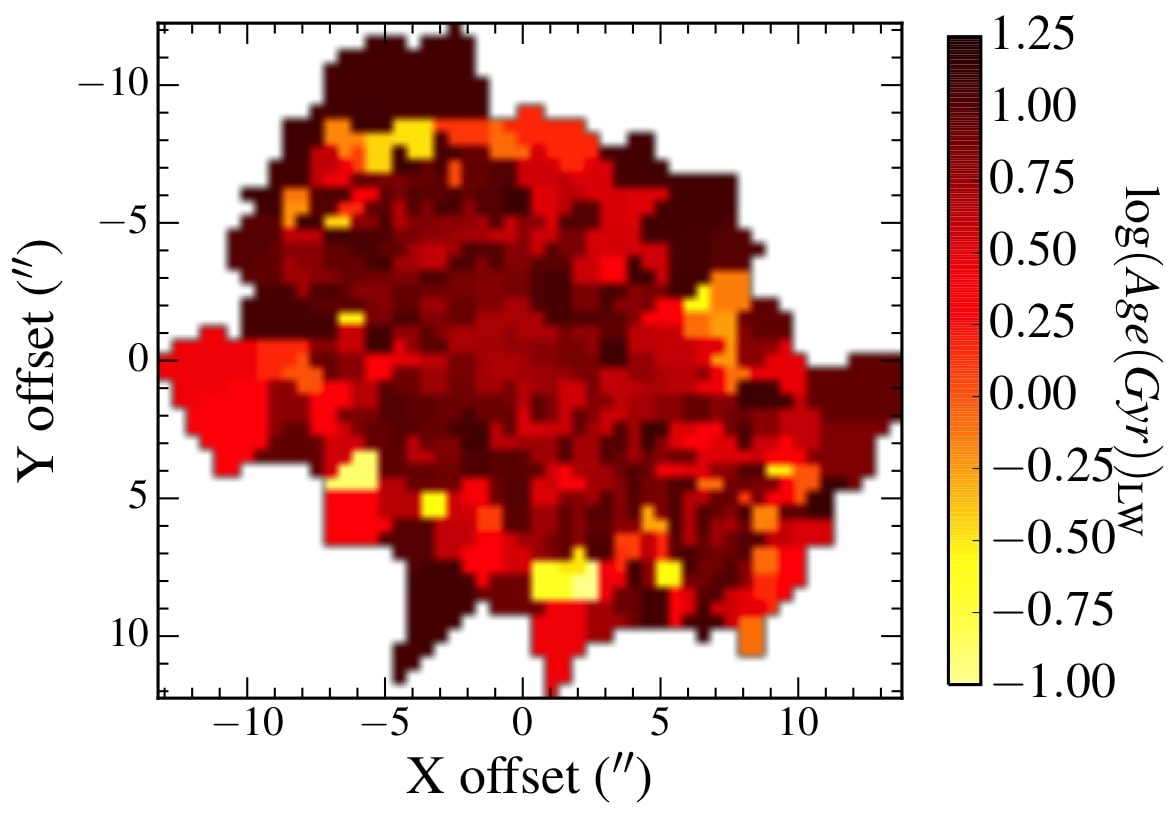}
\includegraphics[height=0.17\textwidth,width=0.223\textwidth, angle=0]{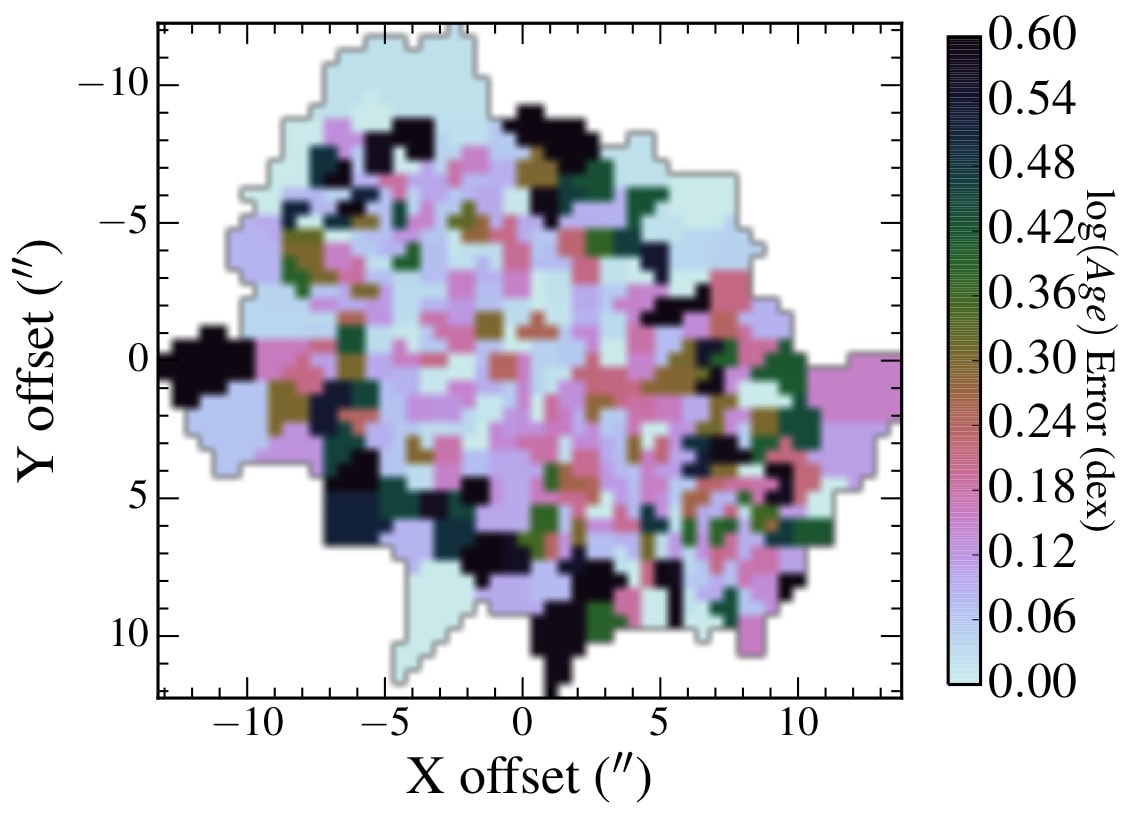}
\includegraphics[height=0.17\textwidth,width=0.223\textwidth, angle=0]{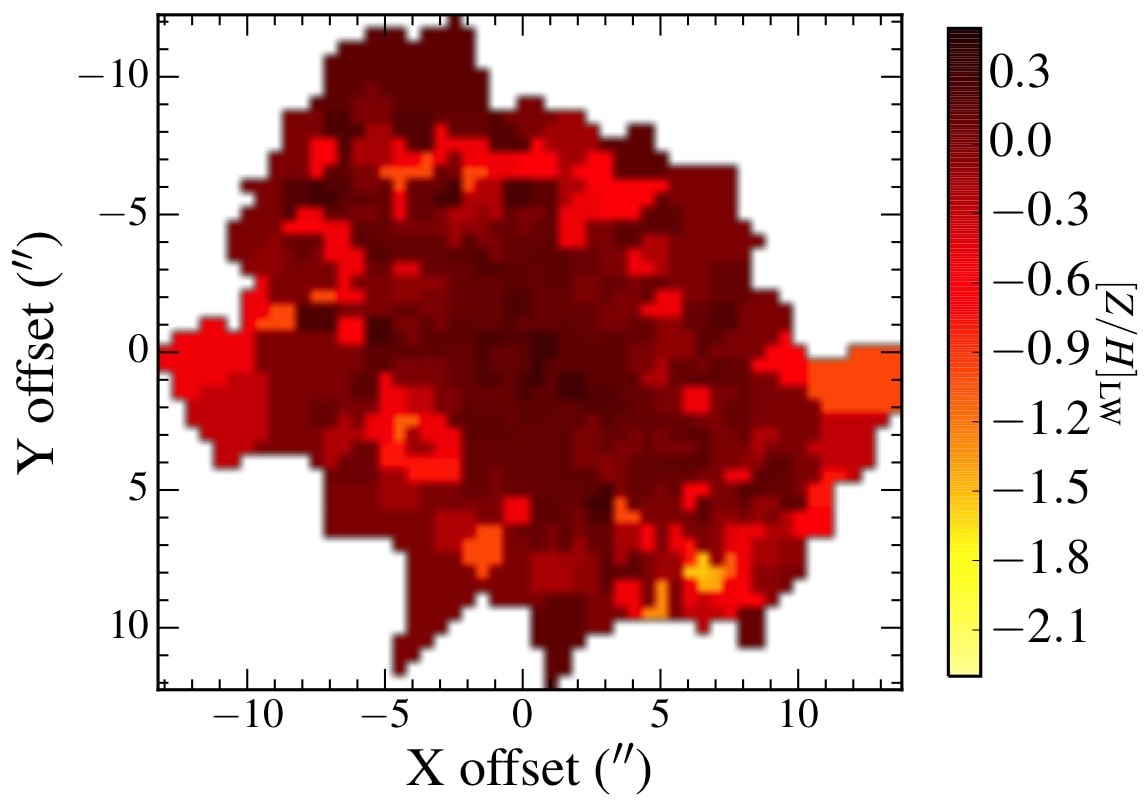}
\includegraphics[height=0.17\textwidth,width=0.223\textwidth, angle=0]{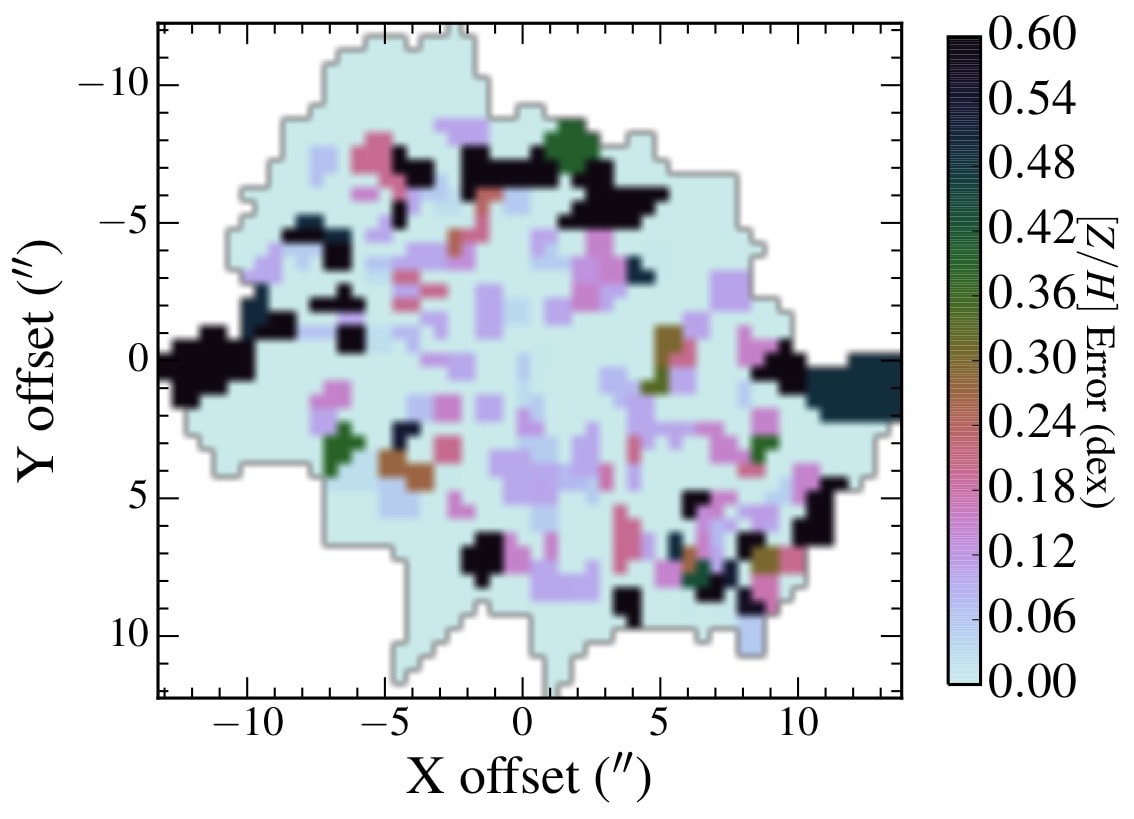}
\includegraphics[height=0.17\textwidth,width=0.223\textwidth, angle=0]{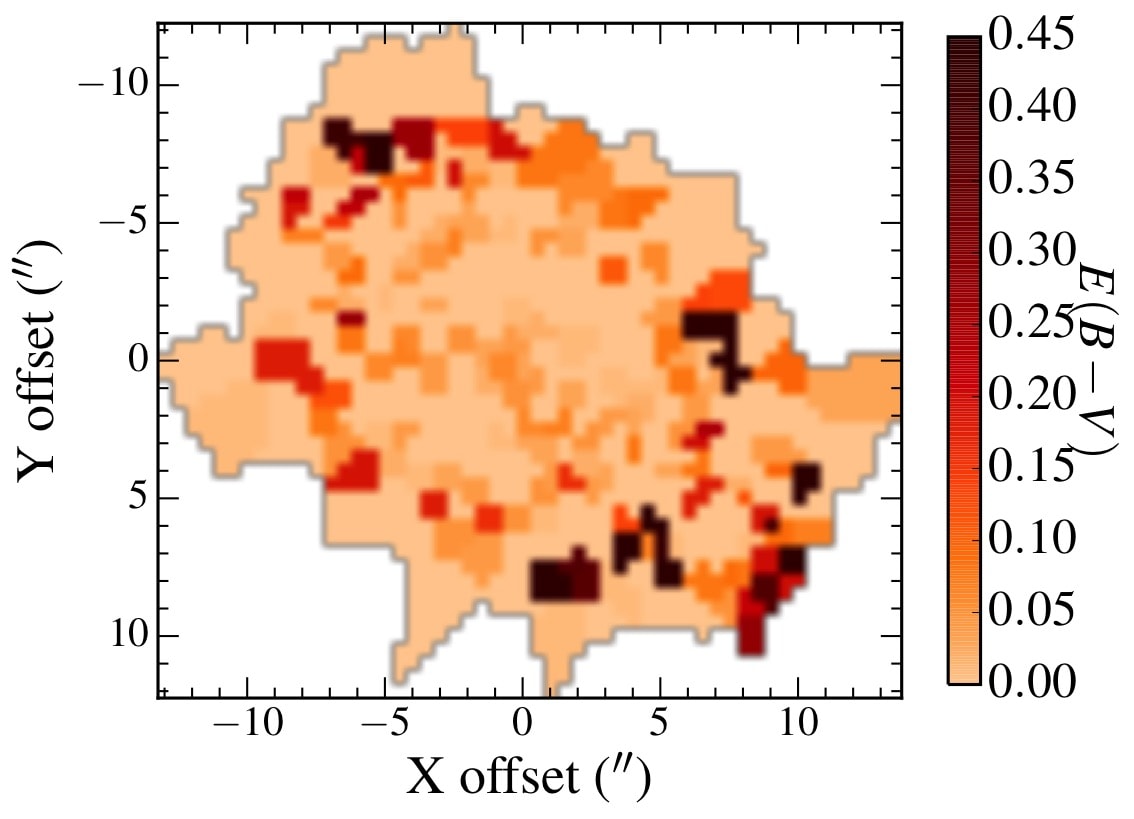}
\includegraphics[height=0.17\textwidth,width=0.223\textwidth, angle=0]{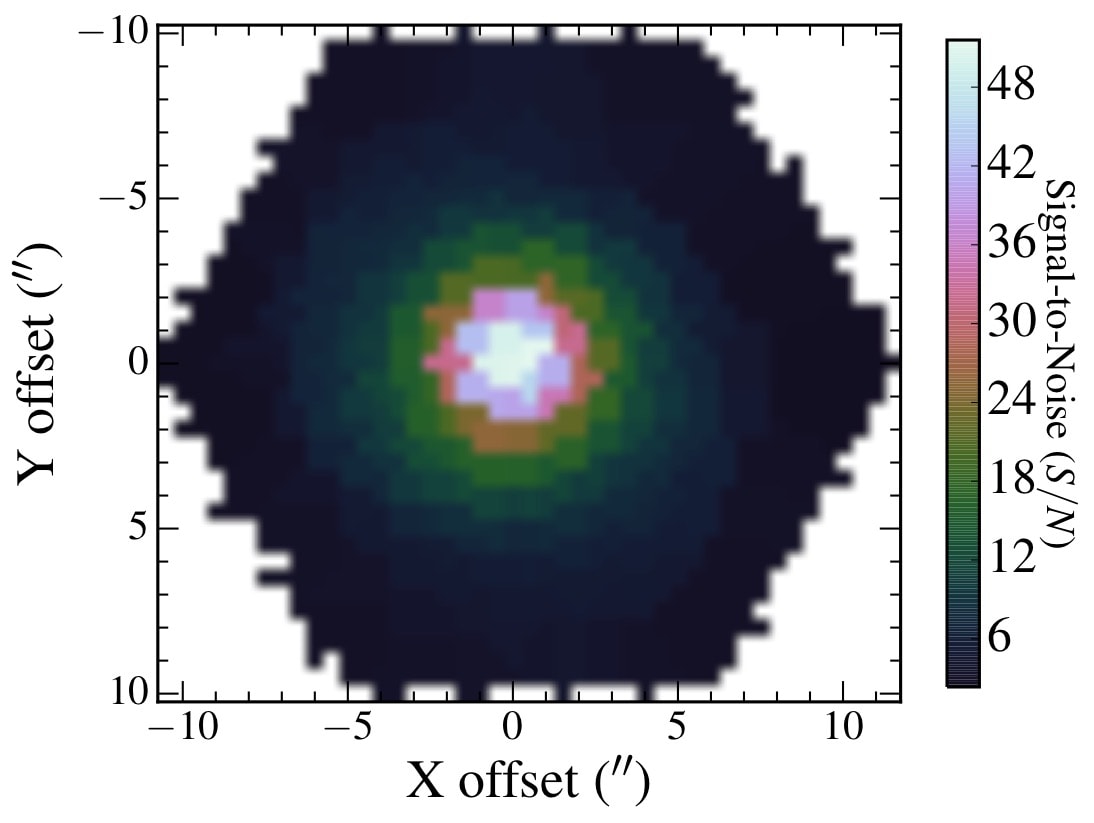}
\includegraphics[height=0.17\textwidth,width=0.223\textwidth, angle=0]{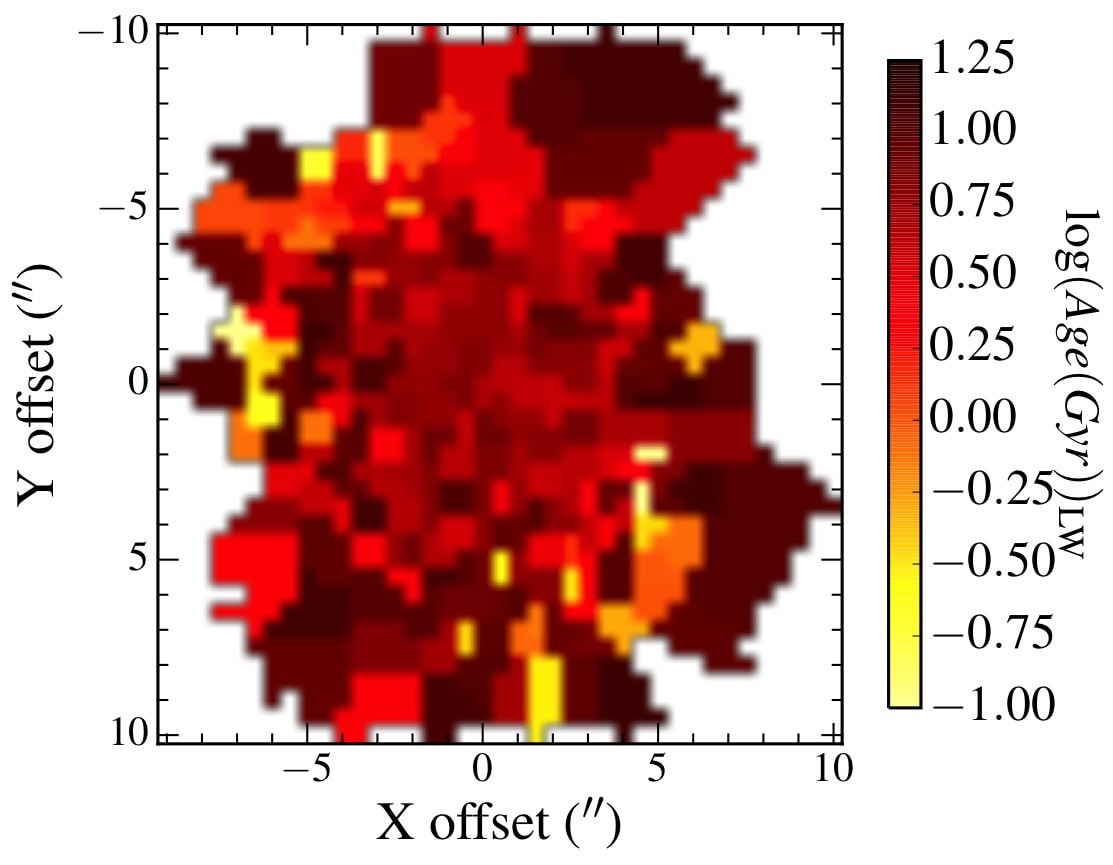}
\includegraphics[height=0.17\textwidth,width=0.223\textwidth, angle=0]{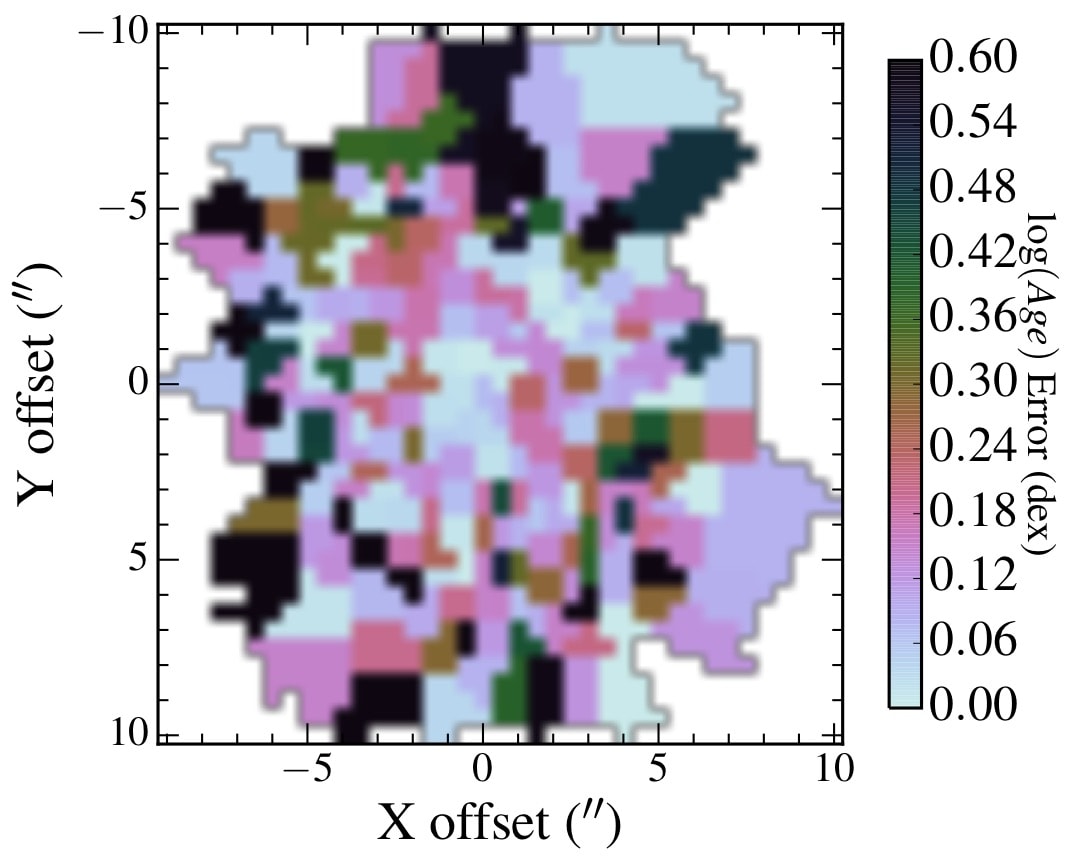}
\includegraphics[height=0.17\textwidth,width=0.223\textwidth, angle=0]{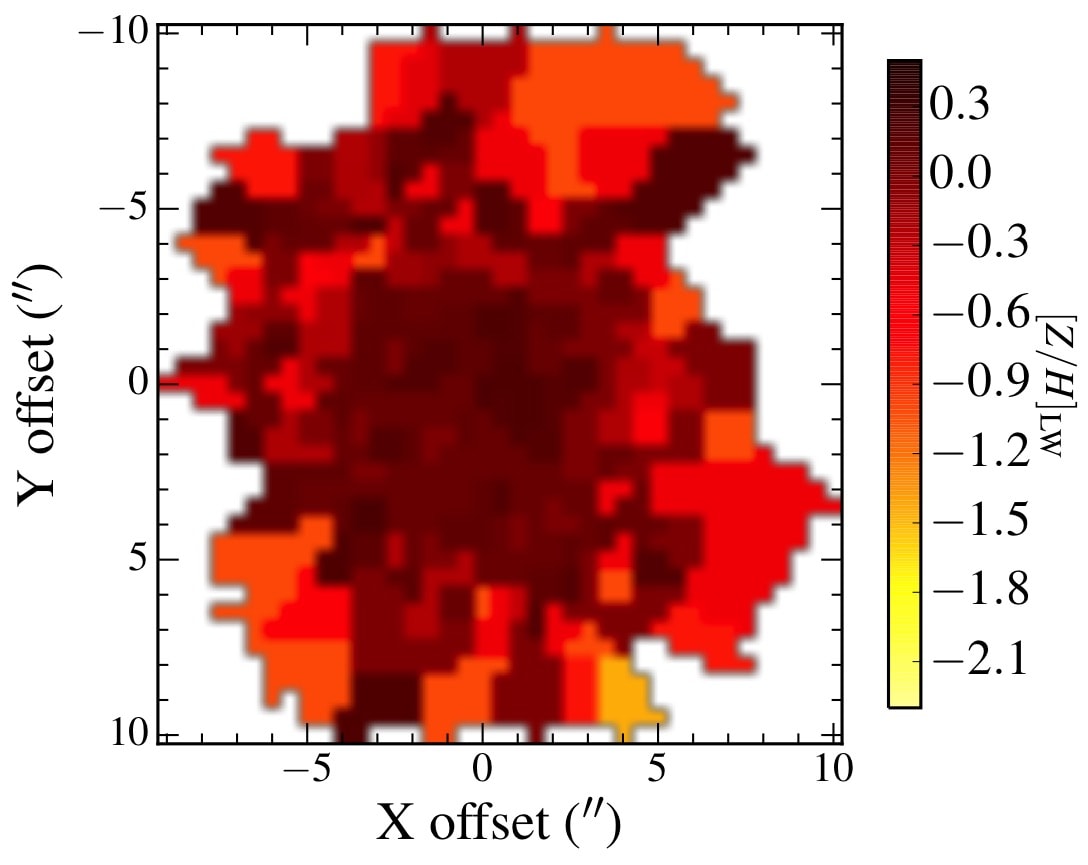}
\includegraphics[height=0.17\textwidth,width=0.223\textwidth, angle=0]{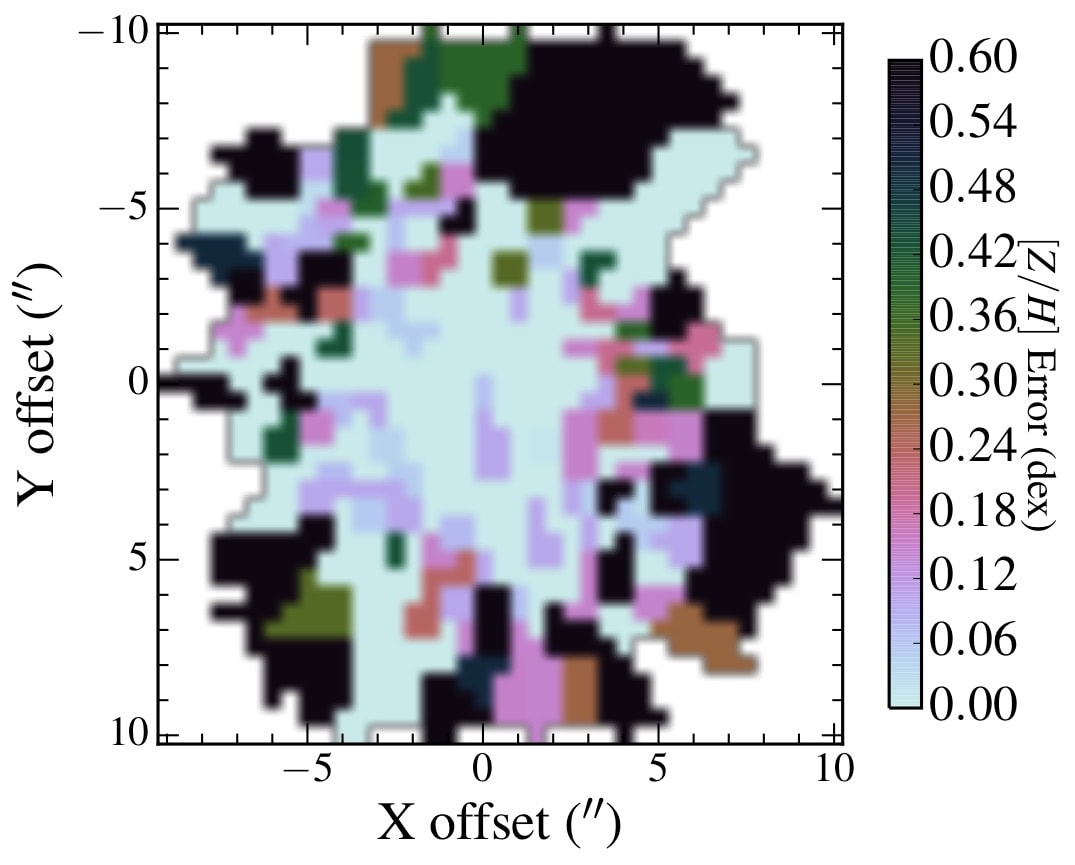}
\includegraphics[height=0.17\textwidth,width=0.223\textwidth, angle=0]{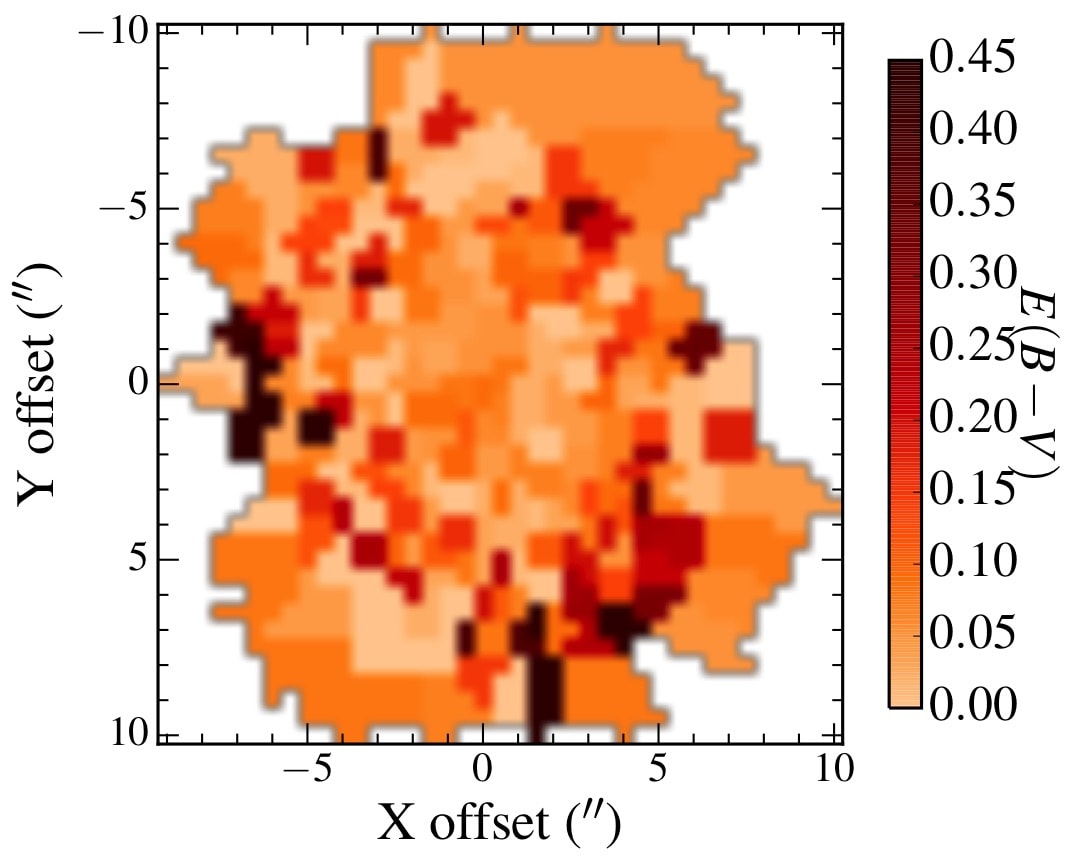}
\includegraphics[height=0.17\textwidth,width=0.223\textwidth, angle=0]{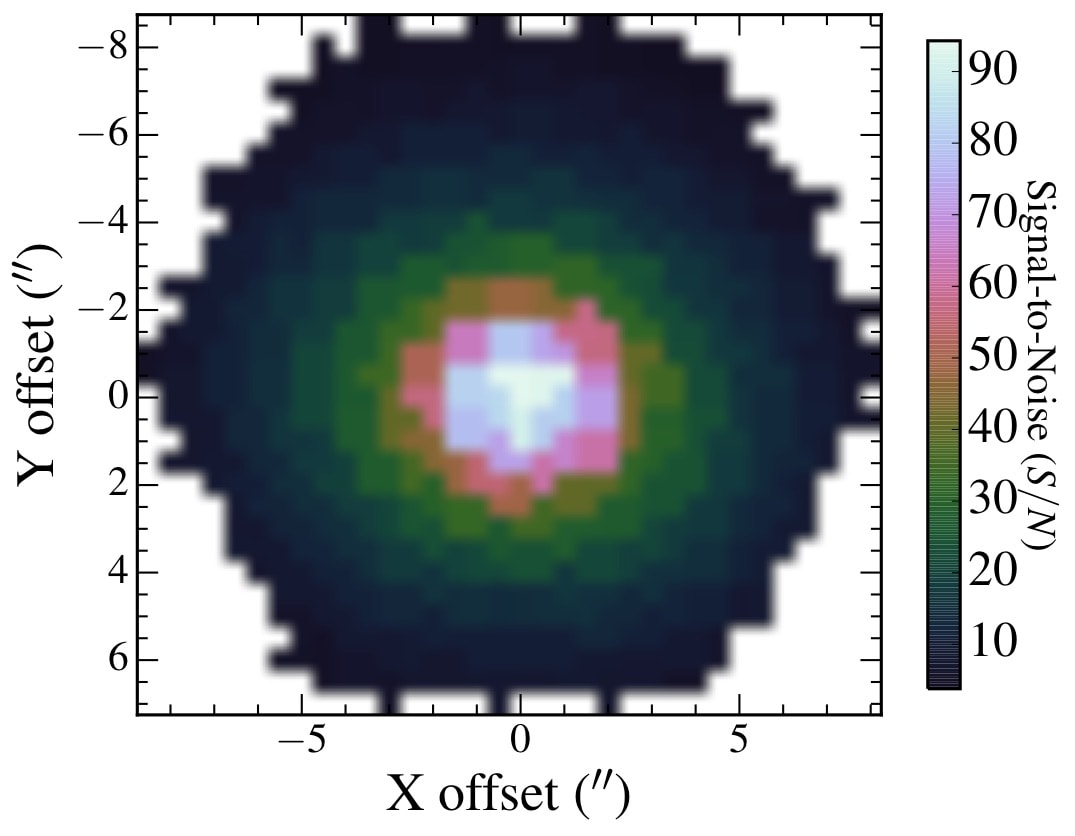}
\includegraphics[height=0.17\textwidth,width=0.223\textwidth, angle=0]{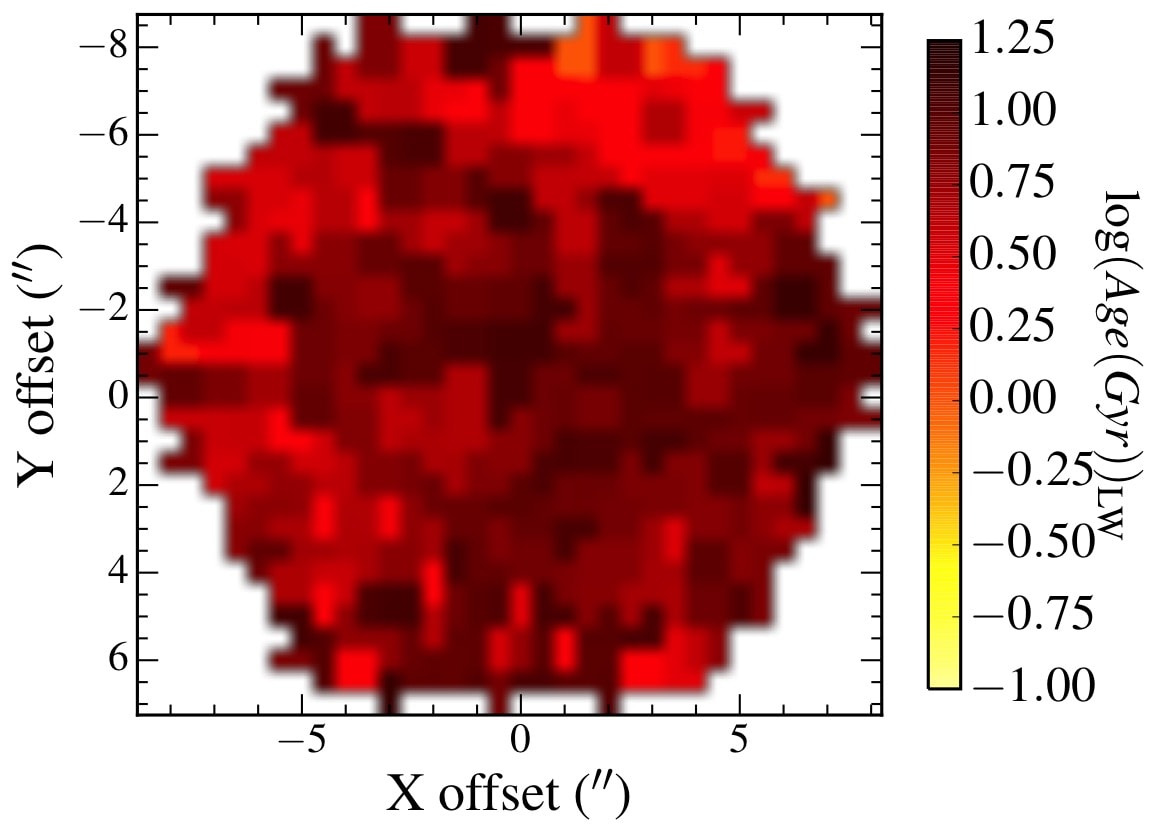}
\includegraphics[height=0.17\textwidth,width=0.223\textwidth, angle=0]{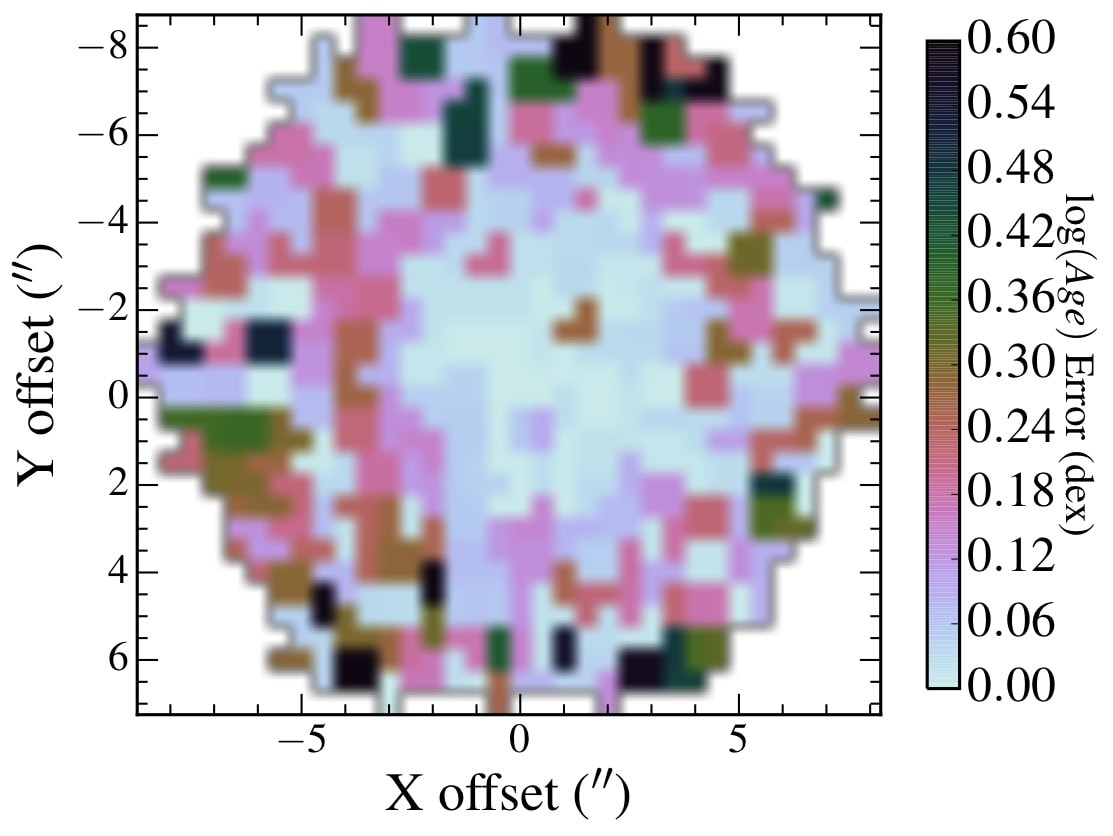}
\includegraphics[height=0.17\textwidth,width=0.223\textwidth, angle=0]{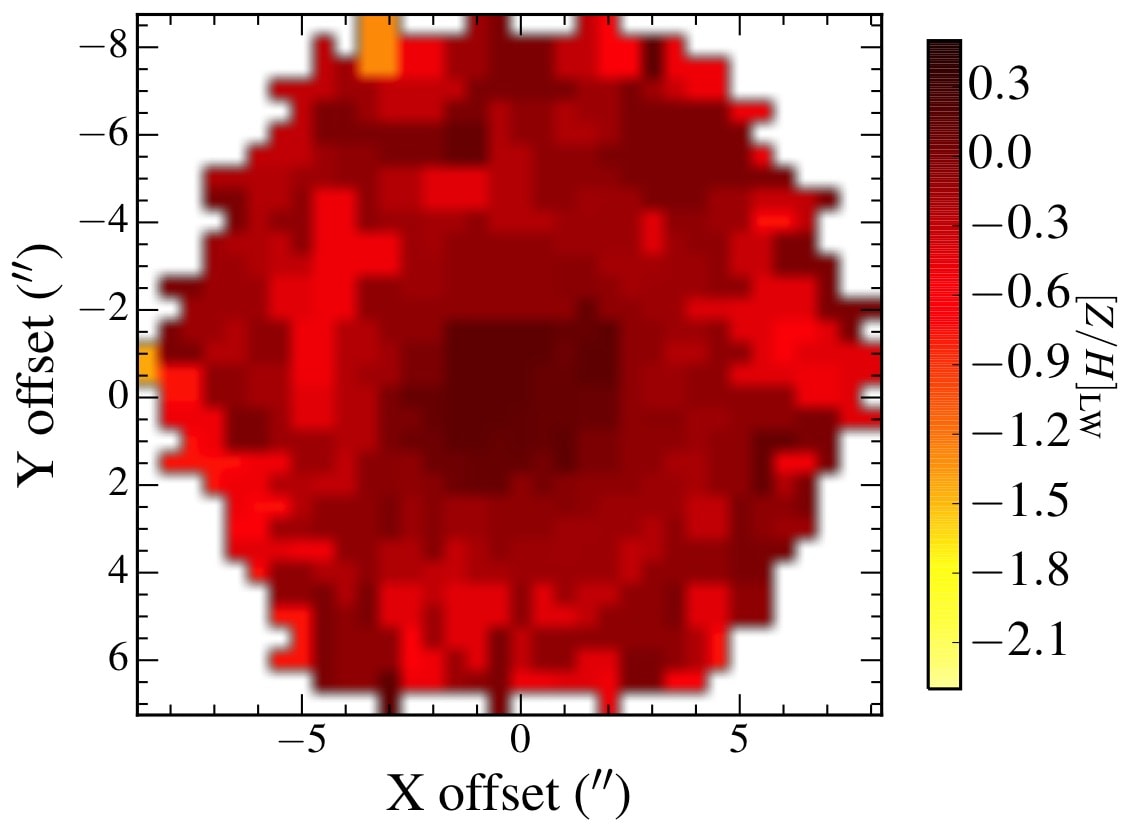}
\includegraphics[height=0.17\textwidth,width=0.223\textwidth, angle=0]{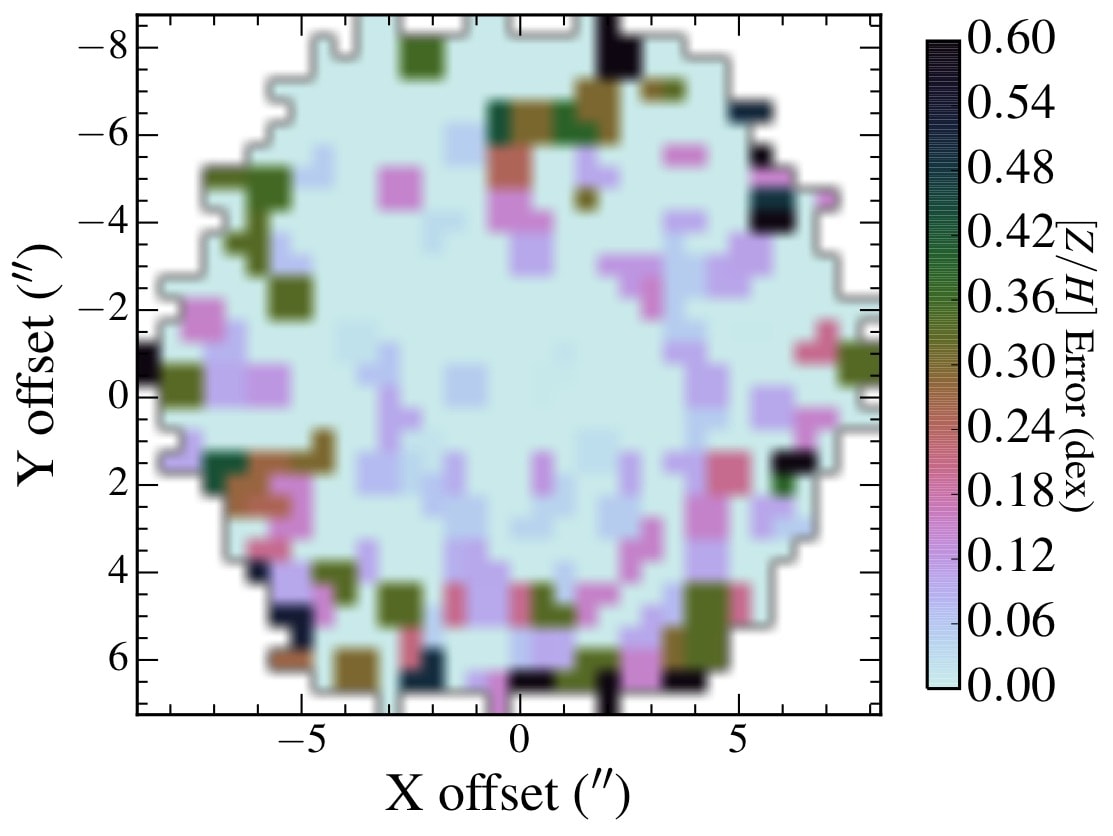}
\includegraphics[height=0.17\textwidth,width=0.223\textwidth, angle=0]{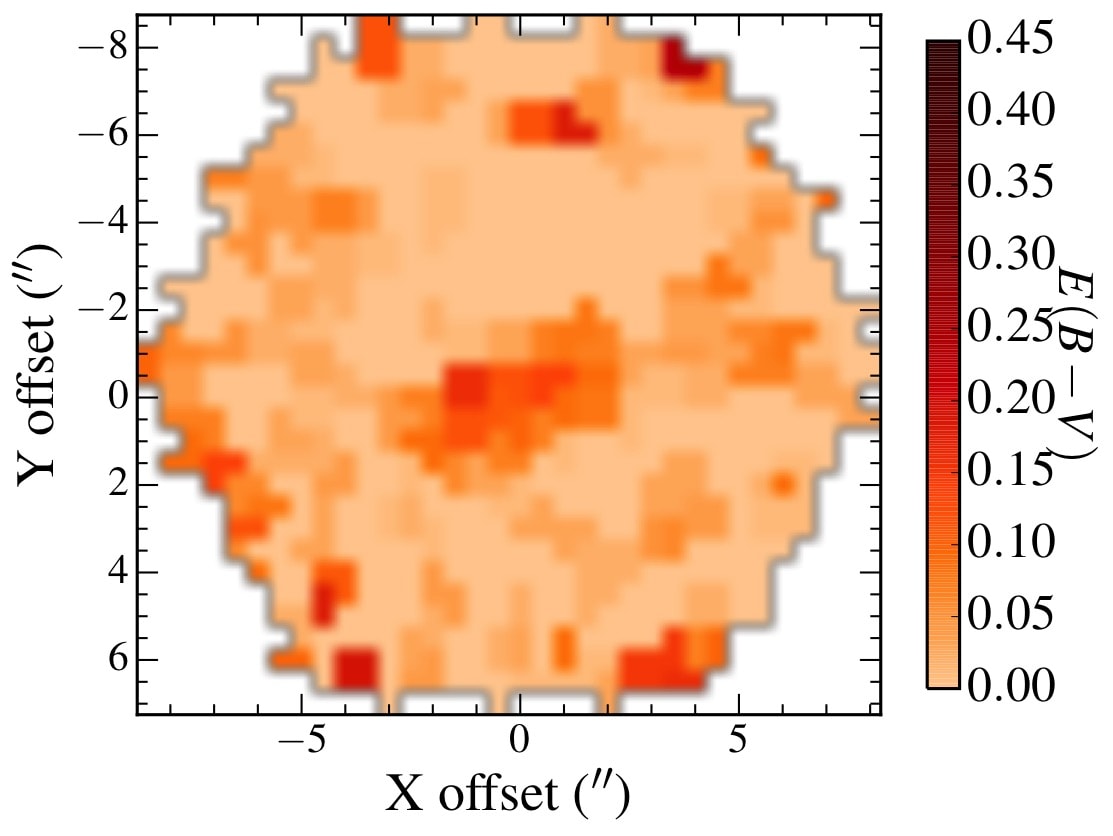}
\includegraphics[height=0.17\textwidth,width=0.223\textwidth, angle=0]{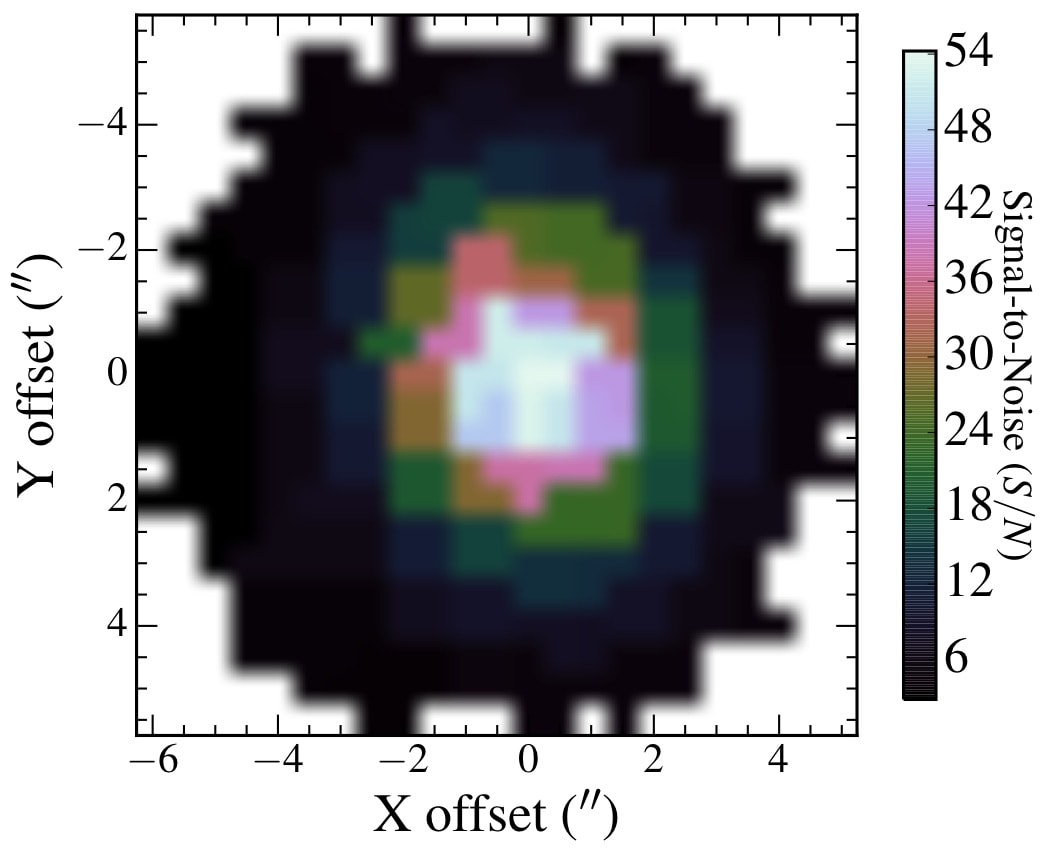}
\includegraphics[height=0.17\textwidth,width=0.223\textwidth, angle=0]{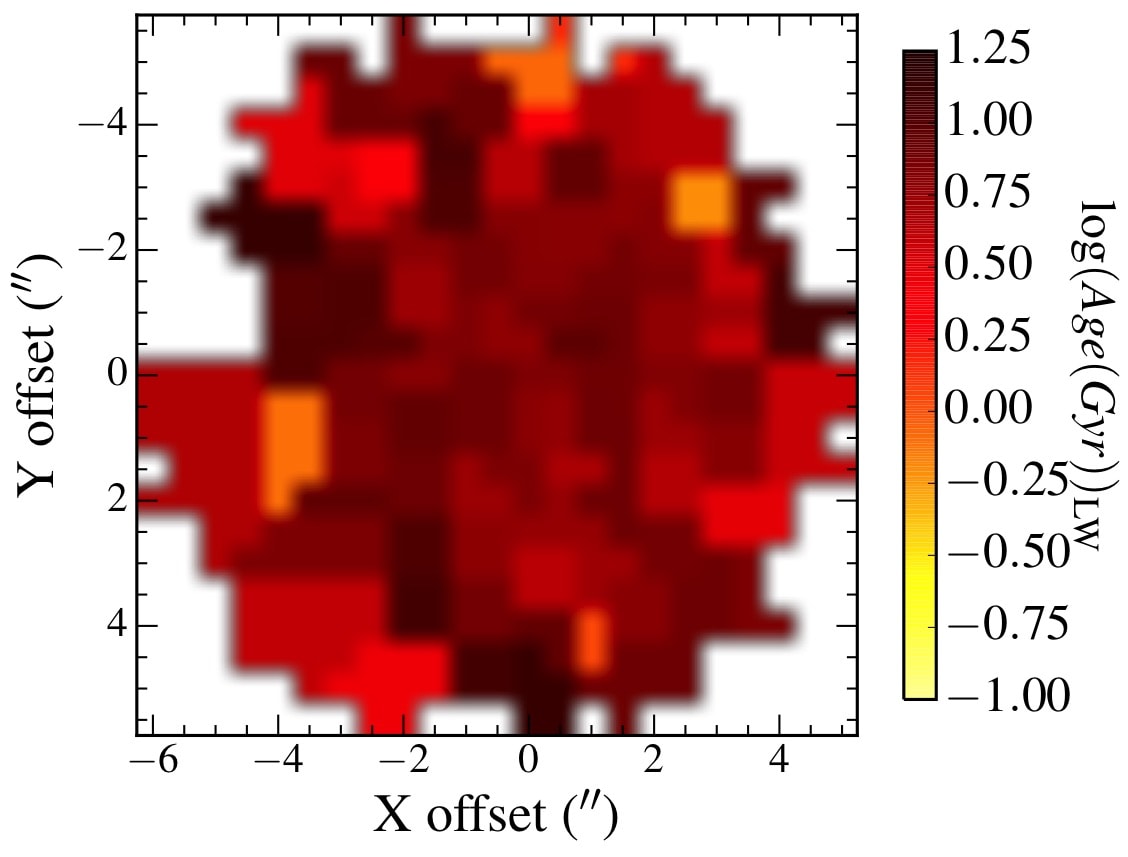}
\includegraphics[height=0.17\textwidth,width=0.223\textwidth, angle=0]{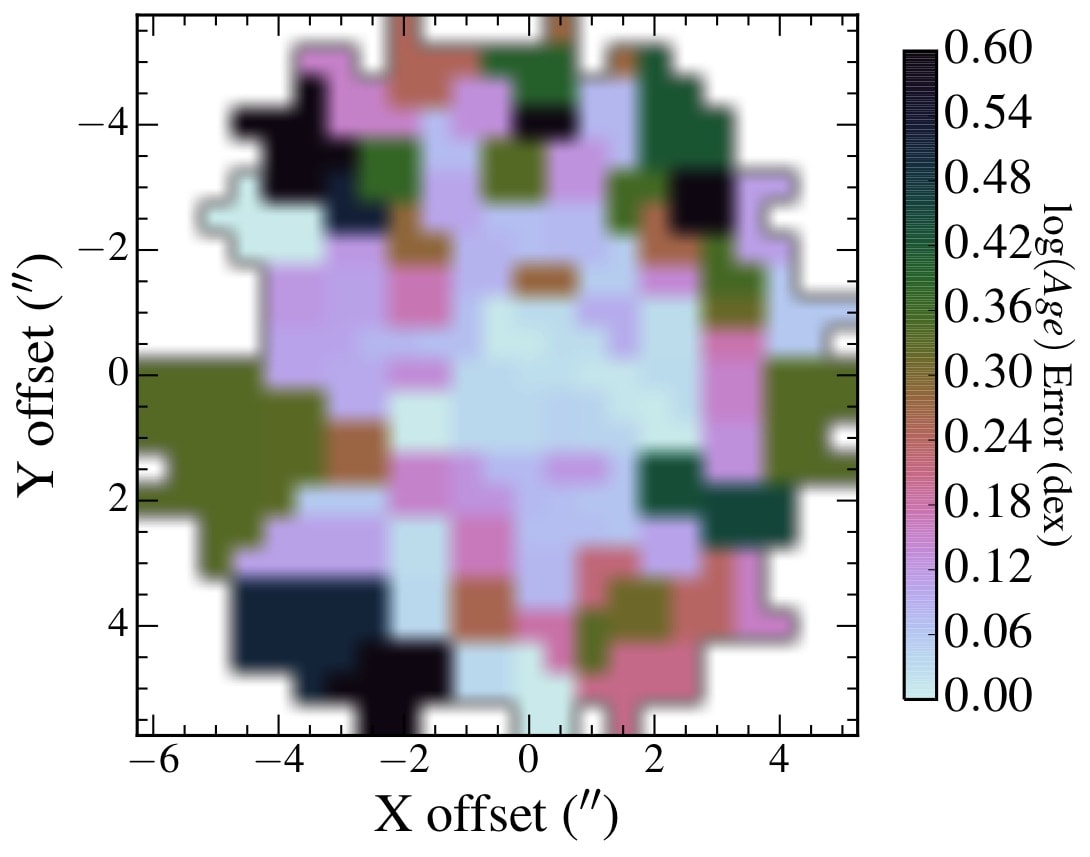}
\includegraphics[height=0.17\textwidth,width=0.223\textwidth, angle=0]{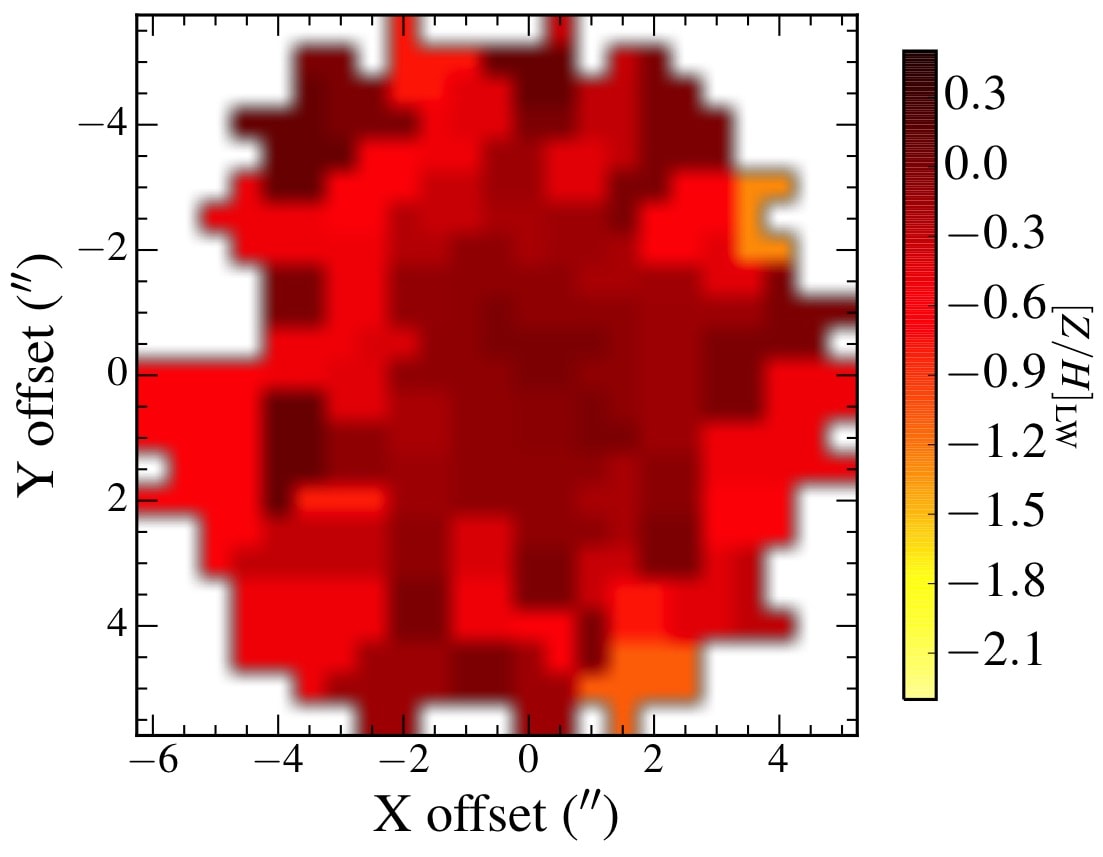}
\includegraphics[height=0.17\textwidth,width=0.223\textwidth, angle=0]{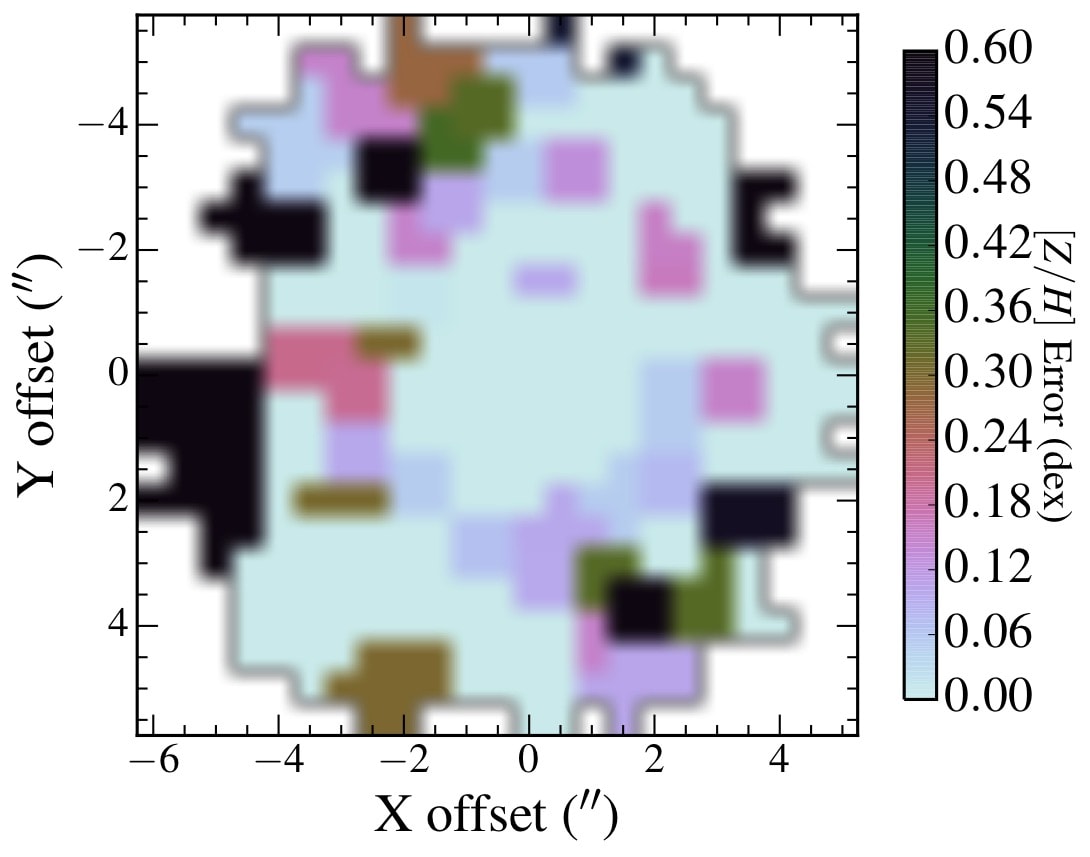}
\includegraphics[height=0.17\textwidth,width=0.223\textwidth, angle=0]{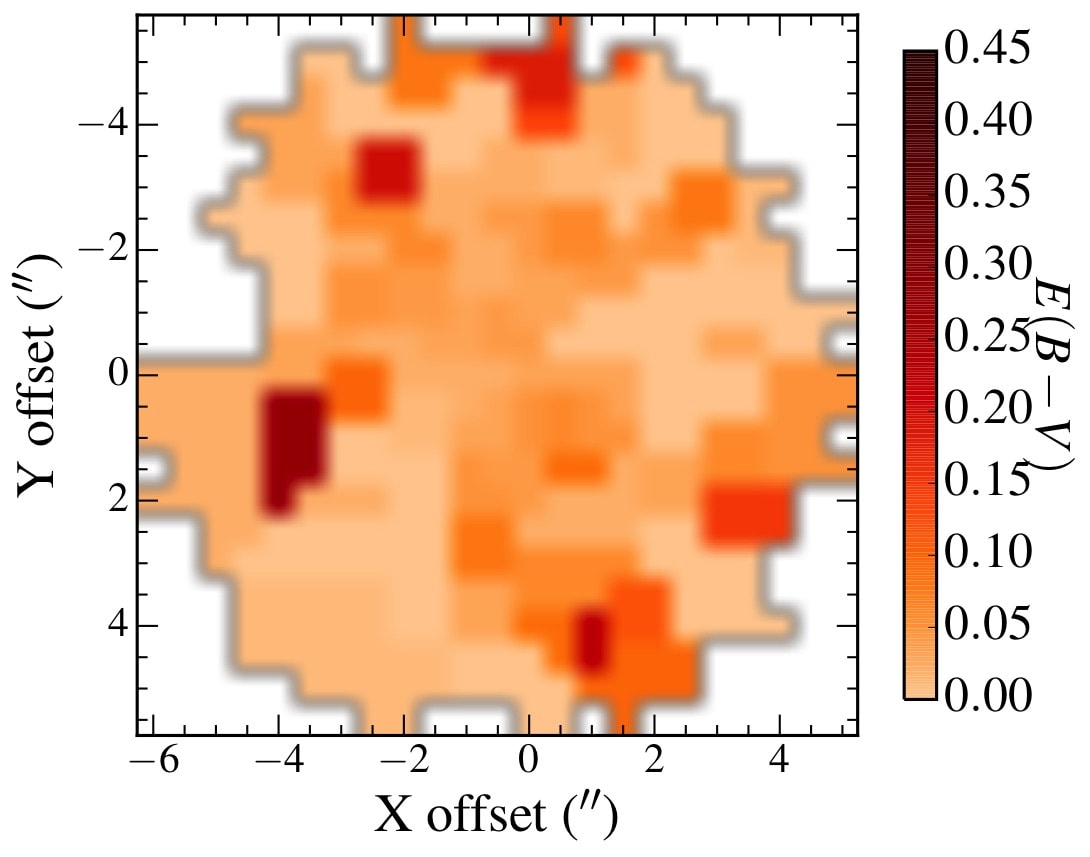}
\caption{Example maps of $S/N$ (Column 1), light-weighted $\log(Age(Gyr))$ with errors (Columns 2 and 3), light-weighted $[Z/{\rm H}]$ with errors (Columns 4 and 5) and $E(B-V)$ values (Column 6) for a Voronoi binning of $S/N=5$. Bins with $S/N < 5$ are masked out as they are not used in our analysis. From top to bottom: MaNGA ID 1-596678 (127 fibre), MaNGA ID 12-84679 (91 fibre), MaNGA ID 1-252070 (61 fibre), MaNGA ID 1-235530 (37 fibre) and MaNGA ID 12-110746 (19 fibre). SDSS colour images of these galaxies are displayed in Figure~\ref{fig:example_gradients}. The masses of the galaxies, in units of $\log(M/M_{\odot})$, are $11.02$, $11.45$, $11.25$, $10.56$ and $9.91$ respectively.}
\label{fig:maps}
\end{figure}
\end{landscape}

Figure~\ref{fig:maps} shows example maps of $S/N$, luminosity-weighted age and metallicity with their corresponding errors, and dust attenuation for the five different IFU bundle sizes in the MaNGA survey. It can be seen that the Voronoi binning to a minimum of $S/N=5$ generally provides a relatively high spatial sampling even for the smaller IFU bundles. The spatial resolution is around 2.5 arcsec, which is somewhat oversampled by the Voronoi cells with $S/N=5$ threshold. The errors on the derived stellar population properties are generally small in the most central parts of the maps where the $S/N$ is greatest, but become larger on the outskirts where $S/N$ begins to drop off. This behaviour can be understood from looking at the running median presented in Figure~\ref{fig:ff_signal-to-noise}. This is key for the derivation of accurate radial gradients across the entire sample. The aim of this figure is to provide a sense of the spatial sampling for the various IFU sizes, and a quantitative assessment of the measurement errors on the stellar parameters as a function of radius is provided below.

\begin{landscape}
\begin{figure}
\raisebox{4.5mm}{\includegraphics[width=0.16\textwidth]{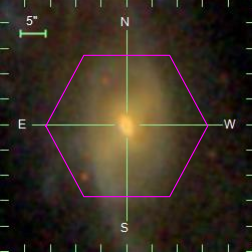}}\hspace{0.01\textwidth}
\includegraphics[width=0.232\textwidth]{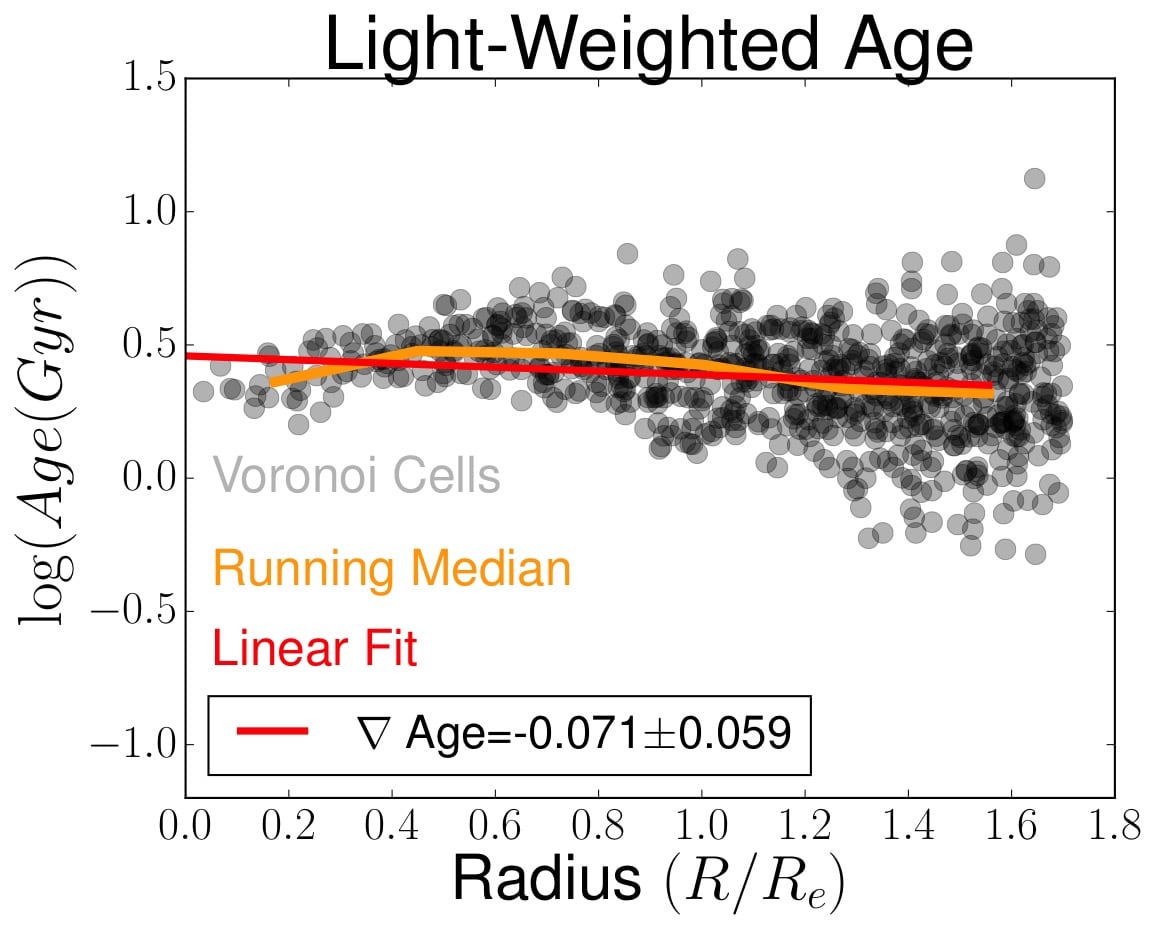}
\includegraphics[width=0.232\textwidth]{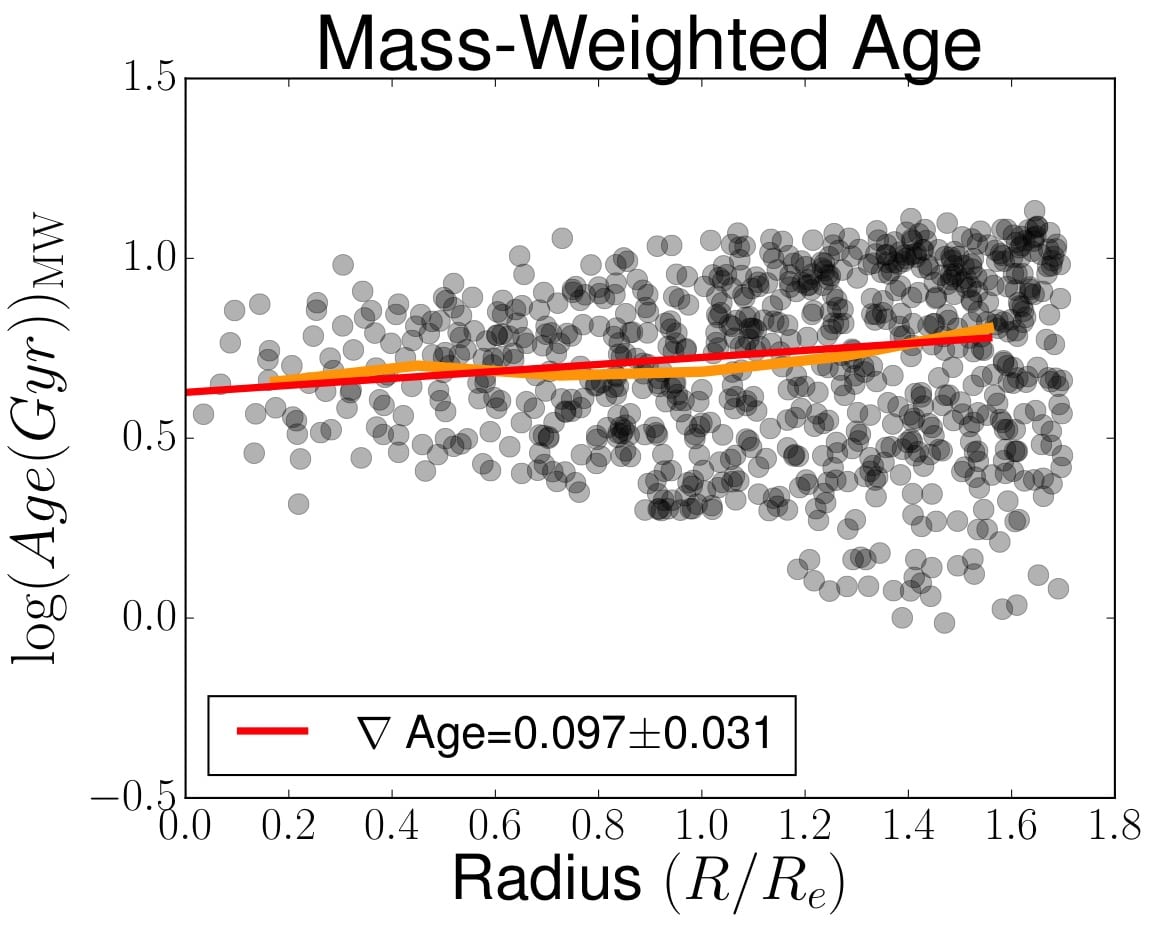}
\includegraphics[width=0.232\textwidth]{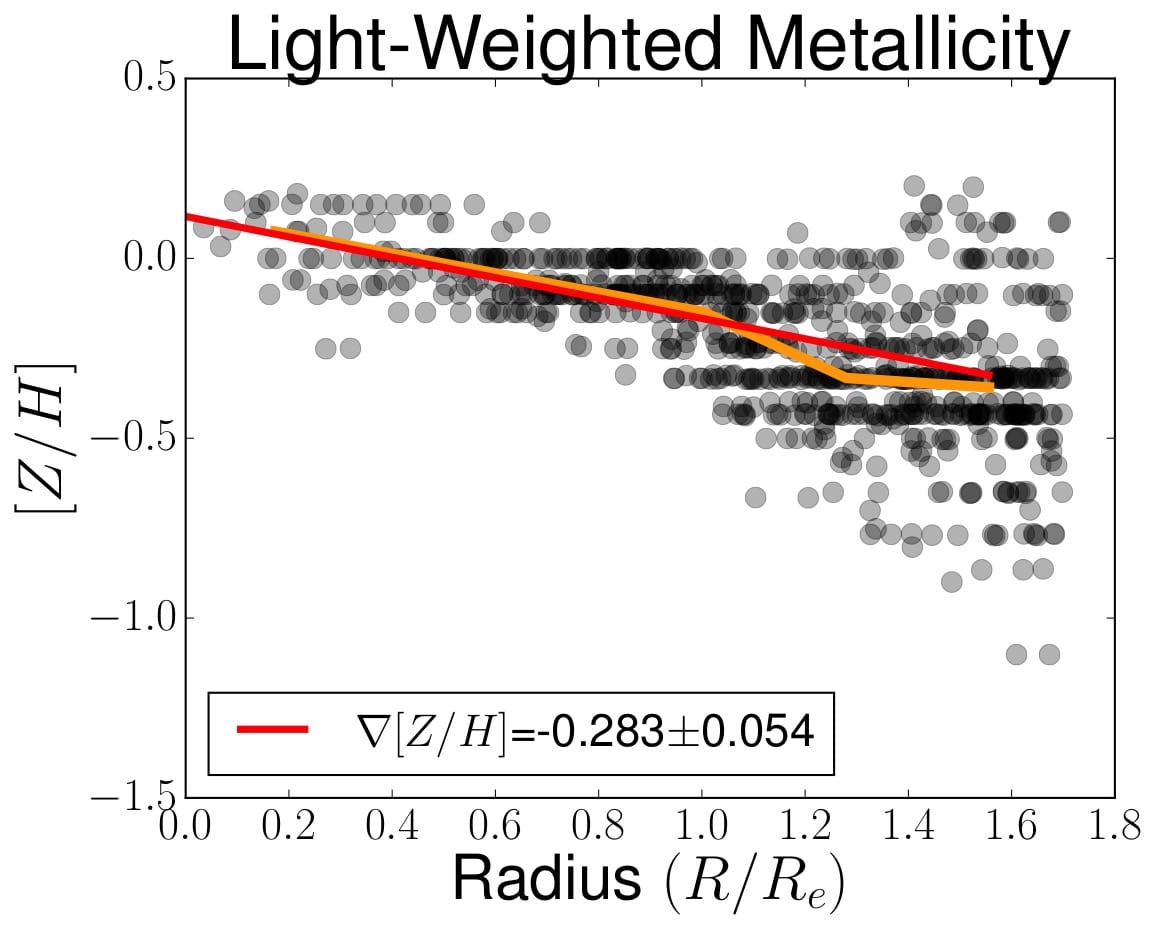}
\includegraphics[width=0.232\textwidth]{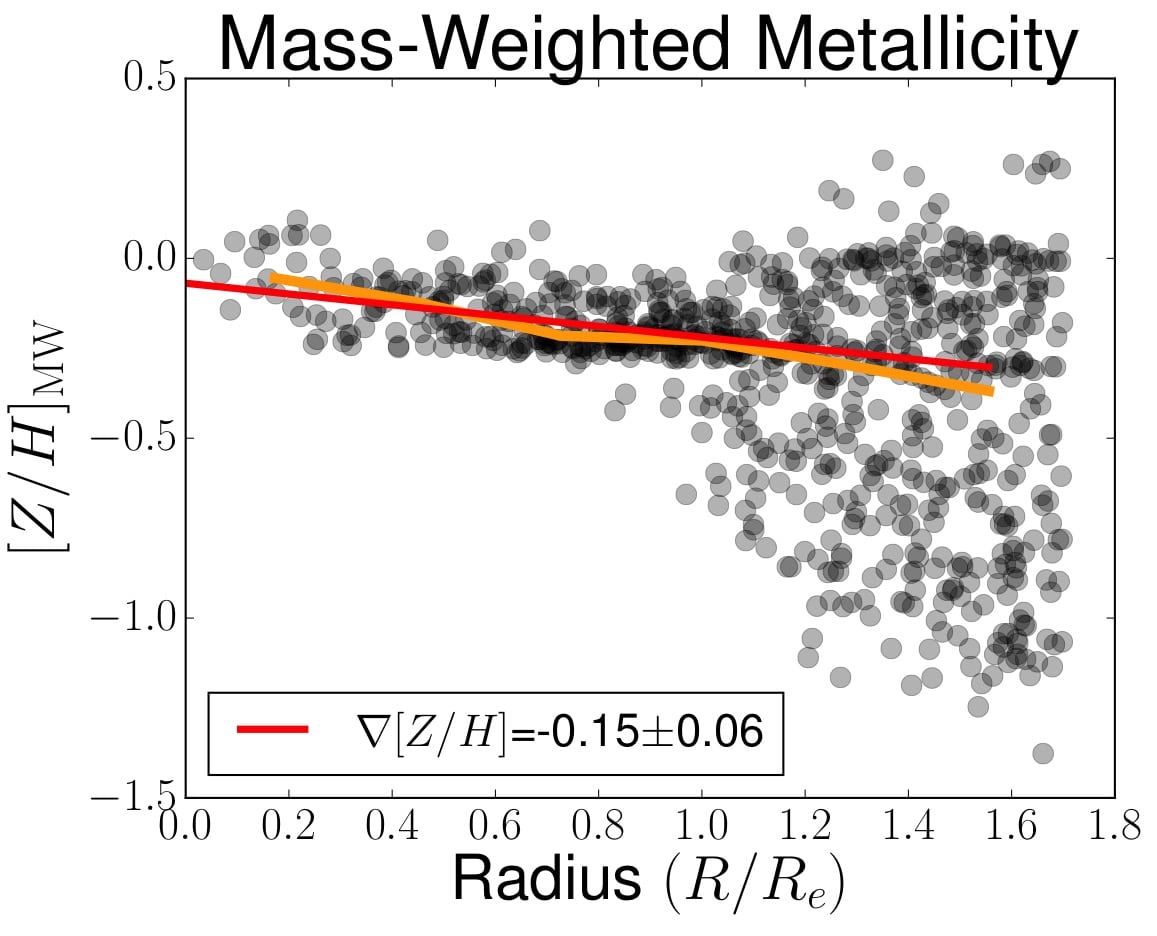}
\includegraphics[width=0.232\textwidth]{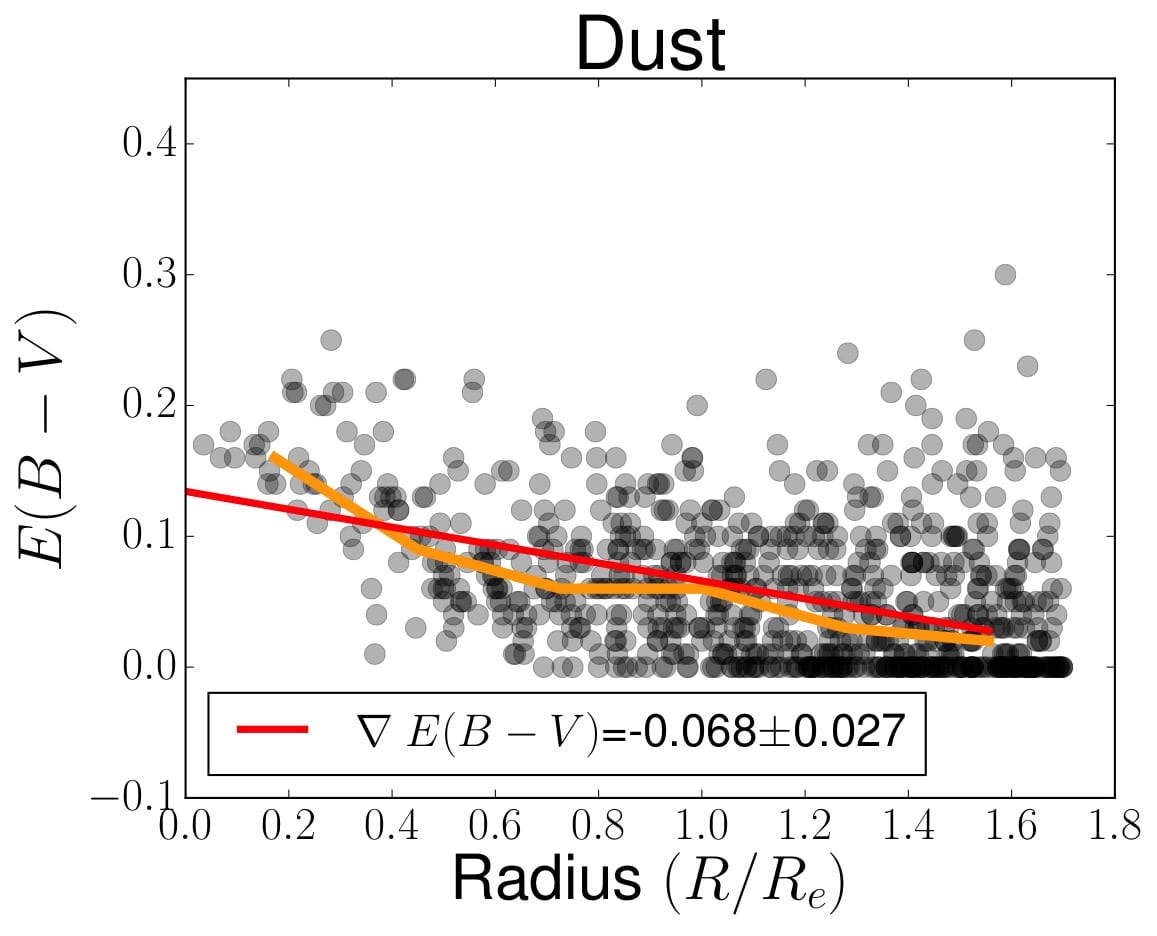}
\raisebox{4.5mm}{\includegraphics[width=0.16\textwidth]{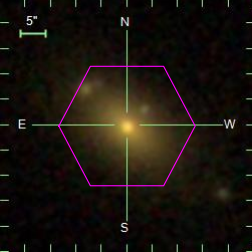}}\hspace{0.01\textwidth}
\includegraphics[width=0.232\textwidth]{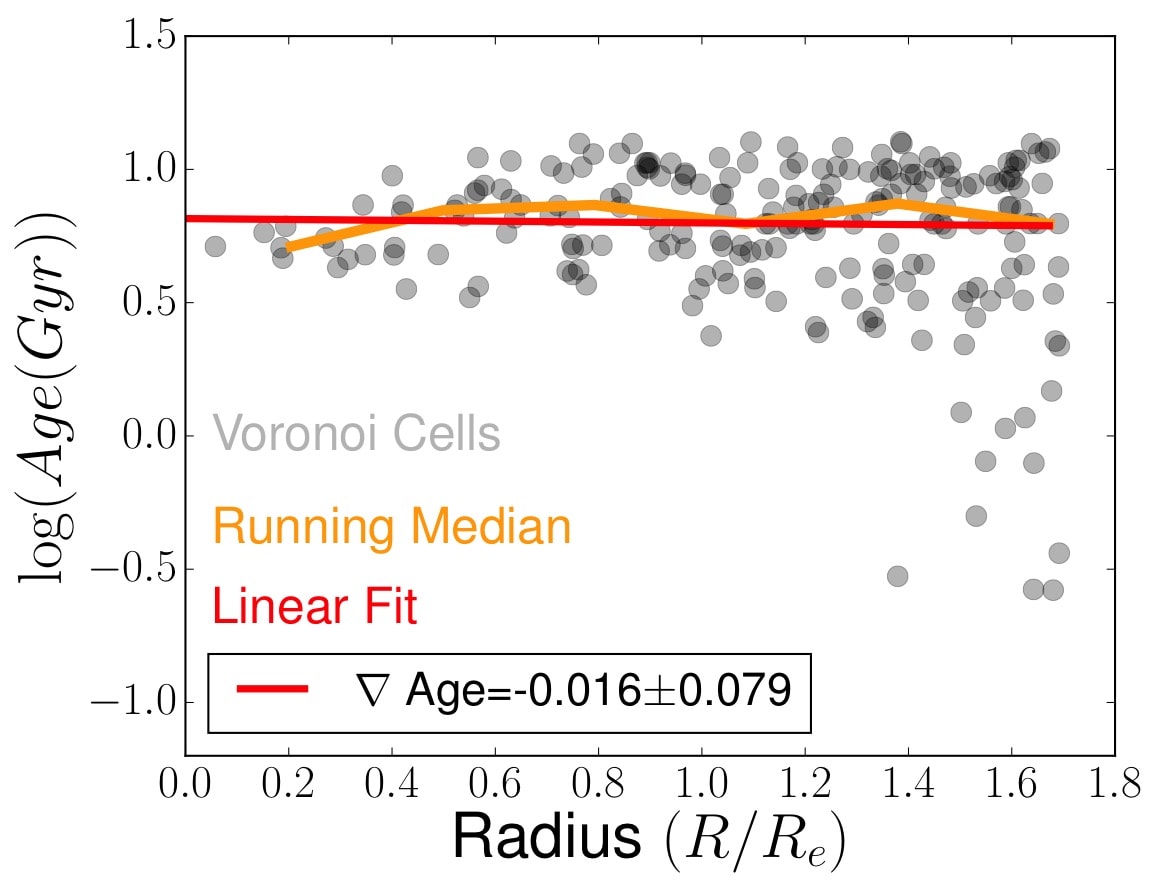}
\includegraphics[width=0.232\textwidth]{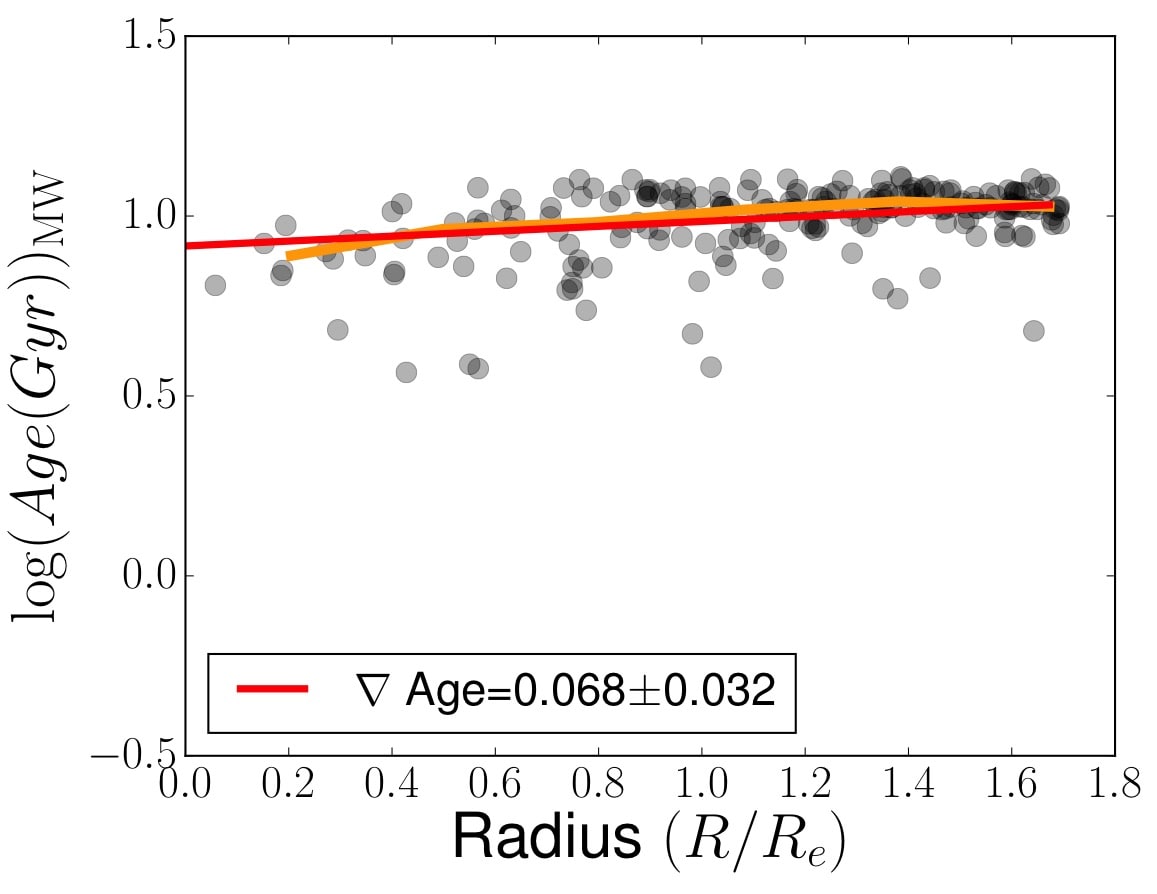}
\includegraphics[width=0.232\textwidth]{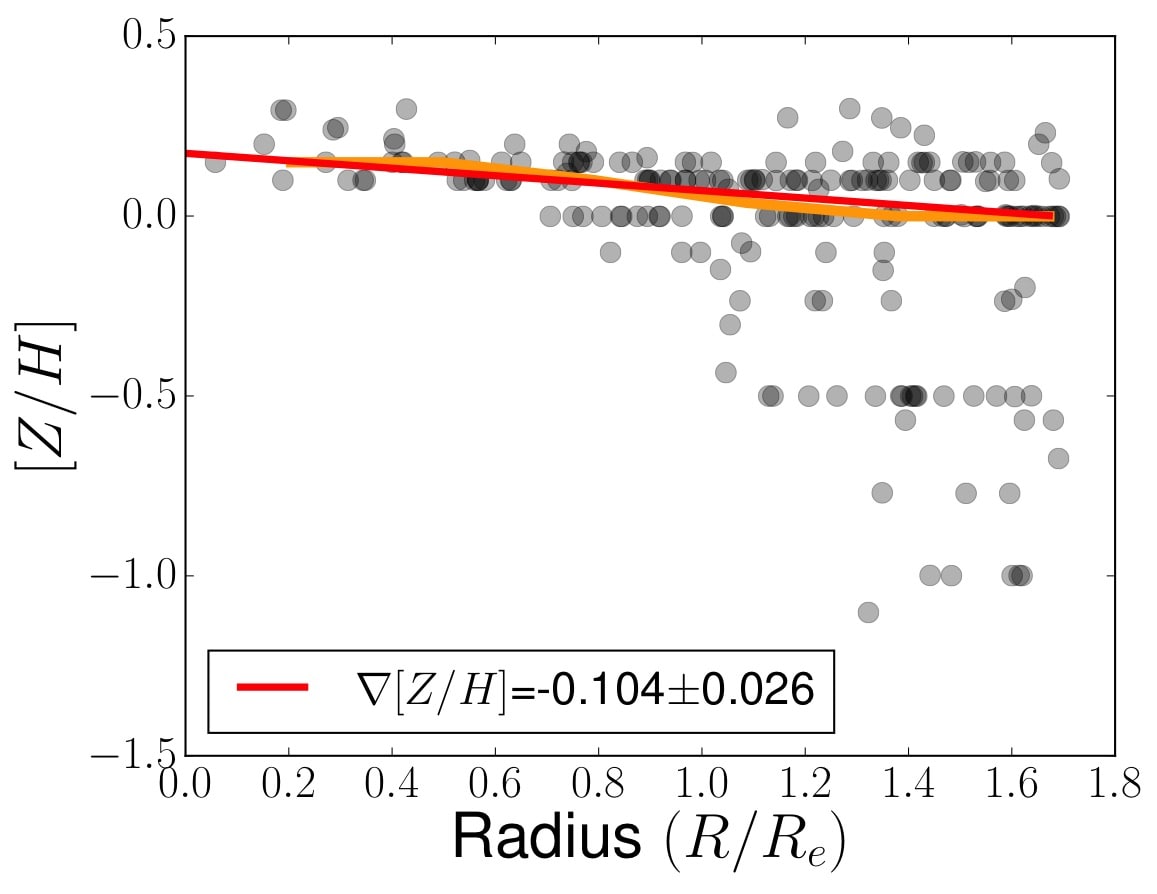}
\includegraphics[width=0.232\textwidth]{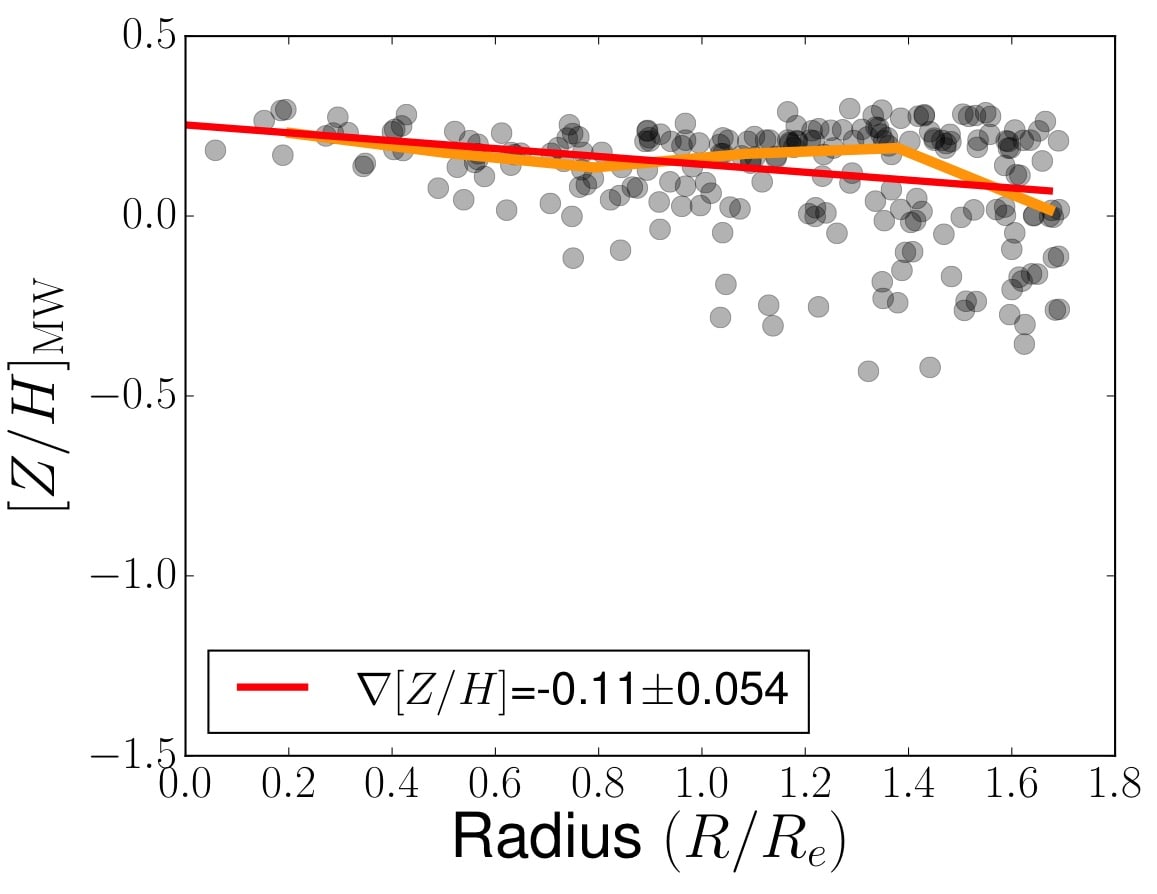}
\includegraphics[width=0.232\textwidth]{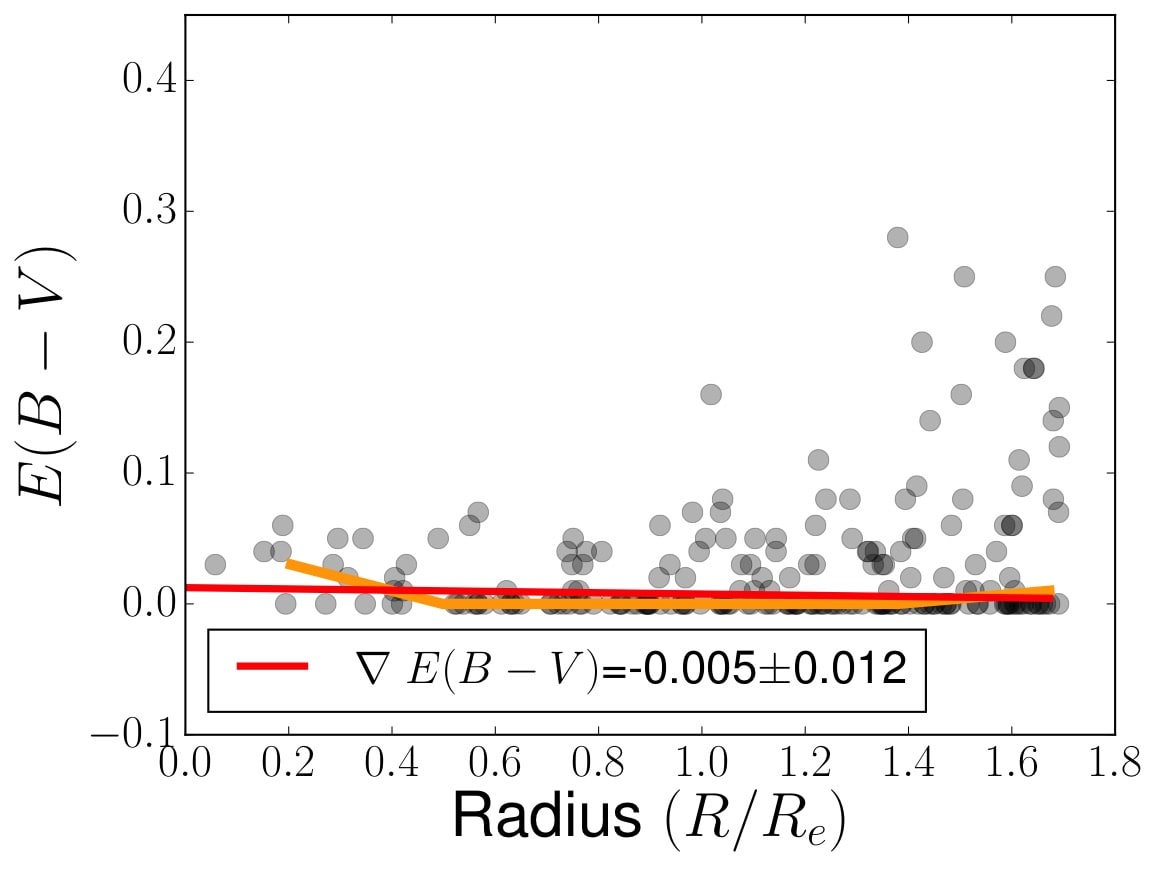}
\raisebox{4.5mm}{\includegraphics[width=0.16\textwidth]{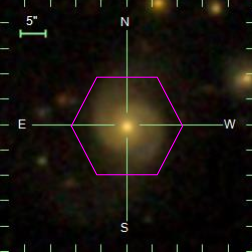}}\hspace{0.01\textwidth}
\includegraphics[width=0.232\textwidth]{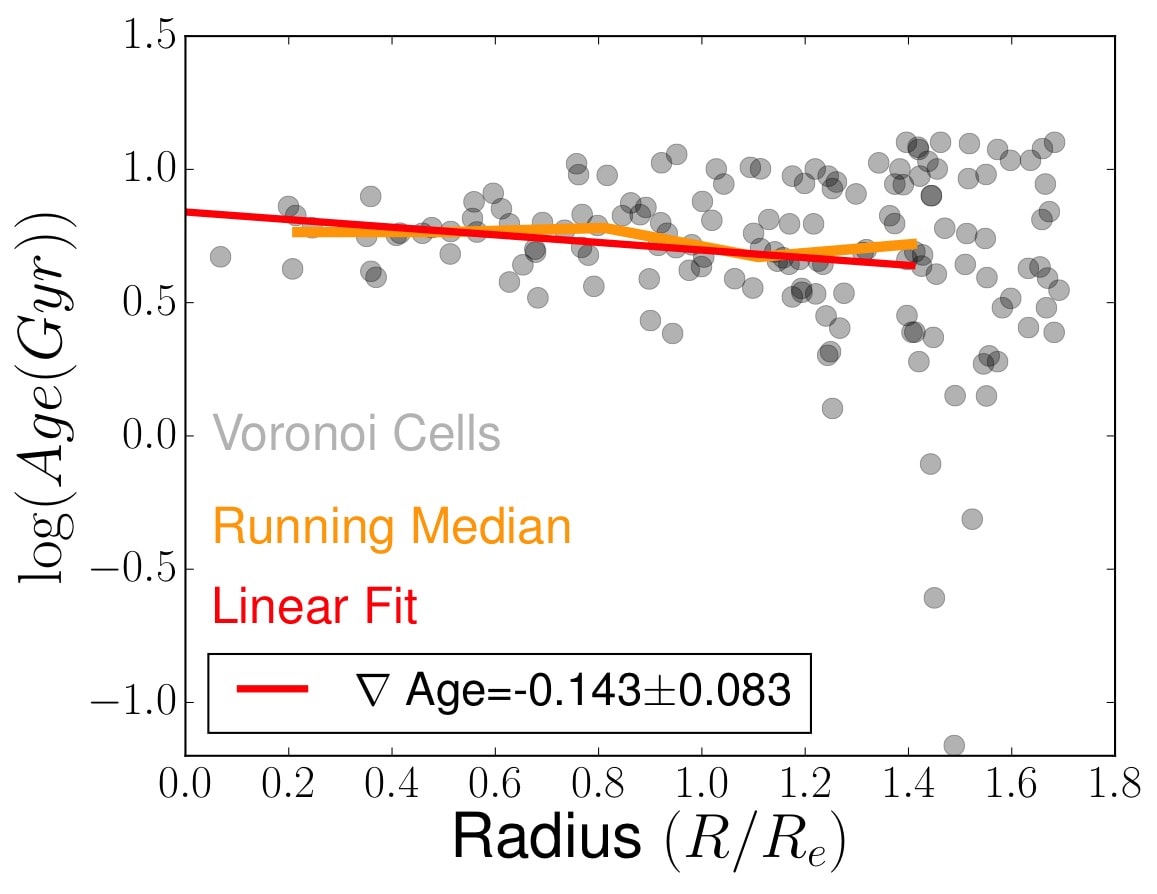}
\includegraphics[width=0.232\textwidth]{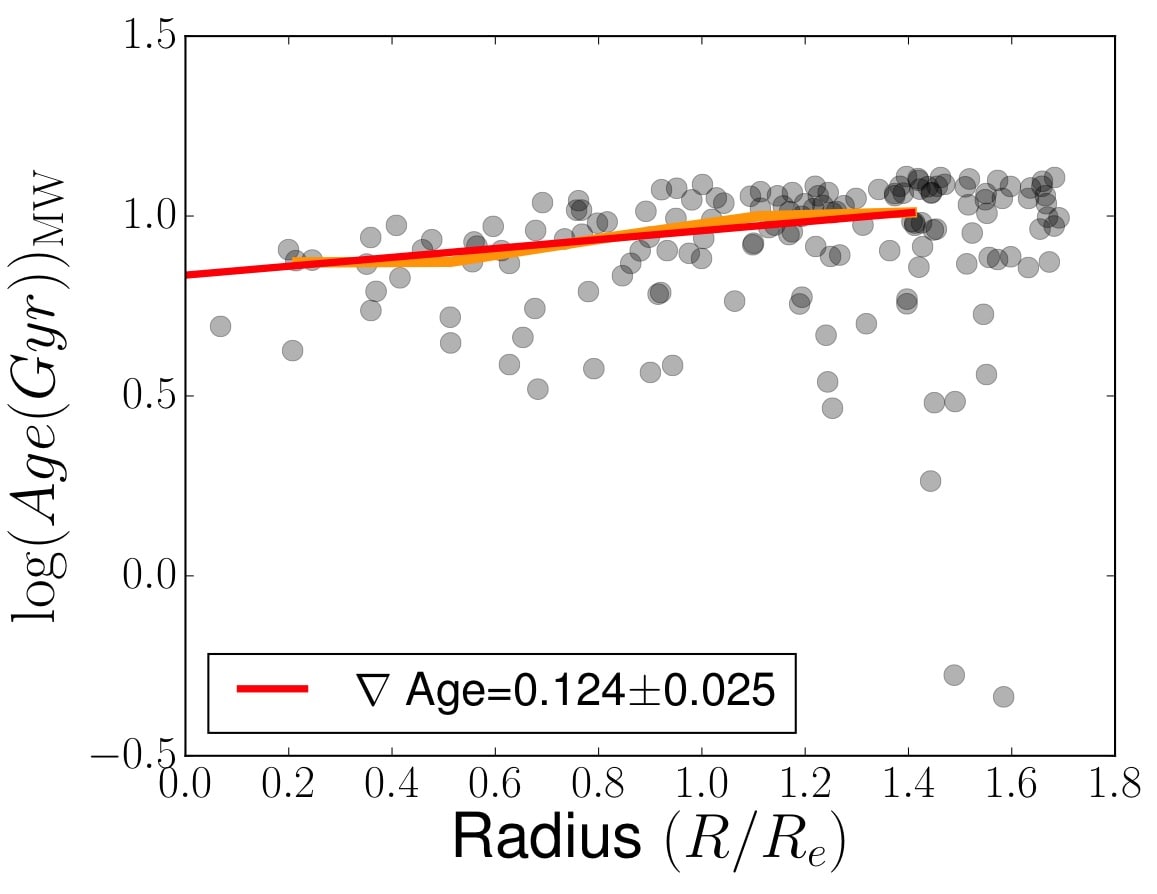}
\includegraphics[width=0.232\textwidth]{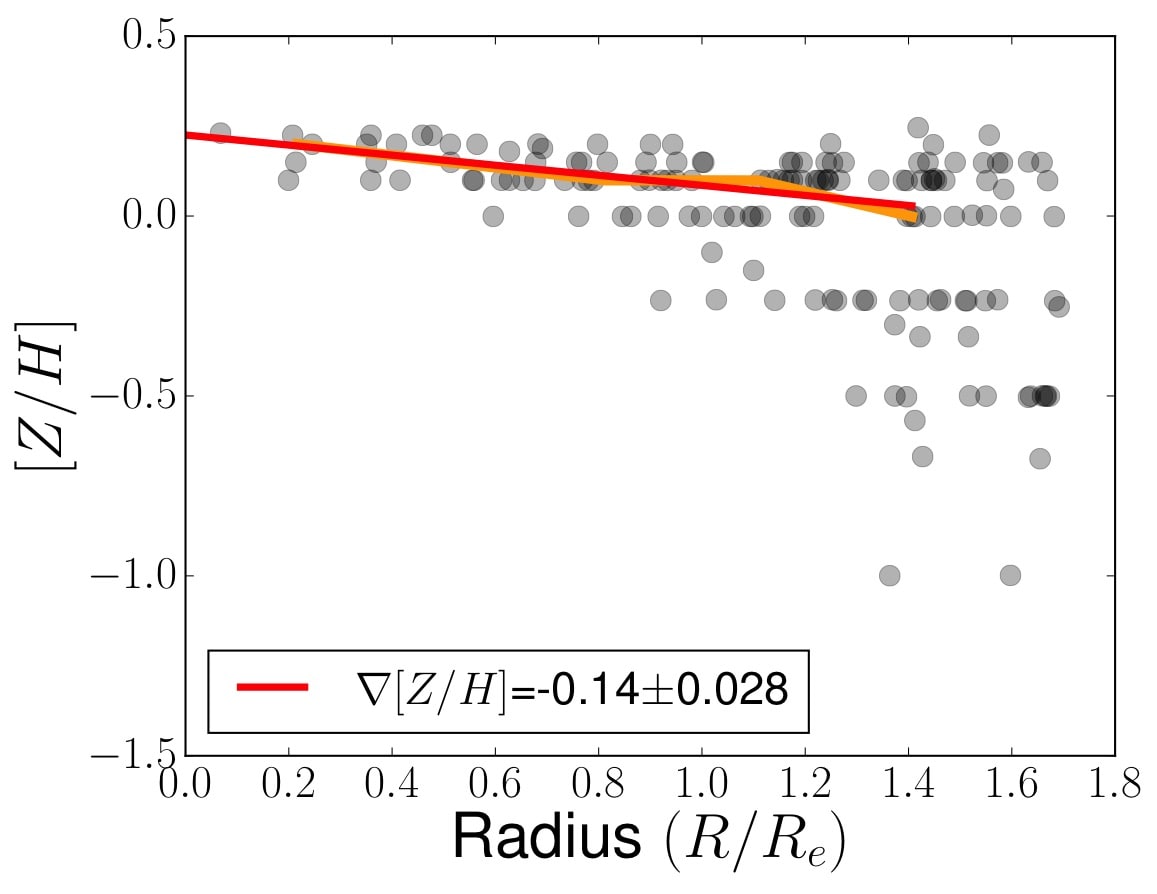}
\includegraphics[width=0.232\textwidth]{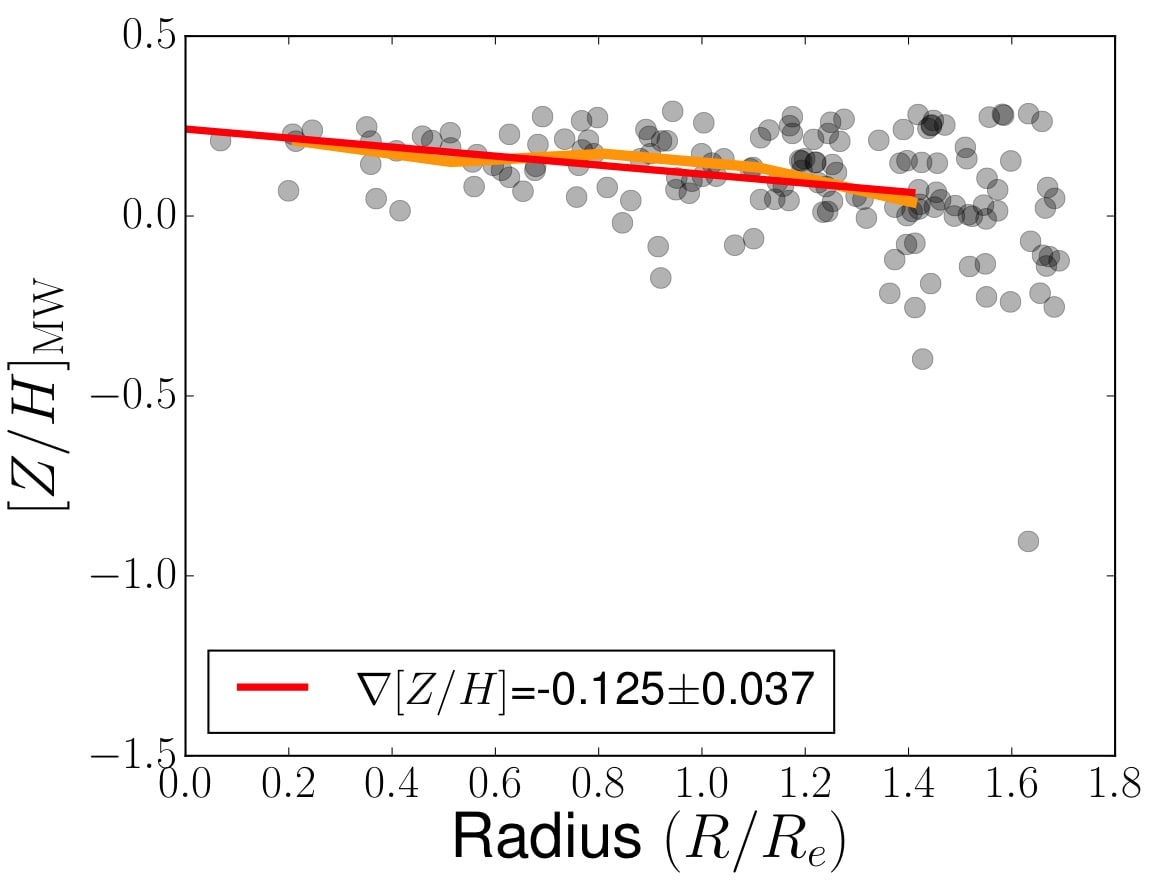}
\includegraphics[width=0.232\textwidth]{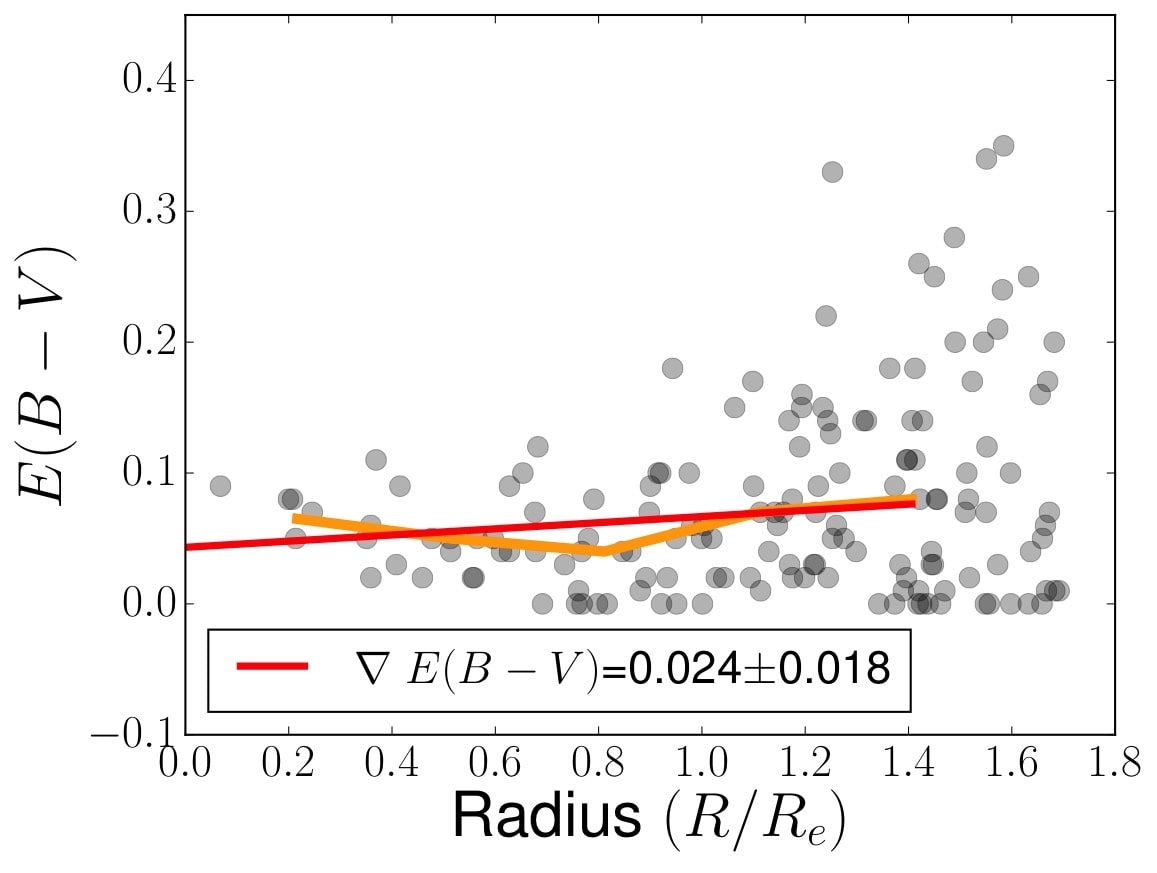}
\raisebox{4.5mm}{\includegraphics[width=0.16\textwidth]{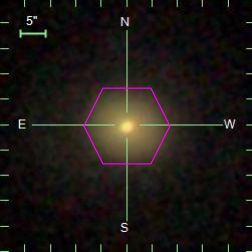}}\hspace{0.01\textwidth}
\includegraphics[width=0.232\textwidth]{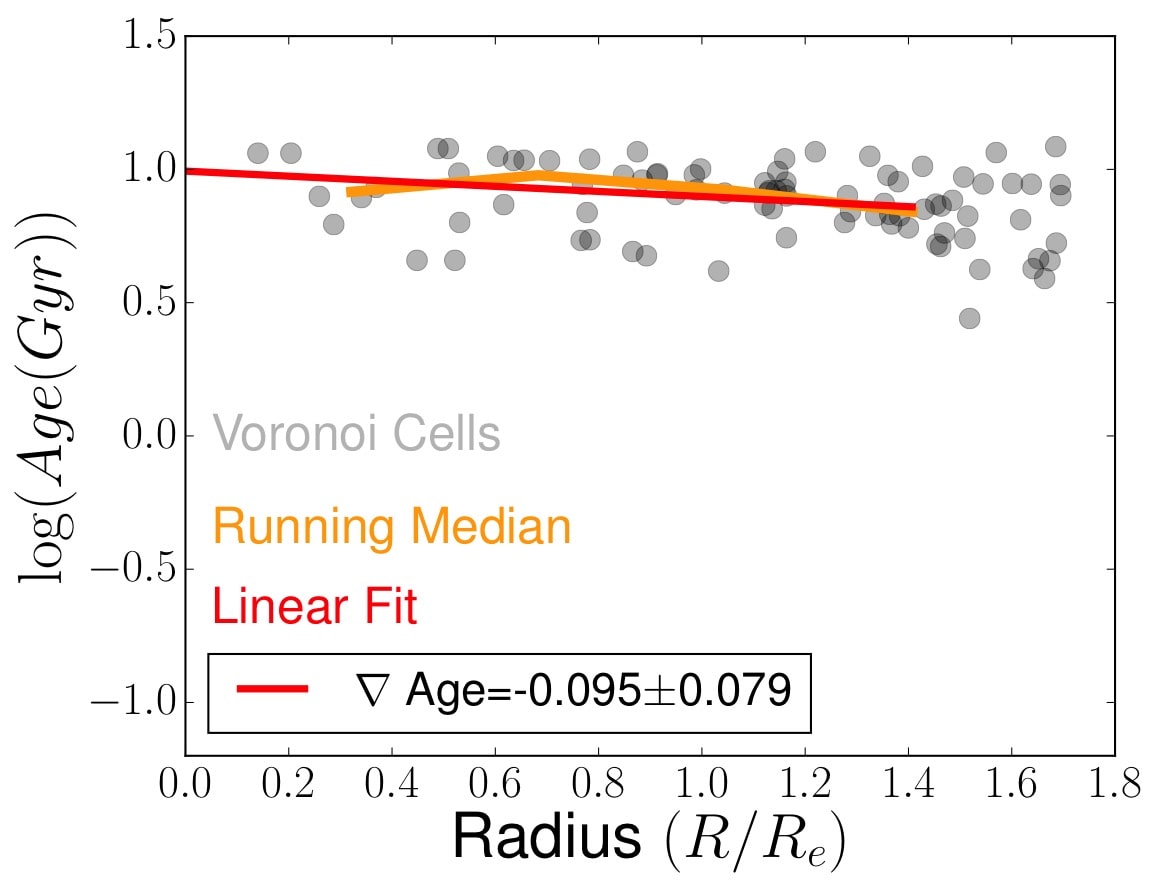}
\includegraphics[width=0.232\textwidth]{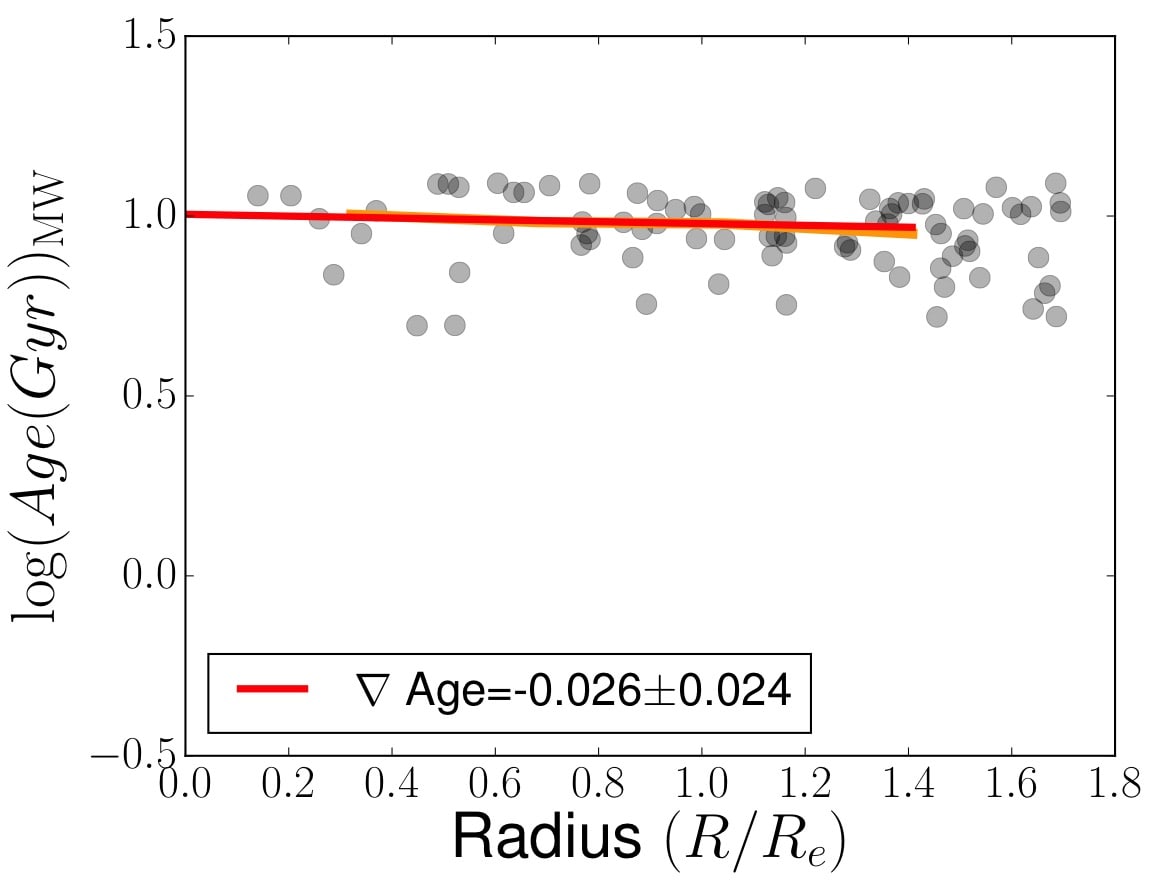}
\includegraphics[width=0.232\textwidth]{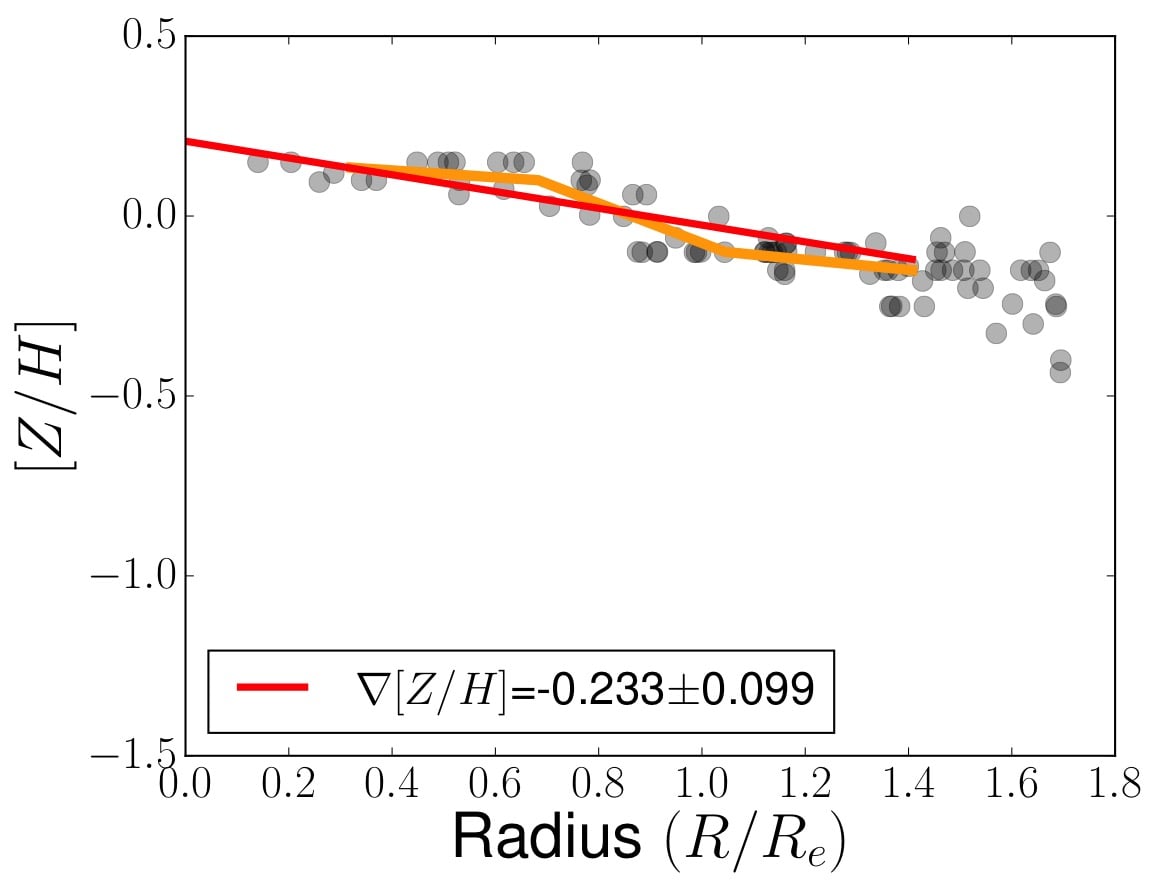}
\includegraphics[width=0.232\textwidth]{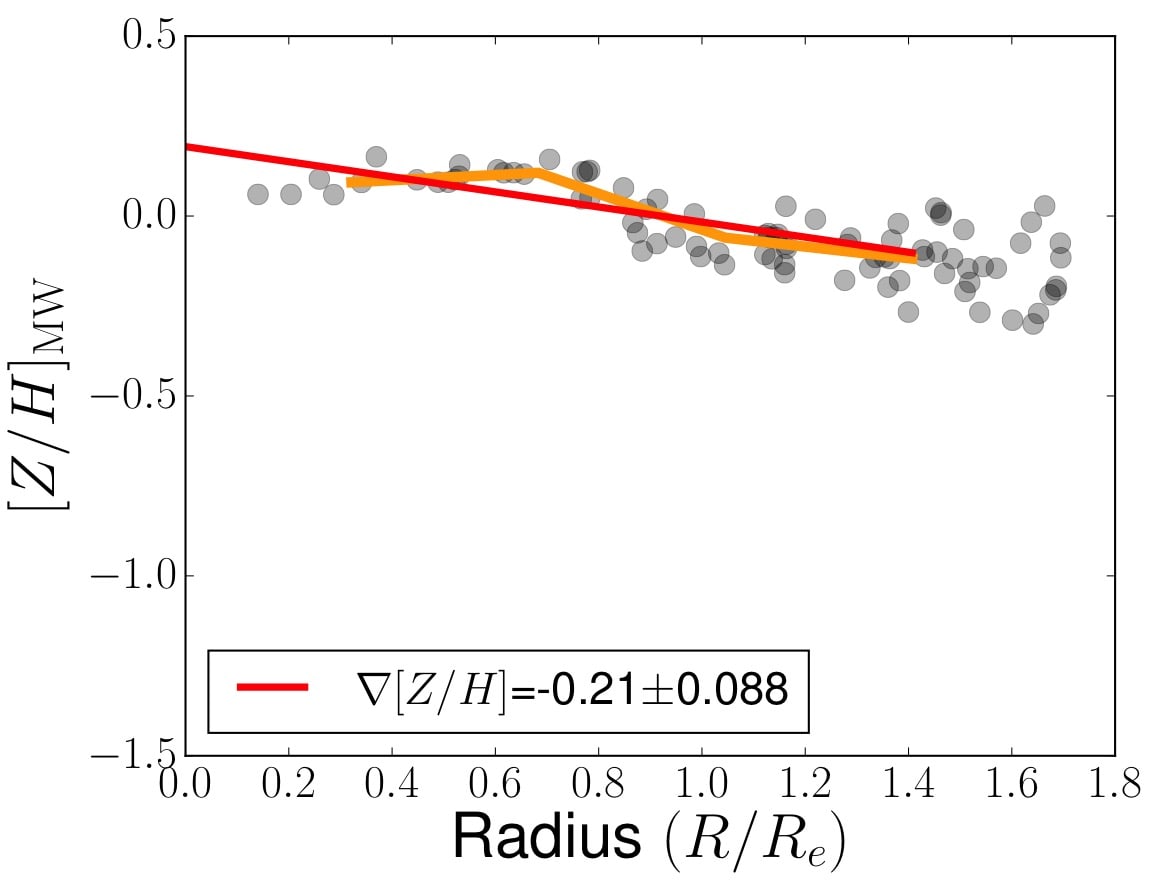}
\includegraphics[width=0.232\textwidth]{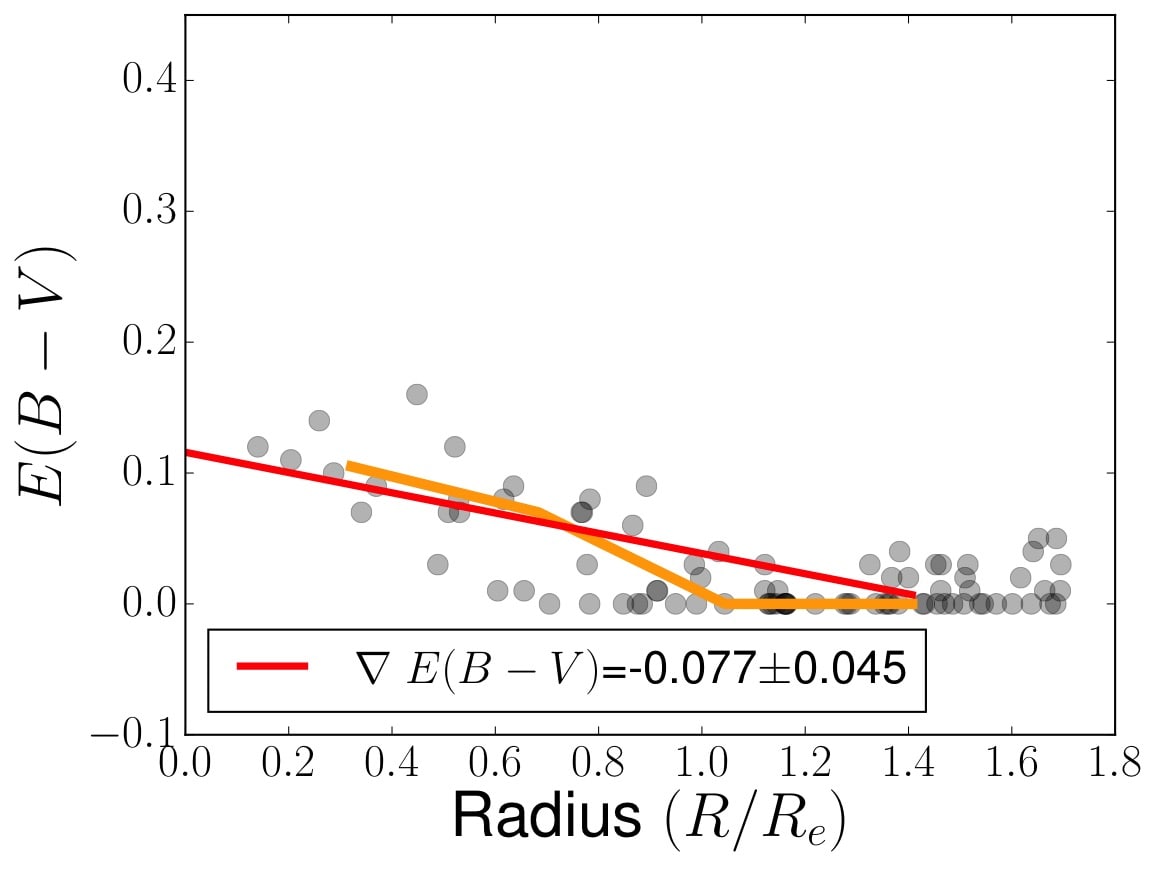}
\raisebox{4.5mm}{\includegraphics[width=0.16\textwidth]{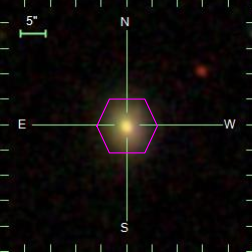}}\hspace{0.01\textwidth}
\includegraphics[width=0.232\textwidth]{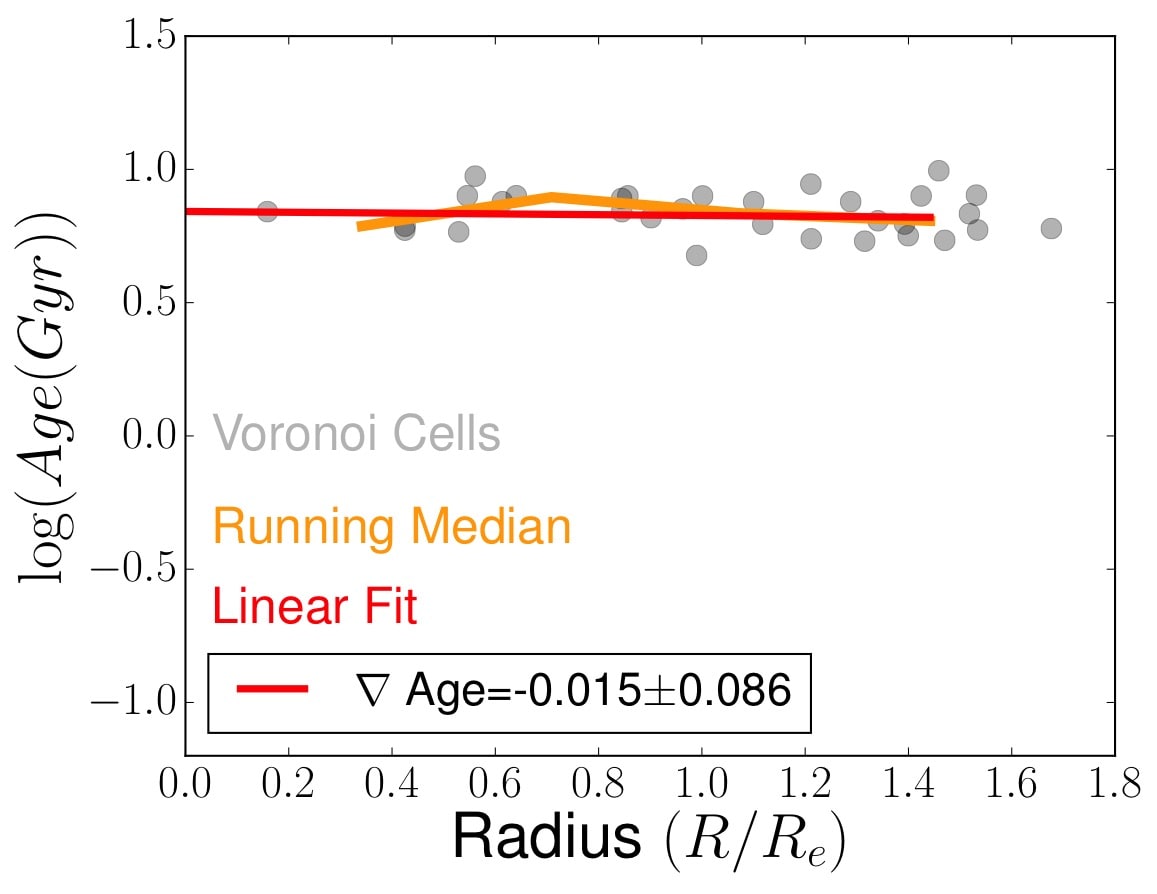}
\includegraphics[width=0.232\textwidth]{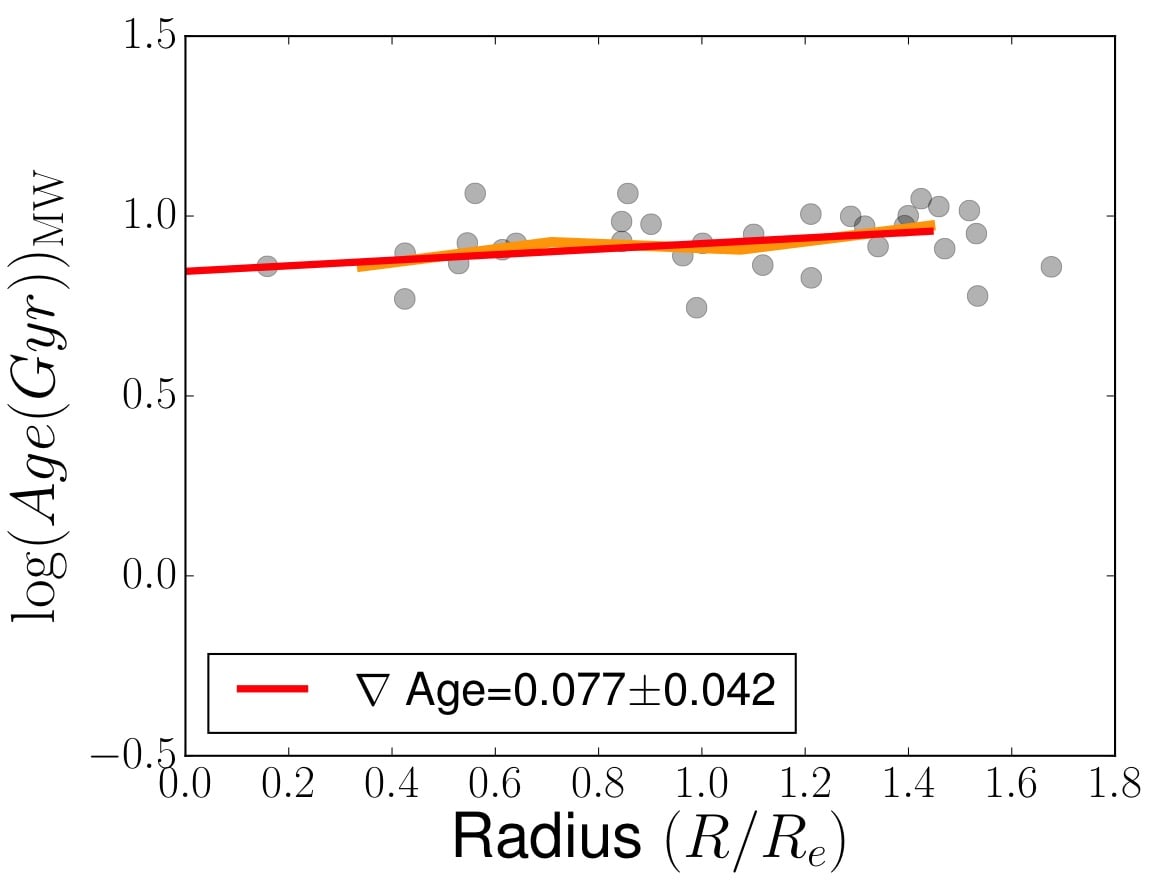}
\includegraphics[width=0.232\textwidth]{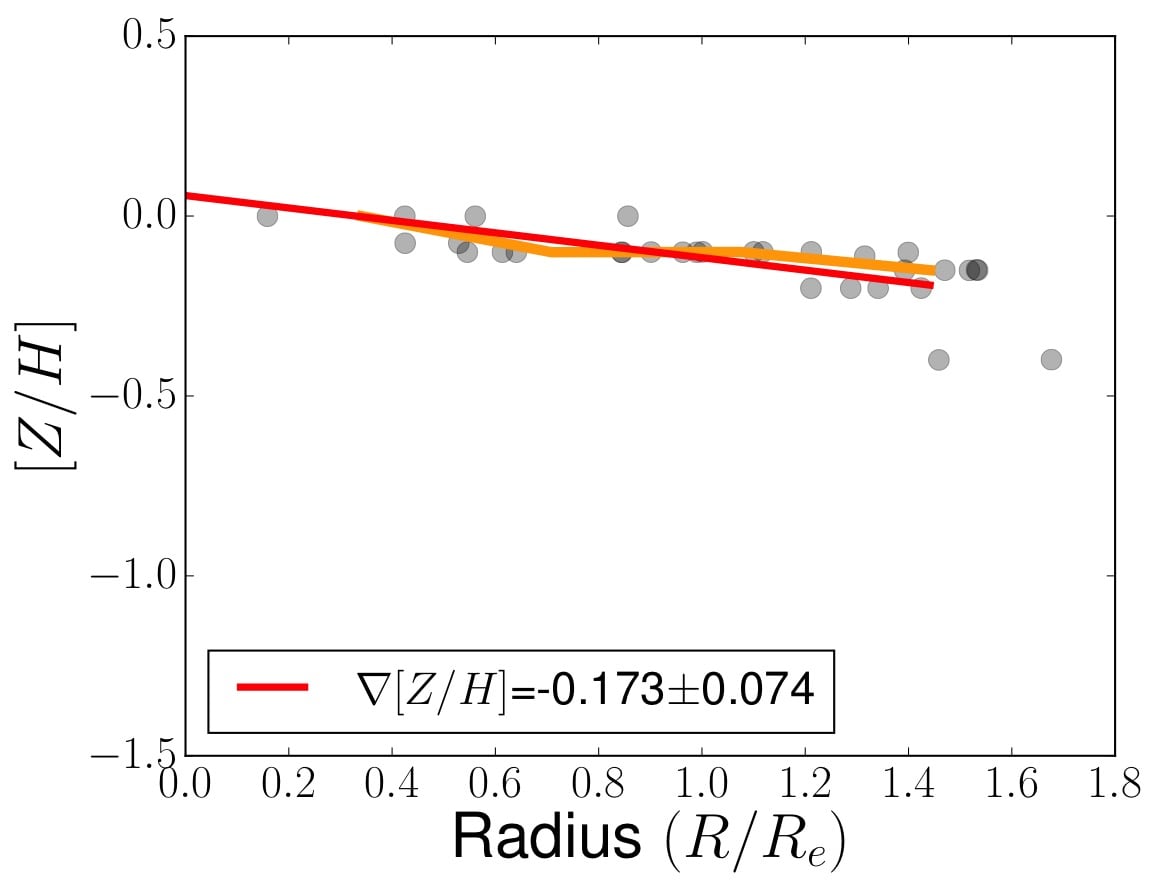}
\includegraphics[width=0.232\textwidth]{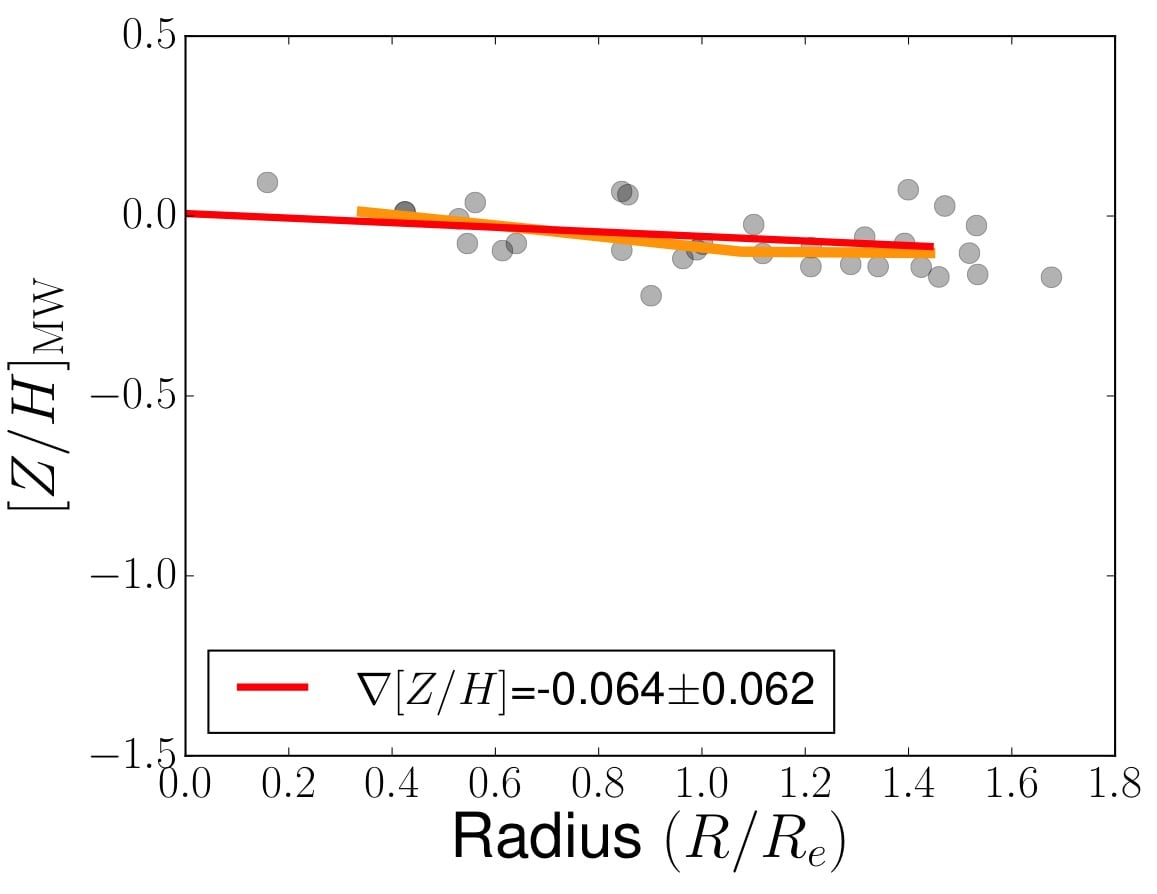}
\includegraphics[width=0.232\textwidth]{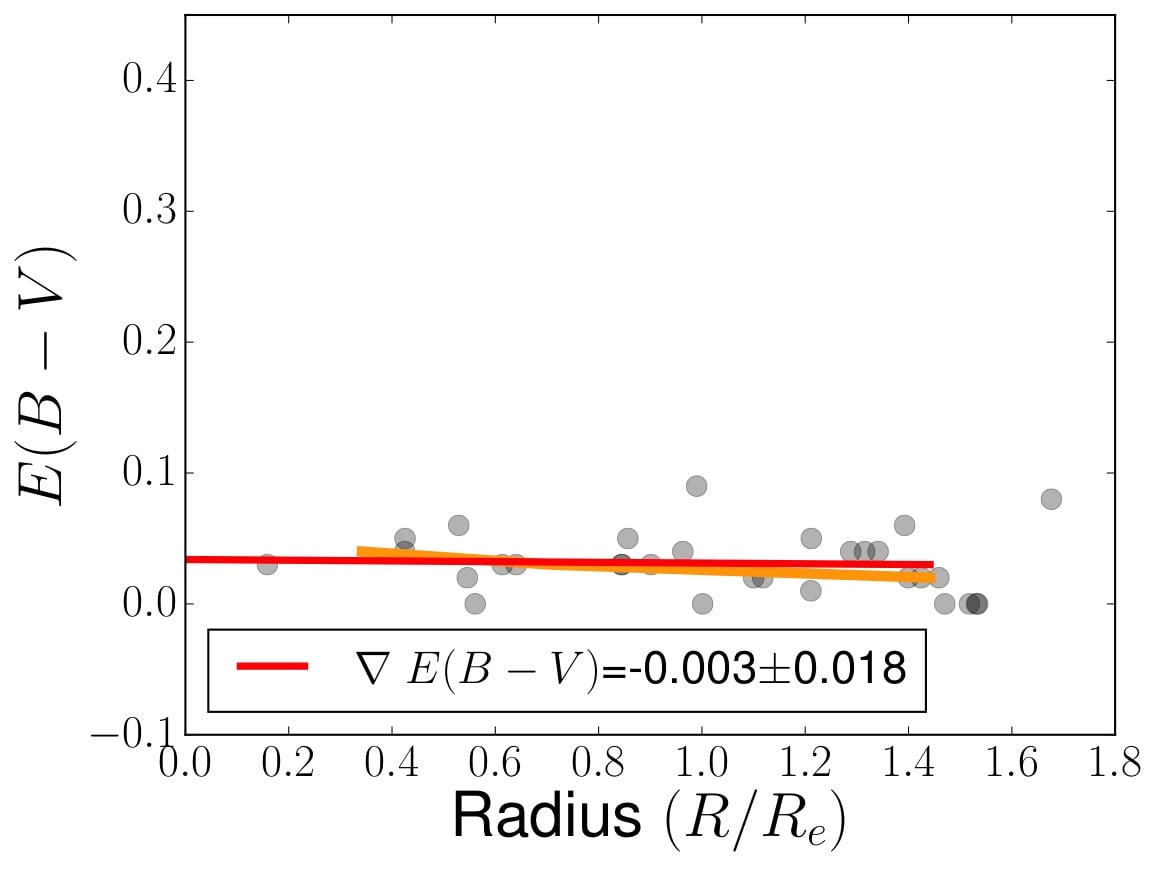}
\caption{Examples of light and mass-weighted radial profiles of age (Columns 2, 3), metallicity (Column 4, 5) and $E(B-V)$ values (Column 6) obtained from full spectral fitting for five galaxies from the MaNGA survey of varying IFU size. Column 1 shows the SDSS image of the galaxy. From top to bottom: MaNGA ID 1-596678 (127 fibre), MaNGA ID 12-84679 (91 fibre), MaNGA ID 1-252070 (61 fibre), MaNGA ID 1-235530 (37 fibre) and MaNGA ID 12-110746 (19 fibre). Grey circles represent the individual Voronoi cells from the DAP data cube, the orange line shows the running median and the red line shows the straight line fit. The value of the stellar population gradient is quoted in the legend.}
\label{fig:example_gradients}
\end{figure}
\end{landscape}

\subsection{Radial Gradients}
\begin{figure}
\centering
\includegraphics[width=0.49\textwidth]{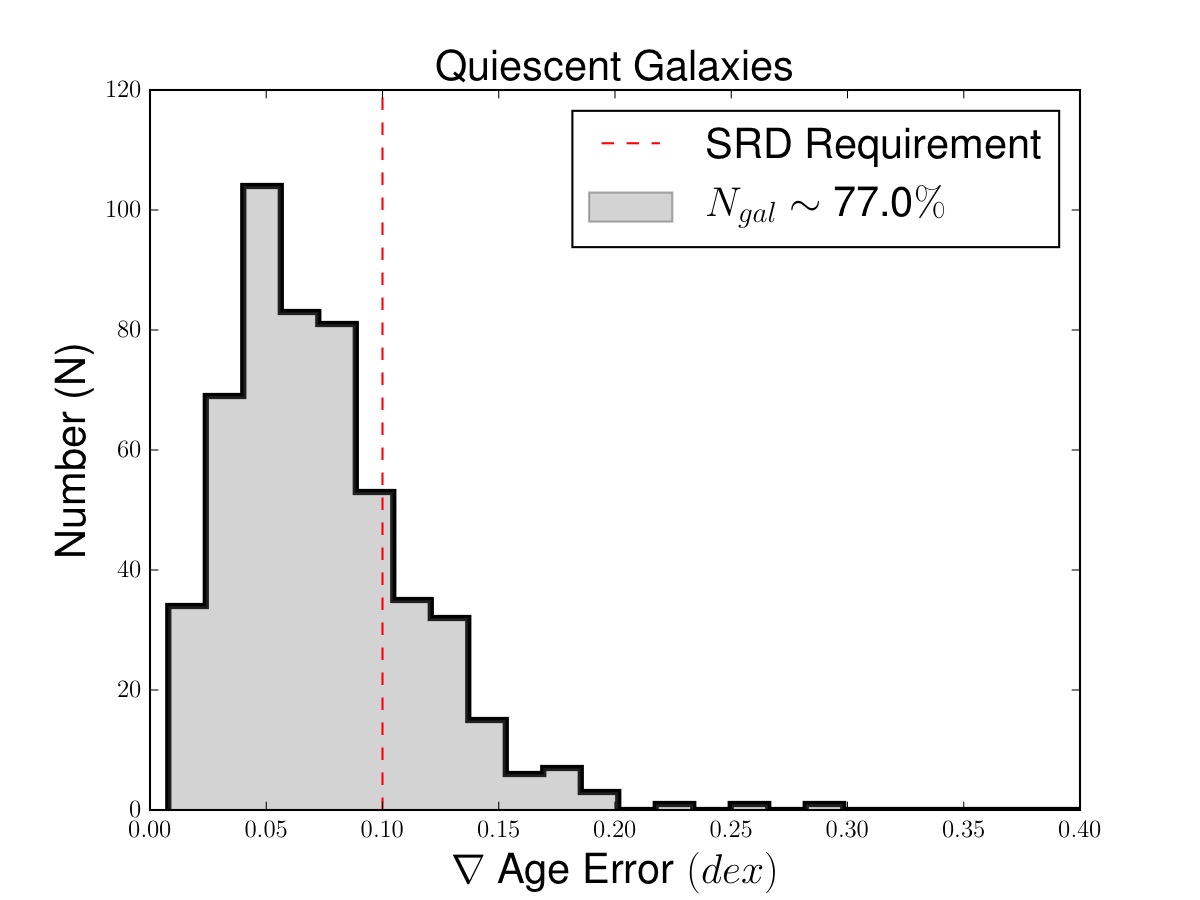}
\includegraphics[width=0.49\textwidth]{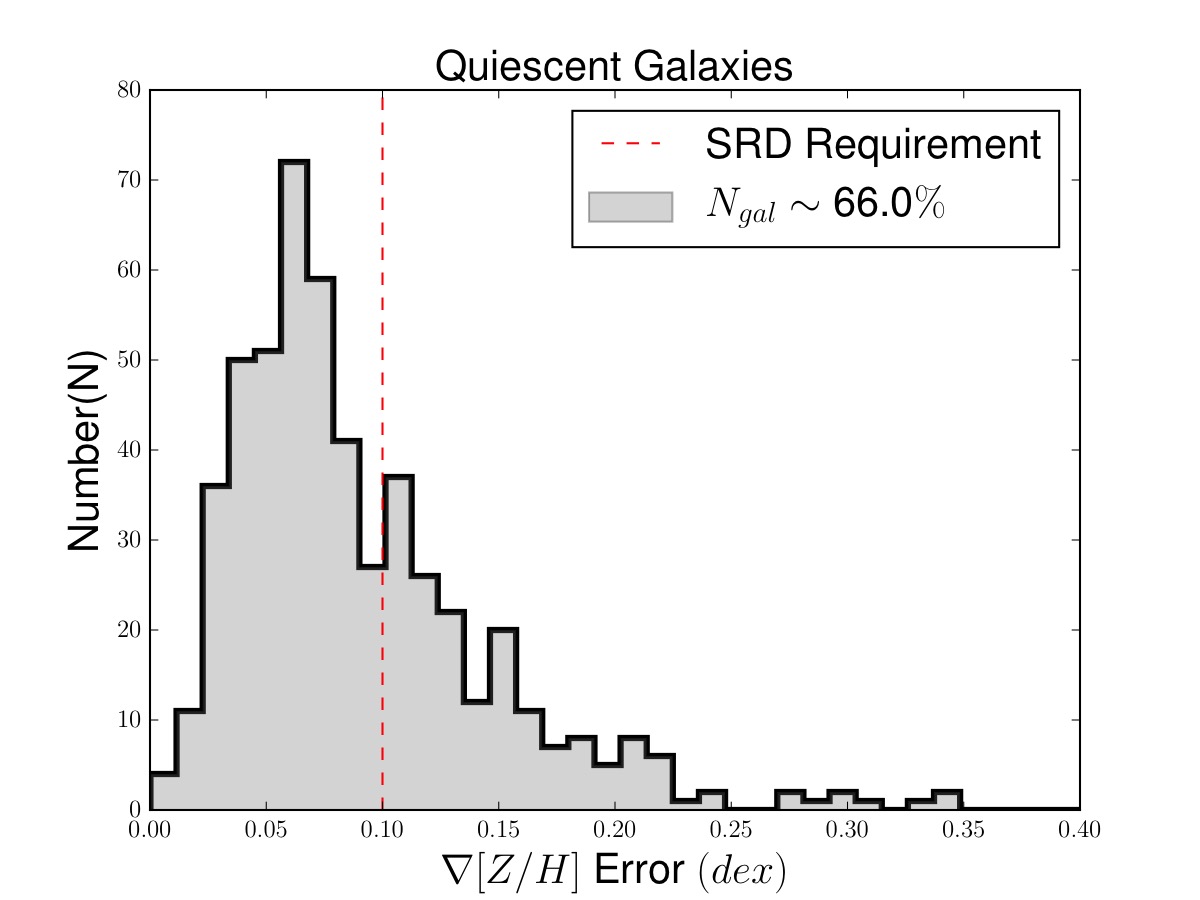}
\caption{Distribution of errors on age (top) and metallicity gradients (bottom). The dashed red line shows the MaNGA science requirement (SRD) of 0.1 dex error on stellar population gradients. Quiescent galaxies are defined as the early-type galaxies used in this paper. $N_{gal}$ represents the percentage of the galaxy sample with an error less than the SRD requirement.}
\label{fig:gradienterrors}
\end{figure}

The on-sky position (relative to the galaxy centre) of each Voronoi cell is used to calculate semi-major axis coordinates, which we then use to define a radius $R$ of the cell. The effective radius $R_{\rm e}$, position angle and ellipticity used in this calculation are adopted from the NSA catalogue. We define the radial gradient of a stellar population property $\theta$ ($\log(Age(Gyr))$, $[Z/H]$) in units of dex/$R_{e}$ as:
\begin{equation}
\nabla \theta= \mathrm{d}\theta/\mathrm{d}R,
\end{equation}
where $R$ is the radius in units of effective radius $R_{\rm e}$. The gradient is measured using least squares linear regression. Errors on the gradients are calculated using a Monte Carlo bootstrap resampling method \citep{recipes}. This technique involves taking an original dataset of size $N$ and creating a synthetic dataset of the same size but which consists of randomly sampled values from the original dataset, and remeasuring the gradient. We iterate this process 1000 times building a distribution of gradient values, and use the standard deviation of this distribution as the $1$-$\sigma$ error on the gradient. Examples of the resulting radial gradients for the same objects as Figure~\ref{fig:maps} for different IFU bundle sizes are shown in Figure~\ref{fig:example_gradients}. Both luminosity-weighted and mass-weighted stellar population parameters are shown in this. It can be seen that spatial sampling is high enough to allow for robust derivations of radial gradients even for the small fibre bundles (bottom rows). It is also reassuring to note that the linear fit (red line) generally agrees well with the running median of stellar population parameter as a function of radius (orange line). The resulting error $\sigma$ in the determination of the stellar population parameter at a radial bin centred around radius $R$ can then be estimated as
\begin{equation}
\sigma (R) = \frac{\sum_{i=1}^{N_{R}} \sigma_{i}^{2}}{\sqrt{N_{R}}},
\end{equation}
where $\sigma_{i}$ is the error for individual Voronoi cells contributing to the radial bin around radius $R$, and $N_{R}$ is the total number of cells in that radial bin. Figure~\ref{fig:dx_r} shows a density plot of the resulting error in age (top panel) and metallicity (bottom panel) as a function of radius for the full galaxy sample. There is a large scatter, but the median error per Voronoi cell increases slightly with increasing radius (orange line) as expected because of the drop in $S/N$. However, at the same time the typical number of cells per radial bin increases with increasing radius with a slight drop beyond a radius of $\sim 1.3\;R_{\rm e}$. The error per radial bin is then $\sigma/\sqrt{N_{R}}$ shown by the yellow line. This final error increases only slightly with radius, as the larger number of cells at large radii compensates for the larger error. The resulting typical errors in age and metallicity at $\sim 1.5\;R_{\rm e}$ are $0.06$ dex and $0.07$ dex respectively, and even lower at smaller radii. We note that some possible contribution of covariance between those cells is neglected here, hence the true error will be slightly larger. We expect the effect to be small, however, because of the generally large spatial separation of most Voronoi cells in each annulus, in particular at large radii.\\
\\
The derived gradients and their errors are quoted in the individual panels of Figure~\ref{fig:example_gradients}. The distribution of the errors in age and metallicity gradients for the early-type galaxies in the present sample are presented in Figure~\ref{fig:gradienterrors}. The median errors in age and metallicity gradient are $0.05$ dex and $0.07$ dex, respectively. The distribution is not symmetric, which implies that the errors of the age and metallicity gradients are below $0.1$ dex for 77\% and 66\% of the sample, respectively. These values are very close to the survey science requirement of 80\% of quiescent galaxies with an error in the age and metallicity gradient below $0.1$ dex \citep{yan2016}.

\begin{figure*}
\includegraphics[width=0.33\textwidth]{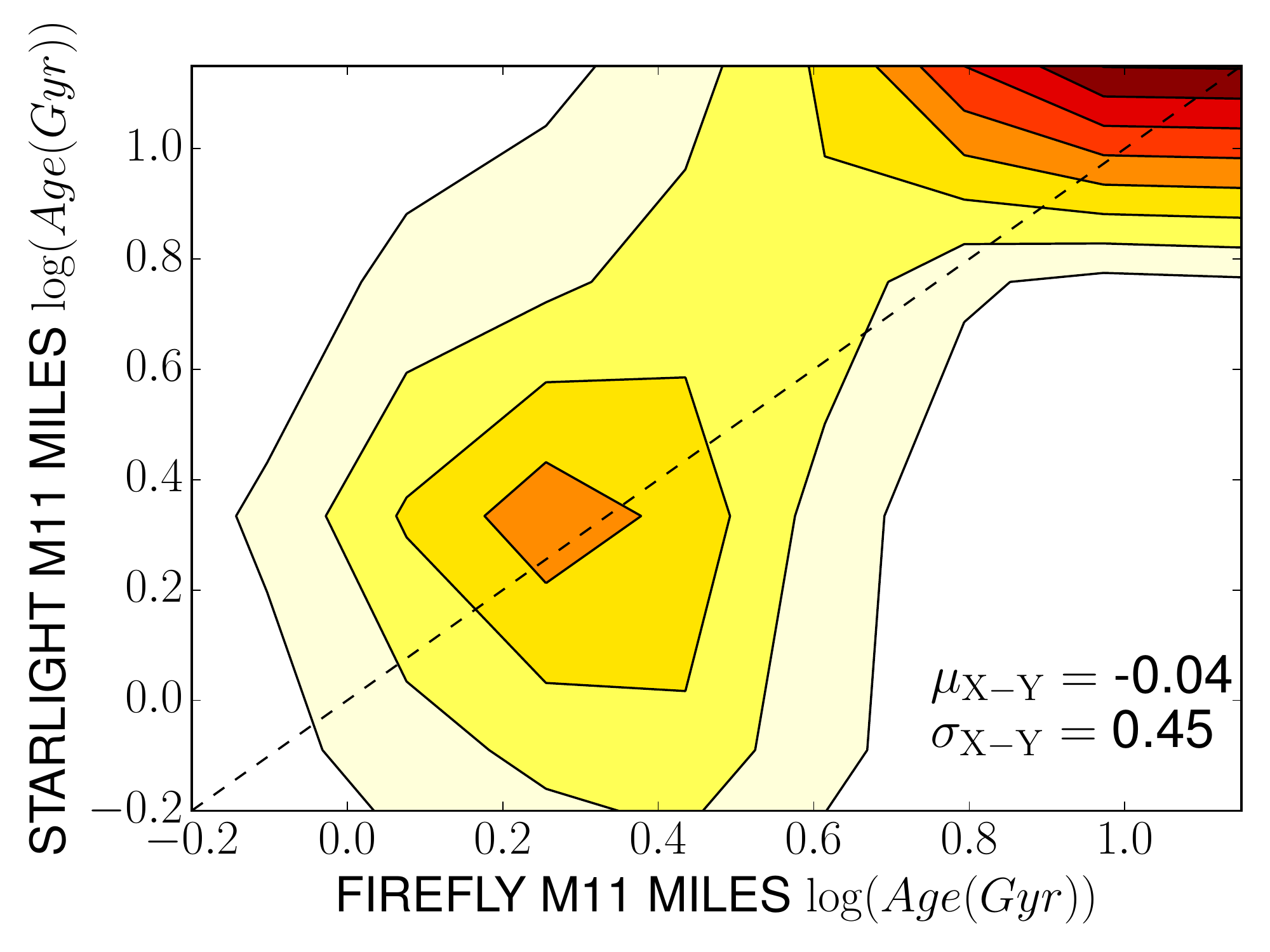}
\includegraphics[width=0.33\textwidth]{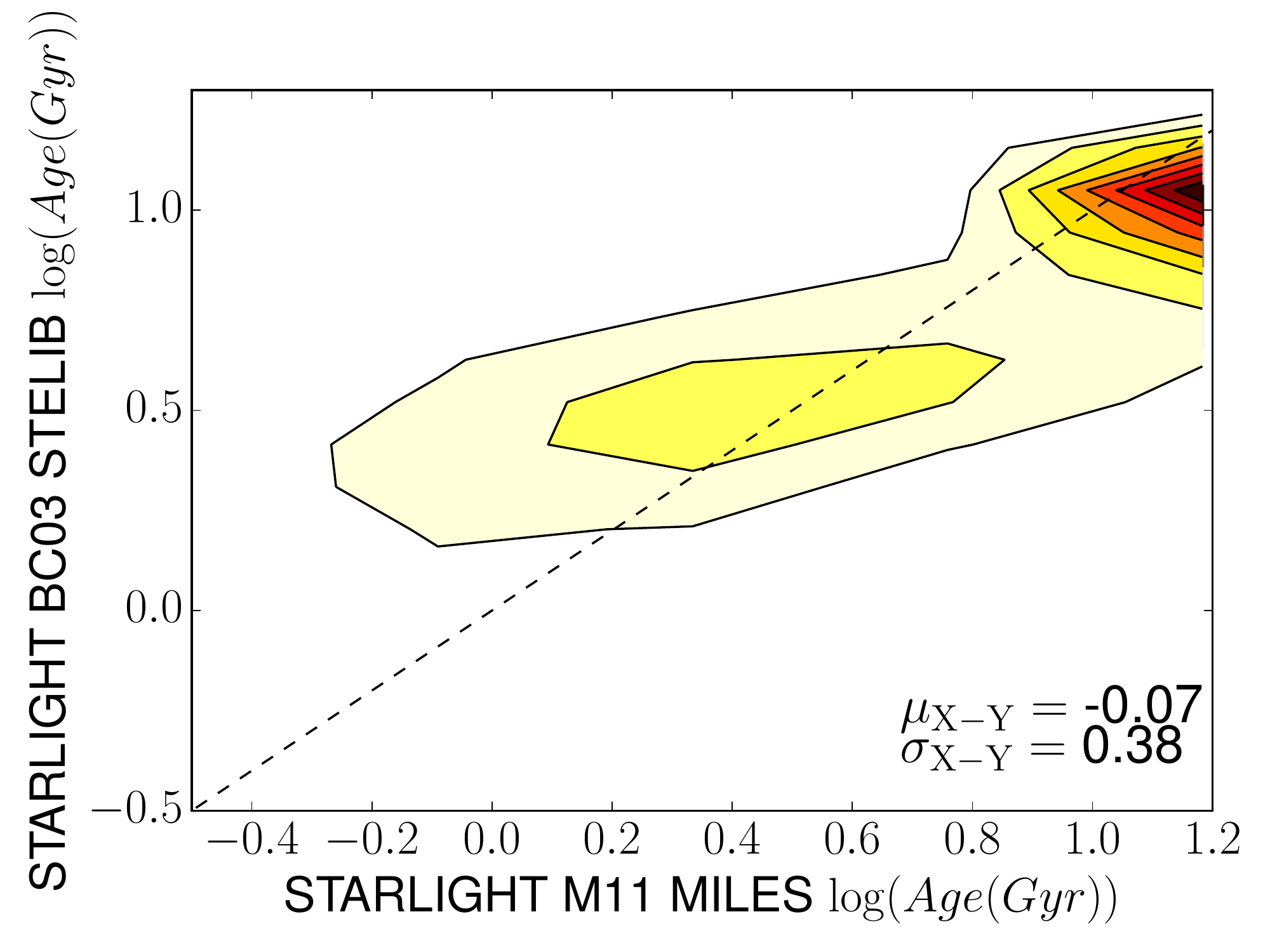}
\includegraphics[width=0.33\textwidth]{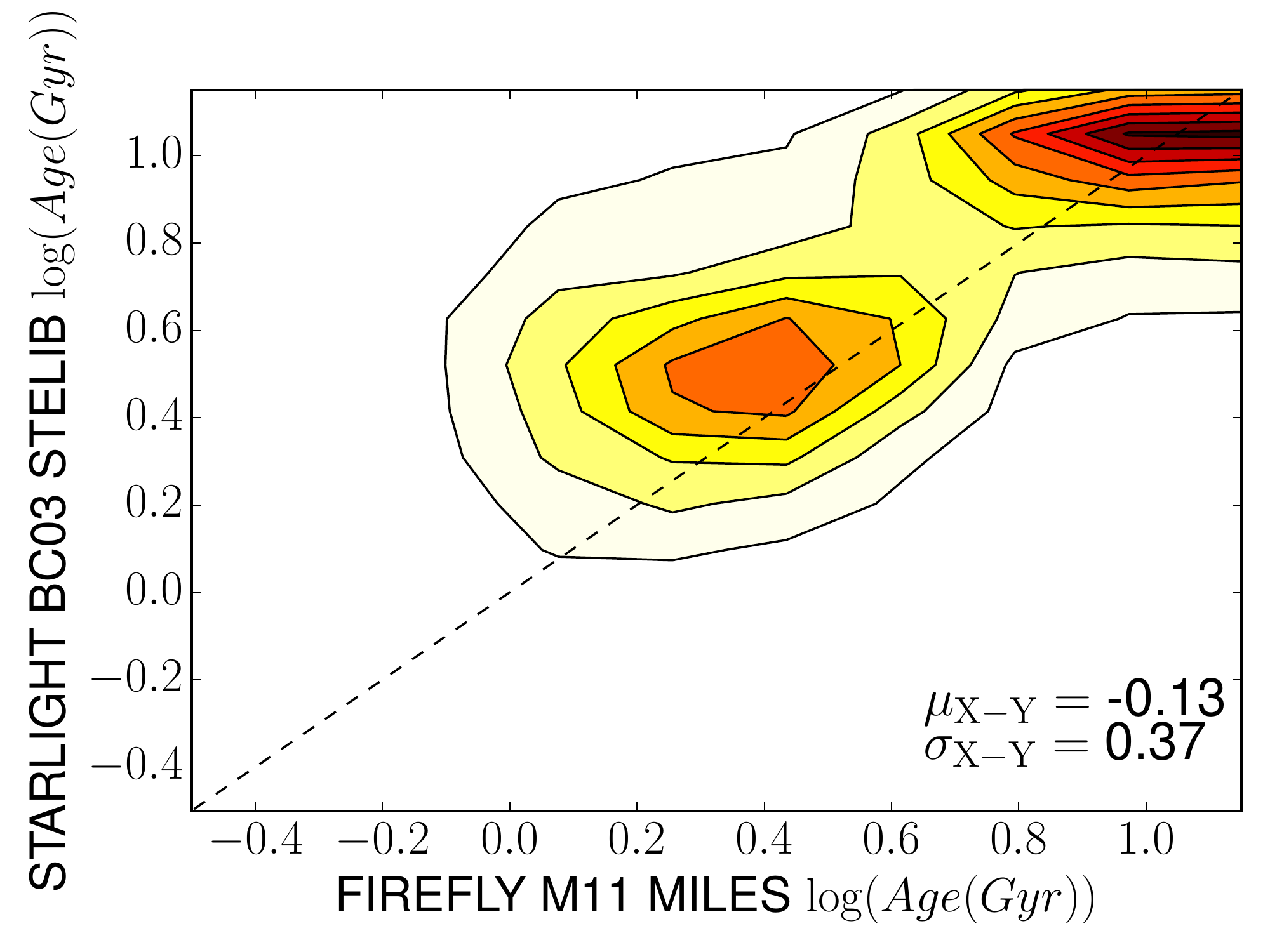}
\includegraphics[width=0.33\textwidth]{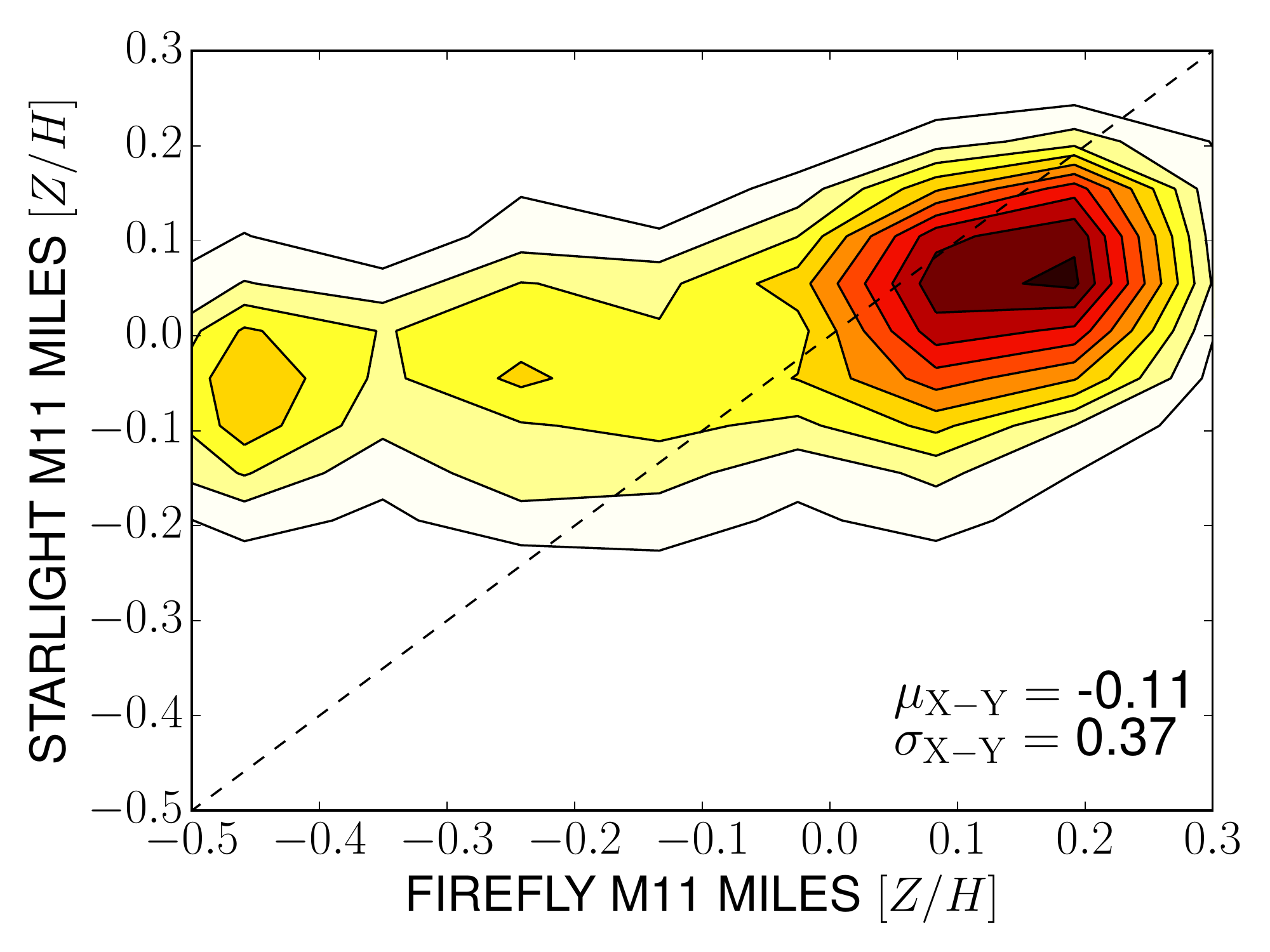}
\includegraphics[width=0.33\textwidth]{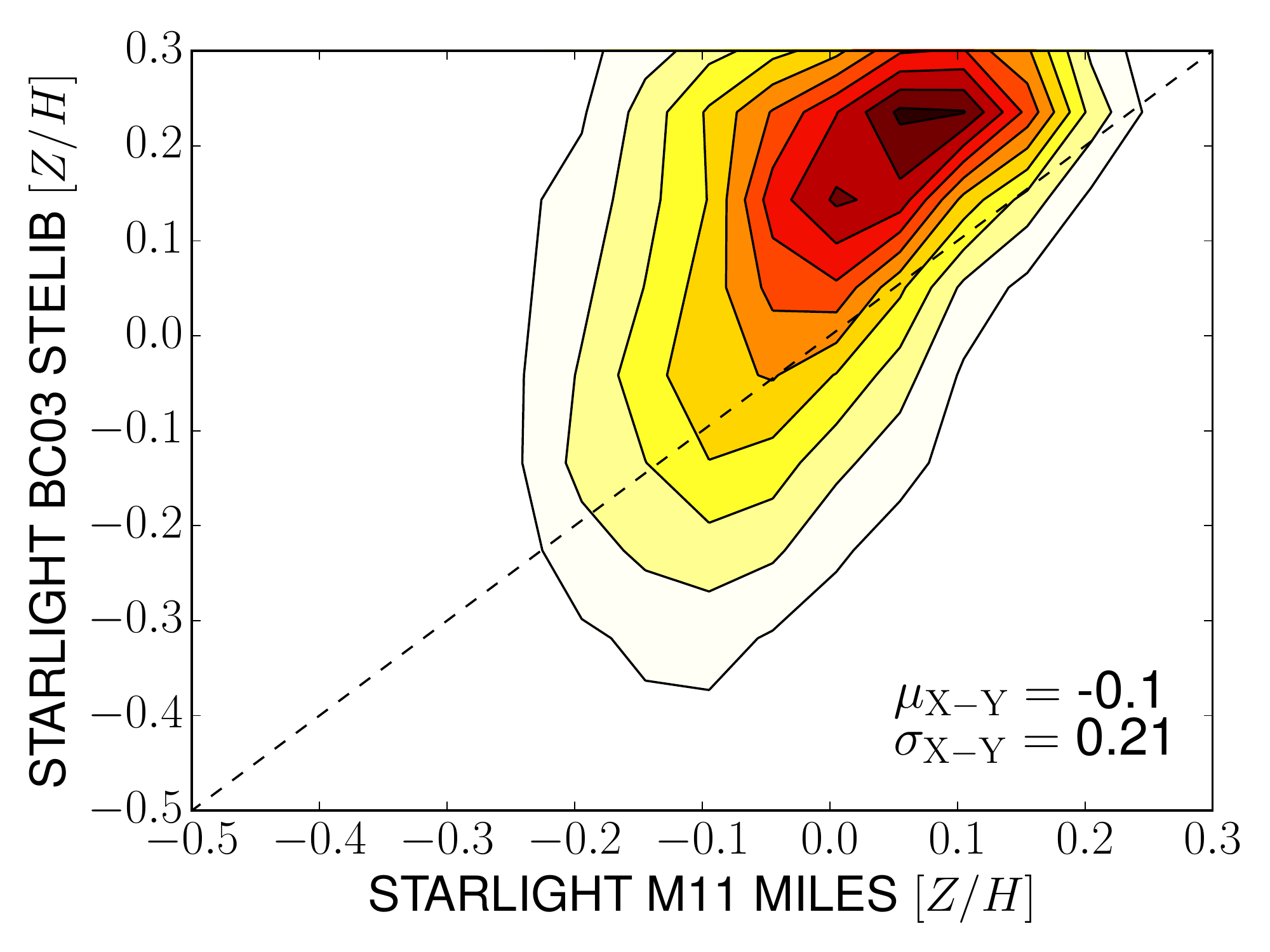}
\includegraphics[width=0.33\textwidth]{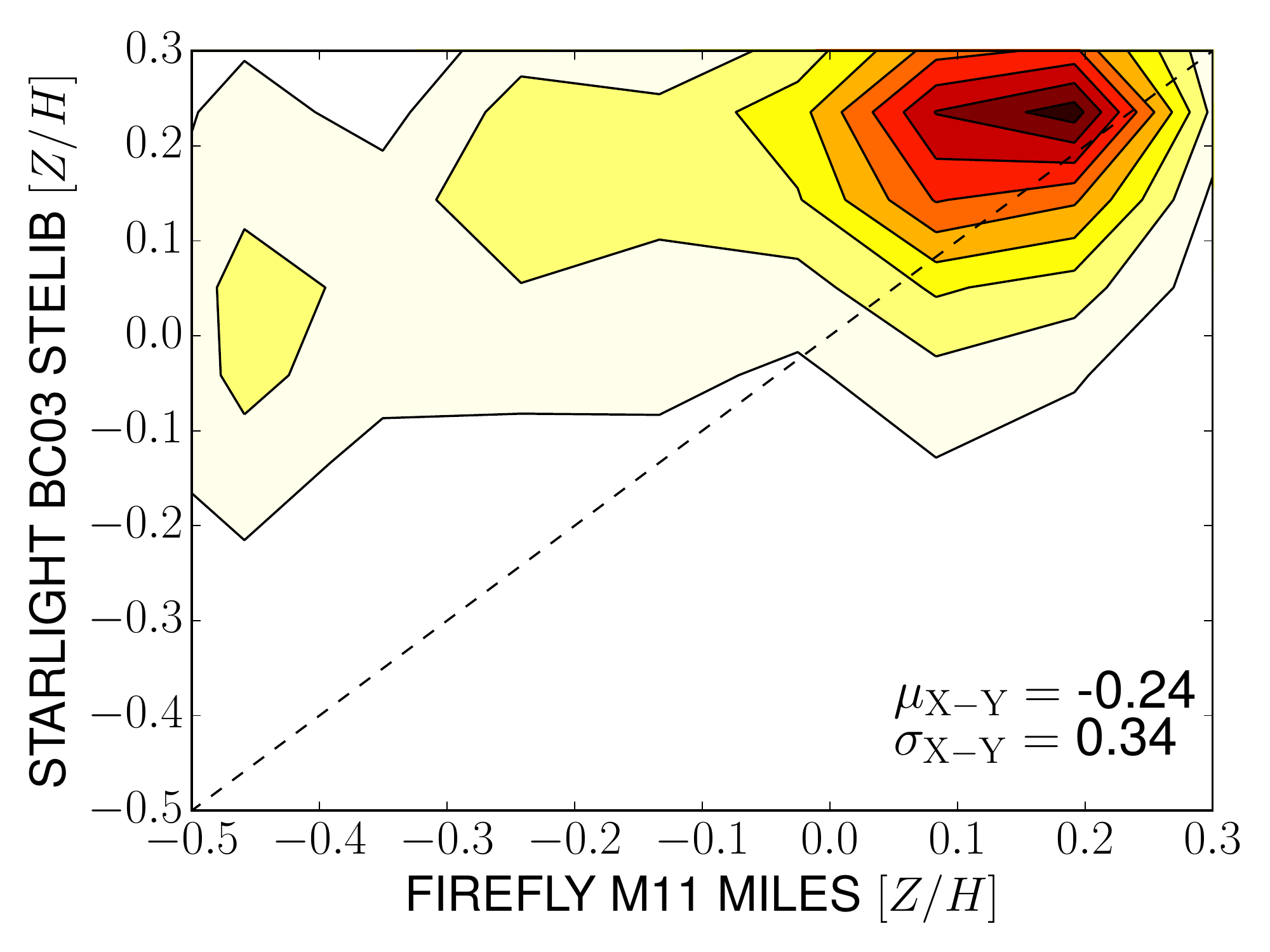}
\caption{Comparison of the light-weighted stellar population parameters age (top row) and metallicity (bottom row), obtained by STARLIGHT and FIREFLY with M11 and BC03 stellar population models for a subset of 30 galaxies from the MaNGA survey. Contours represent the number density of Voronoi cells. The dashed line in each panel is the one-one relation. The values $\mu_{\mathrm{X}-\mathrm{Y}}$ and $\sigma_{\mathrm{X}-\mathrm{Y}}$ represent the median difference and dispersion between the $\mathrm{X}$ and $\mathrm{Y}$ axis, respectively.}
\label{fig:compare_SL_FF_properties}
\end{figure*}
\subsection{Comparison Between Fitting Codes and Stellar Population Models}
\label{sec:comparison}
To test the dependence of our stellar population measurements both on the spectral fitting technique and the underlying stellar population model, we compare to the results obtained with the spectral fitting code STARLIGHT \citep{starlight2005} using both \citet[M11]{maraston2011} and \citet[BC03]{bruzual2003} models for a subset of 30 galaxies. STARLIGHT with BC03 STELIB models is used in \citet{zheng2016}. To ensure our comparison of the codes is a fair one, we adopt the same base stellar population models of M11 with Kroupa IMF that were used in the present FIREFLY fits and we fit over the same wavelength range of 3500-7428$\AA$. A fundamental difference is the treatment of dust: while STARLIGHT assumes a dust reddening law, FIREFLY is parameter free, because it does not fit the continuum shape to constrain the stellar population properties. It should be emphasised, however, that this comparison is very complicated and a more detailed analysis is beyond the scope of this work.\\
\\
The results of this comparison are presented in Figure~\ref{fig:compare_SL_FF_properties}. In general there is a large scatter and some systematic offsets. On the positive side, young and intermediate-age populations are consistently identified by both codes but then there are considerable difference in the exact ages derived. FIREFLY yields younger light-averaged ages for old populations than STARLIGHT (top left-hand panel) by about $\sim 0.1$ dex, but are in reasonable agreement for younger populations leading to an median offset of $\mu_{\mathrm{X}-\mathrm{Y}} = -0.04$ dex and scatter of $\sigma_{\mathrm{X}-\mathrm{Y}} = 0.45$ dex. The choice of stellar population model turns out to produce an effect of comparable size. The BC03 STELIB models yield somewhat younger ages by $\sim 0.08\;$dex for the old population and older ages by $\sim 0.1-0.2$ dex for the younger population (top middle panel). This is a consequence of the different stellar tracks used in the two models: the onset of the red giant evolutionary phase is earlier in M11 models, based on \cite{cassisi1997} tracks, whereas BC03 is based on Padova isochrones \citep{girardi2000} (see \citet{2005MNRAS.362..799M} for a detailed discussion). This conspires to a smaller offset between the ages of old populations derived with STARLIGHT+BC03 in \citet{zheng2016} and in the present work with FIREFLY+M11 (top right-hand panel). Again, there is a large scatter. Young and intermediate-age populations tend to have slightly younger ages when derived with the M11 models, which leads to a systematic offset of $\sim 0.07\;$dex in the comparison between this work and \citet{zheng2016}. Overall, we can conclude that luminosity-weighted ages are affected by systematic offsets between the various codes and underlying stellar population models of the order of $\mu_{\mathrm{X}-\mathrm{Y}} = -0.13$ dex, however the scatter is large, with $\sigma_{\mathrm{X}-\mathrm{Y}} = 0.37$ dex.
\\
\\
The comparison for luminosity-weighted metallicity provides a more complex picture as shown by the bottom panels in Figure~\ref{fig:compare_SL_FF_properties}. STARLIGHT appears to produce systematically lower metallicities for the bulk metal-rich population by $\sim 0.1\;$dex and higher metallicities for the metal-poor population by $\sim 0.3\;$dex (bottom left-hand panel), leading to a median offset $\mu_{\mathrm{X}-\mathrm{Y}} = -0.11$ dex. The bottom middle panel in Figure~\ref{fig:compare_SL_FF_properties} shows an offset of $\mu_{\mathrm{X}-\mathrm{Y}} = -0.1$ dex between the STARLIGHT+M11 and STARLIGHT+BC03 models, with M11 producing lower metallicities. The blend of both of these results leads to an overall difference of $\mu_{\mathrm{X}-\mathrm{Y}} = -0.24$ dex between STARLIGHT+BC03 and FIREFLY+M11, with lower metallicities being obtained by the latter (bottom right panel). The most striking difference between the two fitting codes lies in the distribution of metallicities (bottom left panel). STARLIGHT appears to yield a narrower range of metallicities ($-0.2$ to $0.1$ dex) than FIREFLY ($-0.5$ to $0.3$ dex) and as a consequence, discrepancies can be as large as $0.4$ dex, with FIREFLY obtaining substantially lower metallicities for some spectra than STARLIGHT. This issue clearly requires more investigation and more detailed comparisons with also other spectral fitting codes. However, this work goes beyond the scope of this paper.
\begin{figure*}
\includegraphics[width=0.31\textwidth]{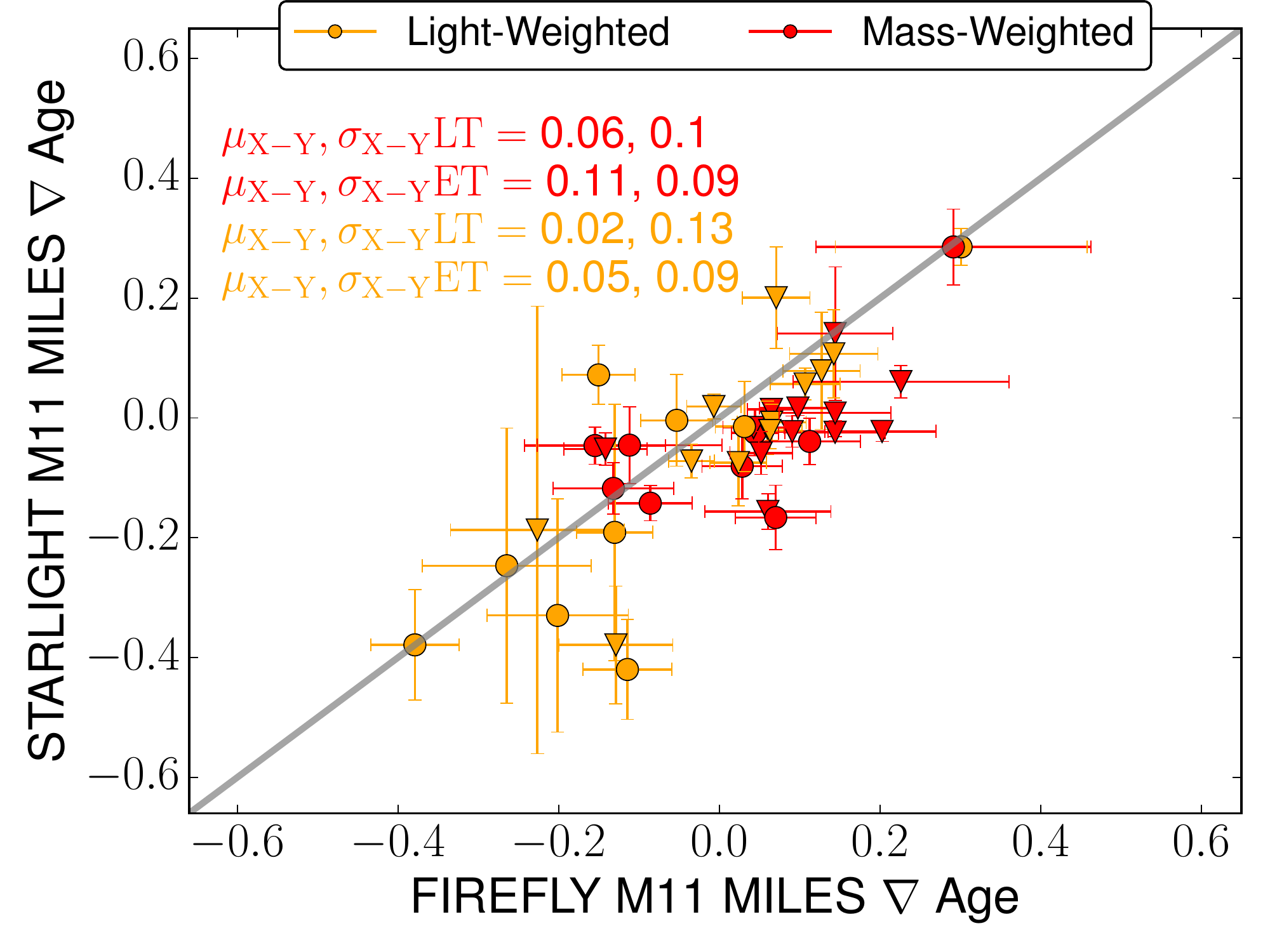}
\includegraphics[width=0.31\textwidth]{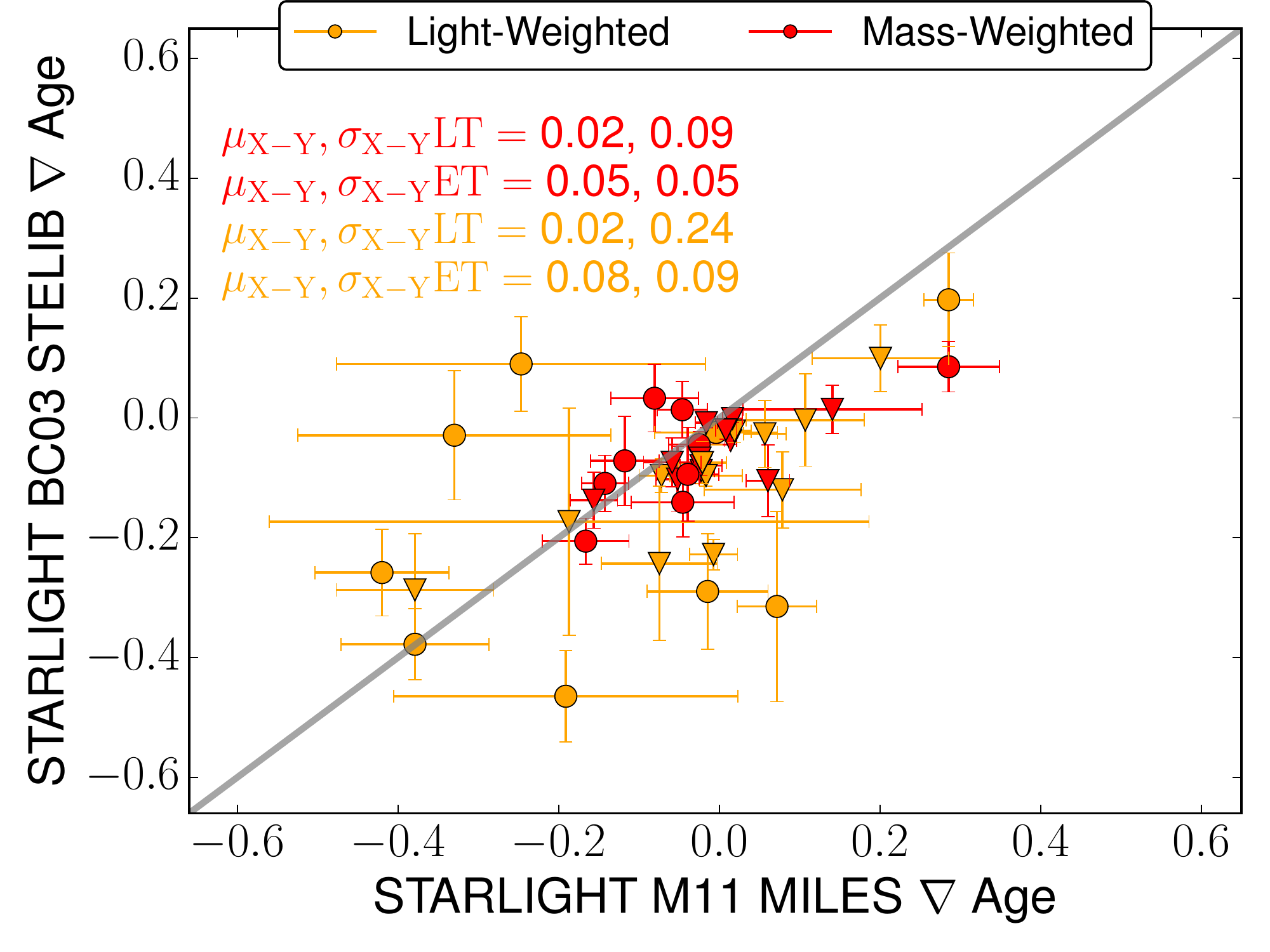}
\includegraphics[width=0.31\textwidth]{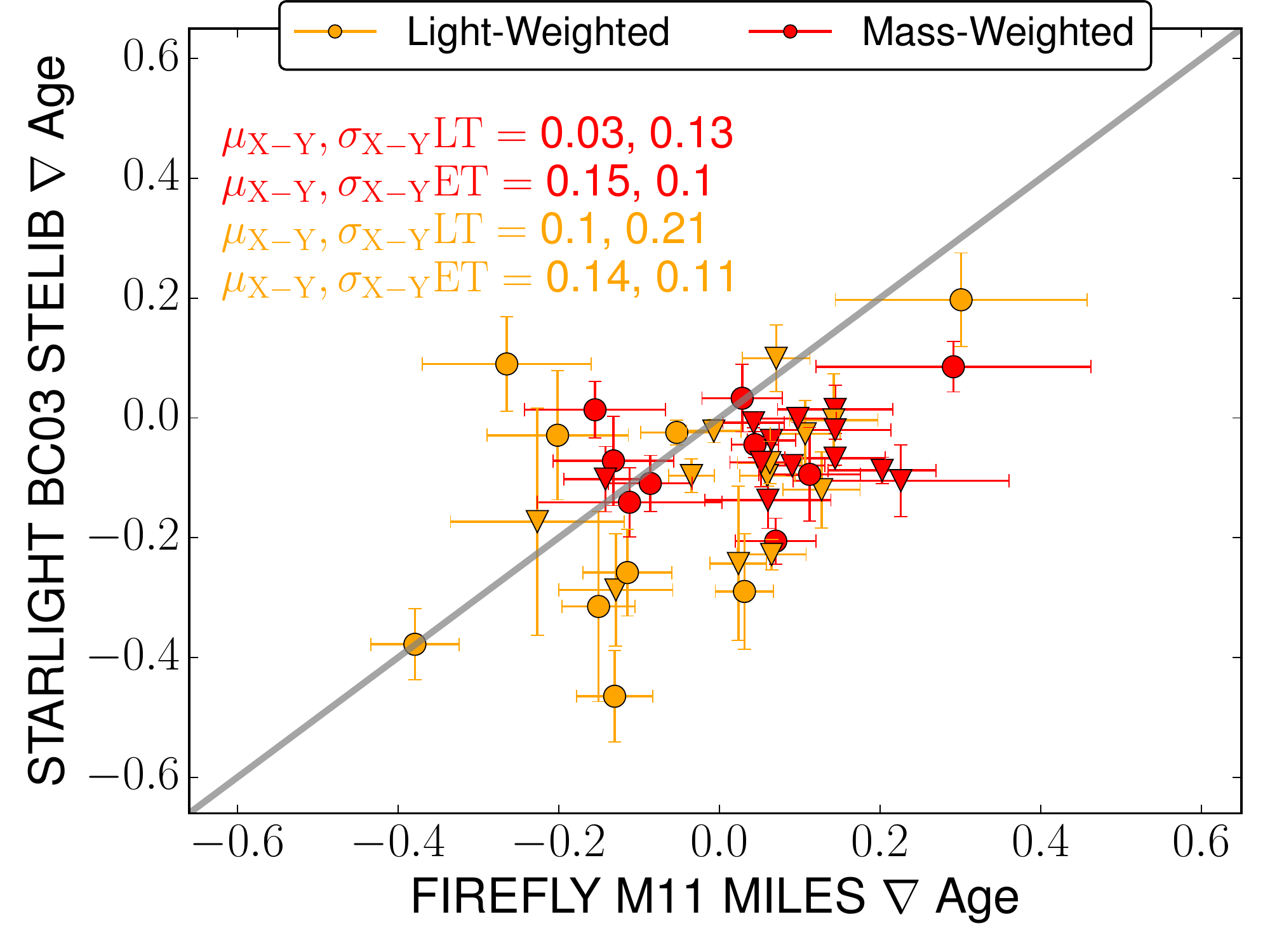}
\includegraphics[width=0.31\textwidth]{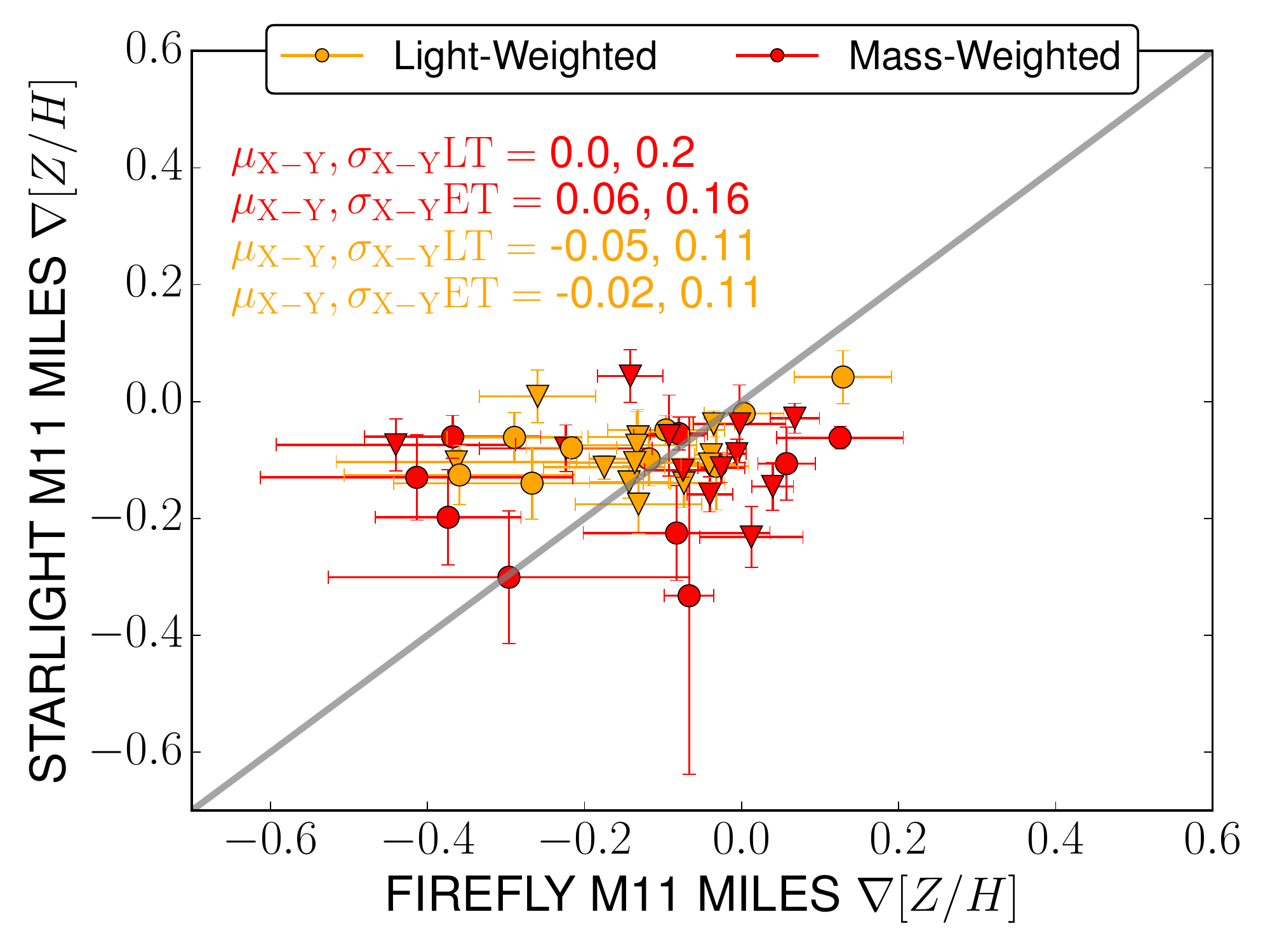}
\includegraphics[width=0.31\textwidth]{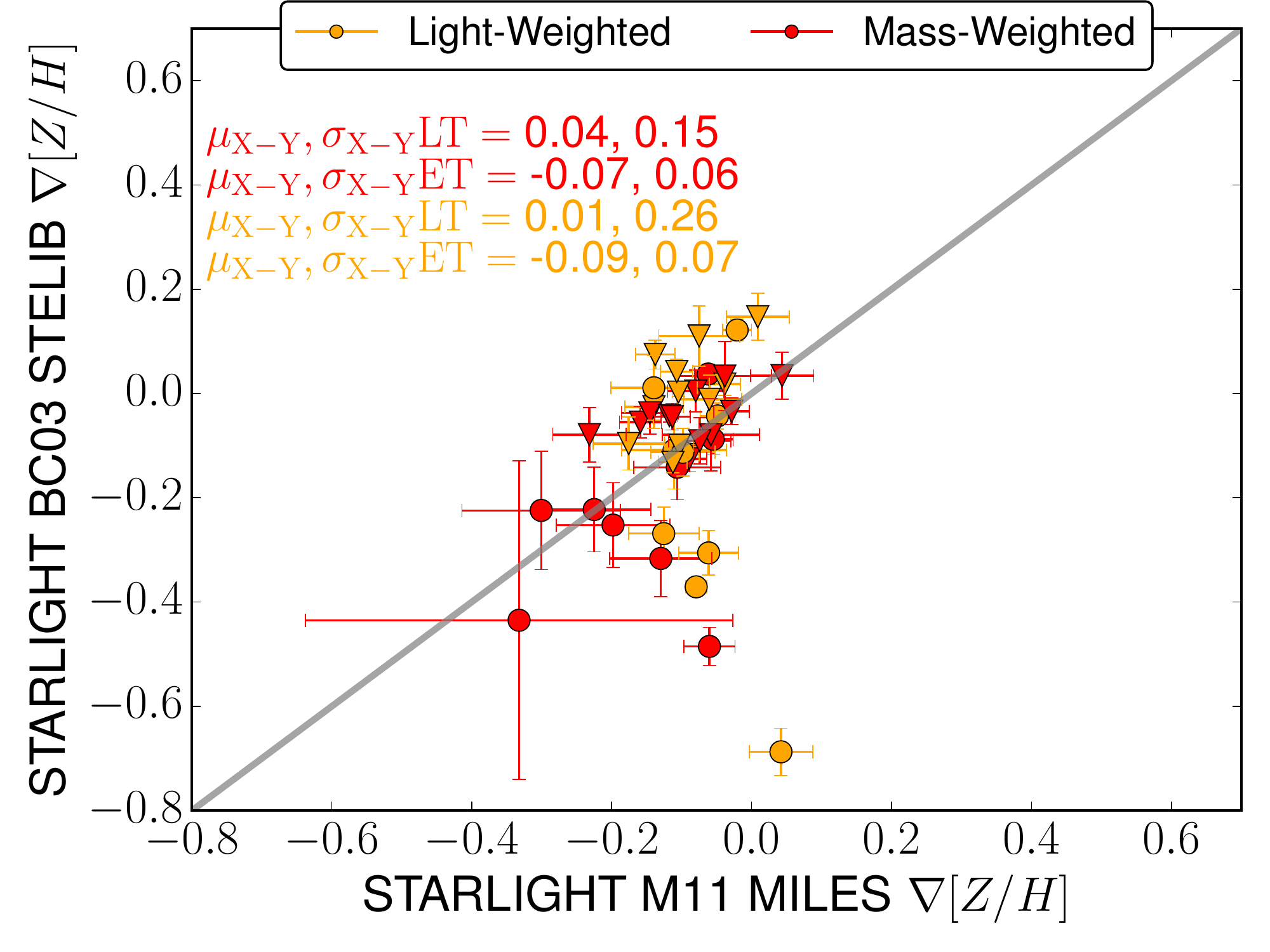}
\includegraphics[width=0.31\textwidth]{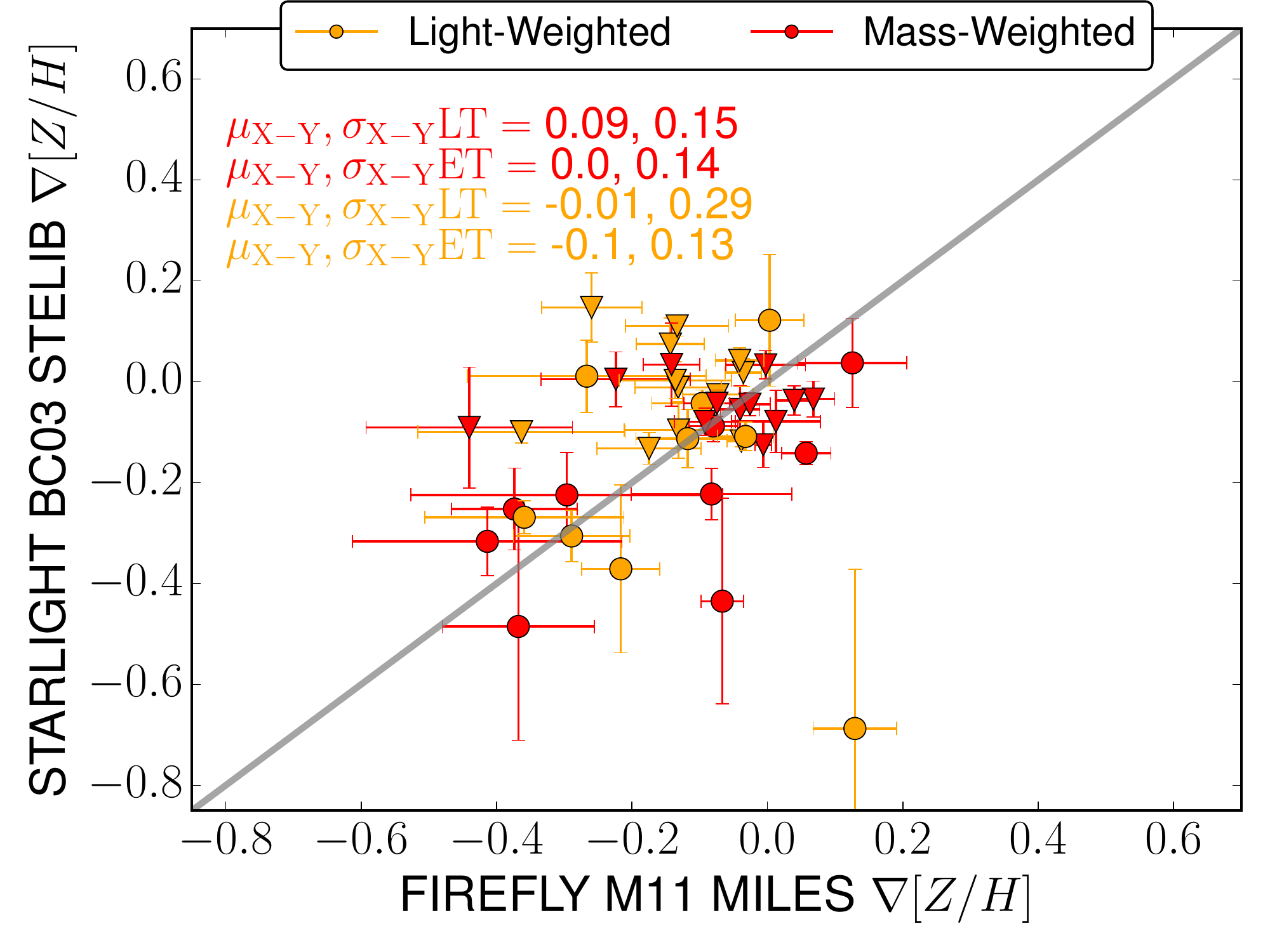}
\caption{Comparison of the stellar population gradients obtained using the output results from the two different spectral fitting codes, FIREFLY and STARLIGHT with two different stellar population models M11 and BC03. Additionally, the same method of obtaining the radial gradient was used and errors were deduced from bootstrap resampling. The grey line shows a 1-1 correspondence between the gradients. The orange and red colours represent light and mass-weighted properties, whereas the circles and triangles represent late-type and early-type galaxies, respectively. Lastly, the values $\mu_{\mathrm{X}-\mathrm{Y}}$ and $\sigma_{\mathrm{X}-\mathrm{Y}}$ represent the median difference and dispersion between the $\mathrm{X}$ and $\mathrm{Y}$ axis as in Figure~\ref{fig:compare_SL_FF_properties}.}
\label{fig:comp_gradients}
\end{figure*}
\\
\\ 
In Figure~\ref{fig:comp_gradients} we compare the final stellar population gradients derived from these three different combinations of fitting code and stellar population model. Here we show both light-averaged (orange symbols) and mass-averaged (red symbols) quantities, as well as separating between early (triangle) and late-type (circle) galaxies. In general, as can be expected from the results in Figure~\ref{fig:compare_SL_FF_properties}, there is quite a large scatter and there are some systematic offsets in the derived gradients. However, it is not necessarily straightforward to relate the discrepancies identified in Figure~\ref{fig:compare_SL_FF_properties} to differences in gradients. The top left panel of Figure~\ref{fig:comp_gradients} shows the comparison between gradients derived using FIREFLY+M11 and STARLIGHT+M11. It can be seen that light-weighted age gradients for STARLIGHT are slightly more negative ($\mu_{\mathrm{X}-\mathrm{Y}} =0.05$ dex/$R_{\rm e}$ for late-types and $\mu_{\mathrm{X}-\mathrm{Y}} =0.02$ dex/$R_{\rm e}$ for early-types) than the ones obtained using FIREFLY, which can be attributed to the lower ages of old (and generally central) populations (top left-hand panel in Figure~\ref{fig:compare_SL_FF_properties}). The mass-weighted age gradients derived from FIREFLY are also more positive by $\mu_{\mathrm{X}-\mathrm{Y}} =0.11$ dex/$R_{\rm e}$ and $\mu_{\mathrm{X}-\mathrm{Y}} =0.06$ dex/$R_{\rm e}$ for early and late-types, respectively. The systematic uncertainty introduced by the fitting method at fixed stellar population model is of the same order of the gradient signal itself, a caveat that needs to be considered when investigating stellar population gradients. \\
\\
The situation becomes more complex when looking at the top right panel in Figure~\ref{fig:comp_gradients}, where the age gradients obtained from FIREFLY+M11 and STARLIGHT+BC03 are compared. The light and mass-weighted age gradients of late-type galaxies obtained by both fitting codes are consistent within the scatter. However, for early-type galaxies there is a marked difference between the two results. Both the light and mass-weighted age gradients produced by FIREFLY+M11 are more positive than STARLIGHT+BC03  ($\mu_{\mathrm{X}-\mathrm{Y}} =0.14$ dex/$R_{\rm e}$ and $\mu_{\mathrm{X}-\mathrm{Y}} =0.15$ dex/$R_{\rm e}$). This leads to the conclusion that depending on the combination of fitting code and underlying stellar population model, the result can range from positive to slightly negative age gradients. The origin of this difference is less clear, but might simply be caused by the fact that ages are generally younger in the FIREFLY+M11 (this work) than the STARLIGHT+BC03 \citep{zheng2016} setup (top right-hand panel in Figure~\ref{fig:compare_SL_FF_properties}). The top right panel showing the age gradients obtained with STARLIGHT+M11 and STARLIGHT+BC03 also shows a large degree of scatter, highlighting further the complexity involved in comparisons such as this one. This significant systematic uncertainty needs to be acknowledged, and future work is needed to assess the origin of these differences and explore possible paths to mitigate them.\\
\\
The bottom panels of Figure~\ref{fig:comp_gradients} show the derived light and mass-weighted metallicity gradients. In general, the metallicity gradient measurements vary quite a bit, but are consistent within the scatter for both early and late-types. \\
\\
We conclude that age and metallicity measurements from full spectral fitting are considerably affected by systematic differences in the full spectral fitting technique and underlying stellar population models. We find offsets of $0.1-0.4\;$dex with a large scatter. This result is not surprising, and similarly large differences between fitting codes and stellar population models have already been highlighted in \citet{wilkinson2015}. However, we also show here that consequently measurements of quantities such as age gradients, are affected by systematic discrepancies of the order of $0.03-0.15\;$dex/$R_{\rm e}$. This implies that the resulting uncertainties are of similar size as the signal being measured, hence a more rigorous and detailed investigation is needed to understand how to address these issues in future studies.

\begin{figure}
\includegraphics[width=0.48\textwidth]{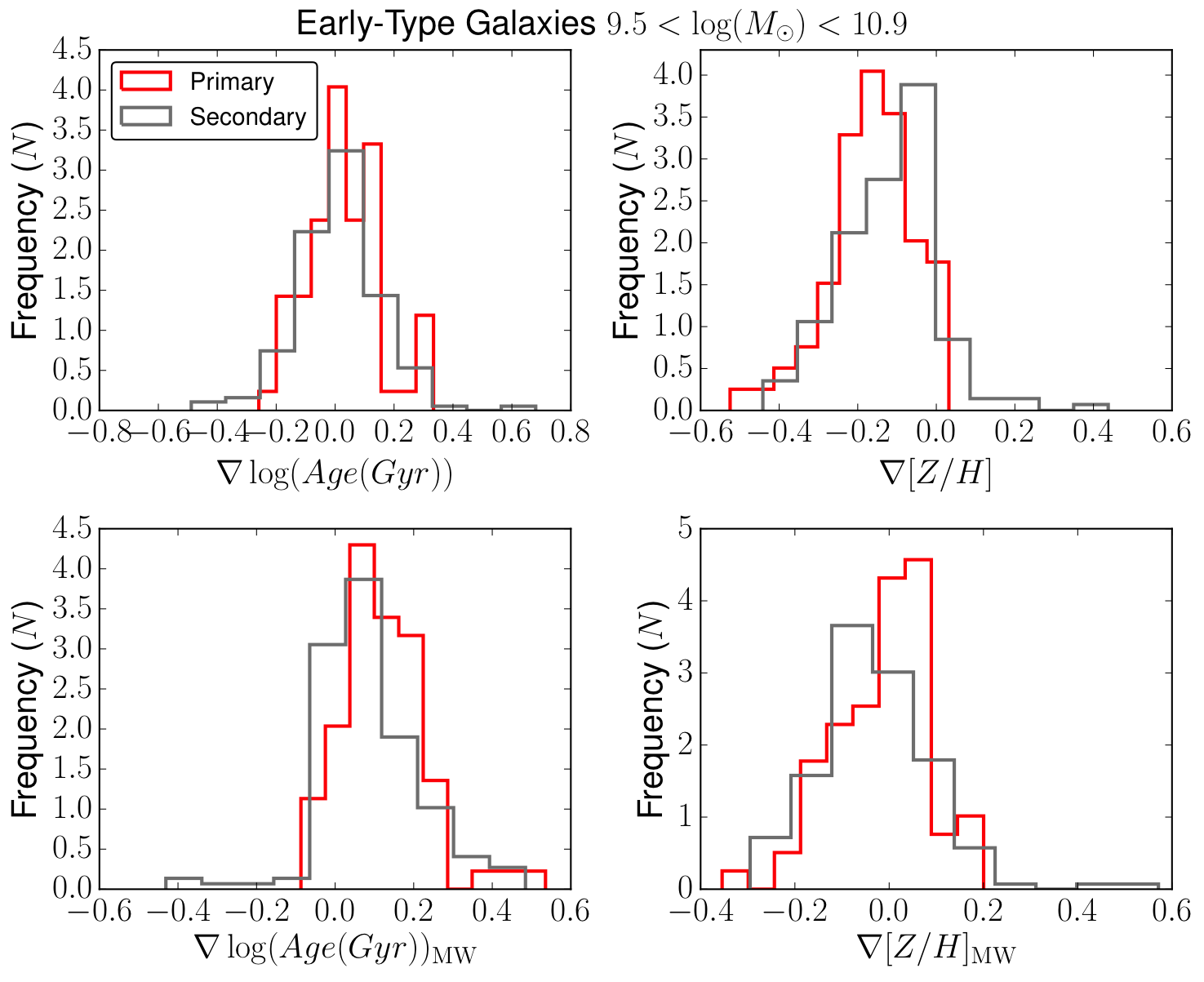}
\includegraphics[width=0.48\textwidth]{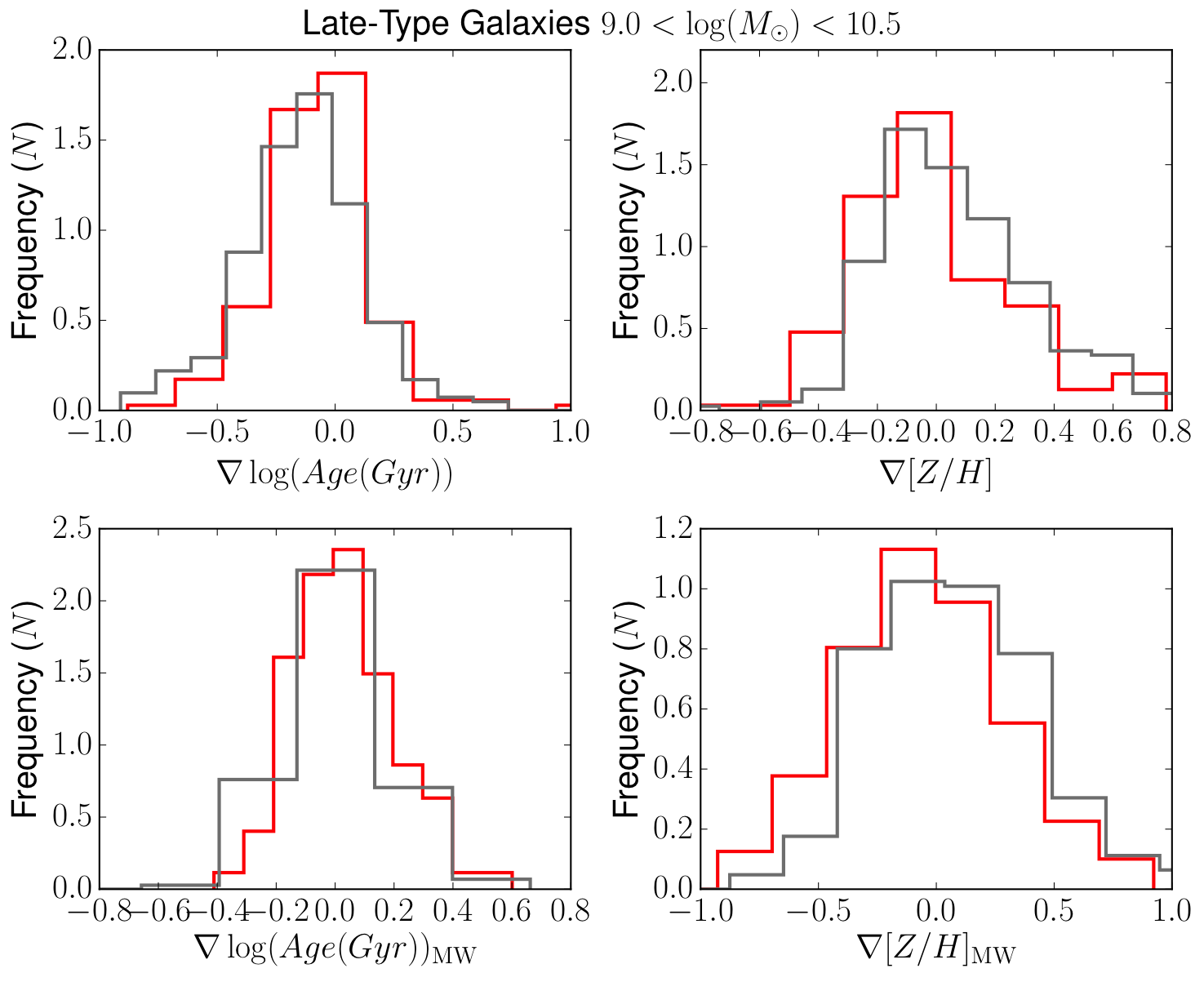}
\caption{Figure investigating the impact of beam smearing on radial gradients. The  panels show the luminosity-weighted age (top left), luminosity-weighted metallicity (top right), mass-weighted age (bottom left) and mass-weighted metallicity gradients (bottom right) for early-types (top plot) and late-types (bottom plot), obtained using the Primary and restricted Secondary MaNGA sample.}
\label{fig:ifu_smearing_1}
\end{figure}

\subsection{Beam Smearing Effects on Radial Gradients}
\begin{figure}
\includegraphics[width=0.48\textwidth]{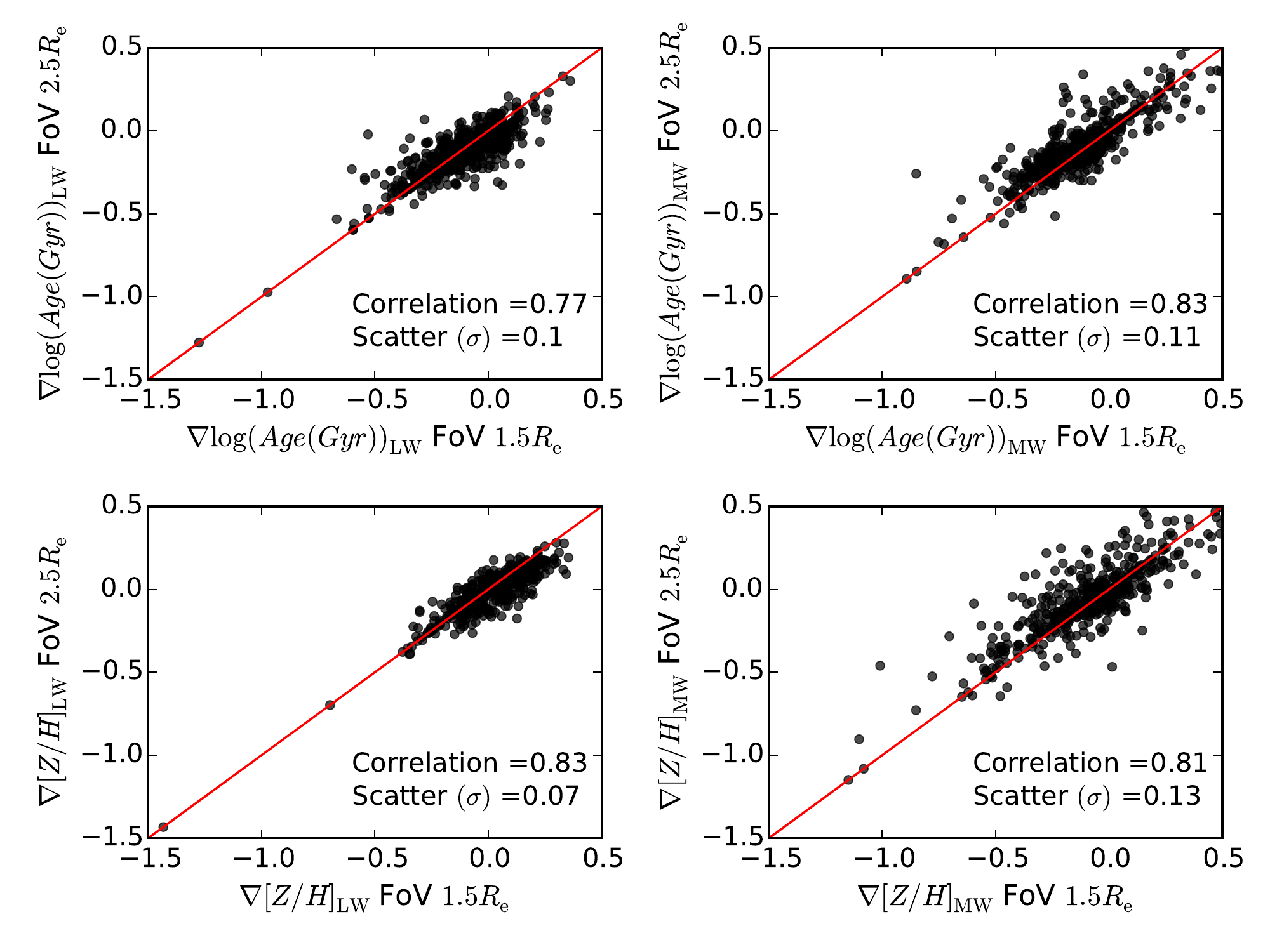}
\caption{Figure showing the impact of the dynamic range on the measured stellar population gradients of luminosity-weighted age (top left), mass-weighted age (top right) , luminosity-weighted metallicity (bottom left) and mass-weighted metallicity gradients (bottom right) for galaxies in the MaNGA secondary sample. Gradients were calculated within $1.5\;R_{\rm e}$ and $2.5\;R_{\rm e}$ for both early and late-type galaxies.}
\label{fig:ifu_smearing_2}
\end{figure}

\begin{table*}
\caption{Median light-weighted and mass-weighted gradients (in dex/$R_{e}$) for both early-type and late-type galaxies for a fixed stellar mass range. The galaxies are split between the Primary MaNGA sample used in this work and the restricted MaNGA Secondary sample to assess the impact of beam smearing investigated in Figure~\ref{fig:ifu_smearing_1}. Errors correspond to the standard deviation of the distribution.}
\centering
\begin{tabular}{c c c c c }
\hline\hline
Galaxy Type & Property & Mass Bin & Primary (dex/$R_{e}$) & Restricted Secondary (dex/$R_{e}$) \\ [0.5ex]
\hline
Early-Type & Light-Weighted Age & $9.5 < \log(M/M_{\odot}) < $ 10.9 & $0.01 \pm$ 0.13 & $0.02 \pm$ 0.14\\
 &  Light-Weighted $[Z/H]$ & &  $-0.14 \pm$ 0.11 & $-0.12 \pm$ 0.12 \\
 &  Mass-Weighted Age &   &  $0.10 \pm$ 0.10 & $0.07 \pm$ 0.12 \\
 &  Mass-Weighted $[Z/H]$ &   & $-0.01 \pm$ 0.11 & $-0.04 \pm$ 0.13 \\
 
Late-Type & Light-Weighted Age & $9.0 < \log(M/M_{\odot}) < $ 10.5 & $-0.07 \pm$ 0.20 & $-0.12 \pm$ 0.28\\
 &  Light-Weighted $[Z/H]$ & &  $-0.03 \pm$ 0.29 & $0.03 \pm$ 0.27\\
   &  Mass-Weighted Age &   &  $0.02 \pm$ 0.17 & $-0.01 \pm$ 0.30 \\
 &  Mass-Weighted $[Z/H]$ &   & $-0.06 \pm$ 0.30 & $0.06 \pm$ 0.36 \\[2ex]
\hline
\end{tabular}
\label{table:ifu_table}
\end{table*}
\begin{table}
\caption{Table explaining the approximate number of beams per IFU for both Primary and Secondary MaNGA galaxy samples within $1.5\;R_{\rm e}$. This highlights the fact that Secondary sample has $\sim$1/3 less spatial resolution than the Primary sample at this given radius.}
\centering
\begin{tabular}{c c c c }
\hline\hline
Bundle & Diameter (arcsec) & Primary  & Secondary  \\  [0.5ex]
 & & ($\mathrm{N}_{\mathrm{Beams}}$)  & ($\mathrm{N}_{\mathrm{Beams}}$) \\ [0.5ex]
\hline
19-fibre &  12$\arcsec$ & 4.8 & 2.9 \\
37-fibre &  17$\arcsec$ & 6.8 & 4.1 \\
61-fibre &  22$\arcsec$ & 8.8 & 5.3 \\
91-fibre &  27$\arcsec$ & 10.8 & 6.5 \\
127-fibre &  32$\arcsec$ & 12.8 & 7.7 \\[2ex]
\hline
\end{tabular}
\label{table:ifu_smearing}
\end{table}
Beam smearing is a cause for concern for most IFU surveys as it can have the effect of diluting the measured radial gradients. By comparing stellar population gradients for the MaNGA Primary and Secondary galaxy samples\footnote{It is important to note that the Secondary sample used in this comparison is not part of the 721 galaxies used in this work.}, we can assess the possible impact of beam smearing on the gradients we present in this paper. \\
\\
The Field of View (FoV) of the IFUs, in effective radius, are $1.5\;R_{\rm e}$ for the Primary and $2.5\;R_{\rm e}$ for the Secondary sample, respectively. If we only consider the Secondary samples radial extent out to $1.5\;R_{\rm e}$ we can calculate the number of beams ($N_{\mathrm{Beams}}$) per FoV in the following manner:
\begin{displaymath}
   N_{\mathrm{Beams}}= \left\{
     \begin{array}{lr}
       d_{\mathrm{IFU}}/d_{\mathrm{B}} & : \text{Primary} \\
       (d_{\mathrm{IFU}}\cdot(1.5/2.5))/d_{\mathrm{B}} & : \text{Restricted Secondary} 
     \end{array}
   \right.
\end{displaymath} 
where $d_{\mathrm{IFU}}$ is the diameter of the IFU in arcsecs and $d_{\mathrm{B}}$ is the typical beam size in MaNGA, which is $2.5\;$arcsec. The values obtained for the five different IFU bundles are presented in Table~\ref{table:ifu_smearing}. From this Table it is clear to see that the Secondary sample (when sampled out to $1.5\;R_{\rm e}$ only) has $\sim$1/3 less spatial resolution than the Primary sample in each IFU. If beam smearing was having a significant effect, this change in spatial resolution would lead to flatter radial gradients in the restricted Secondary sample compared to the Primary sample. 
\\
\\
Figure~\ref{fig:ifu_smearing_1} shows the light and mass-weighted stellar population gradients for early-type (top four panels) and late-type galaxies (bottom four panels) obtained for a subset of galaxies in a fixed mass range (early-types: $9.5 < \log(M_{\odot}) <  10.9$, $-18.5 < M_{i} <  -21.6$ and late-types: $9.0 < \log(M_{\odot}) <  10.5$, $-17.4 < M_{i} <  -21.0$). The median and standard deviations of the distributions are also summarised in Table~\ref{table:ifu_table}.
From Table~\ref{table:ifu_table}, we can see that the median gradients obtained from the distributions in Figure~\ref{fig:ifu_smearing_1} are consistent for both early and late-type galaxies. This tells us that the restricted Secondary sample, which has lower spatial resolution, has comparable gradients to what we present in this work and provides us with confidence that the effect of beam smearing must be quite small in our work and does not affect the gradients in any significant way. \\
\\
We can however, conduct one further exercise to consolidate our conclusion regarding beam smearing. Again using the Secondary sample, we can investigate the effect that the radial dynamic range used in measuring stellar population gradients, has on the derived gradient value. This is done by measuring the radial gradients on Secondary sample galaxies out to two different radii ($1.5\;R_{\rm e}$ and $2.5\;R_{\rm e}$)\footnote{Note, this exercise can not be done with the Primary galaxy sample as their radial extent is only observed out to $\sim 1.5\;R_{\rm e}$.} and comparing the results. The results of this analysis can be seen in Figure~\ref{fig:ifu_smearing_2}. Overall, the gradients are fairly consistent when measured using these two different dynamic ranges, with high correlations (ranging between $0.77$ to $0.83$) and relatively low scatter (ranging between $0.07$ and $0.13$) between the derived gradients. This once again emphasises that the effect of beam smearing must be extremely small in our work and does not affect the gradients presented in this paper in any significant way. \\
\\
Our results obtained in this Section are slightly different compared to a preliminary study based on P-MaNGA data by \citet{wilkinson2015}, who found mild evidence of beam smearing on the gradients. However, the sample size for their study was very small (17 galaxies) and the MaNGA data reduction has significantly improved since the P-MaNGA data. Although our results suggest no significant beam smearing, in order to truly quantify the effect of beam smearing simulations would be necessary, however this is beyond the scope of this paper.

\section{Results}
\begin{table*}
\caption{Light and mass-weighted median stellar population properties at the centre of the galaxy and at $1\;R_{\rm e}$, split by galaxy mass. These numbers are representative of the plots shown in Figure~\ref{fig:gradients}.}
\centering
\begin{tabular}{c c c c c c}
\hline\hline
Property & Mass Bin & Early-types & Early-types & Late-types & Late-types \\
& & Value at Centre & Value at $1\;R_{\rm e}$ & Value at Centre & Value at $1\;R_{\rm e}$ \\ 
& & (dex) & (dex) & (dex) & (dex) \\ [0.5ex]
\hline
$\log(Age(Gyr))_{\mathrm{LW}}$ & $\log(M/M_{\odot})< 9.935$ & $0.56$ & $0.60$ & $0.44$ & $0.26$ \\
$[Z/H]_{\mathrm{LW}}$ & & $-0.24$ & $-0.32$ & $-0.73$ & $-0.65$ \\
$\log(Age(Gyr))_{\mathrm{MW}}$  & & $0.84$ & $0.90$ & $0.77$ & $0.70$ \\
$[Z/H]_{\mathrm{MW}}$  &  & $-0.28$& $-0.32$ & $-1.07$ & $-1.12$ \\
$\log(Age(Gyr))_{\mathrm{LW}}$ & $9.935 <\log(M/M_{\odot})< 10.552$  & $0.60$ & $0.53$ & $0.45$ & $0.28$ \\
$[Z/H]_{\mathrm{LW}}$ & & $-0.10$ & $-0.15$ & $-0.49$ & $-0.58$ \\
$\log(Age(Gyr))_{\mathrm{MW}}$  & & $0.80$ & $0.87$ & $0.73$ & $0.68$\\
$[Z/H]_{\mathrm{MW}}$ &  & $-0.18$& $-0.16$ & $-0.77$ & $-1.01$\\
$\log(Age(Gyr))_{\mathrm{LW}}$ & $10.552 <\log(M/M_{\odot})< 11.054$ & $0.71$ & $0.64$ & $0.43$ & $0.29$ \\
$[Z/H]_{\mathrm{LW}}$ & & $0.08$ & $-0.06$& $-0.33$ & $-0.43$ \\
$\log(Age(Gyr))_{\mathrm{MW}}$ & & $0.82$ & $0.91$ & $0.68$ & $0.70$\\
$[Z/H]_{\mathrm{MW}}$ &  & $0.06$& $-0.04$ & $-0.48$ & $-0.80$\\
$\log(Age(Gyr))_{\mathrm{LW}}$ &  $\log(M/M_{\odot})> 11.054$ & $0.81$ & $0.75$ & $0.65$ & $0.34$ \\
$[Z/H]_{\mathrm{LW}}$& & $0.15$& $-0.01$& $-0.01$ & $-0.33$ \\
$\log(Age(Gyr))_{\mathrm{MW}}$ & & $0.91$ & $0.99$ & $0.75$ & $0.81$\\
$[Z/H]_{\mathrm{MW}}$&  & $0.16$& $0.03$ & $-0.06$ & $-0.30$\\
\hline
\end{tabular}
\label{table:one_re_values}
\end{table*} 

We now turn to presenting the resulting stellar population gradients for the full sample of 505 early-type and 216 late-type galaxies and their dependence on galaxy mass and morphological type. Firstly, we present dependencies of the stellar population properties, age and metallicity, as a function of radius in the four quartiles of the mass distribution corresponding to the following four mass bins: $\log(M/M_{\odot})<9.935$, $9.935<\log(M/M_{\odot})<10.552$, $10.552<\log(M/M_{\odot})<11.054$, and $\log(M/M_{\odot})>11.054$. Information regarding the breakdown of galaxies into these mass bins can be found in Table~\ref{table:sample_classification}. We then calculate the median and its distribution of age and metallicity in radial bins (see orange line in Figure~\ref{fig:example_gradients}) for these four mass bins, considering both light and mass-weighted stellar population properties, as well as $E(B-V)$. In a final step, we present the full star formation histories in the age-metallicity plane as a function of galaxy mass and radius.

\subsection{Median Stellar Population Properties}
\begin{figure*}
 \centering
 \textbf{\Large{Early-Types}}\par\medskip
\includegraphics[width=1.0\textwidth]{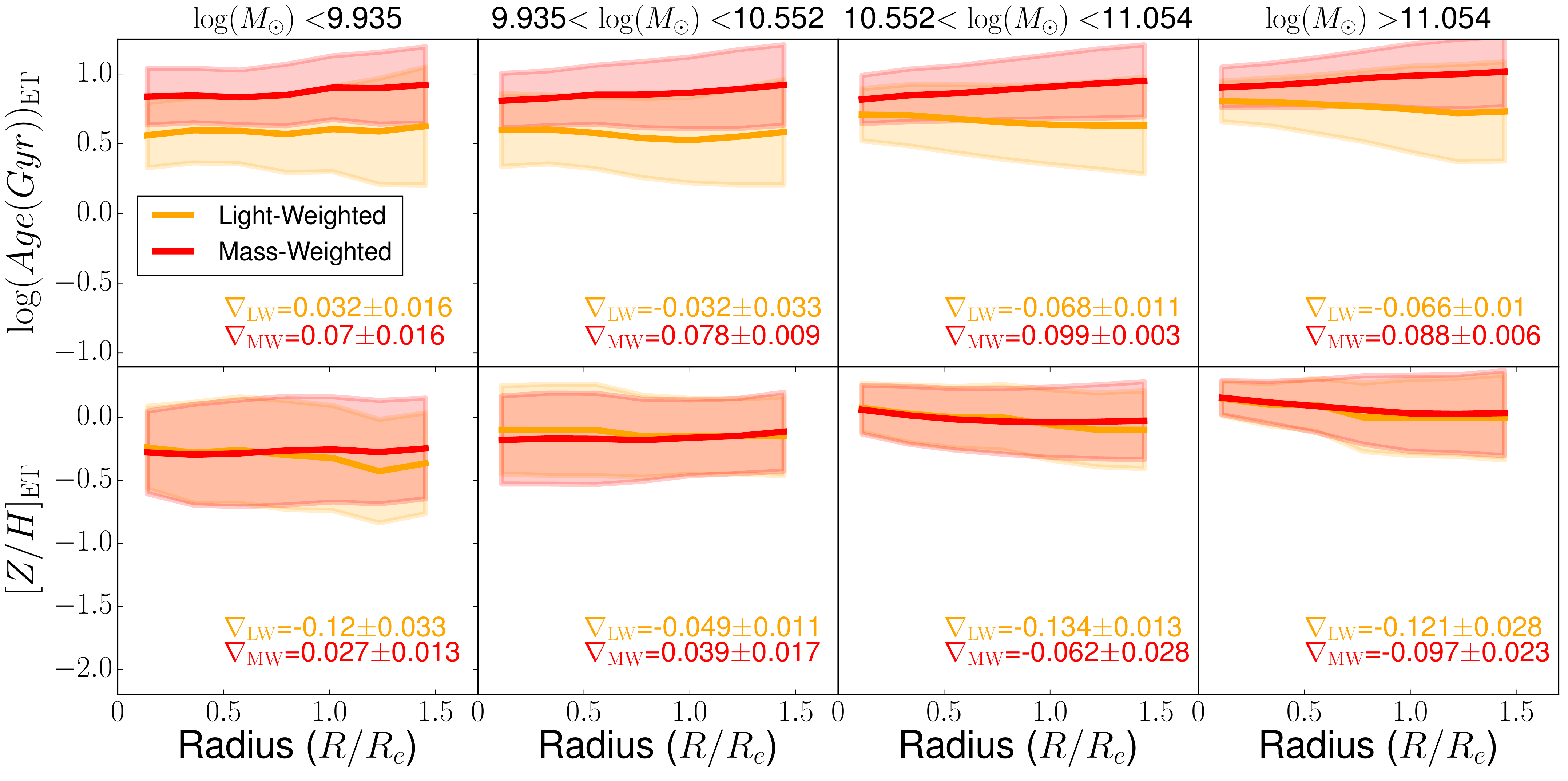}
 \centering
 \textbf{\Large{Late-Types}}\par\medskip
\includegraphics[width=1.0\textwidth]{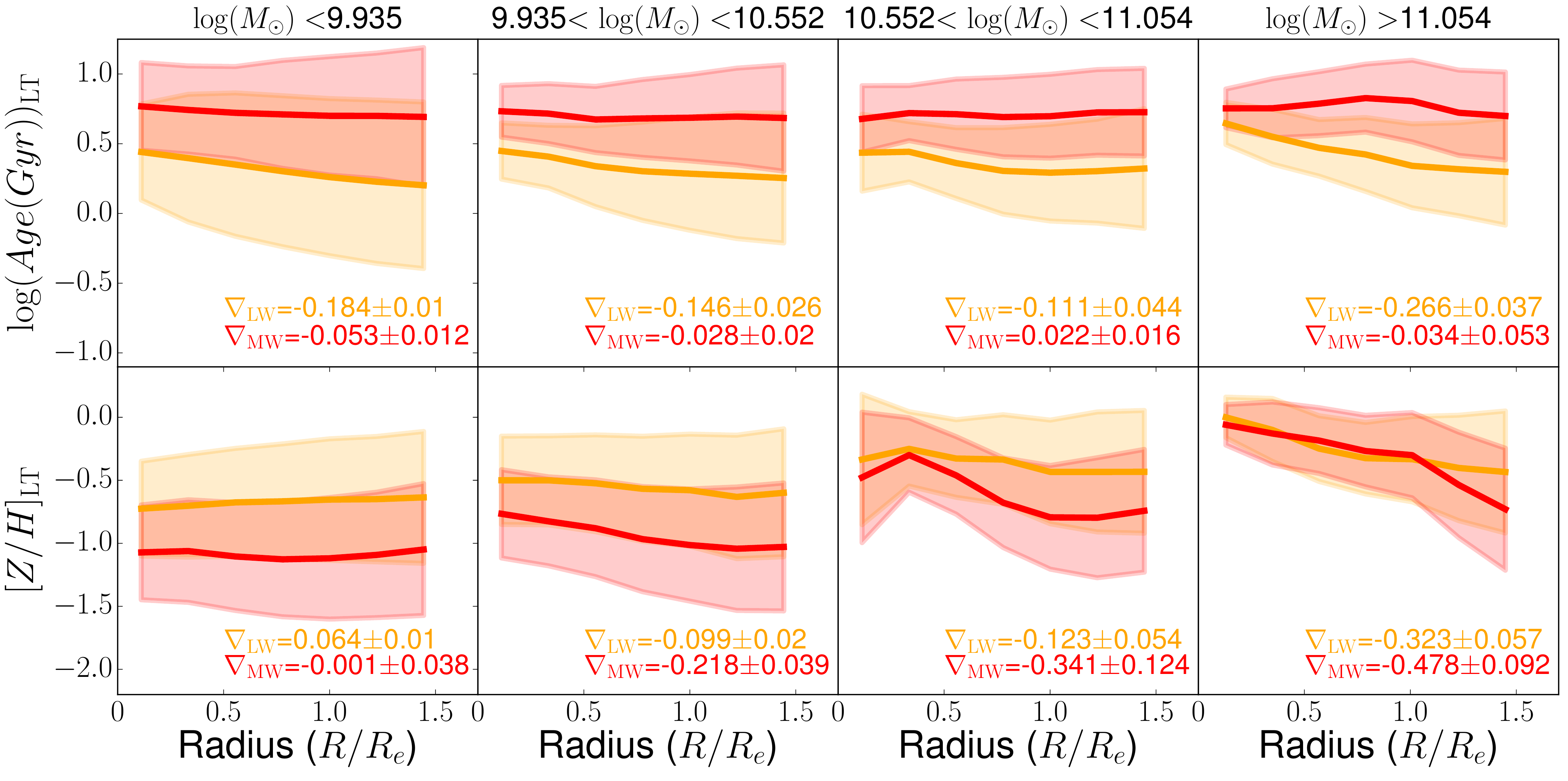}
\caption{Median stellar population parameters age (top rows) and metallicity (bottom rows) as a function of radius in four different bins of galaxy mass (see labels) for the full sample containing 505 early-type and 216 late-type galaxies. The top panels are early-type galaxies and the bottom panels are late-type galaxies, respectively. Light-weighted and mass-weighted quantities are shown as orange and red lines, respectively. The shaded area is the $1$-$\sigma$ width of the distribution around the median value. The resulting fitted gradients and their bootstrapped errors are quoted in each panel.}
\label{fig:gradients}
\end{figure*}

Figure~\ref{fig:gradients} shows the median stellar population parameters, age and metallicity, as a function of radius in the four different bins of galaxy mass (see labels) and Table~\ref{table:one_re_values} shows the derived stellar population properties at $1\;R_{\rm e}$ for both early and late-type galaxies used in this work.

\begin{figure*}
\includegraphics[width=0.49\textwidth]{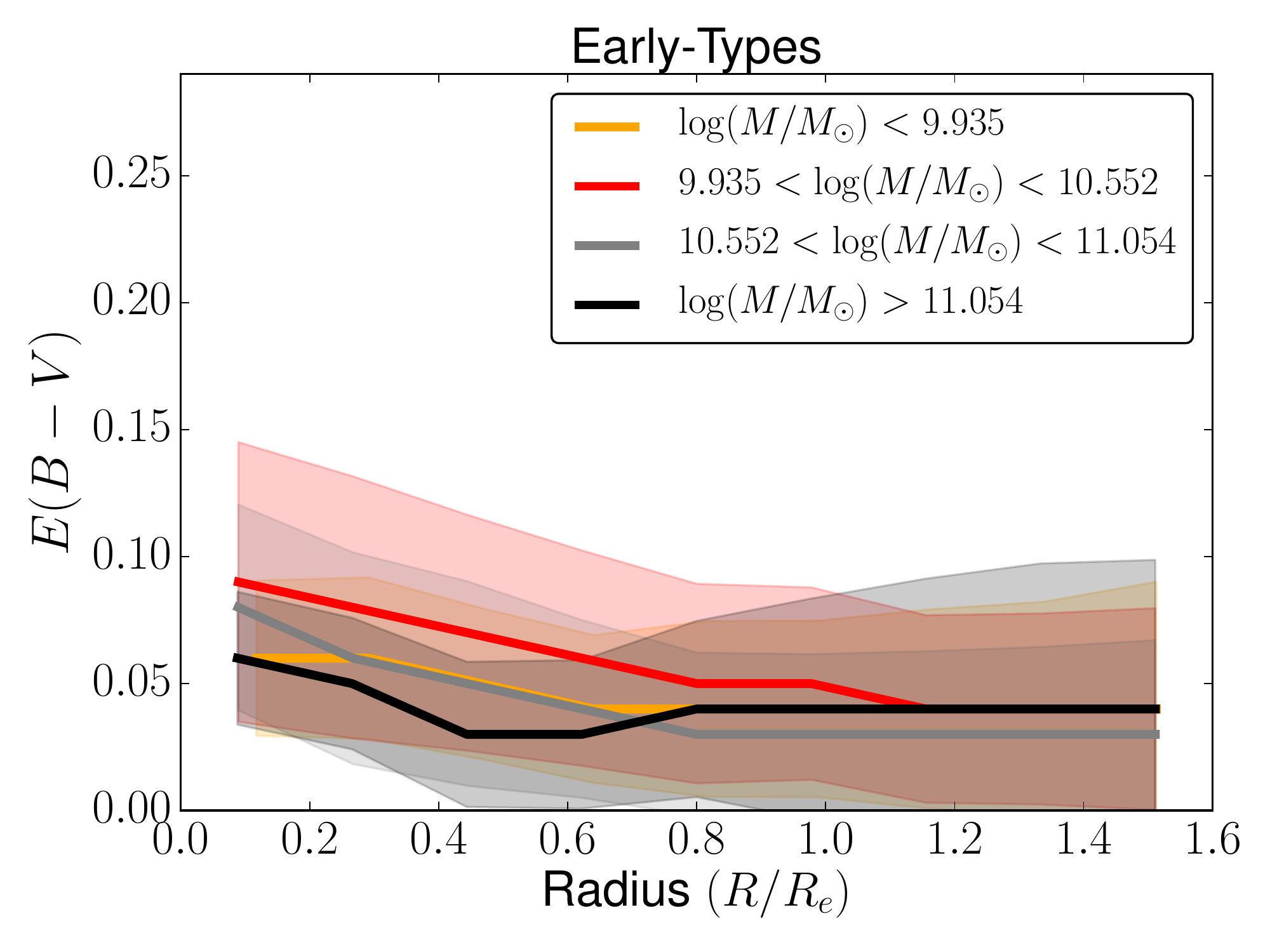}
\includegraphics[width=0.49\textwidth]{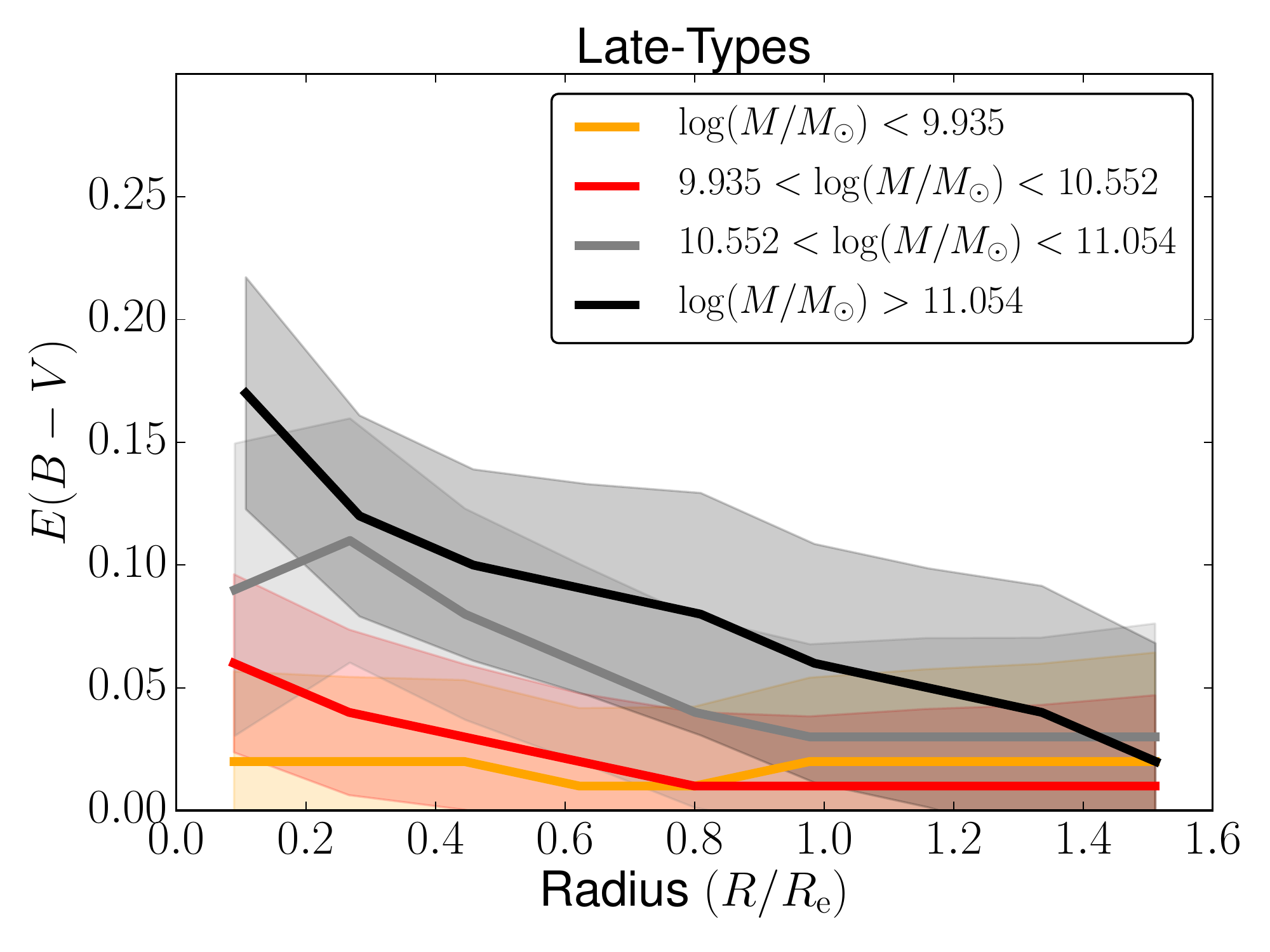}
\caption{$E(B-V)$ values derived from FIREFLY as a function of radius in four different bins of galaxy mass (see legend) for the sample of 505 early-type galaxies (left panel) and 216 late-type galaxies (right panel). The thick line represents that median and the the shaded region is the $1$-$\sigma$ width of the distribution around the median value.}
\label{fig:dust_radius}
\end{figure*}

\begin{table*}
\caption{Table showing the breakdown of number and stellar mass ($\log(M/M_{\odot})$) of galaxies in each of the four quartiles of the mass distribution used in this work. $\sigma$ shows the standard deviation of the galaxy masses relating to galaxies inside each mass bin.}
\centering
\begin{tabular}{c c c c c c}
\hline\hline
Mass Bin & Number (N) & Number (N) & Number (N) & Mean Stellar Mass & Standard Deviation \\
 & & Early-types & Late-types & $\log(M/M_{\odot})$ & ($\sigma$) \\ [0.5ex]
\hline
$\log(M/M_{\odot})< 9.935$ & $160$ & $54$ & $106$  & $9.58$ & $\pm 0.26$ \\ 
$9.935 <\log(M/M_{\odot})< 10.552$& $167$ & $96$ & $71$ & $10.30$ &  $\pm 0.19$ \\ 
$10.552 <\log(M/M_{\odot})< 11.054$ & $192$ & $162$ & $30$  & $10.78$ &  $\pm 0.15$ \\ 
$\log(M/M_{\odot})> 11.054$ & $202$ & $197$ & $5$ & $11.41$ & $\pm 0.25$ \\ 
\hline
\end{tabular}
\label{table:sample_classification}
\end{table*}

\subsubsection{Age Gradients}
Early-type galaxies generally exhibit relatively shallow age gradients. The light-weighted age gradients (orange lines) are either not significant within 2-$\sigma$ (lower mass bins), or very small (higher mass bins). This is consistent with previous results in the literature \citep[][see also Introduction]{saglia2000,mehlert2003,kuntschner2010,rawle2010,gonz2015}. However, there is a marked difference between light and mass-weighted measurements. The light-averaged ages are systematically younger than the mass-weighted ones at all radii, ranging from $\sim 0.27$ dex in the lowest mass bin to $\sim 0.17$ dex in the highest mass bins. This result is as expected, simply because younger stellar populations are brighter \citep[e.g.][]{greggio1997}. Very interestingly, there is a difference in gradient between light and mass-weight, and the mass-weighted median age does show some radial dependence with positive gradients at all galaxy masses. The gradient is very shallow for galaxies with $\log(M/M_{\odot})<9.935$ ($0.07\pm 0.016\;$dex/$R_{\rm e}$), but increases with increasing galaxy mass to a gradient of about $\sim 0.1\;$dex/$R_{\rm e}$ in galaxies with $\log(M/M_{\odot}) > 10.552$. In other words, mass-weighted age increases with galactic radius. This implies that stellar populations in early-type galaxies, even though generally old, are slightly younger in their centres pointing to an "outside-in" progression of star formation. We will provide a more detailed assessment of this difference in Section~\ref{sec:sfh}.\\
\\
Late-type galaxies exhibit very different age gradients compared to early-types. The light-weighted age gradients are negative and with statistically significant slopes scattering around $\sim 0.2\;$dex/$R_{\rm e}$ for all galaxy masses. This is consistent with recent results in the literature \citep{sanchez2014,gonz2015} and points to older stellar populations in the bulge and more prominent star formation activity at large radii, in line with "inside-out" formation scenarios \citep{li2015}. However, the mass-weighted age gradients are flat, independent of galaxy mass, as also found by \citet{sanchez2014}. This shows that any excess of star formation in the outskirts of galaxy discs compared to the centre must be small and does not contribute significantly to the overall mass budget. Hence, while galaxy discs appear to form "inside-out" in contrast to early-types, this pattern is still surprisingly weak.
\begin{figure*}
\includegraphics[width=0.33\textwidth]{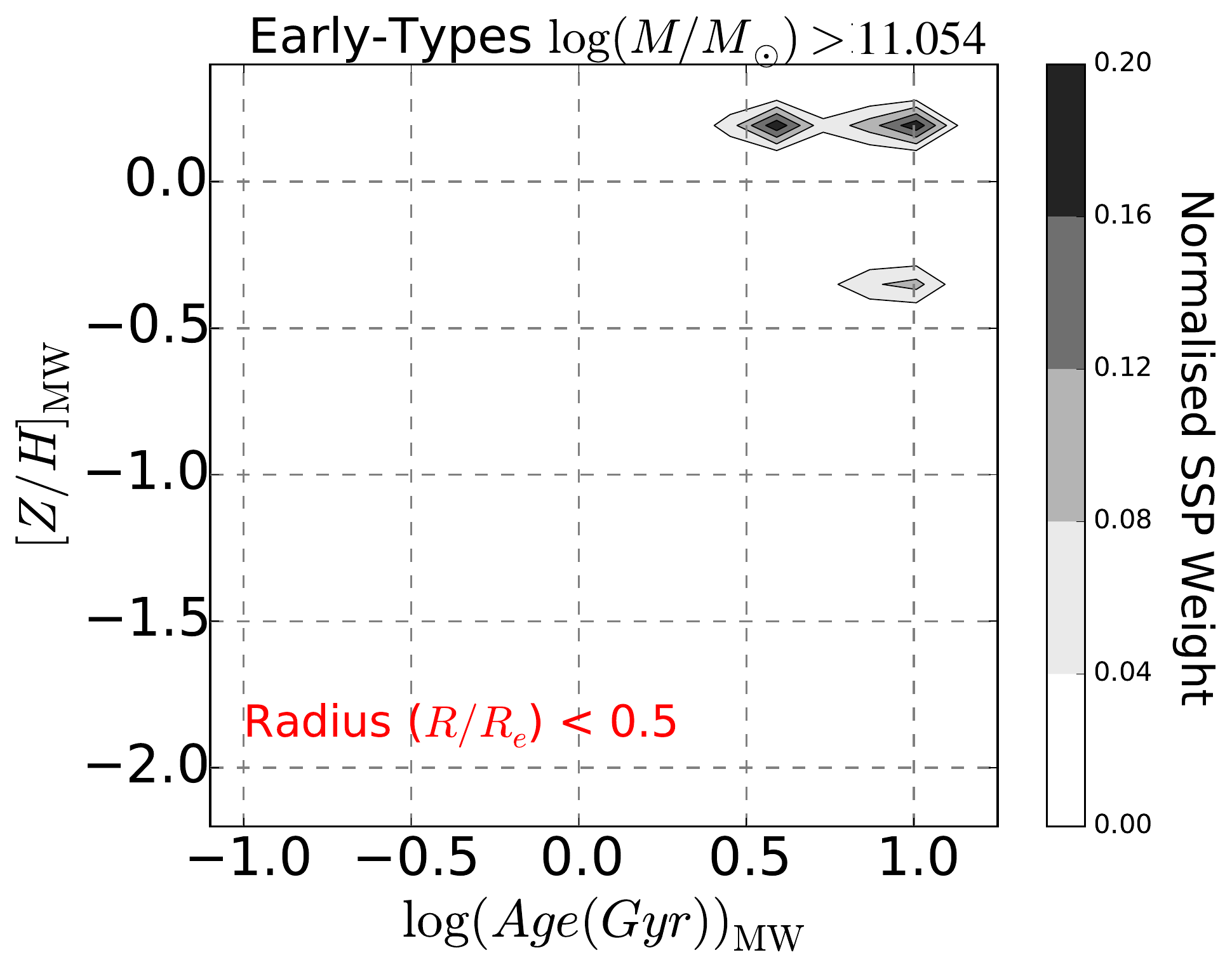}
\includegraphics[width=0.33\textwidth]{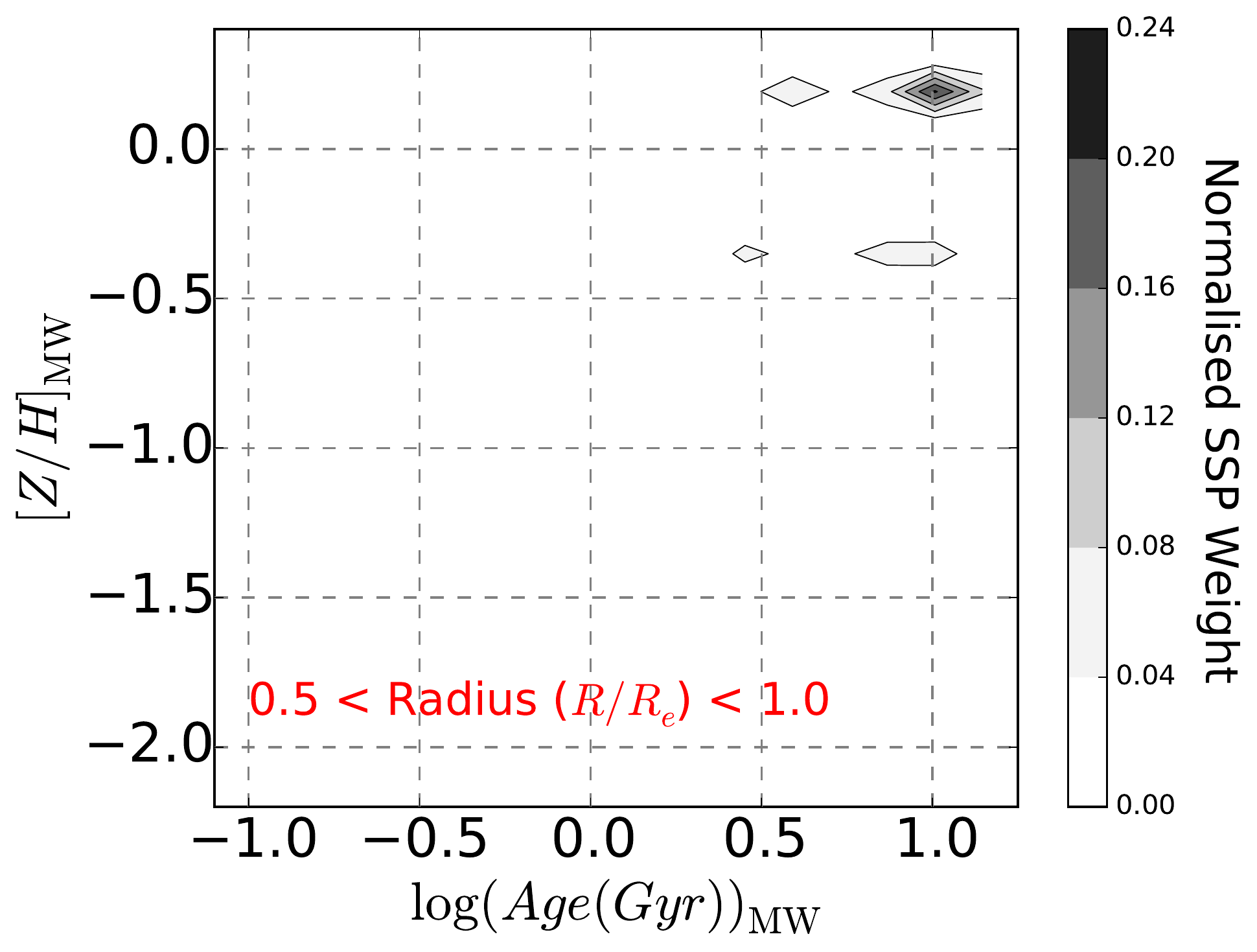}
\includegraphics[width=0.33\textwidth]{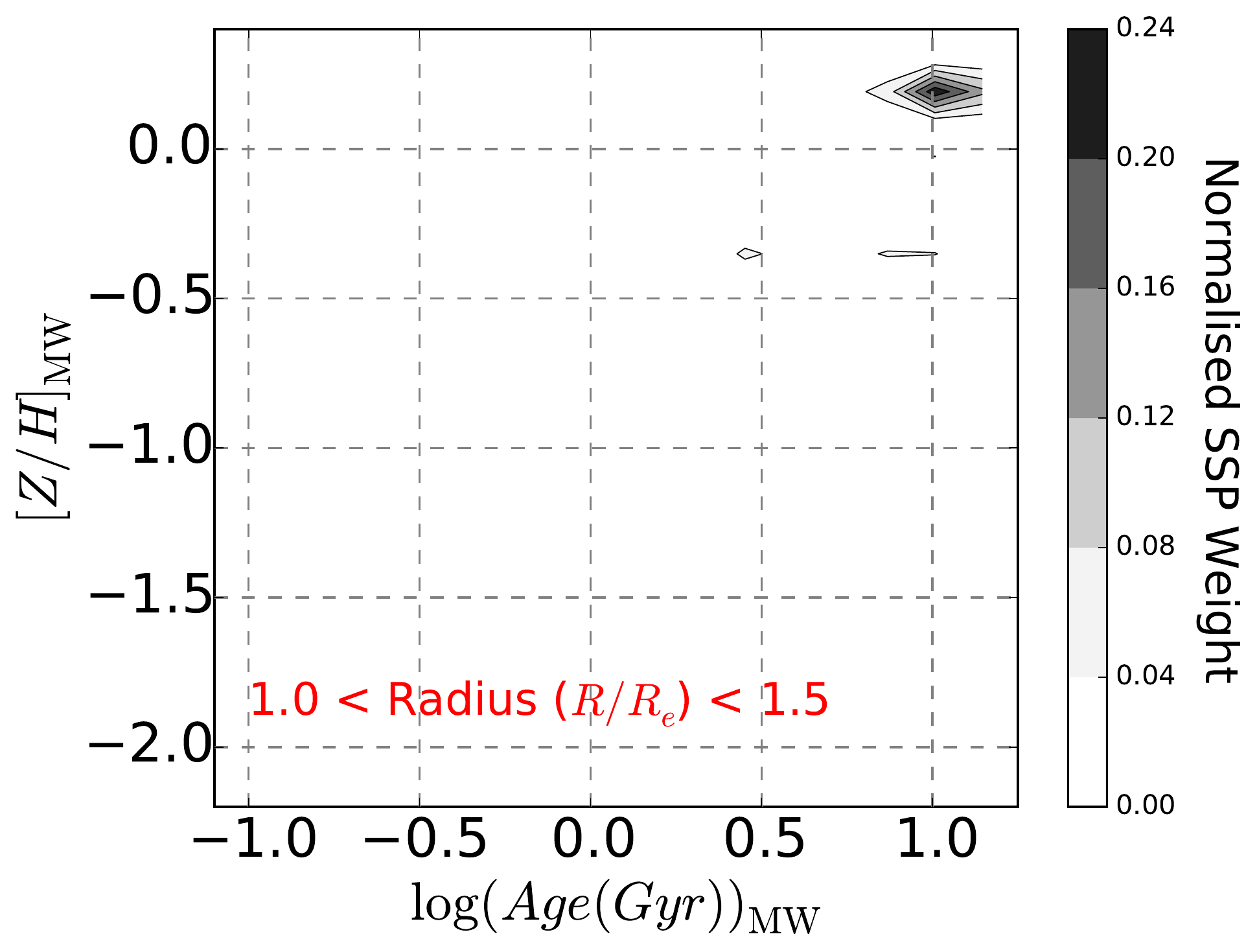}
\includegraphics[width=0.33\textwidth]{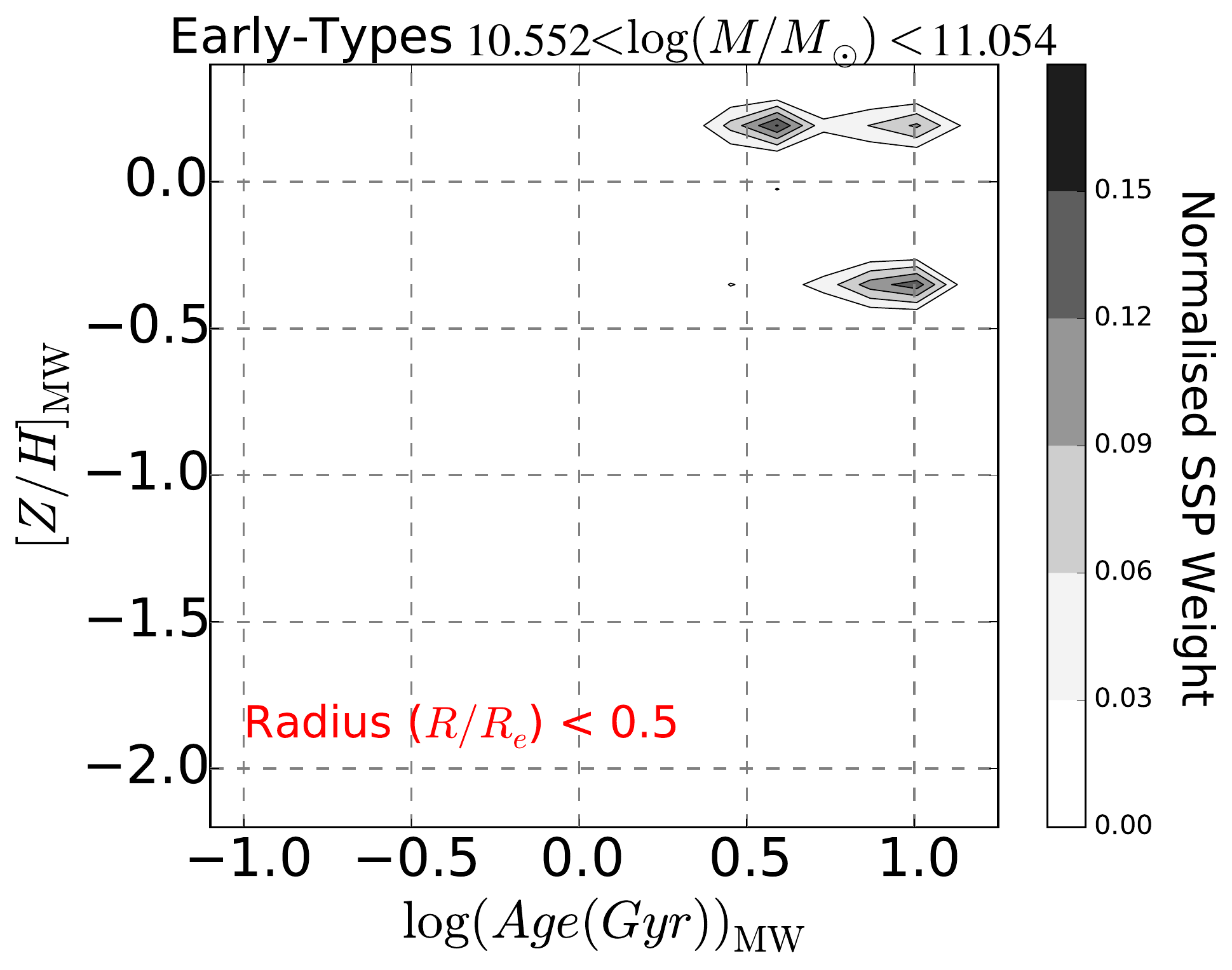}
\includegraphics[width=0.33\textwidth]{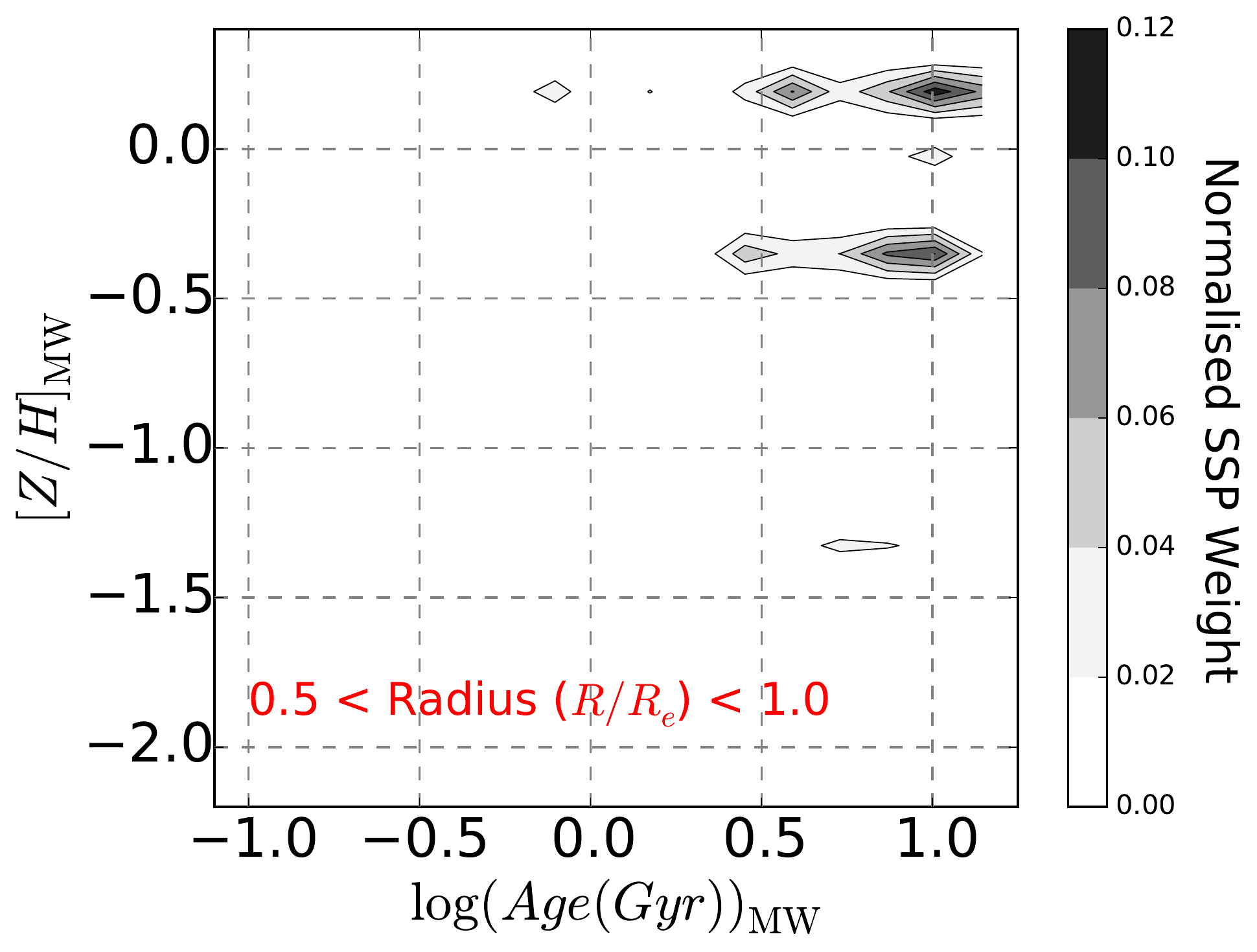}
\includegraphics[width=0.33\textwidth]{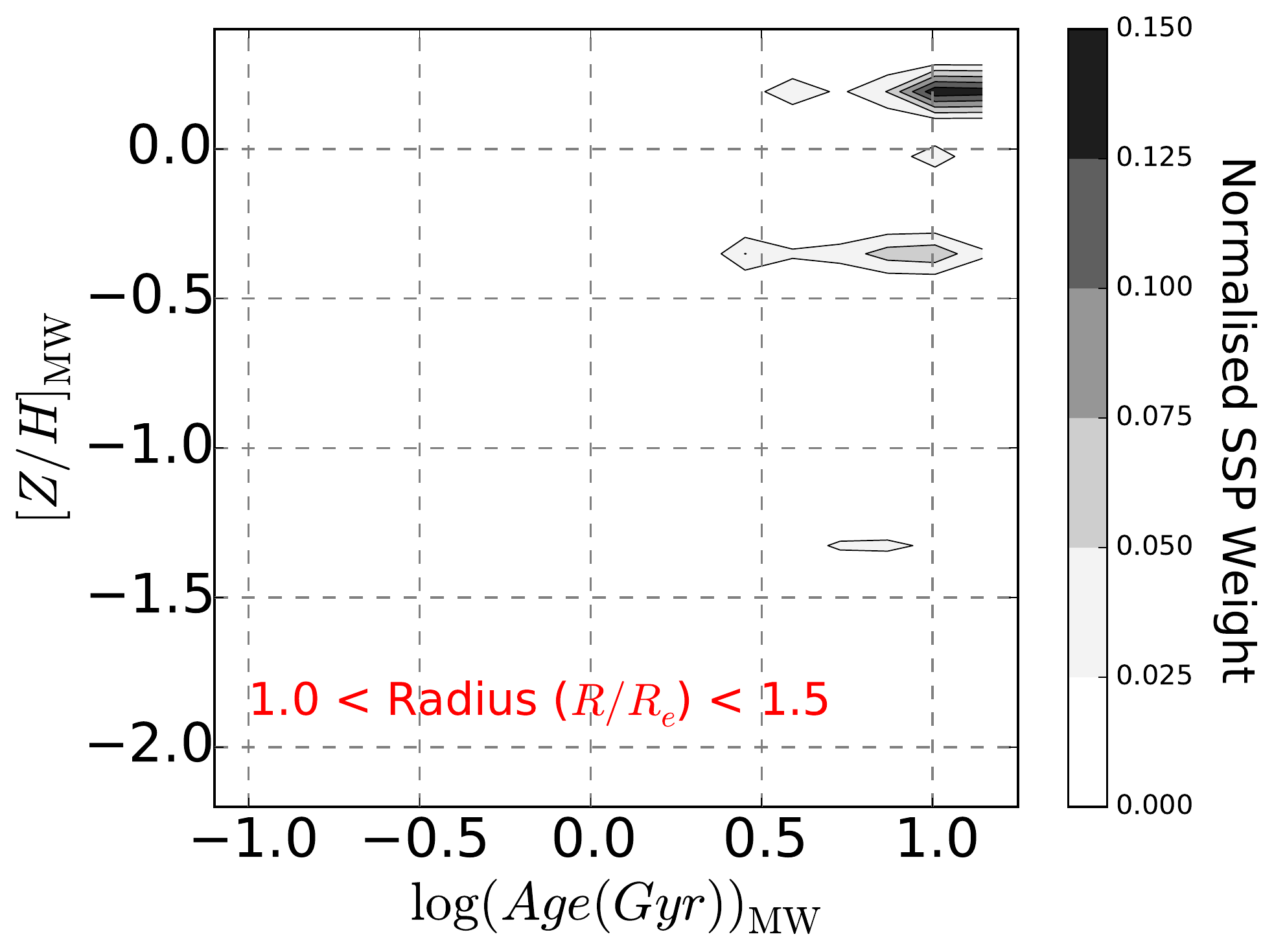}
\includegraphics[width=0.33\textwidth]{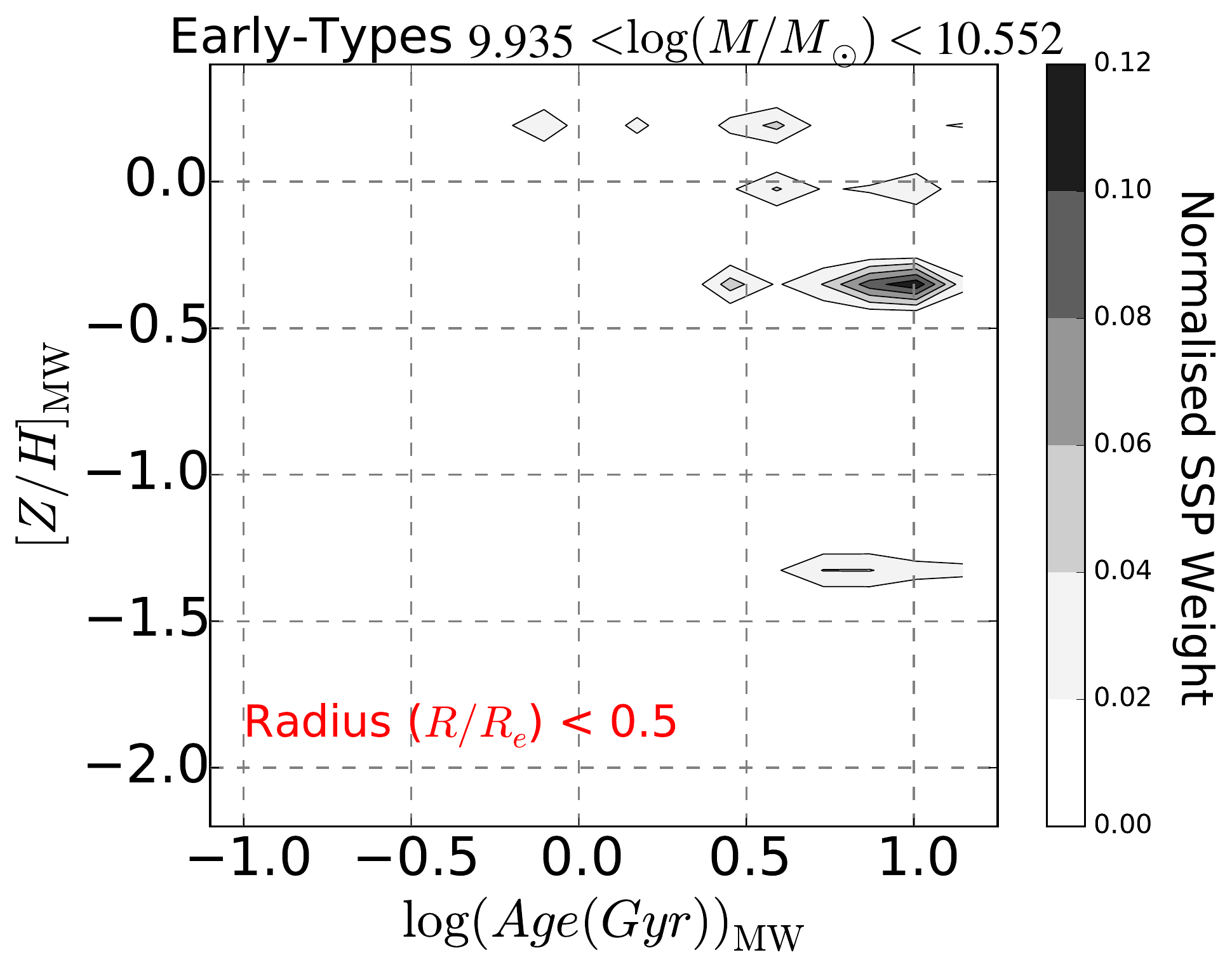}
\includegraphics[width=0.33\textwidth]{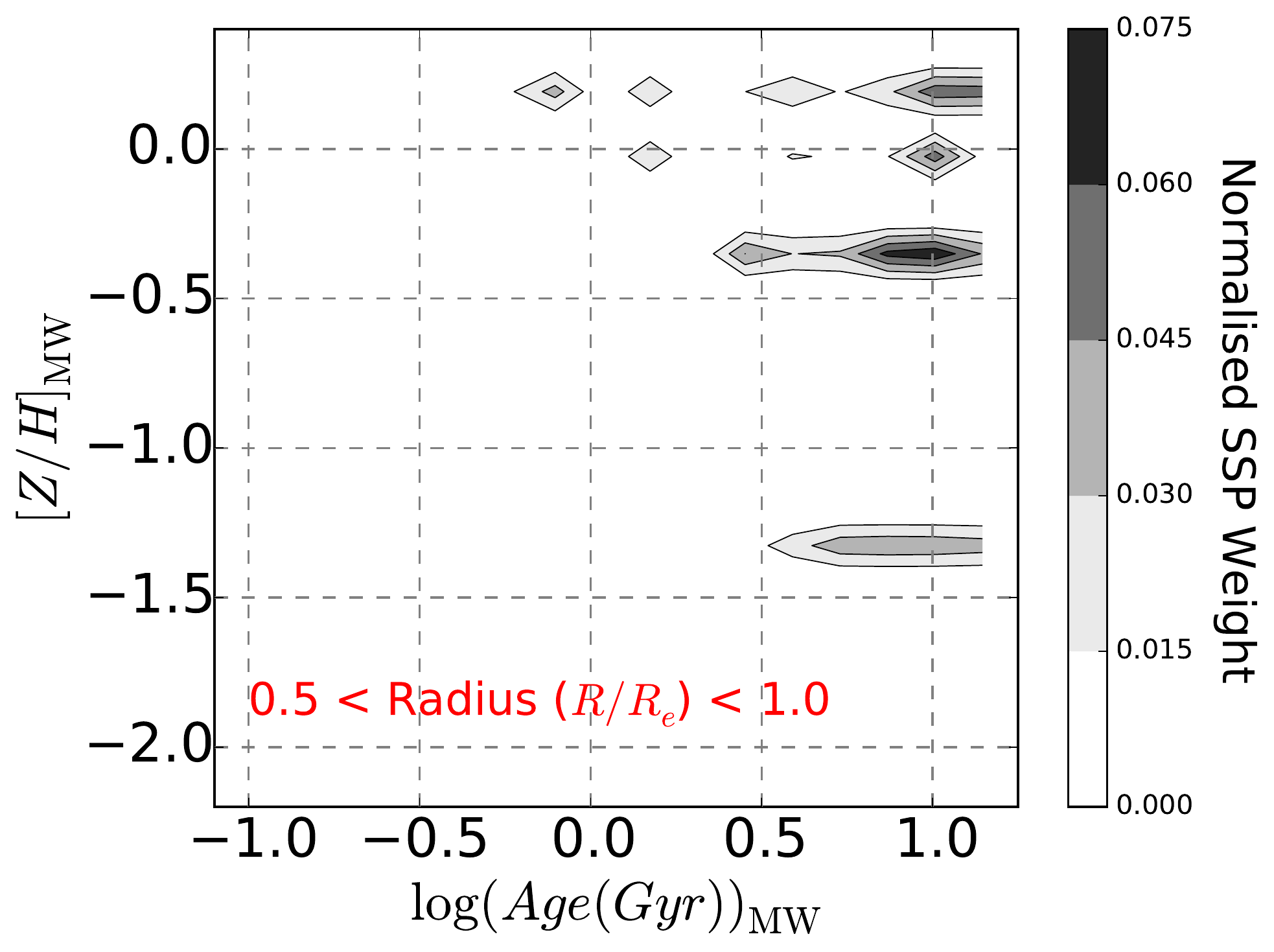}
\includegraphics[width=0.33\textwidth]{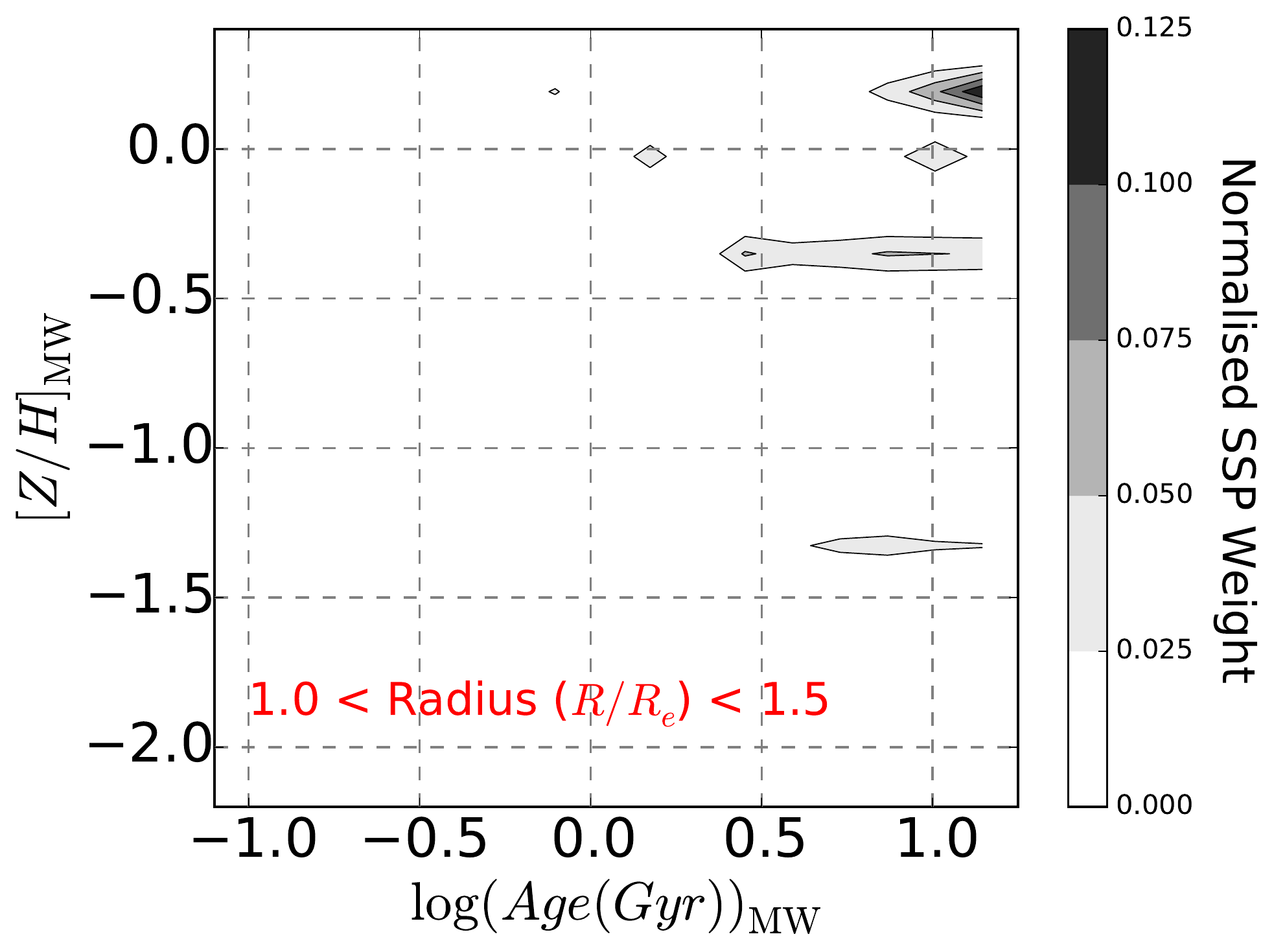}
\includegraphics[width=0.33\textwidth]{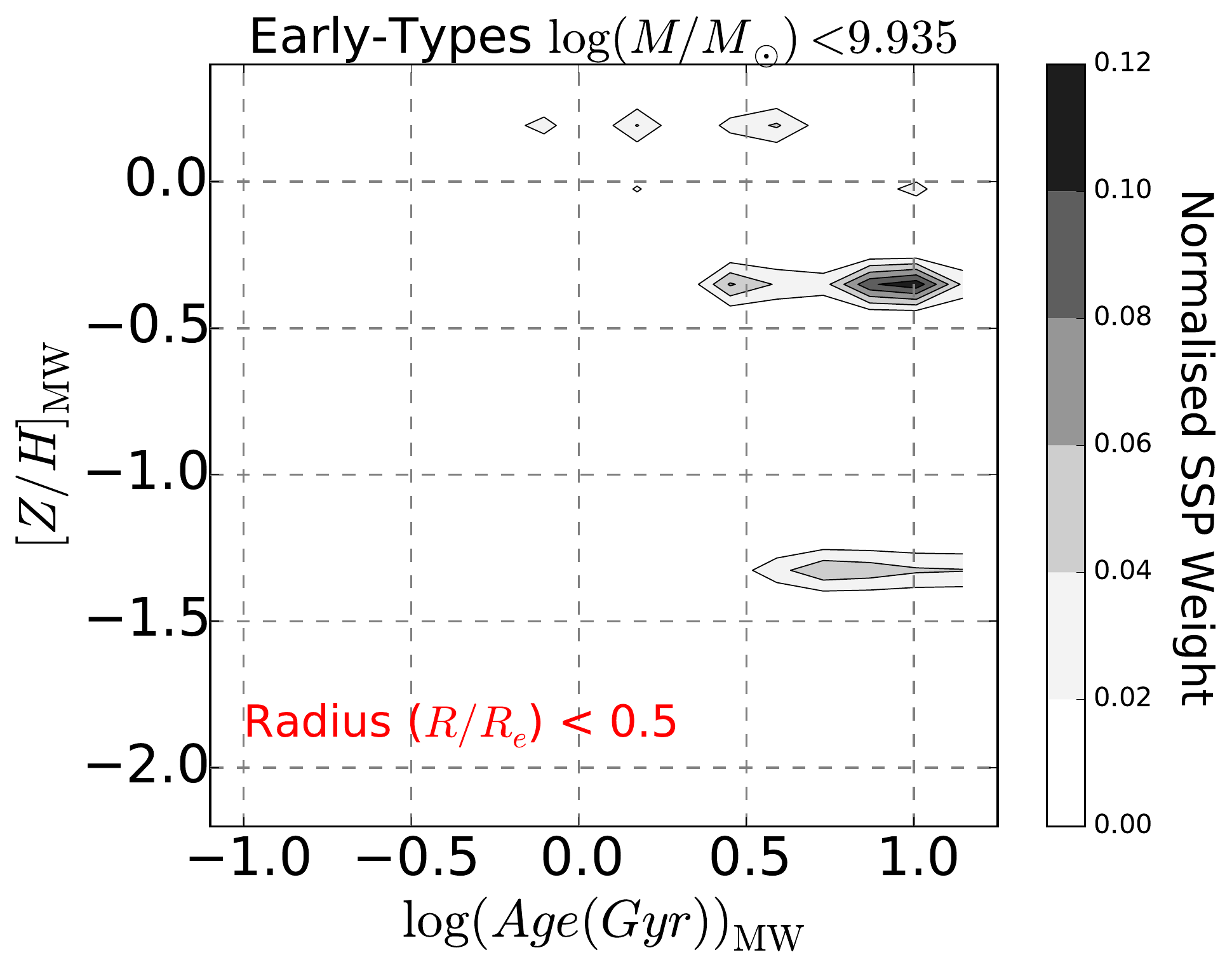}
\includegraphics[width=0.33\textwidth]{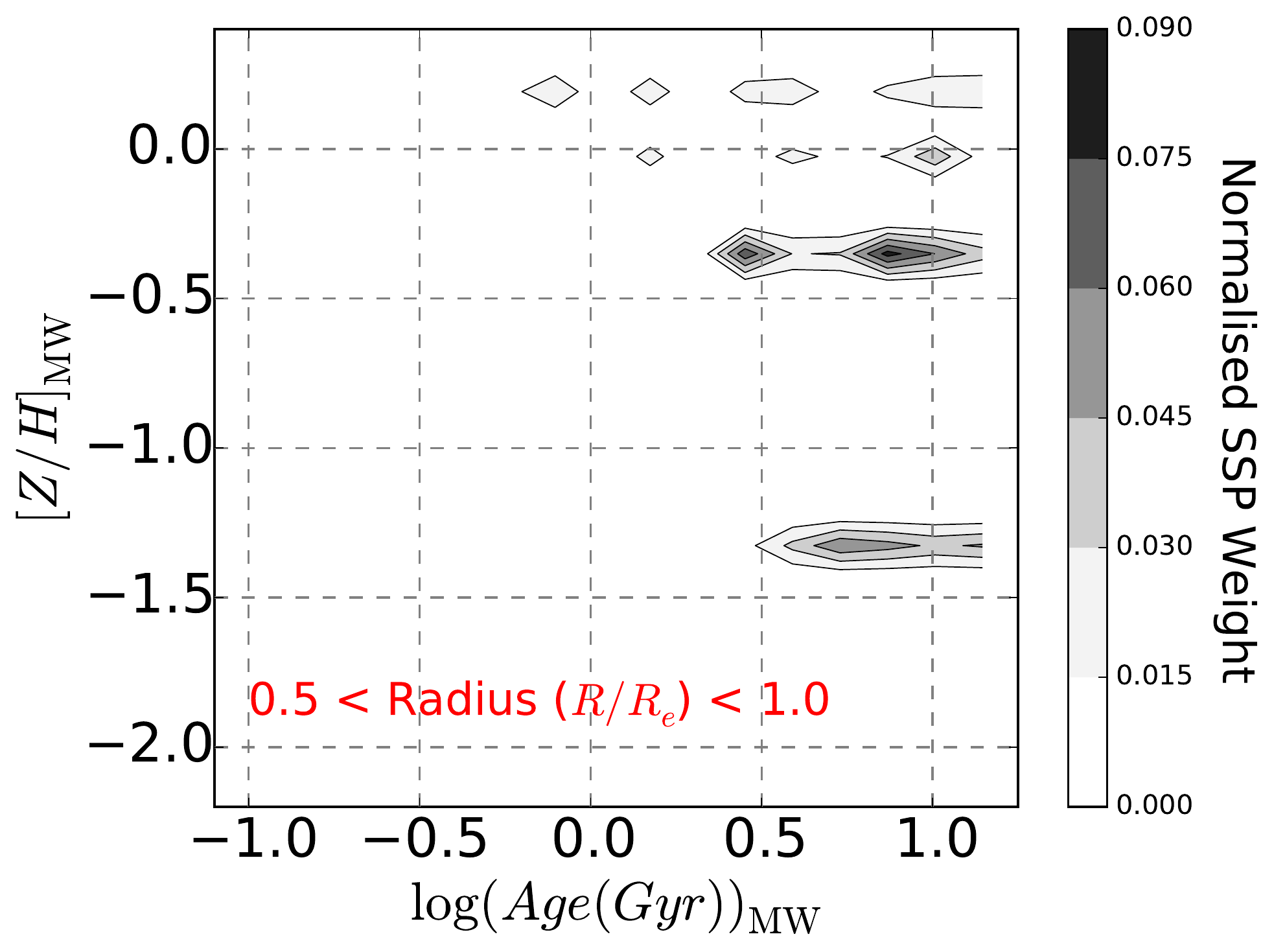}
\includegraphics[width=0.33\textwidth]{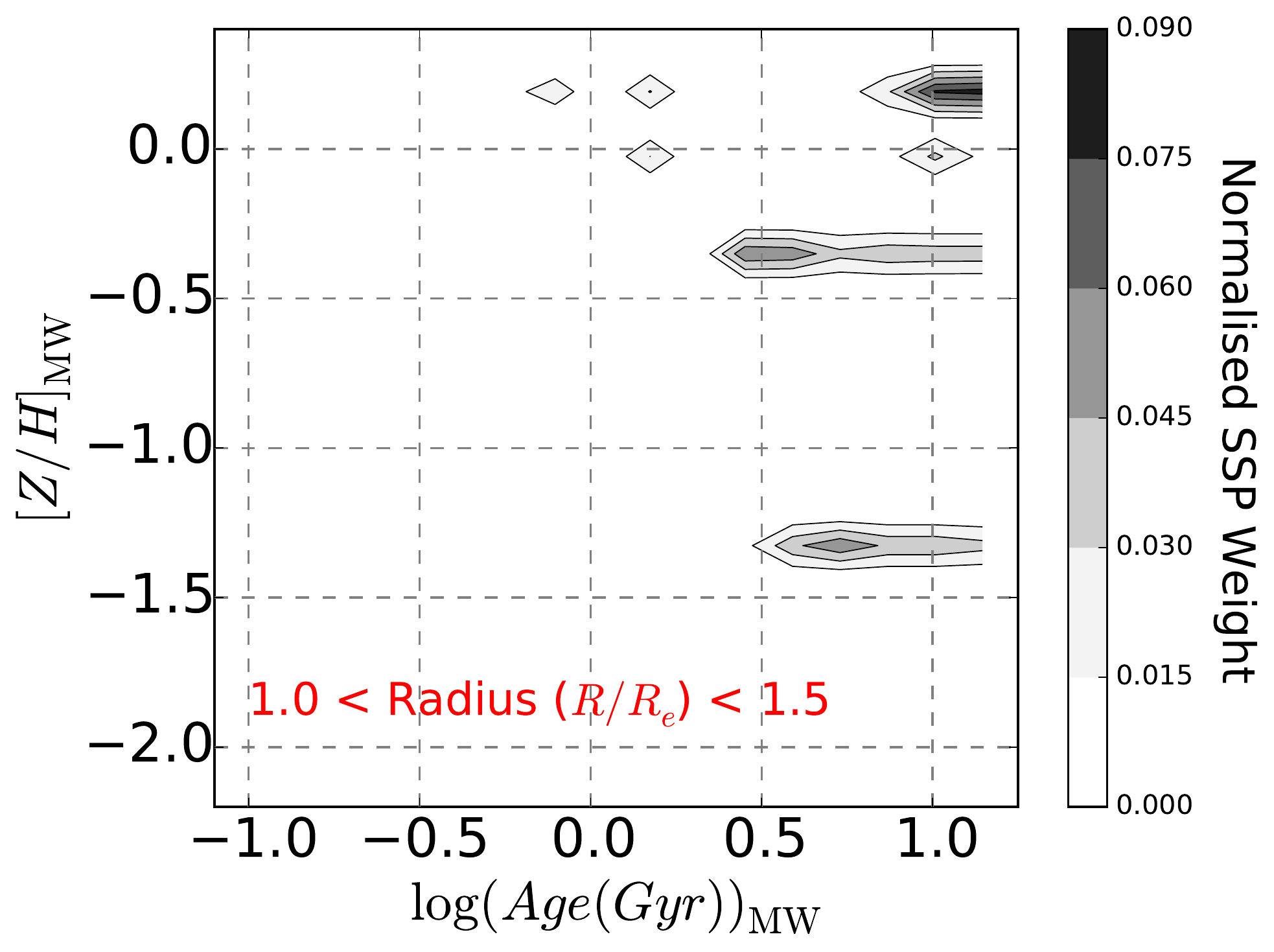}
\caption{Star formation and metal enrichment histories for early-type galaxies as function of galaxy mass and radius. The density scale indicates the relative {\em mass}-weights of the stellar populations in the spectral fit in age-metallicity space. The columns are three radial bins with radius increasing from left to right (see labels), rows are the four mass bins with mass increasing from bottom to top (see labels). See Figure~\ref{fig:sfh_et_lw} for light-weighted quantities. For more details see Section 4.2}
\label{fig:sfh_et_mw}
\end{figure*}
\begin{figure*}
\includegraphics[width=0.33\textwidth]{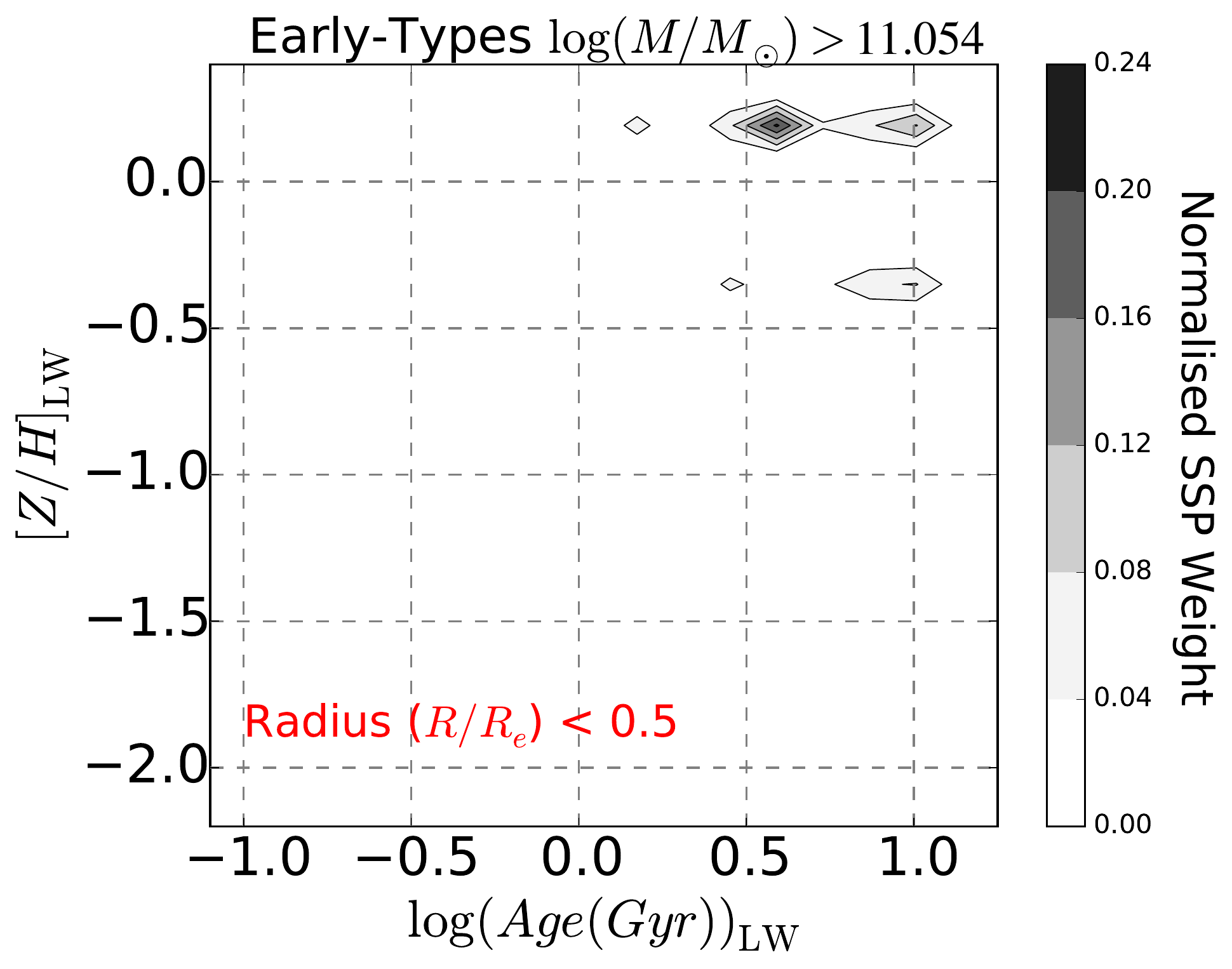}
\includegraphics[width=0.33\textwidth]{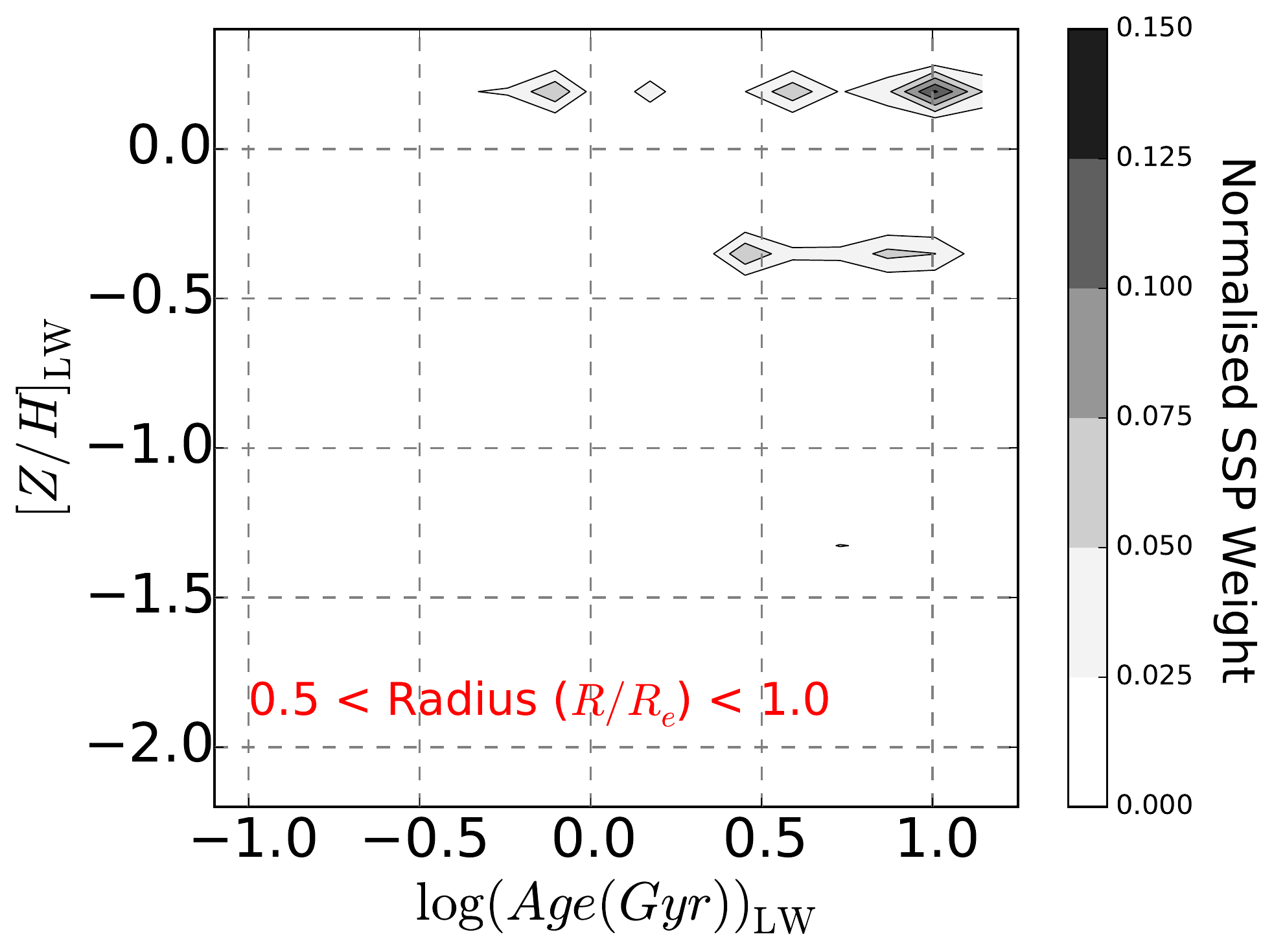}
\includegraphics[width=0.33\textwidth]{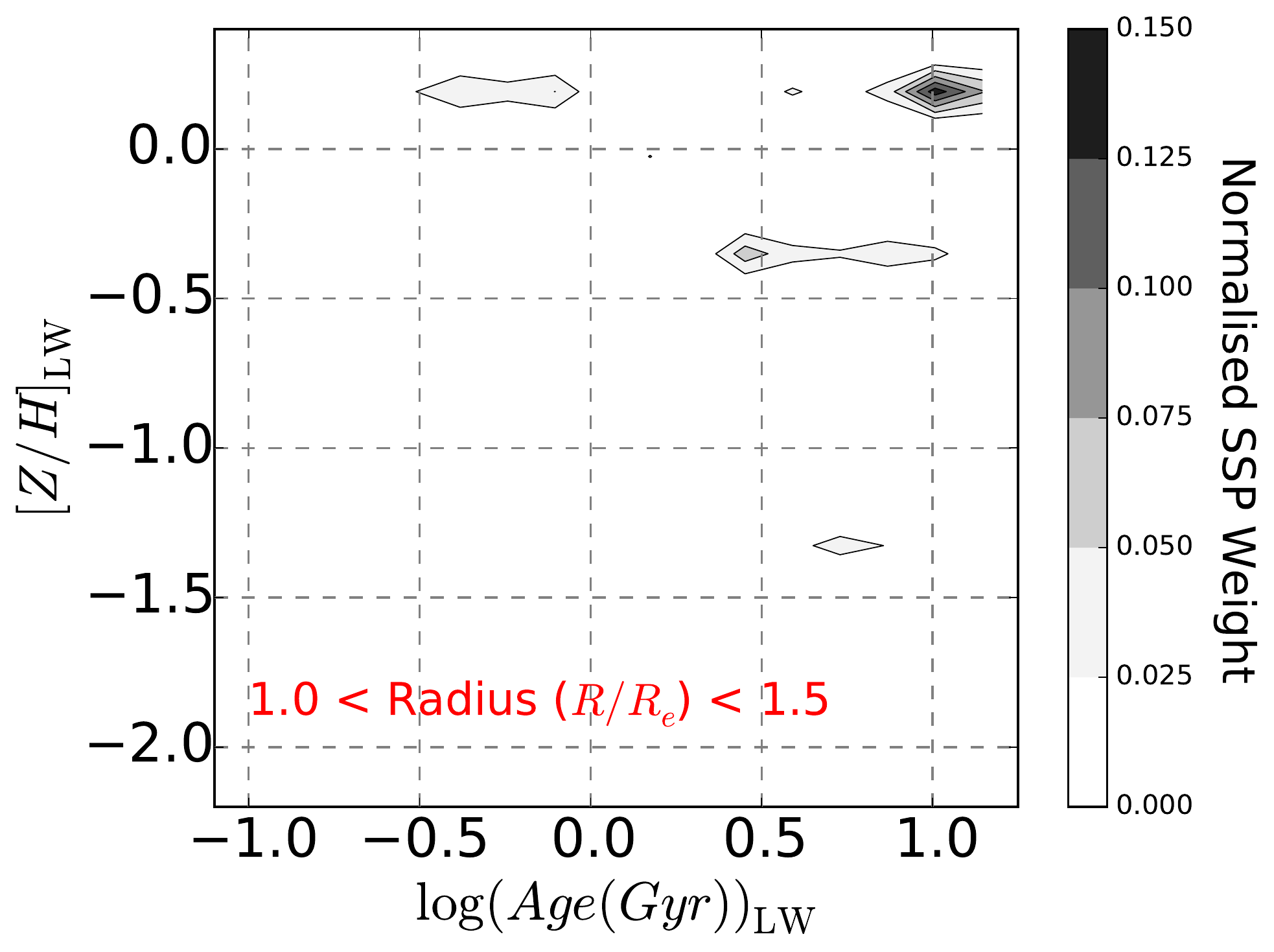}
\includegraphics[width=0.33\textwidth]{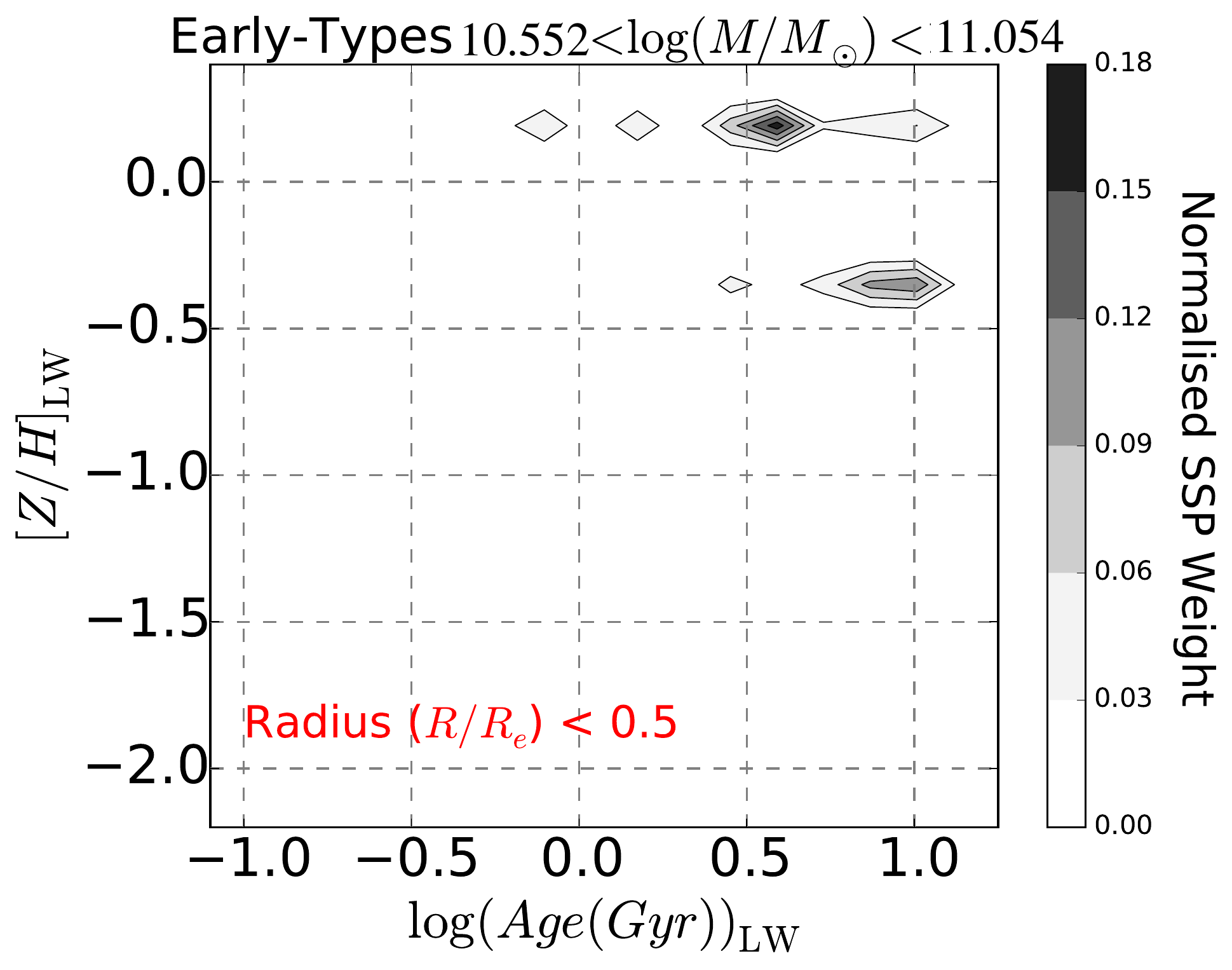}
\includegraphics[width=0.33\textwidth]{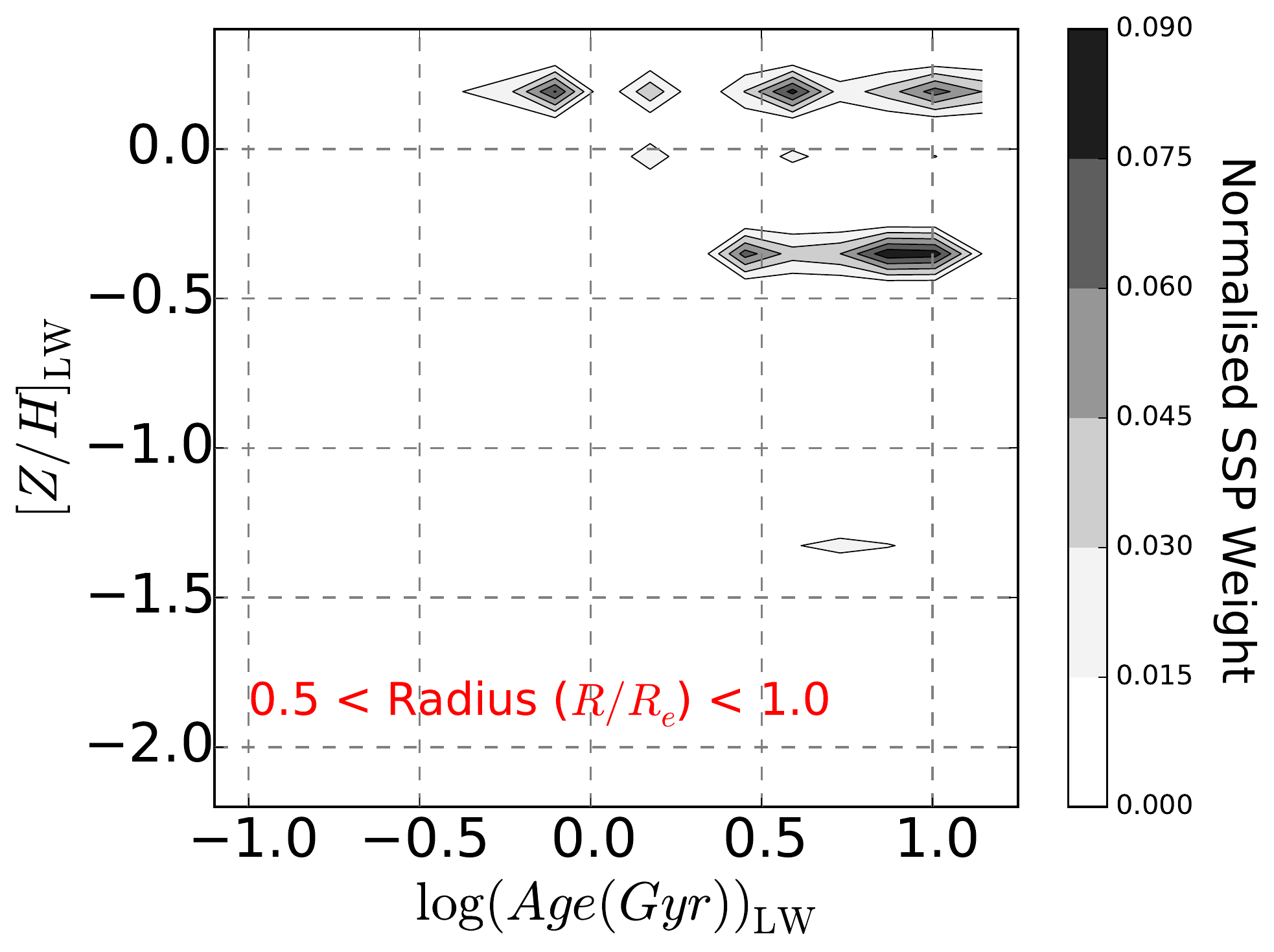}
\includegraphics[width=0.33\textwidth]{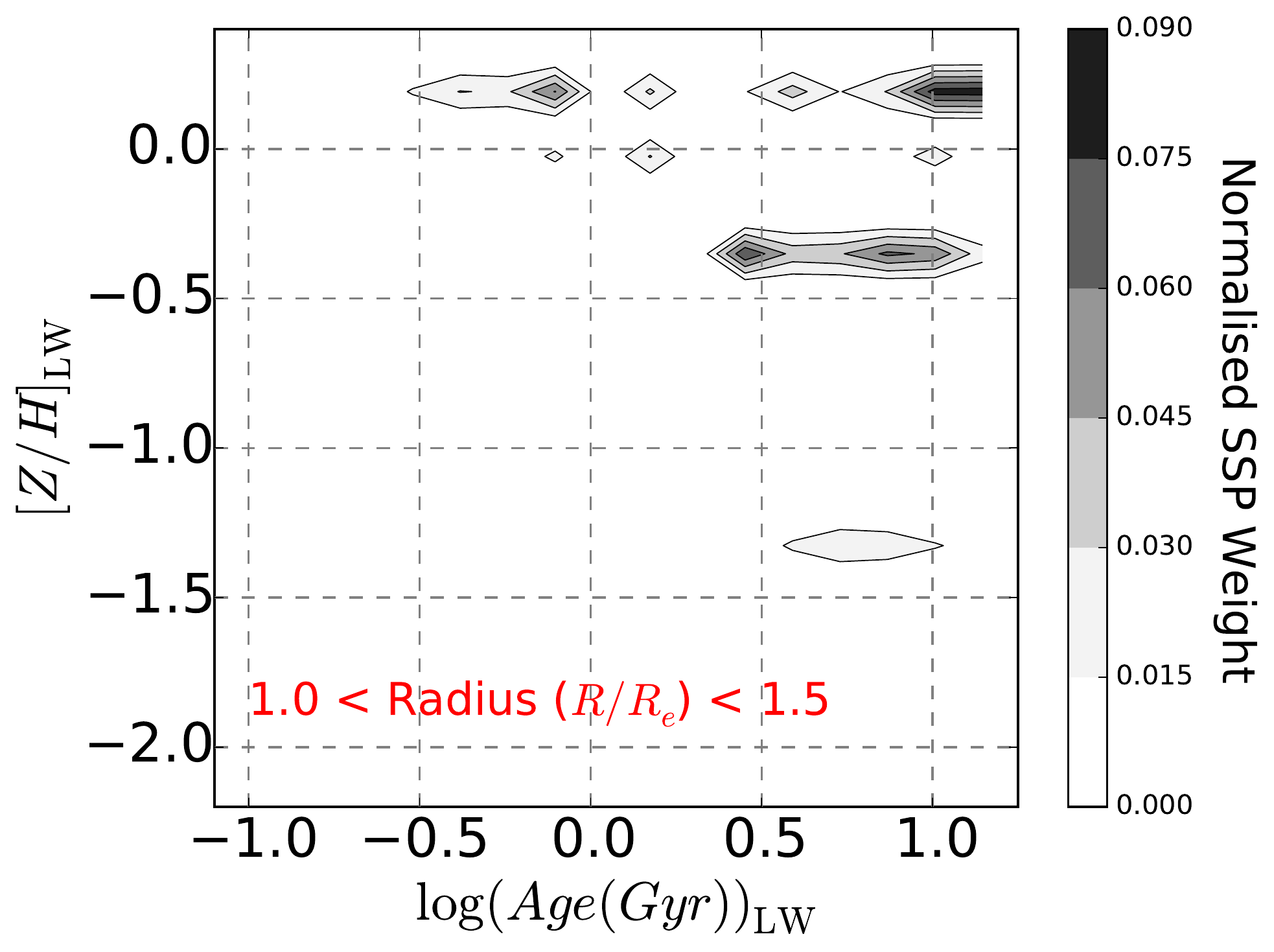}
\includegraphics[width=0.33\textwidth]{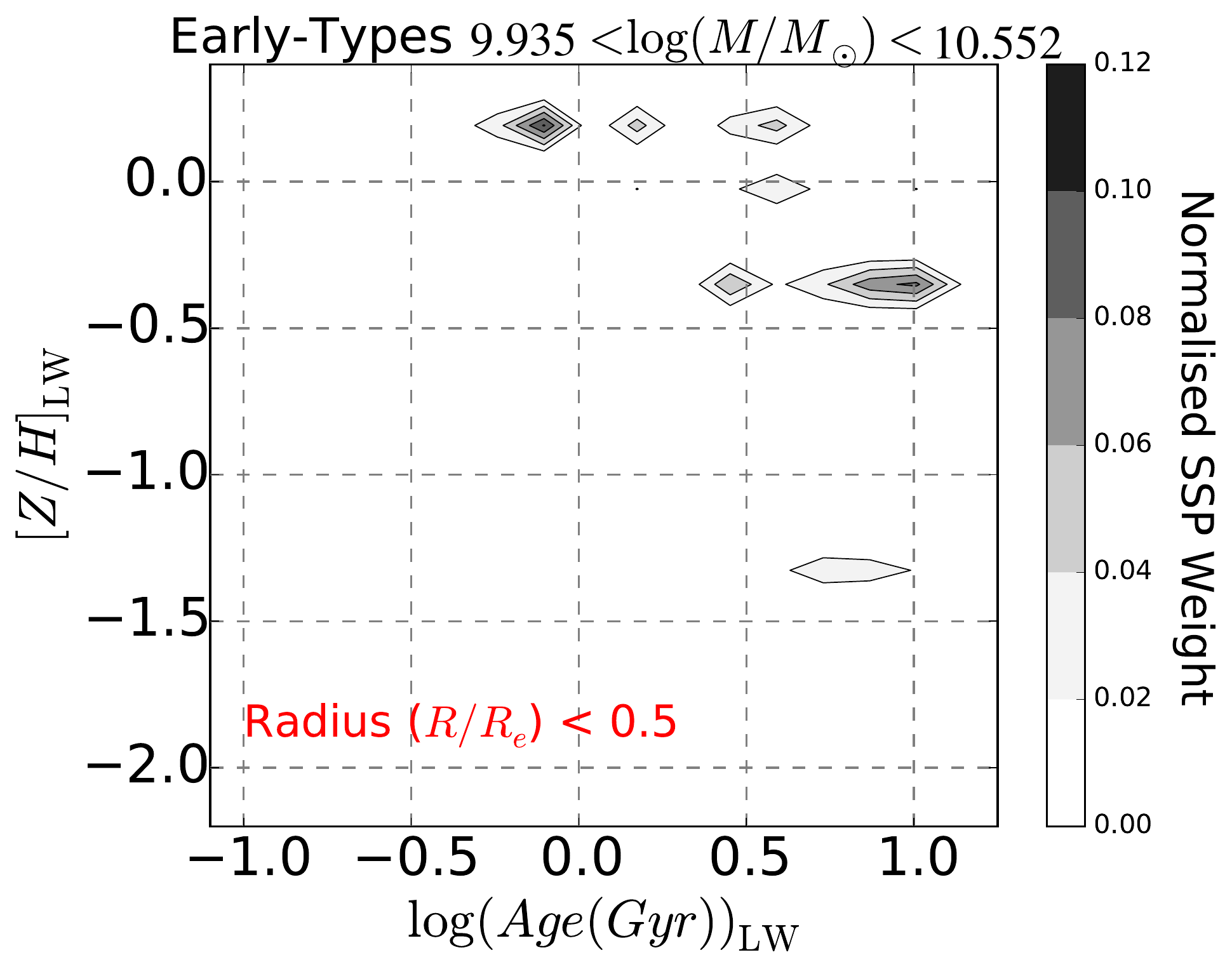}
\includegraphics[width=0.33\textwidth]{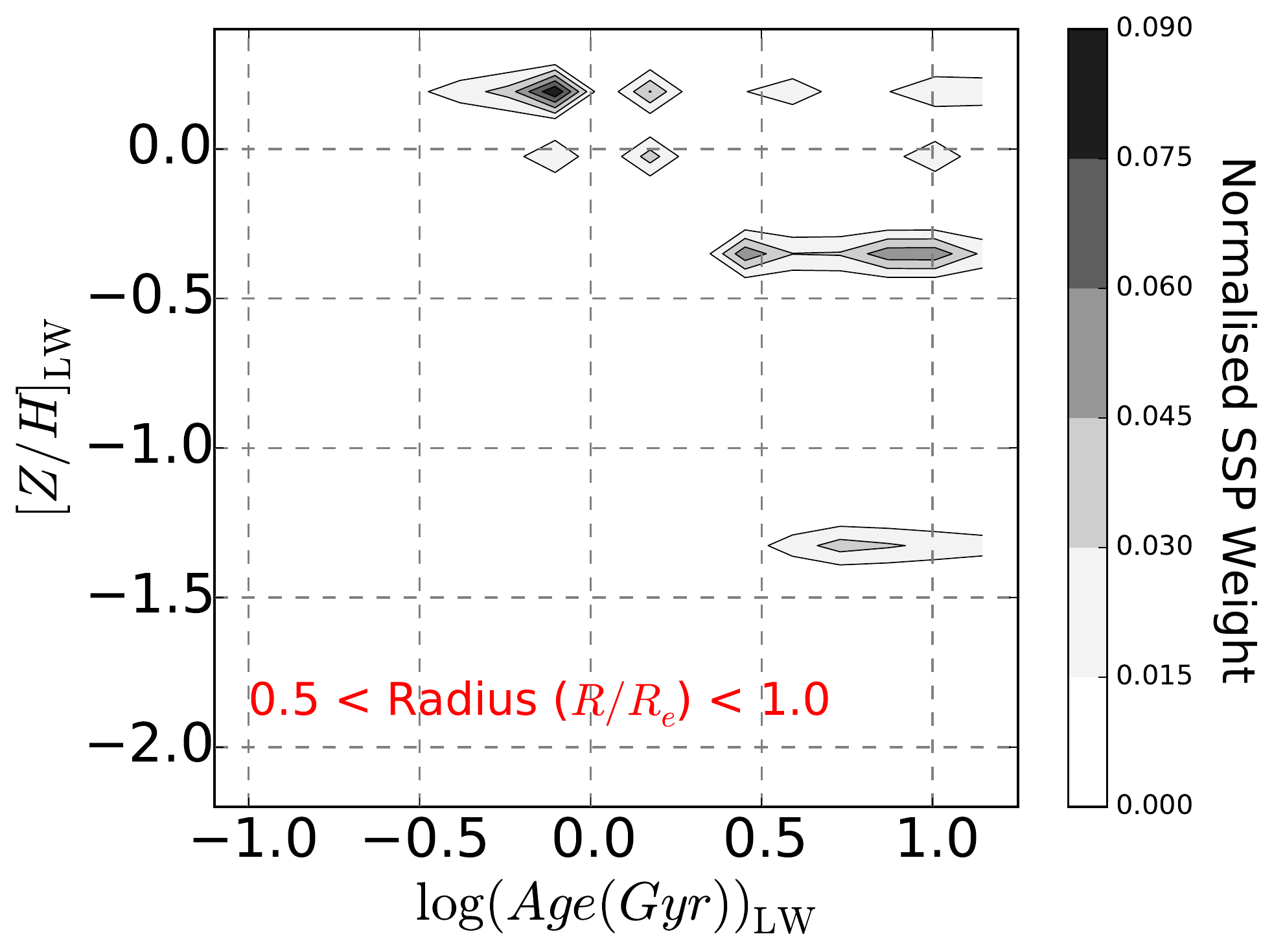}
\includegraphics[width=0.33\textwidth]{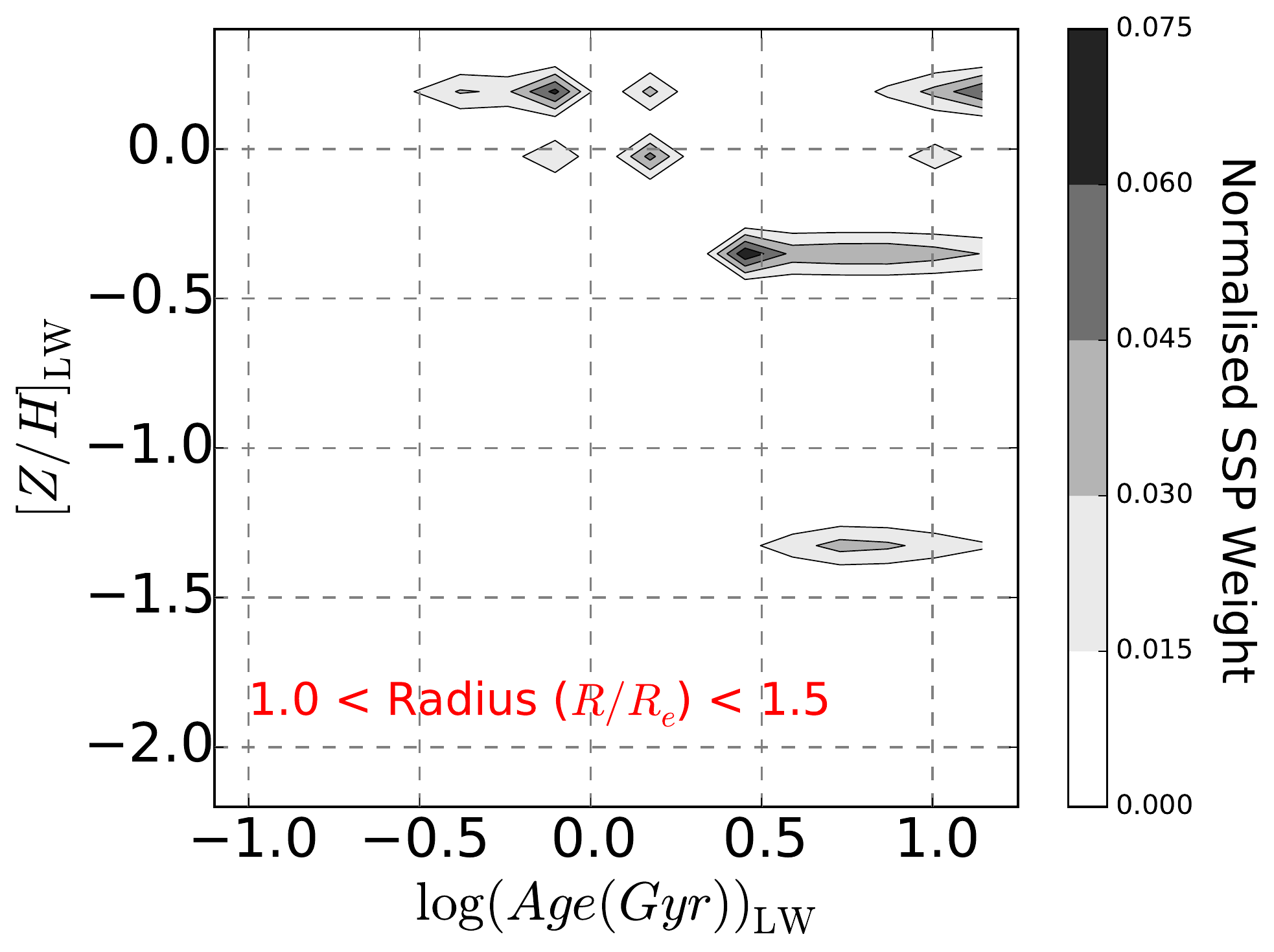}
\includegraphics[width=0.33\textwidth]{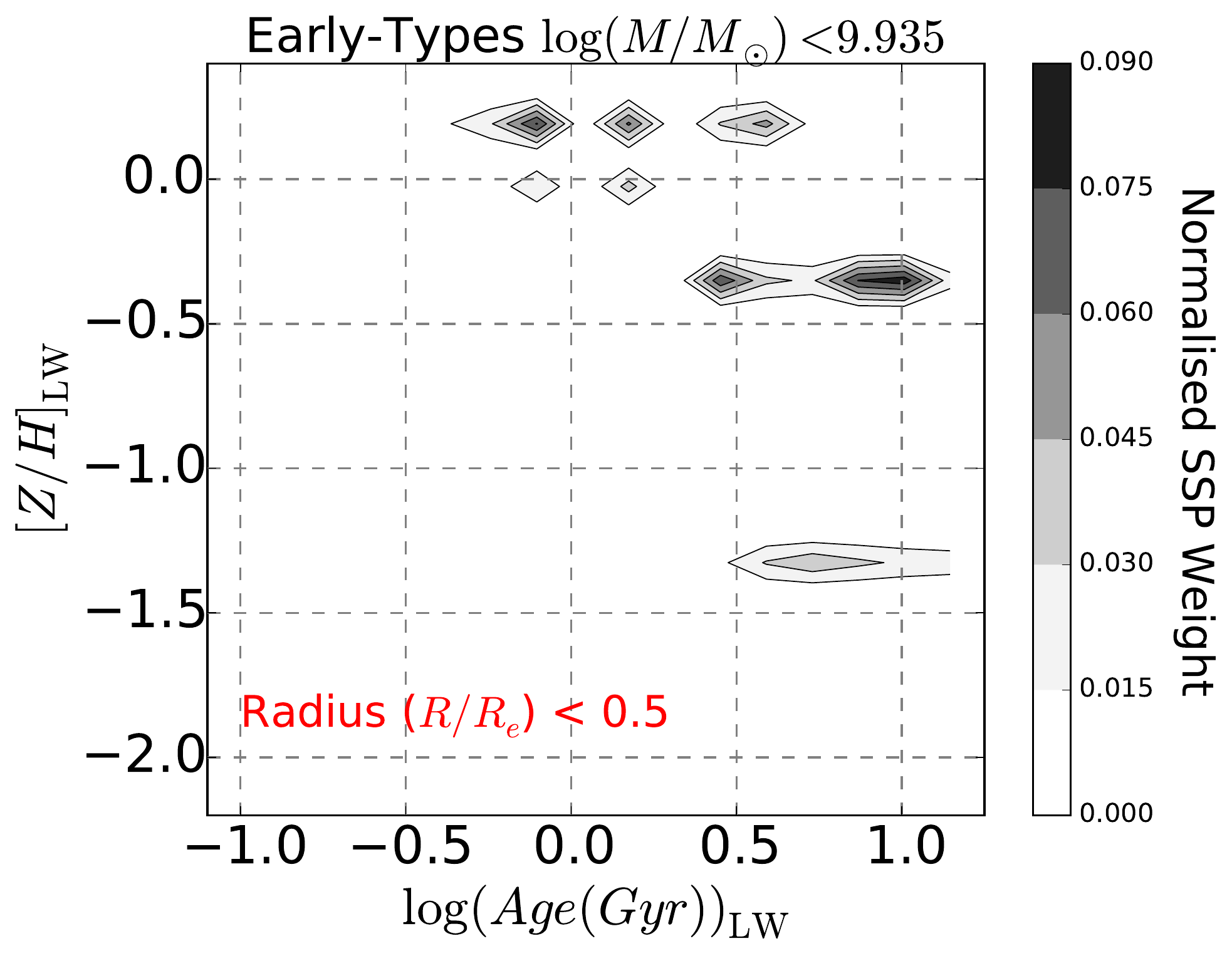}
\includegraphics[width=0.33\textwidth]{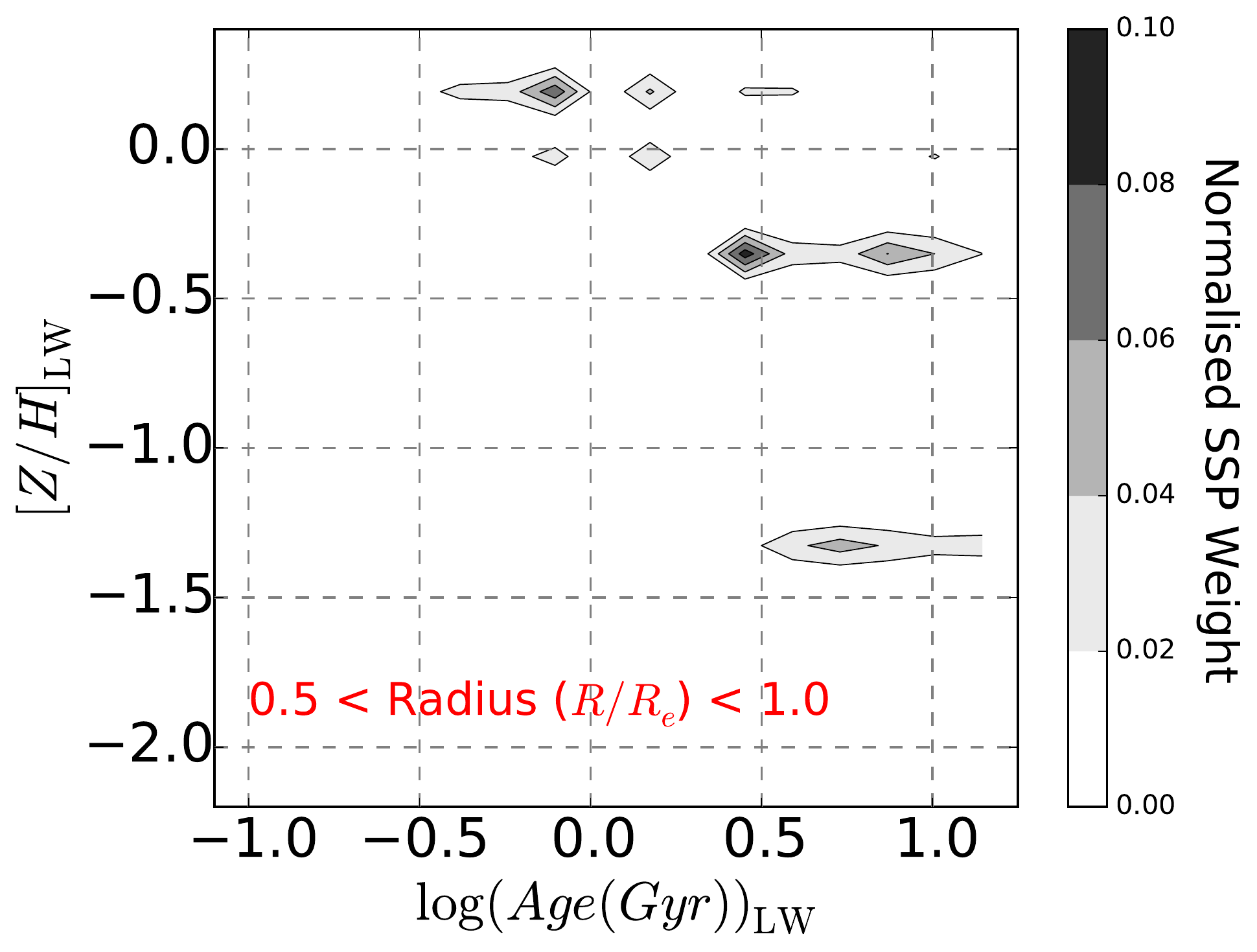}
\includegraphics[width=0.33\textwidth]{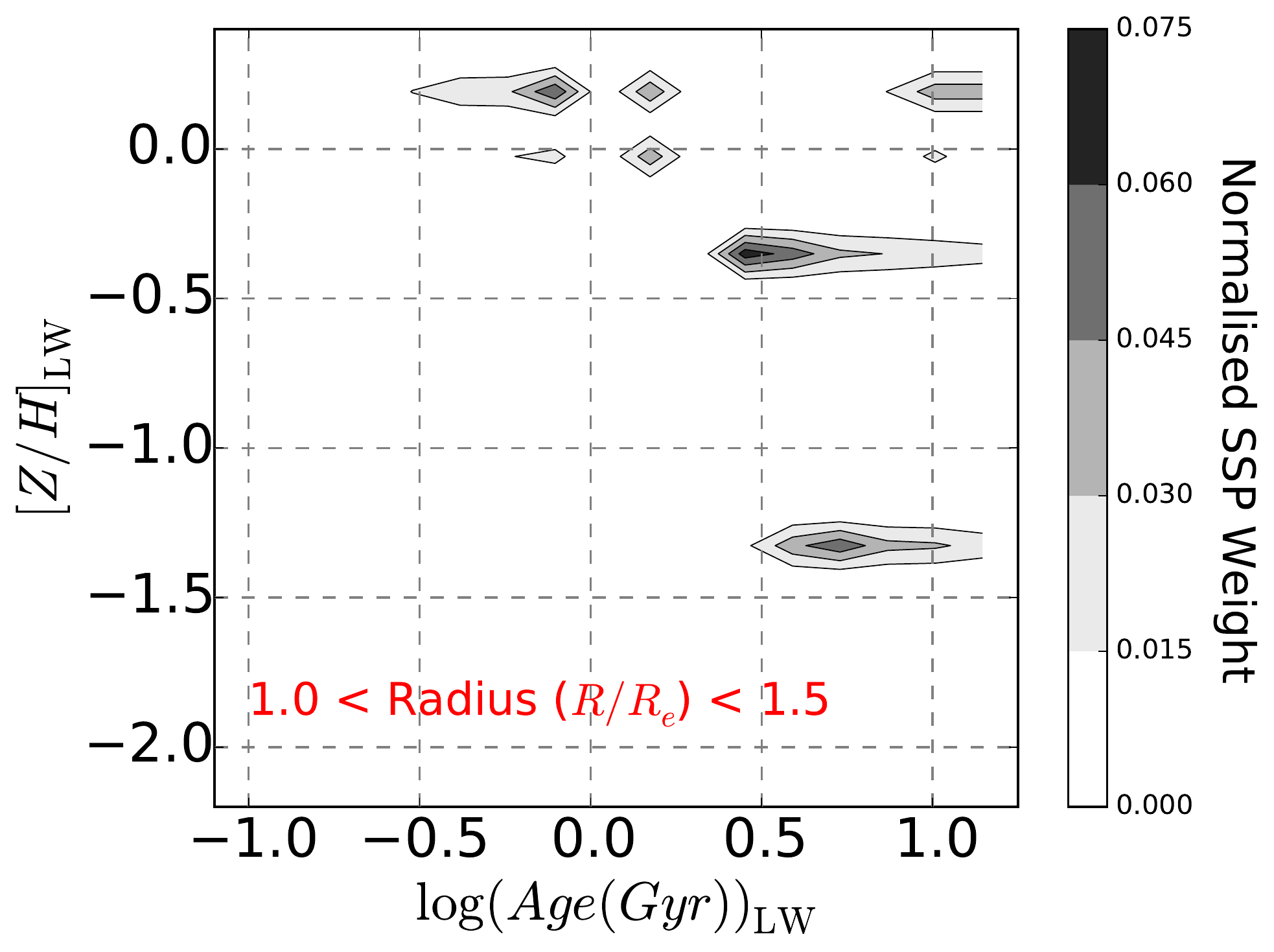}
\caption{Star formation and metal enrichment histories for early-type galaxies as function of galaxy mass and radius. The density scale indicates the relative {\em luminosity}-weights of the stellar populations in the spectral fit in age-metallicity space. The columns are three radial bins with radius increasing from left to right (see labels), rows are the four mass bins with mass increasing from bottom to top (see labels). See Figure~\ref{fig:sfh_et_mw} for mass-weighted quantities. For more details see Section 4.2}
\label{fig:sfh_et_lw}
\end{figure*}
\begin{figure*}
\includegraphics[width=0.33\textwidth]{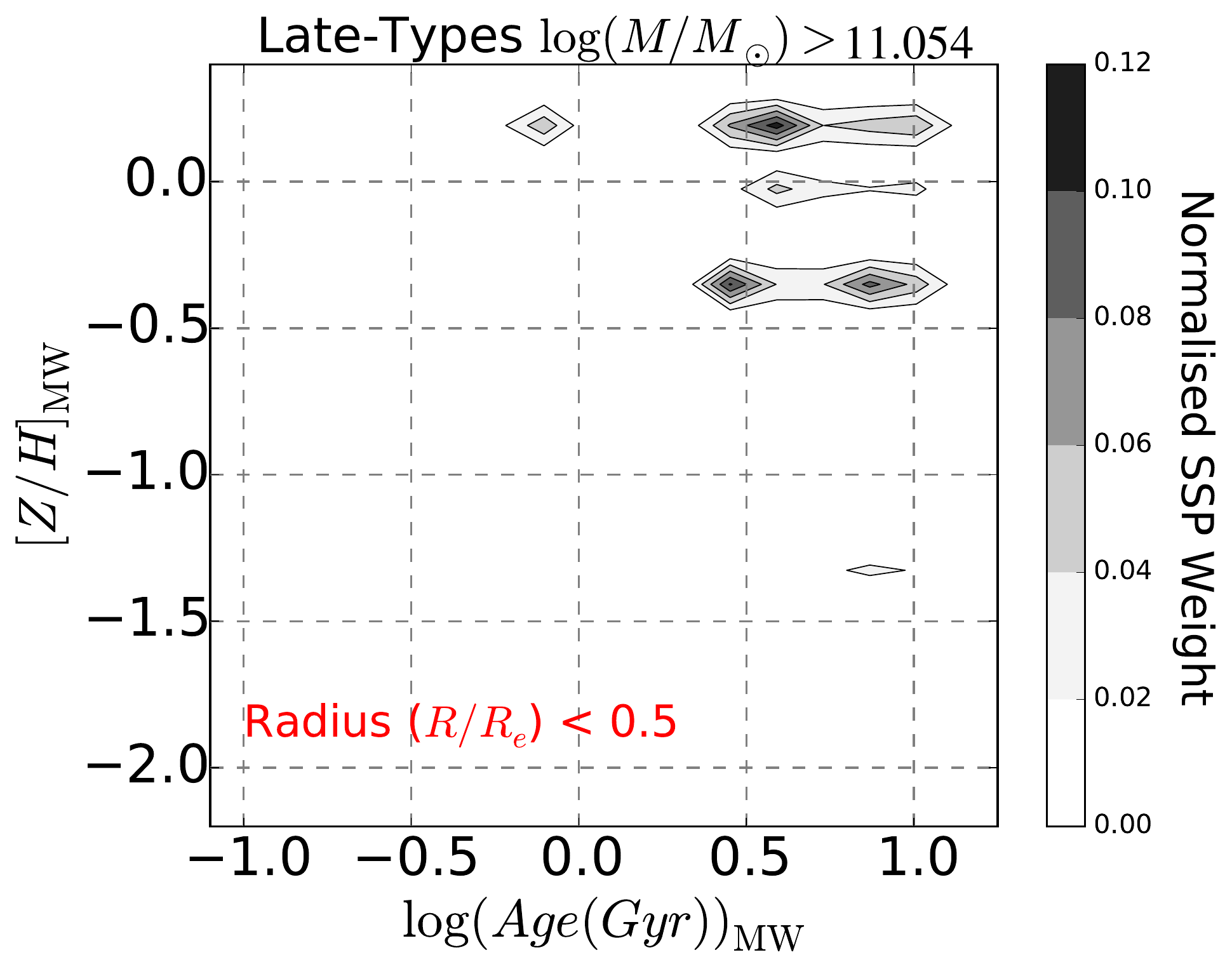}
\includegraphics[width=0.33\textwidth]{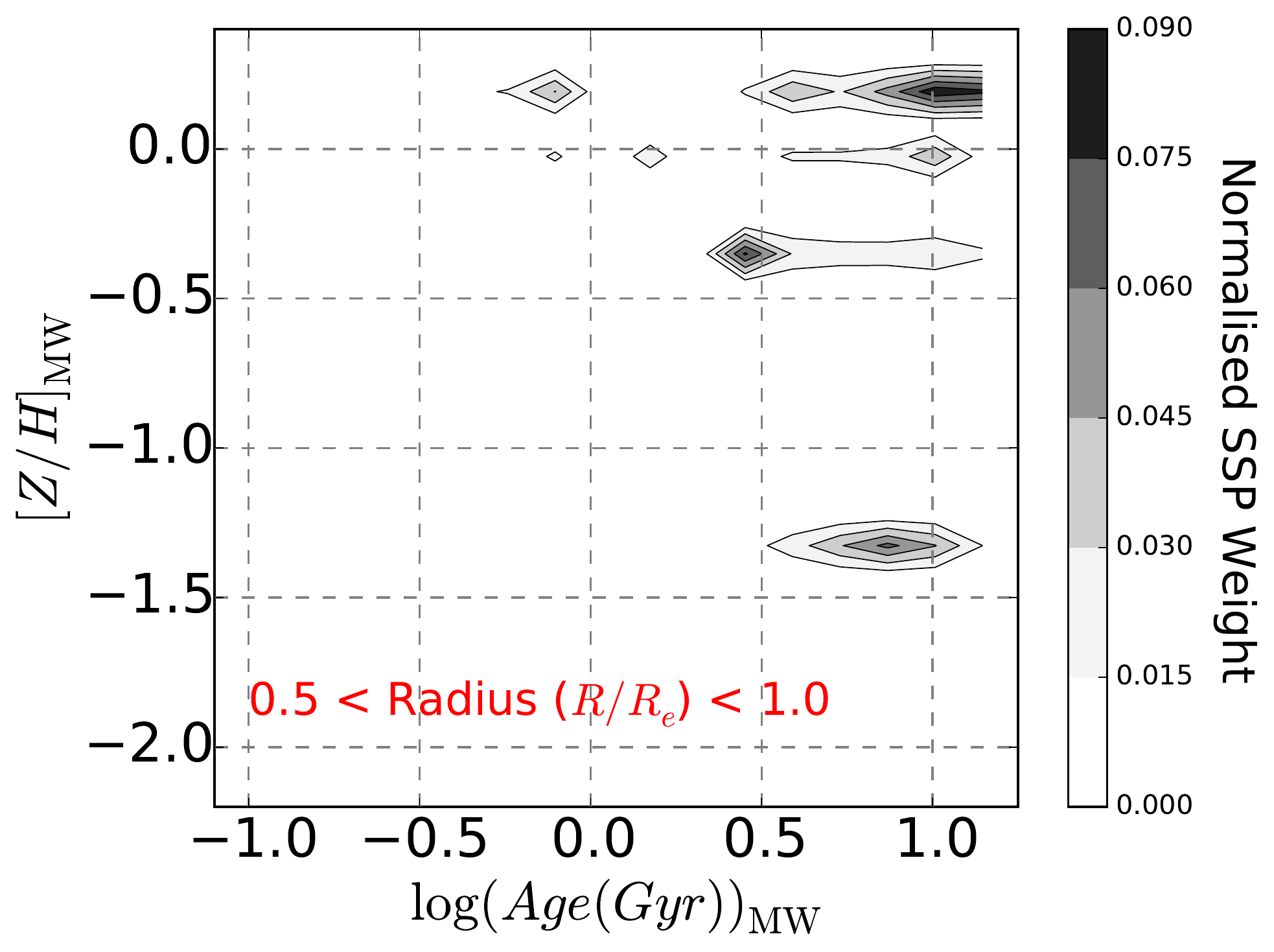}
\includegraphics[width=0.33\textwidth]{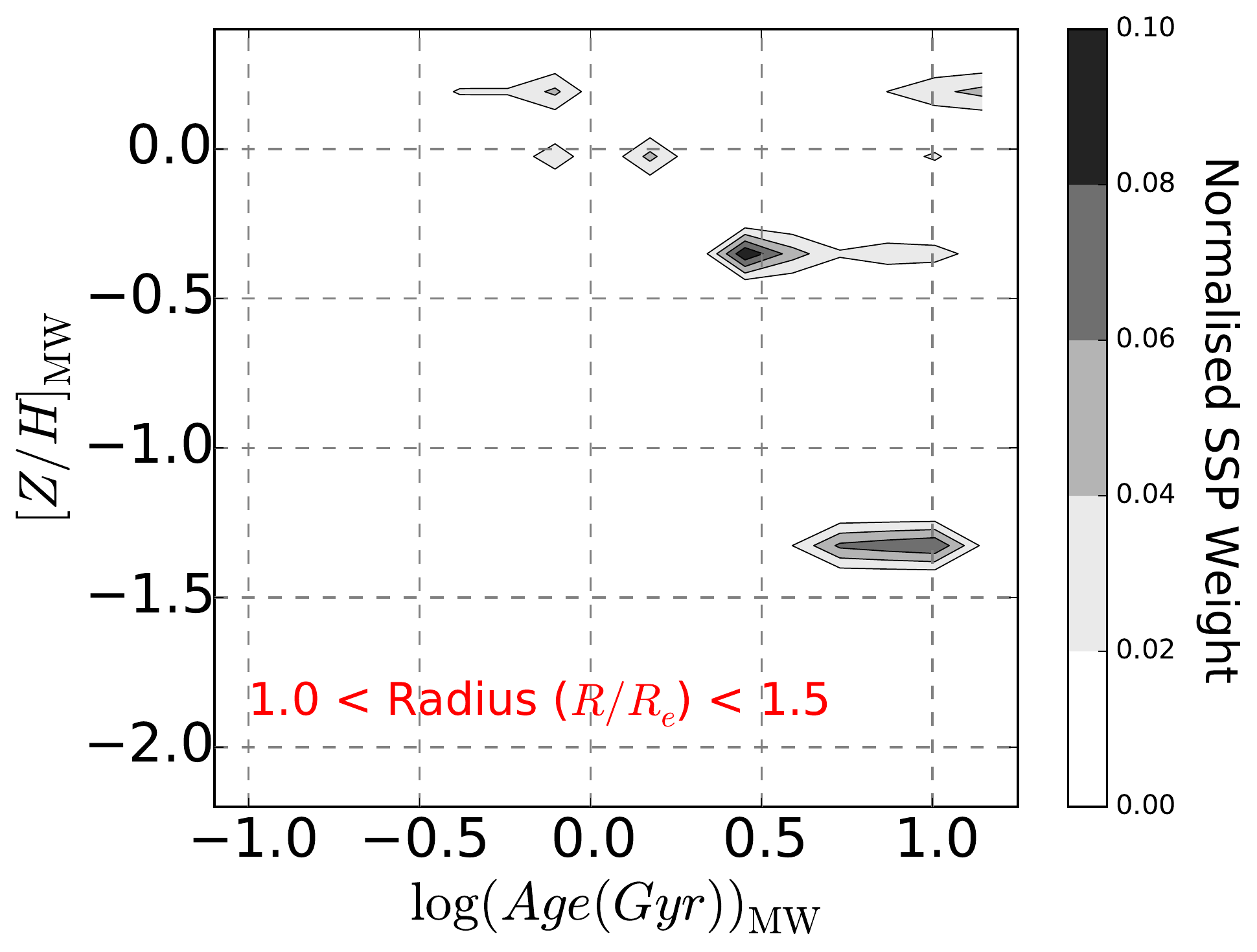}
\includegraphics[width=0.33\textwidth]{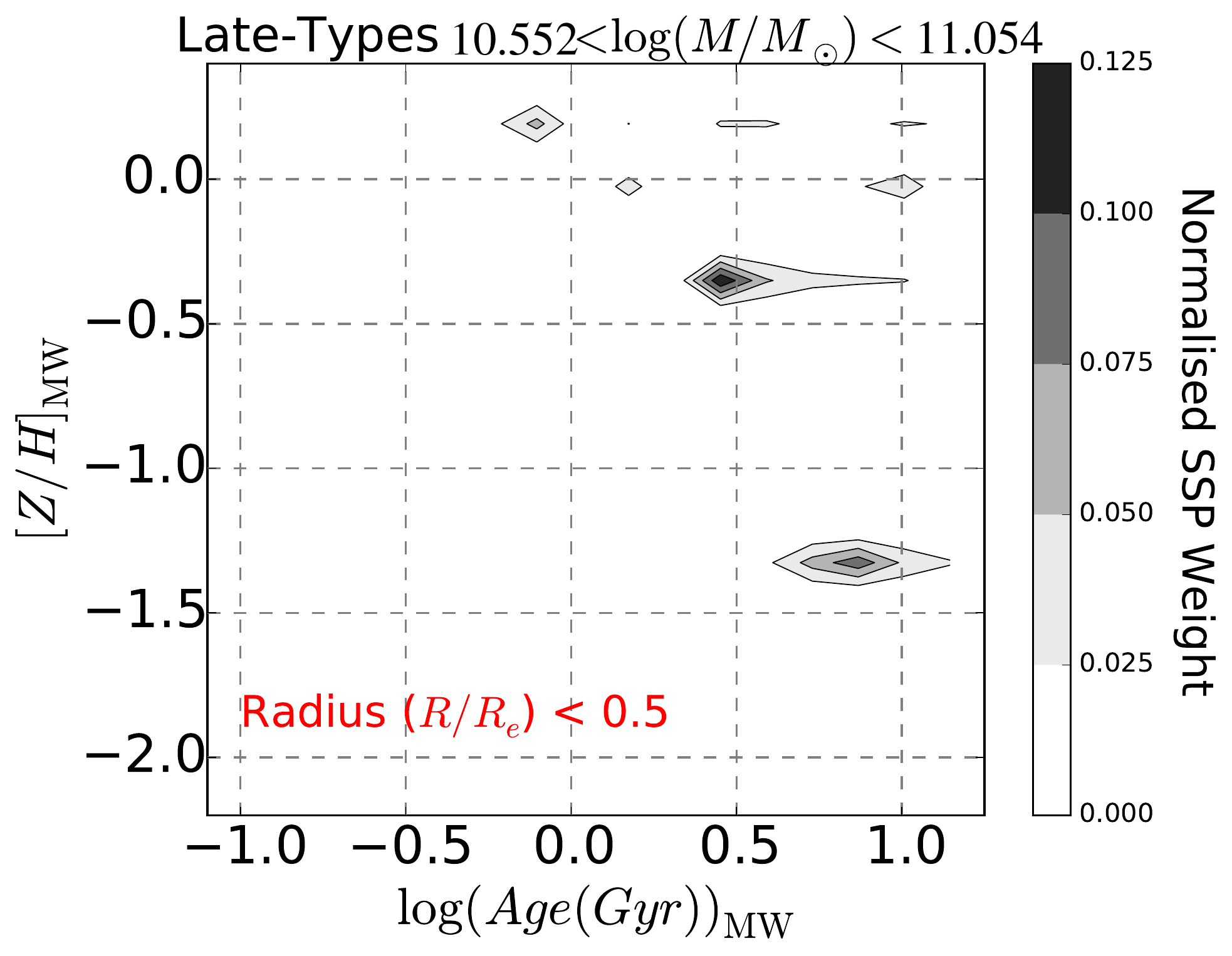}
\includegraphics[width=0.33\textwidth]{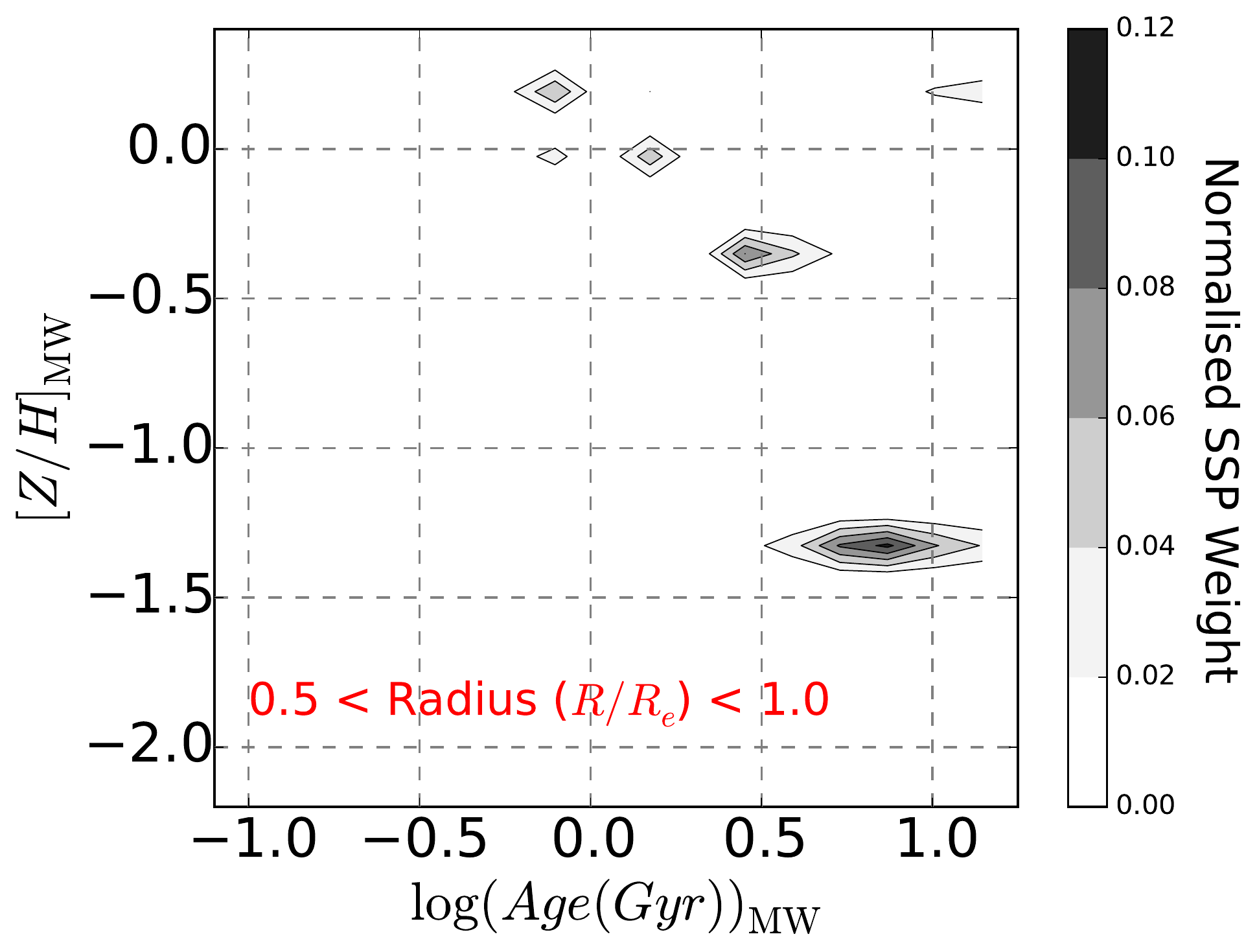}
\includegraphics[width=0.33\textwidth]{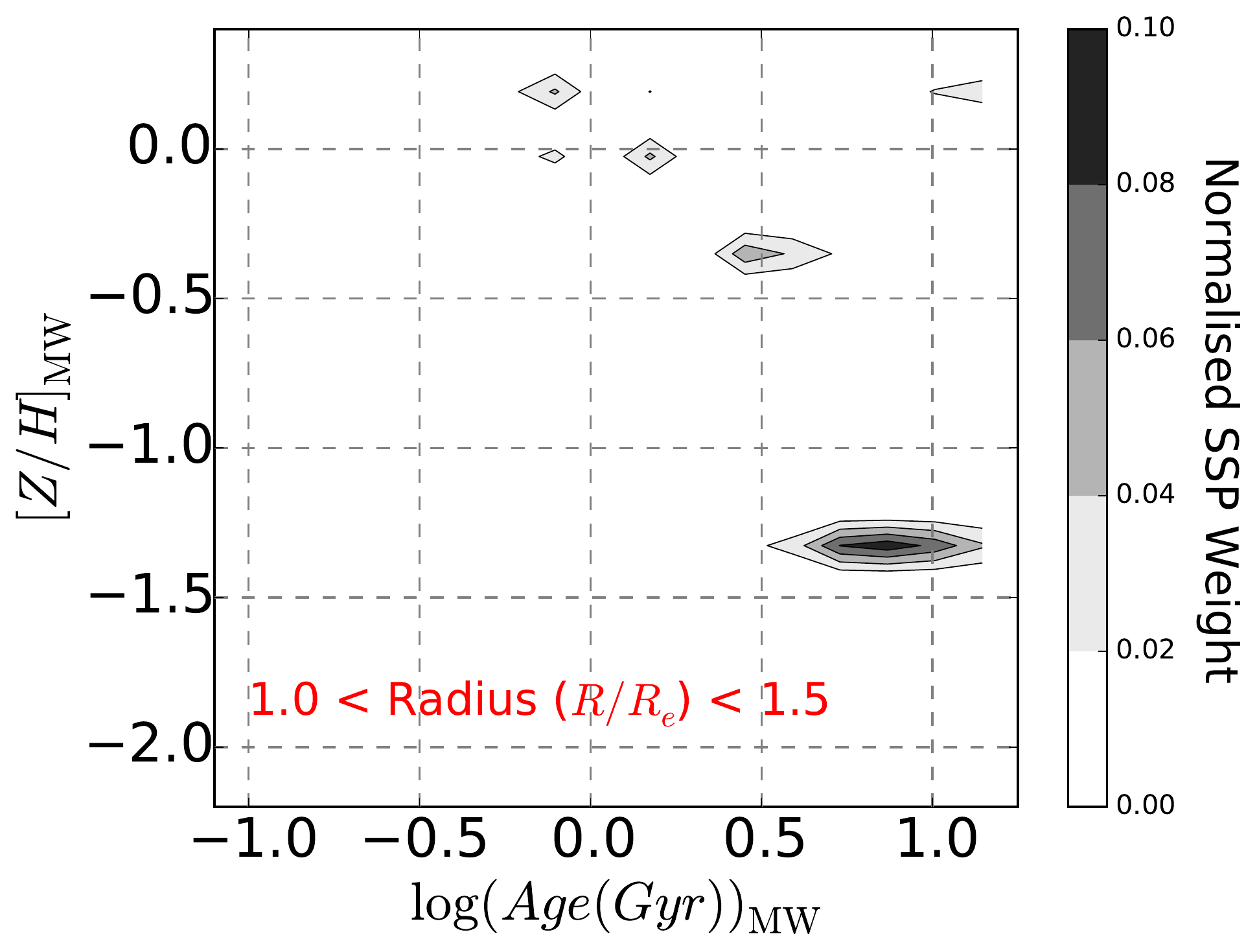}
\includegraphics[width=0.33\textwidth]{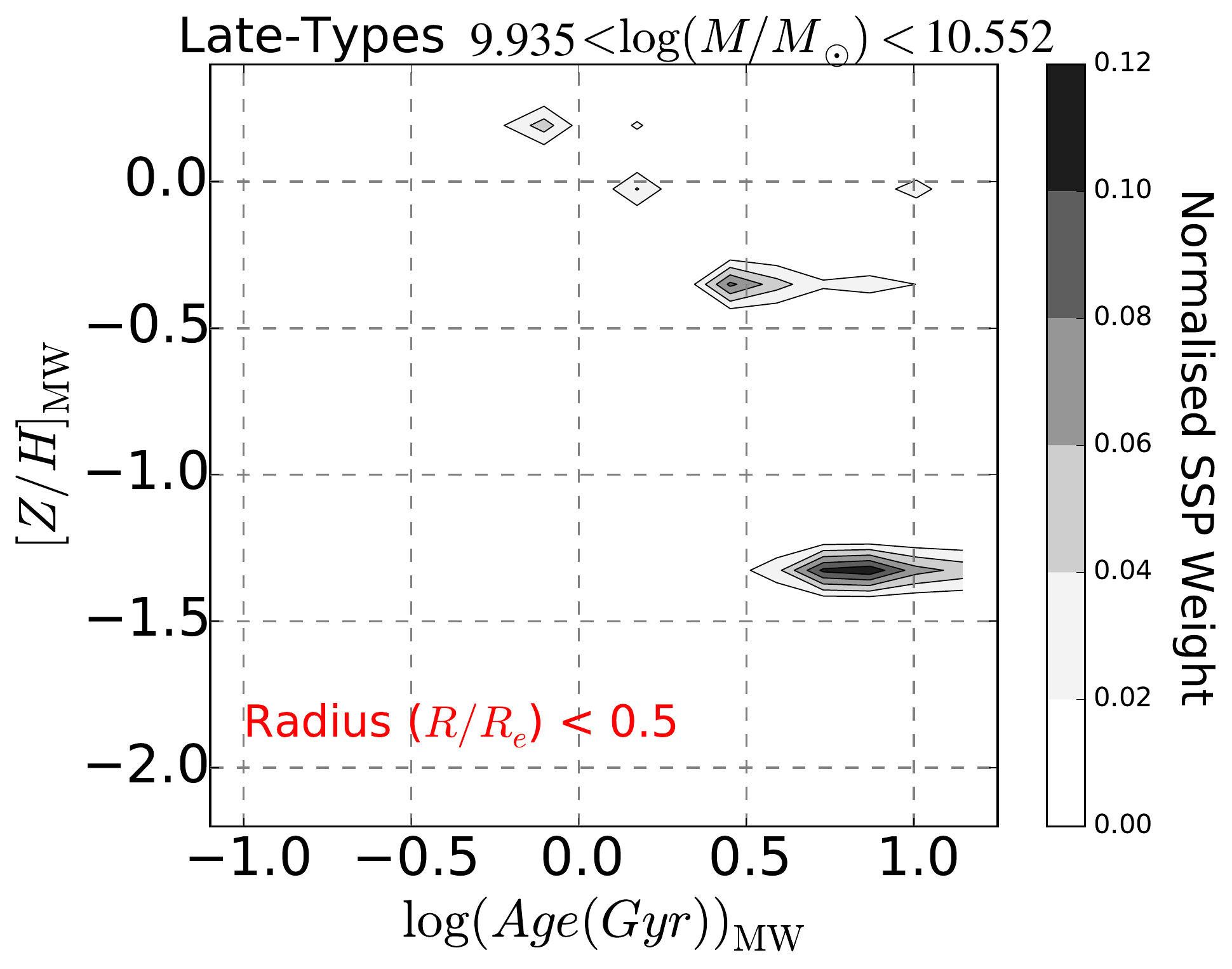}
\includegraphics[width=0.33\textwidth]{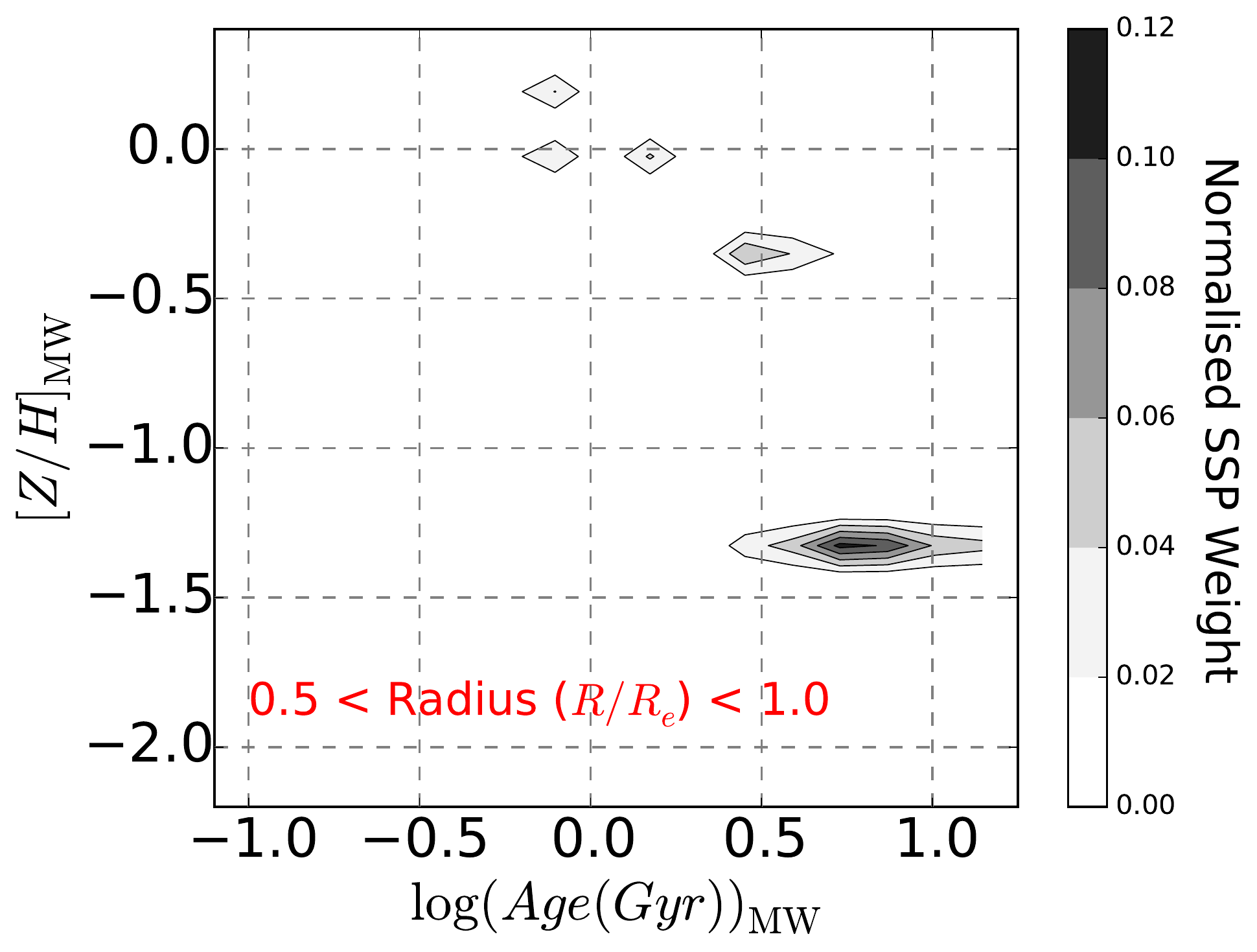}
\includegraphics[width=0.33\textwidth]{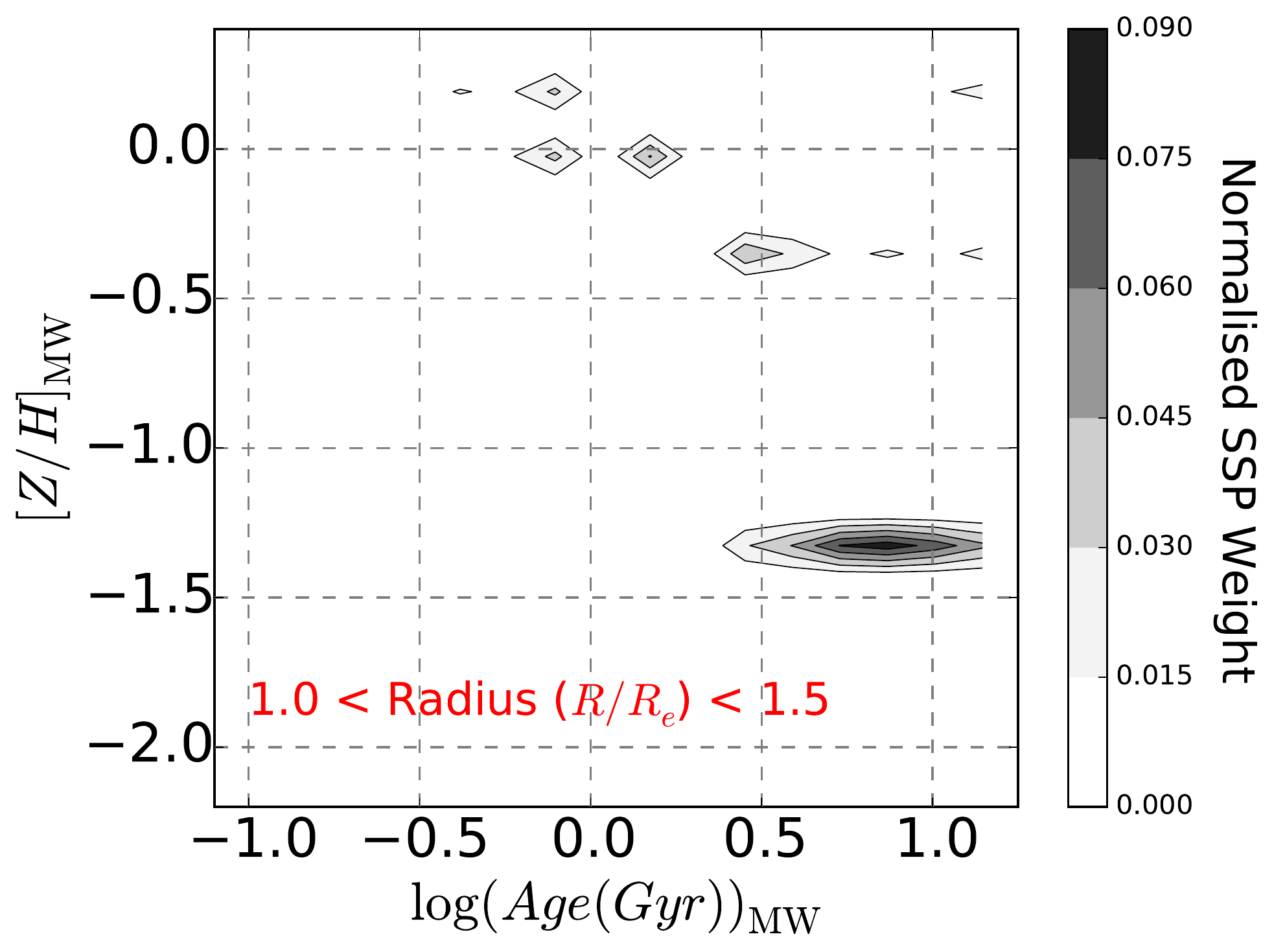}
\includegraphics[width=0.33\textwidth]{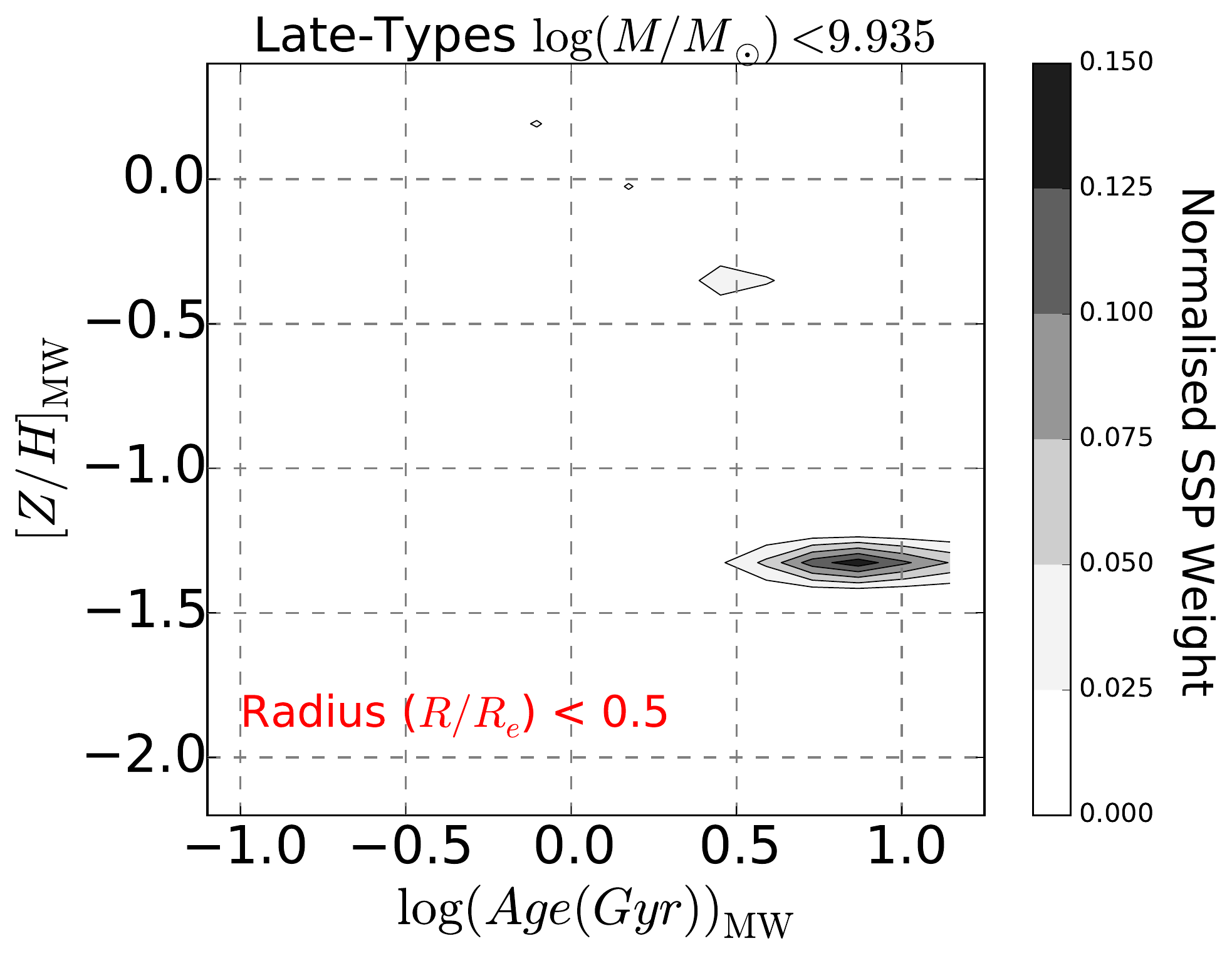}
\includegraphics[width=0.33\textwidth]{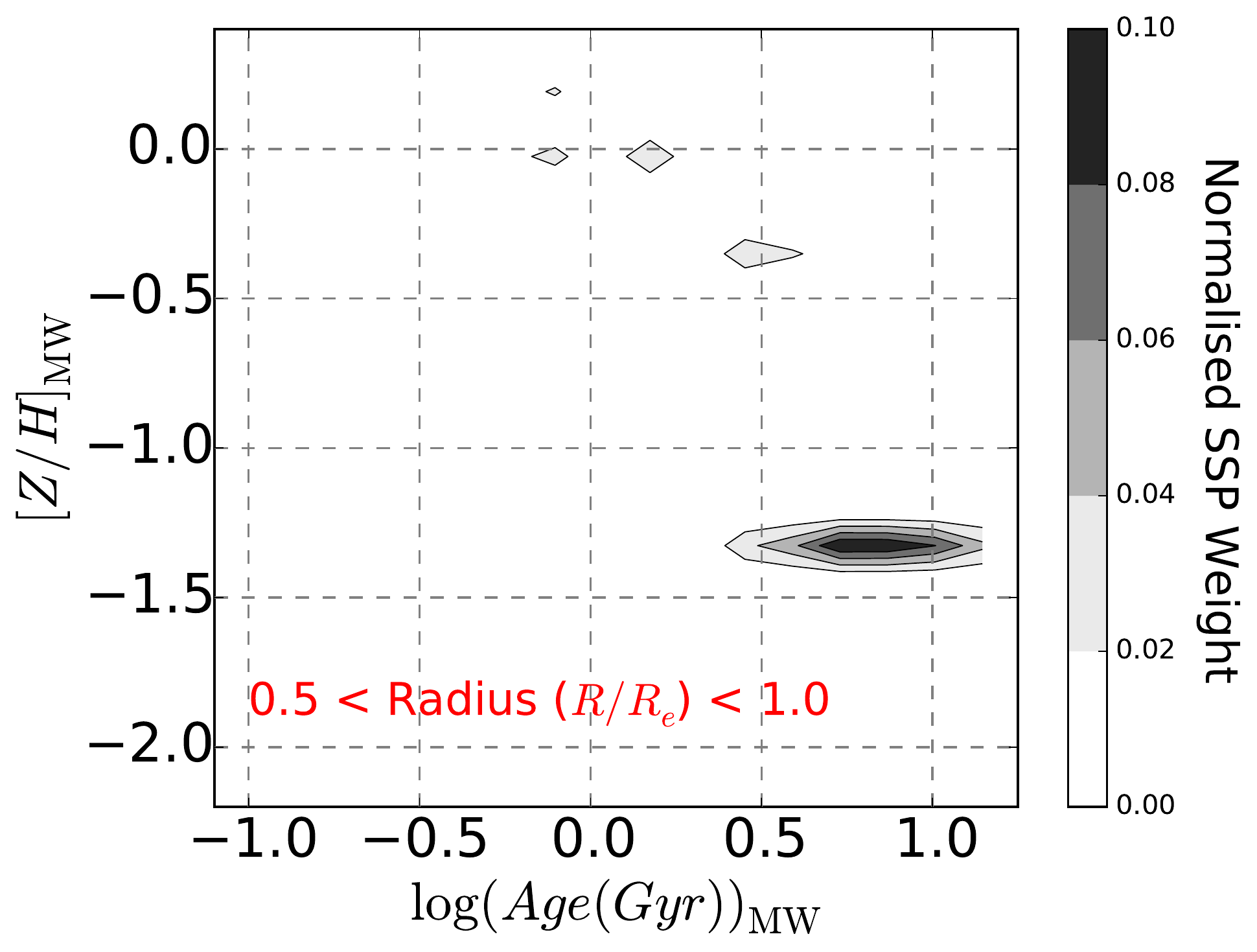}
\includegraphics[width=0.33\textwidth]{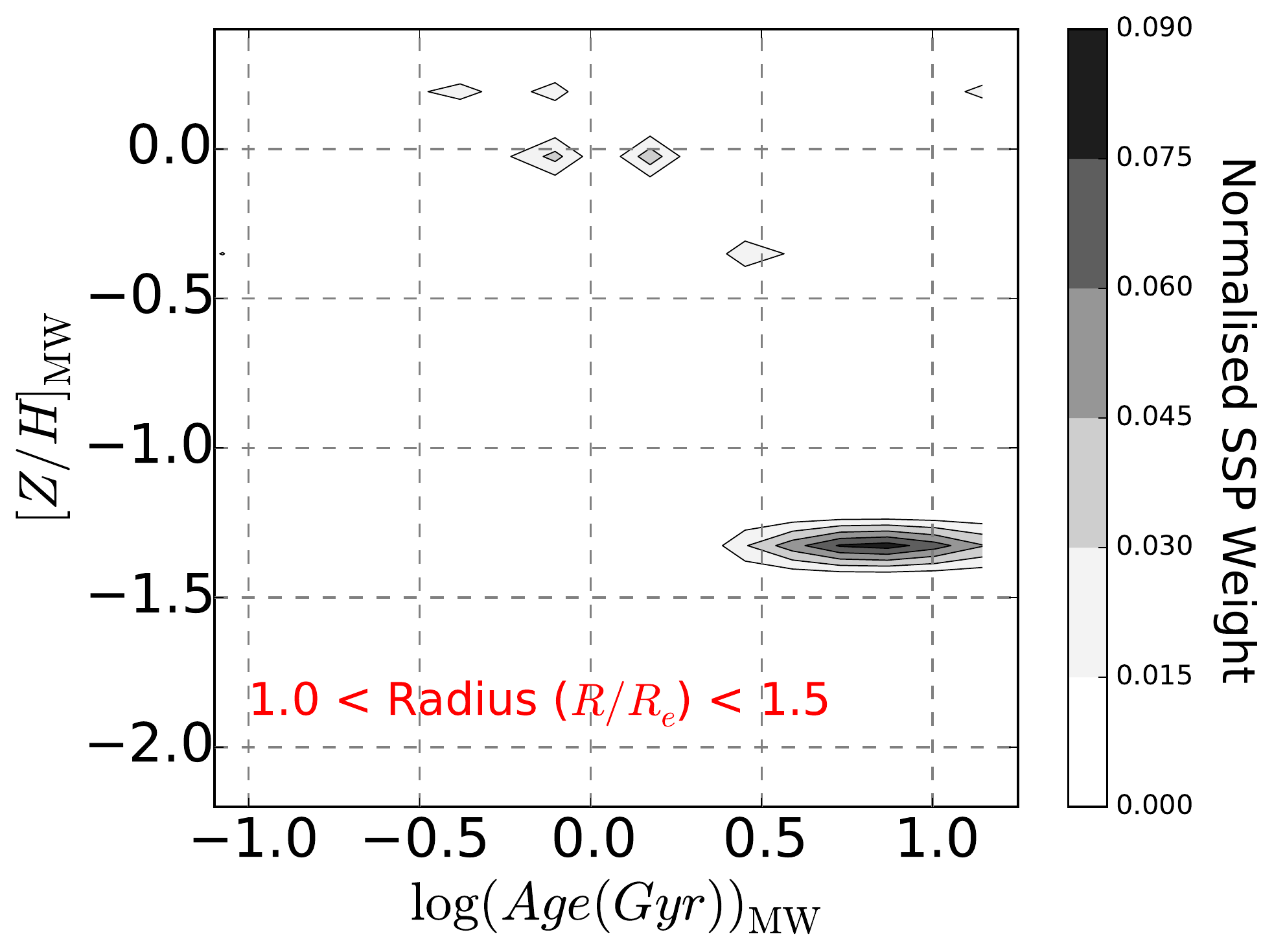}
\caption{Star formation and metal enrichment histories for late-type galaxies as function of galaxy mass and radius. The density scale indicates the relative {\em mass}-weights of the stellar populations in the spectral fit in age-metallicity space. The columns are three radial bins with radius increasing from left to right (see labels), rows are the four mass bins with mass increasing from bottom to top (see labels). See Figure~\ref{fig:sfh_lt_lw} for light-weighted quantities. For more details see Section 4.2}
\label{fig:sfh_lt_mw}
\end{figure*}
\begin{figure*}
\includegraphics[width=0.33\textwidth]{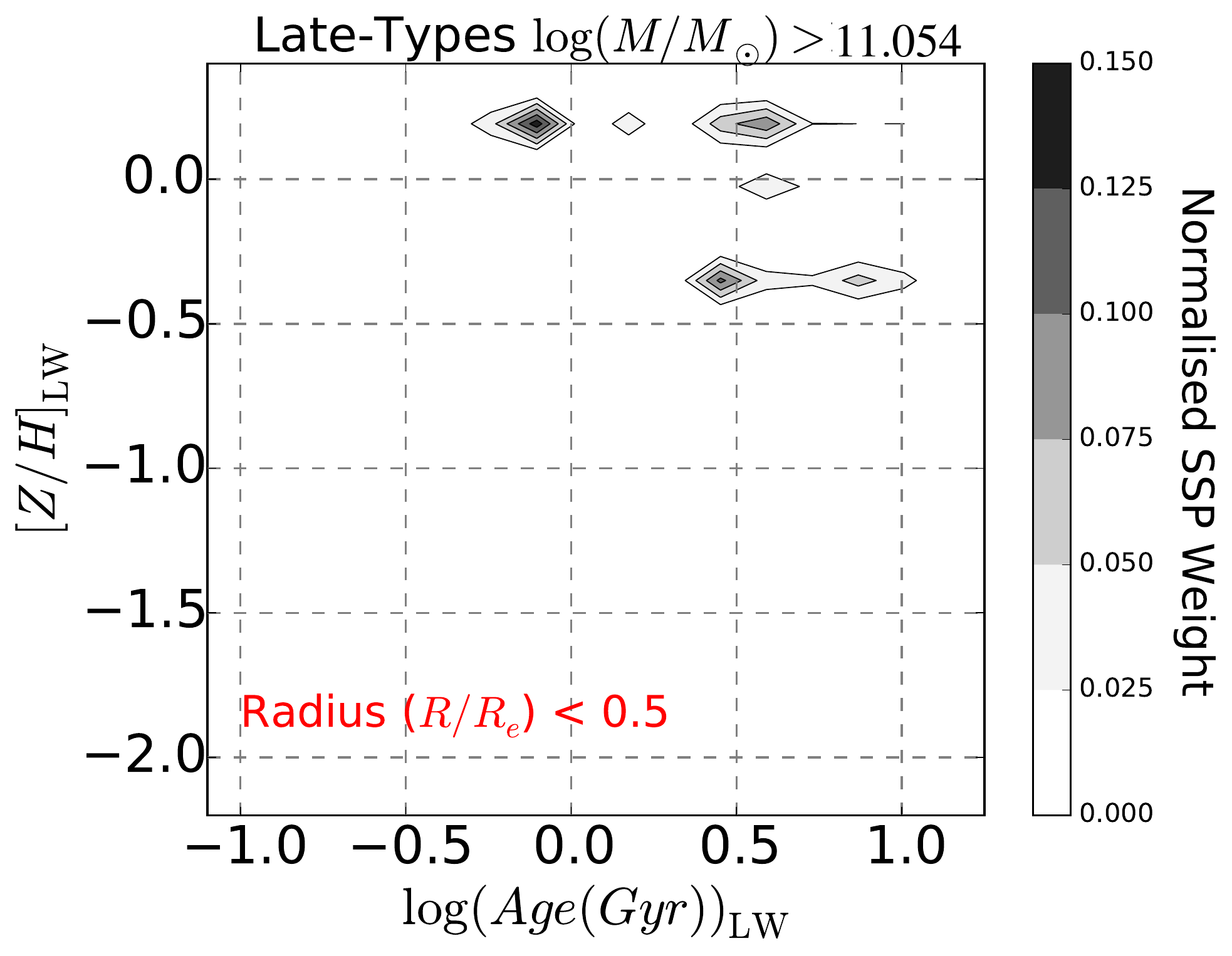}
\includegraphics[width=0.33\textwidth]{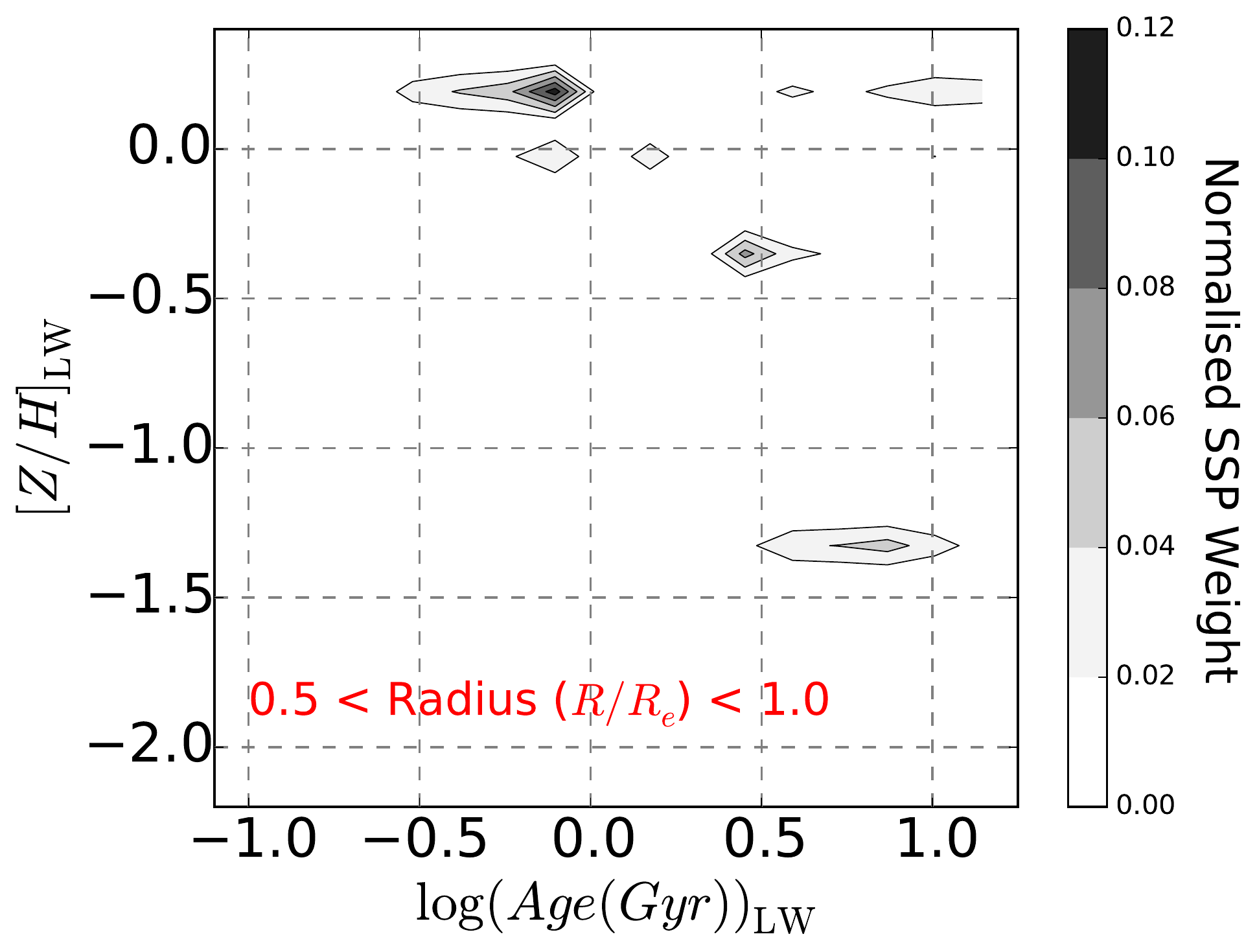}
\includegraphics[width=0.33\textwidth]{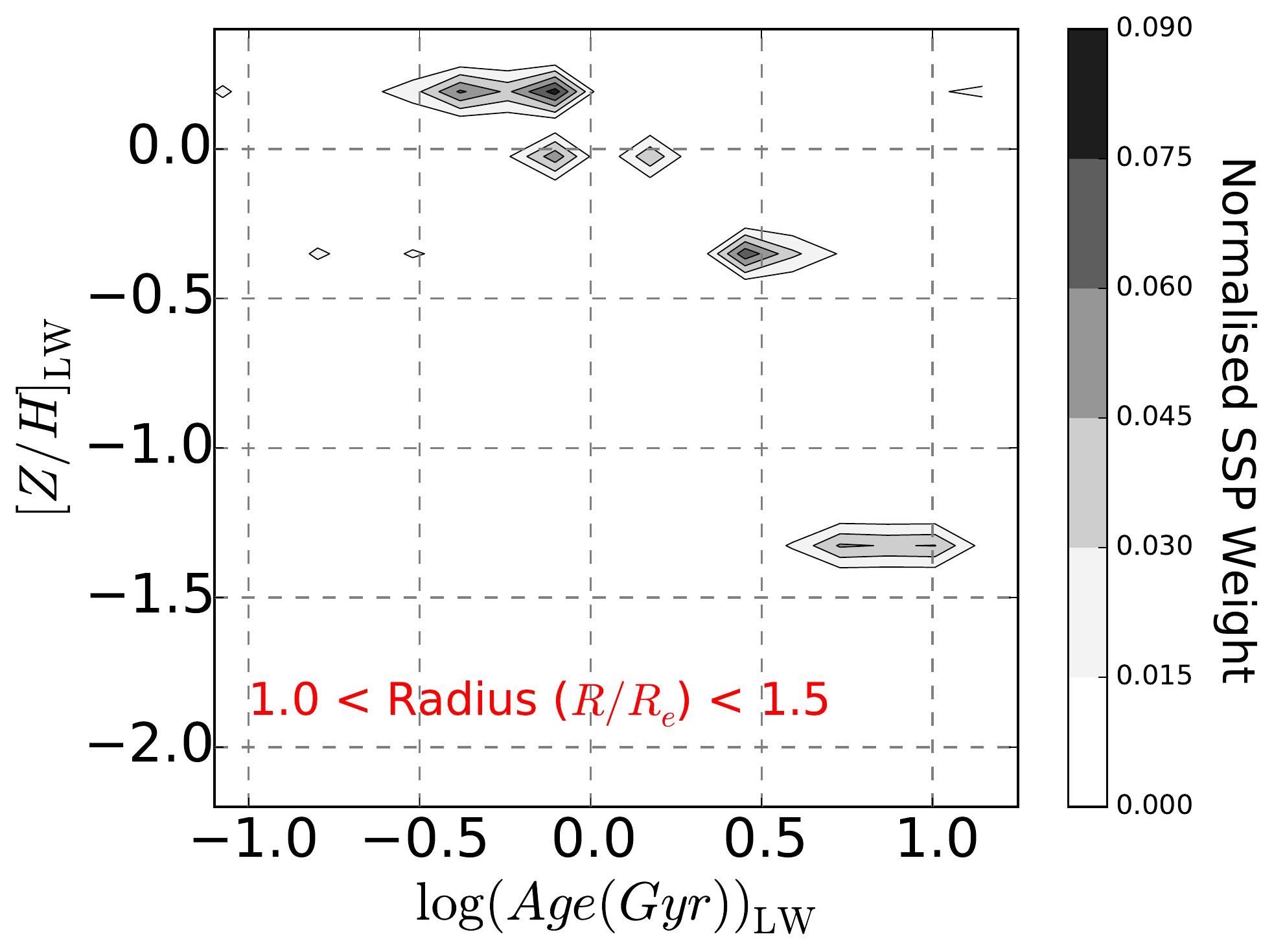}
\includegraphics[width=0.33\textwidth]{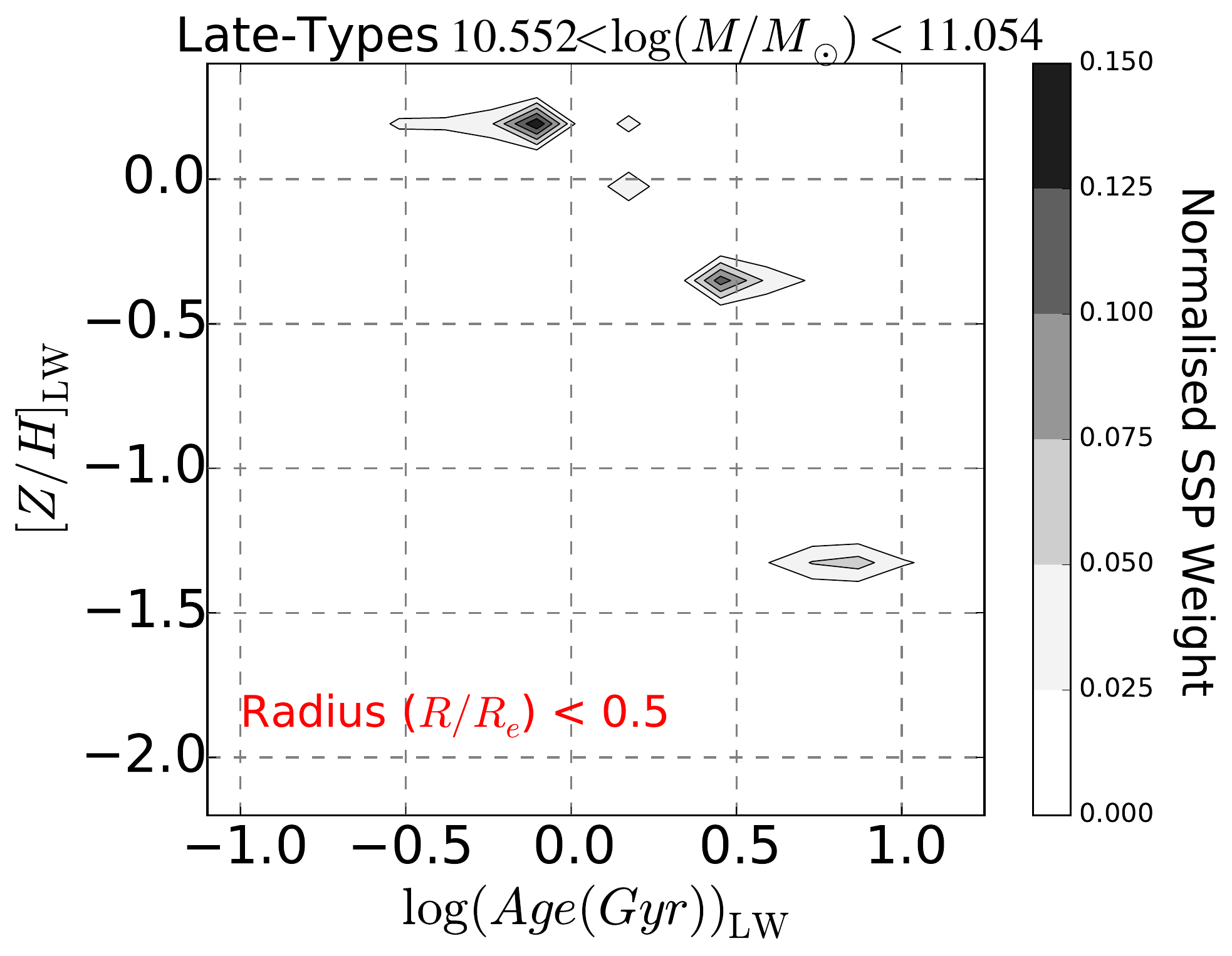}
\includegraphics[width=0.33\textwidth]{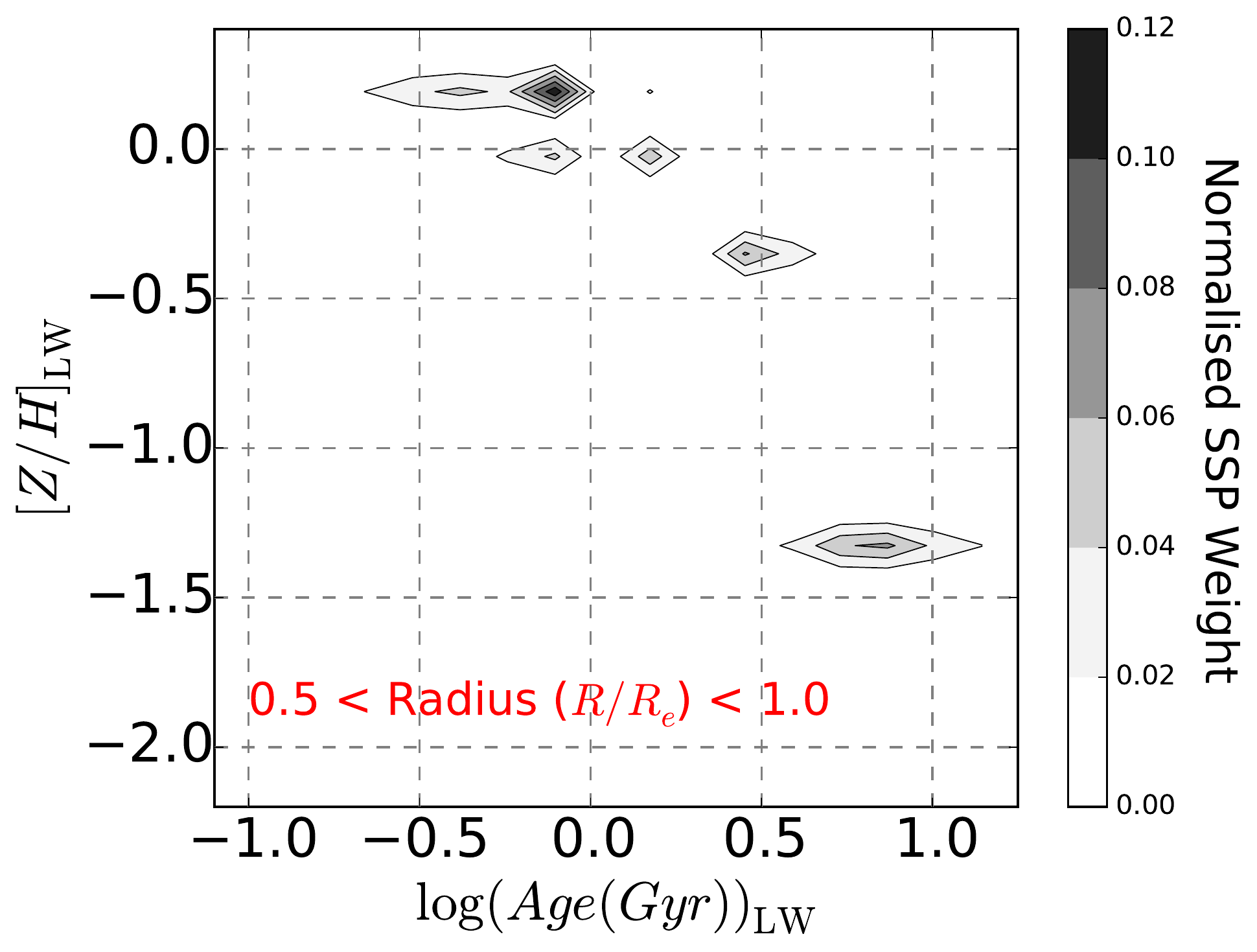}
\includegraphics[width=0.33\textwidth]{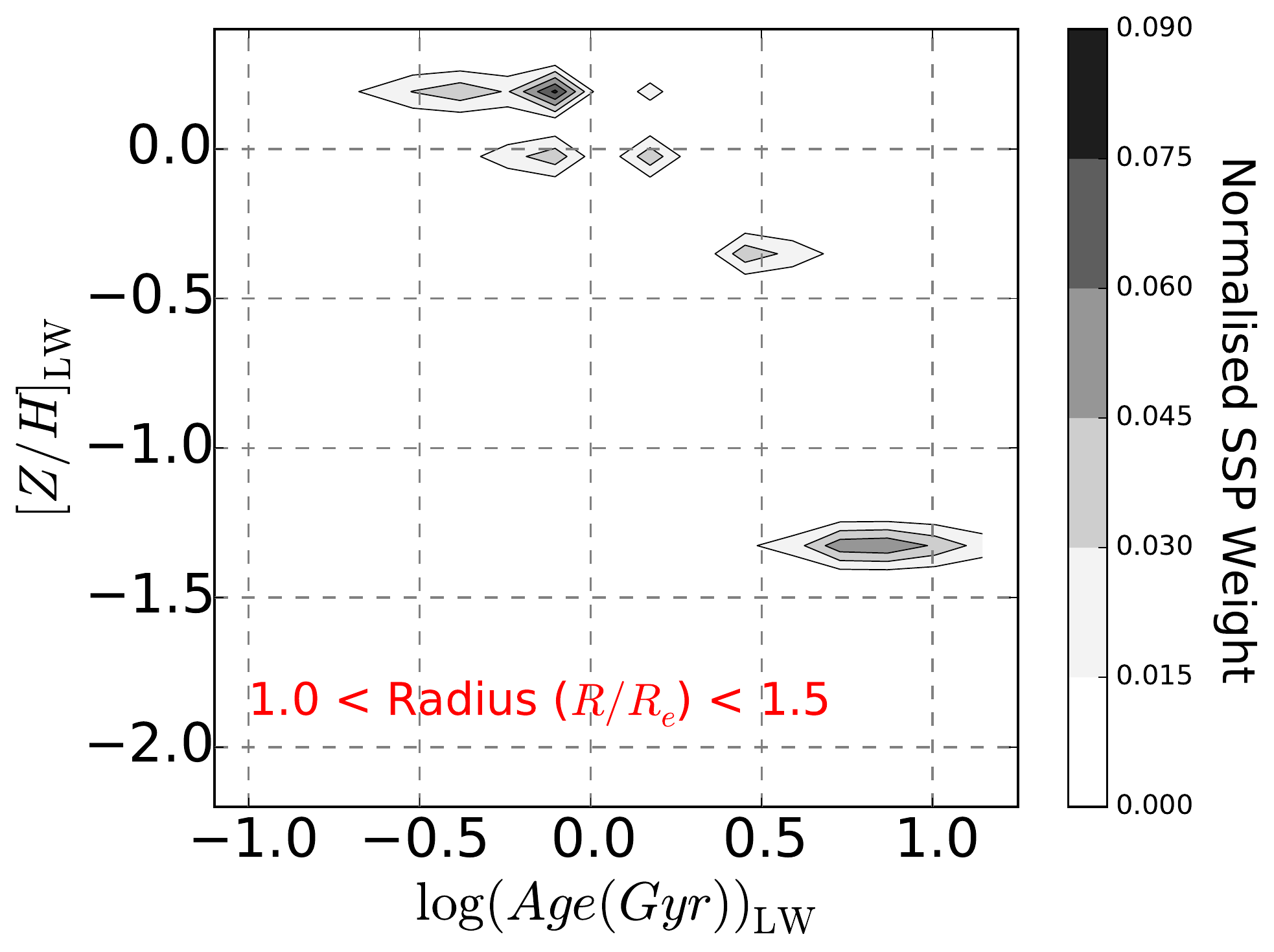}
\includegraphics[width=0.33\textwidth]{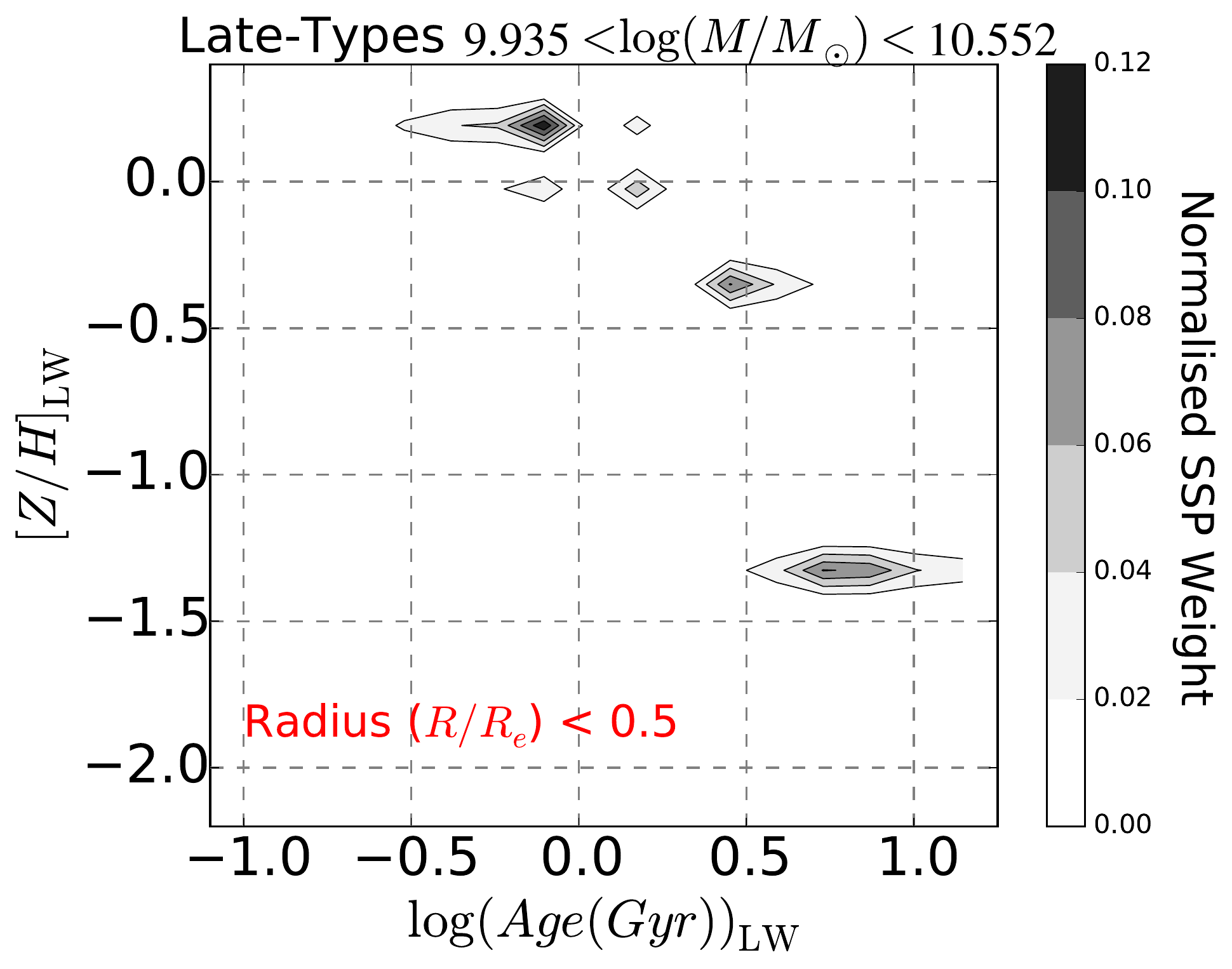}
\includegraphics[width=0.33\textwidth]{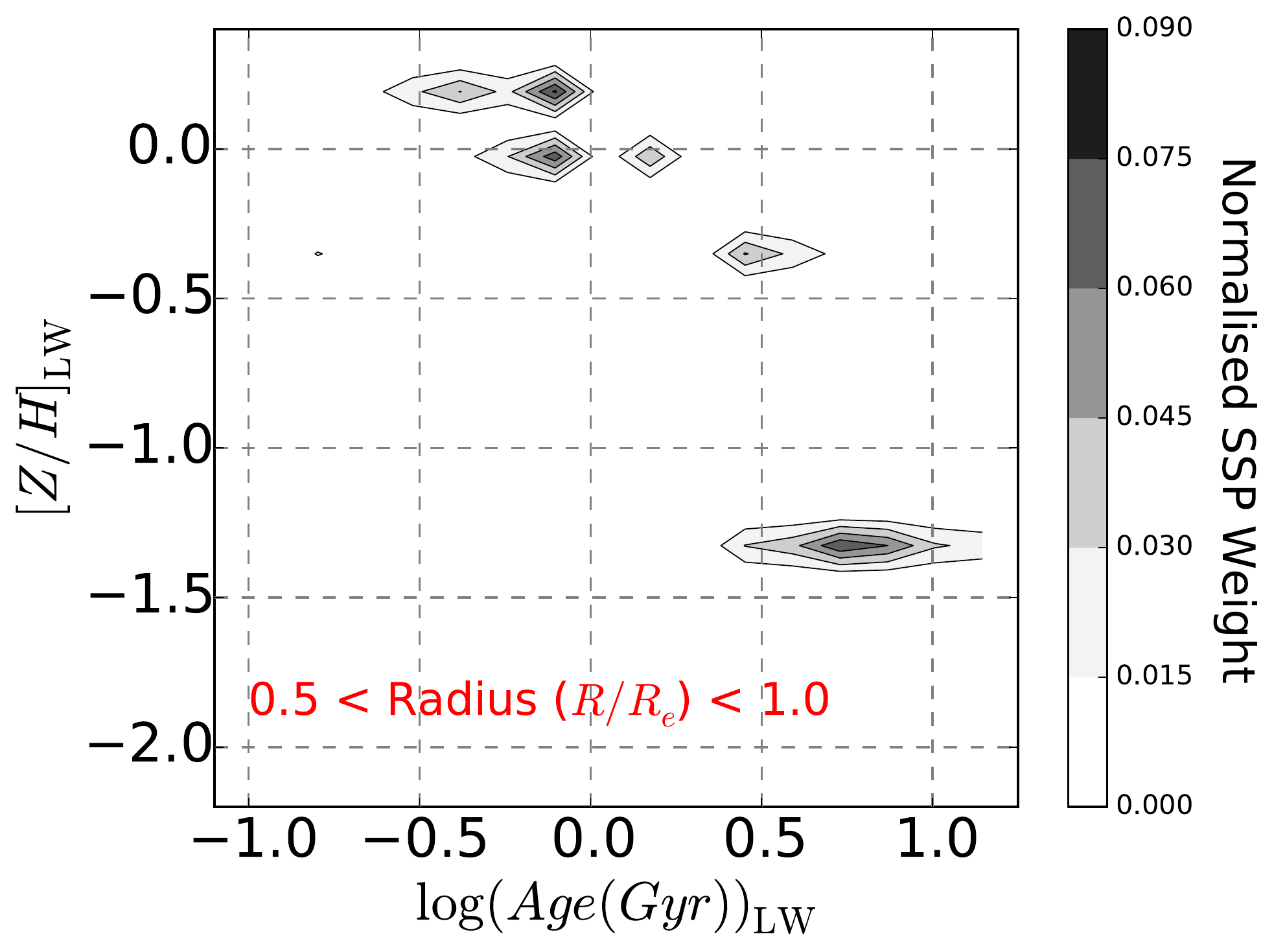}
\includegraphics[width=0.33\textwidth]{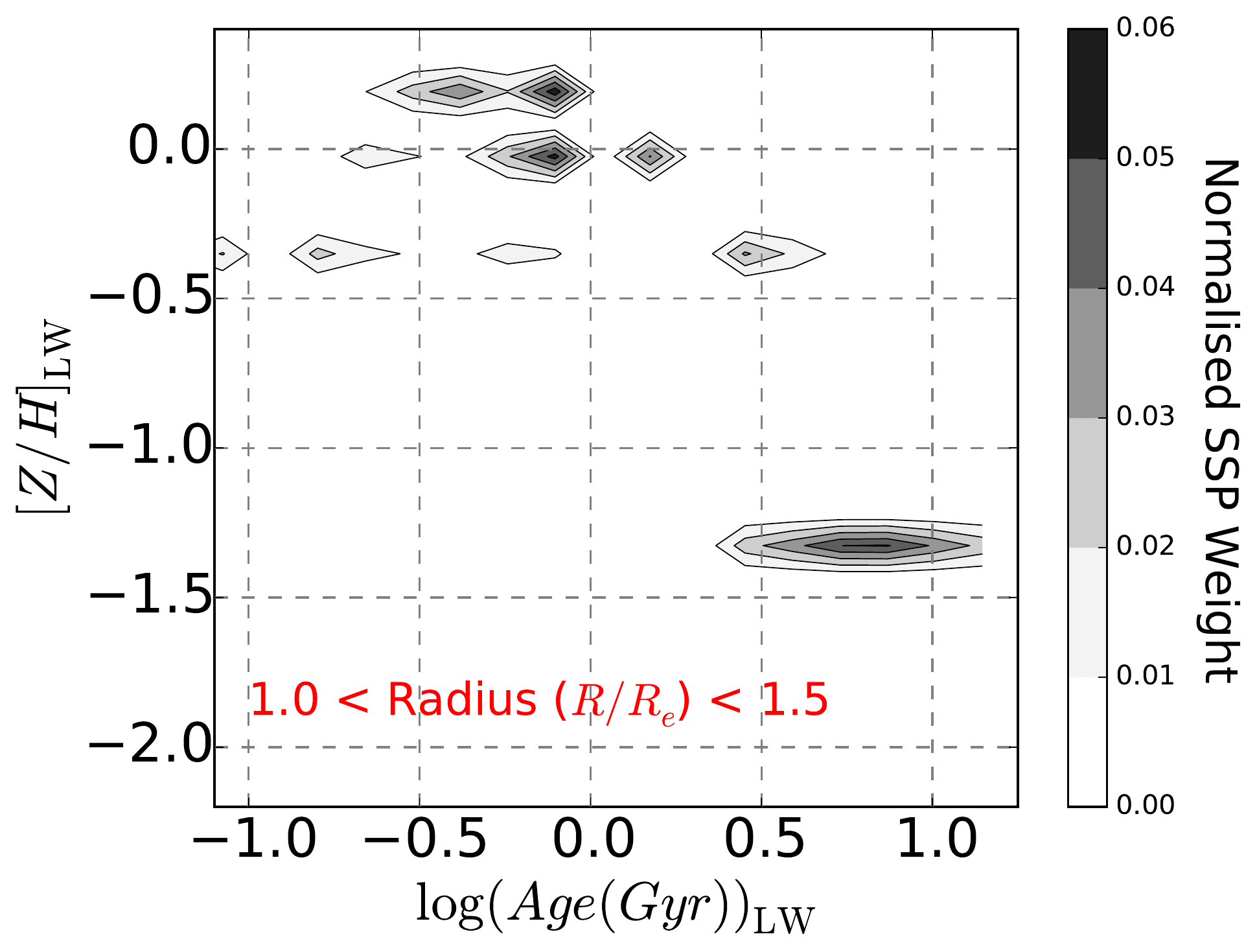}
\includegraphics[width=0.33\textwidth]{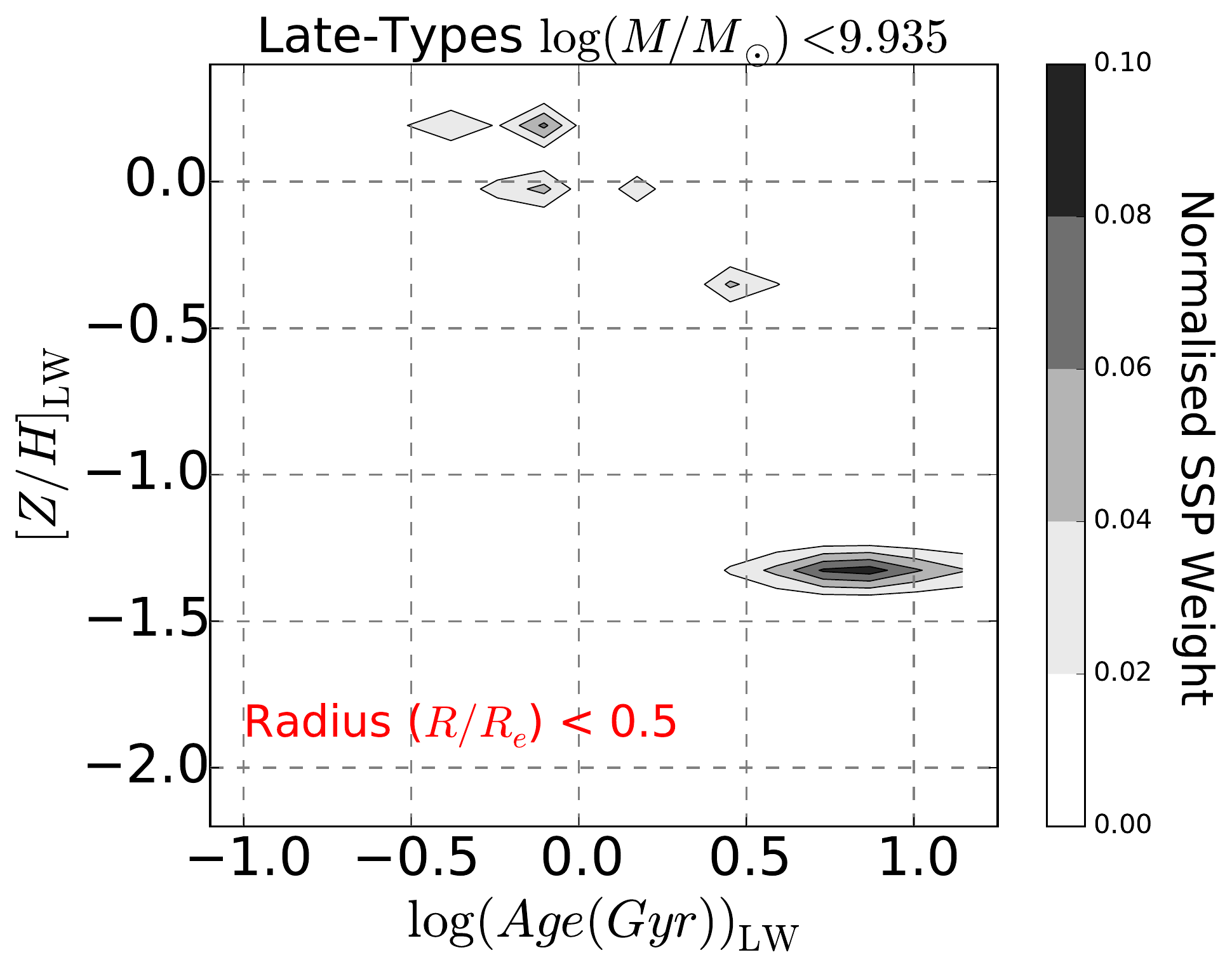}
\includegraphics[width=0.33\textwidth]{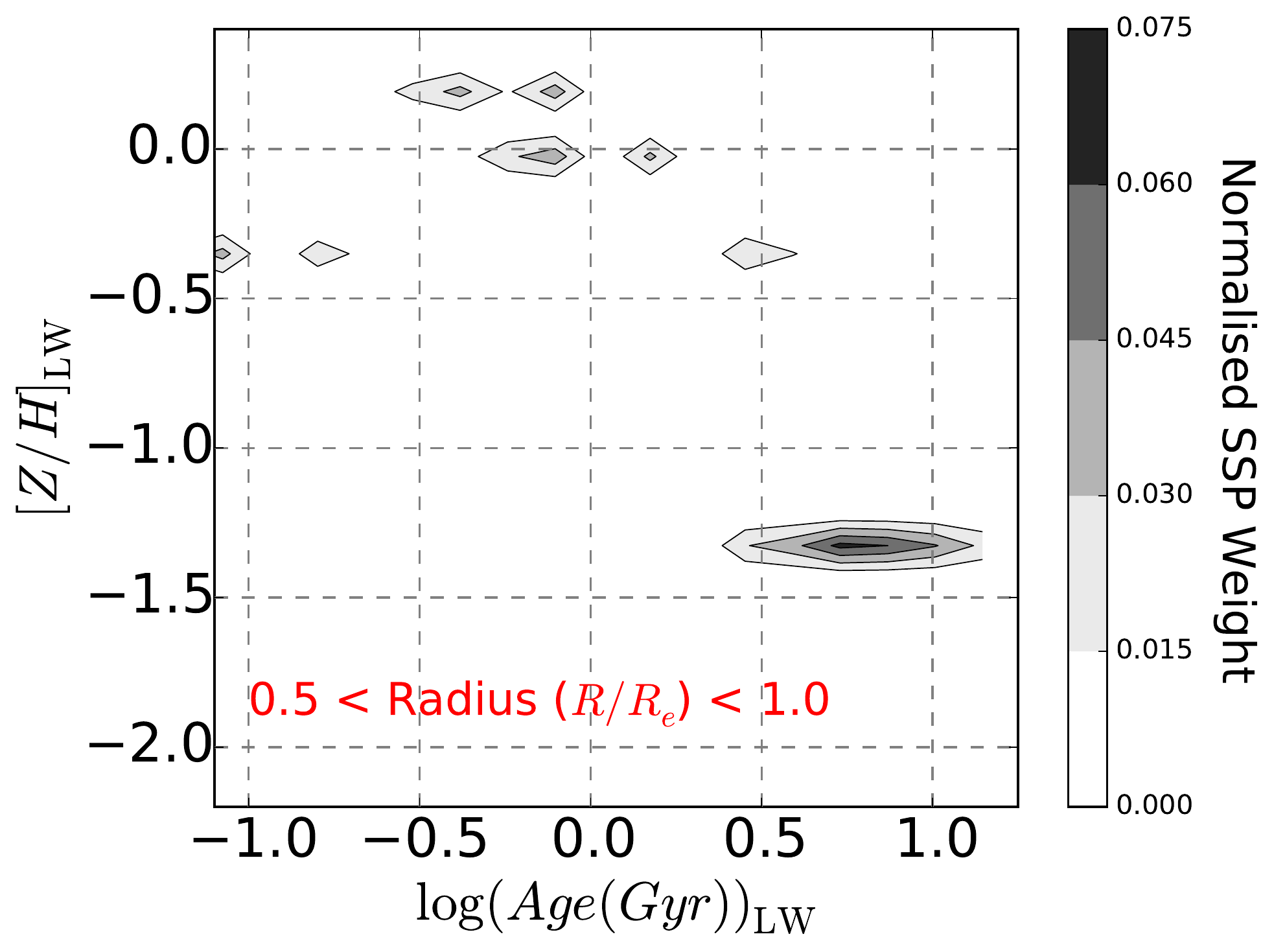}
\includegraphics[width=0.33\textwidth]{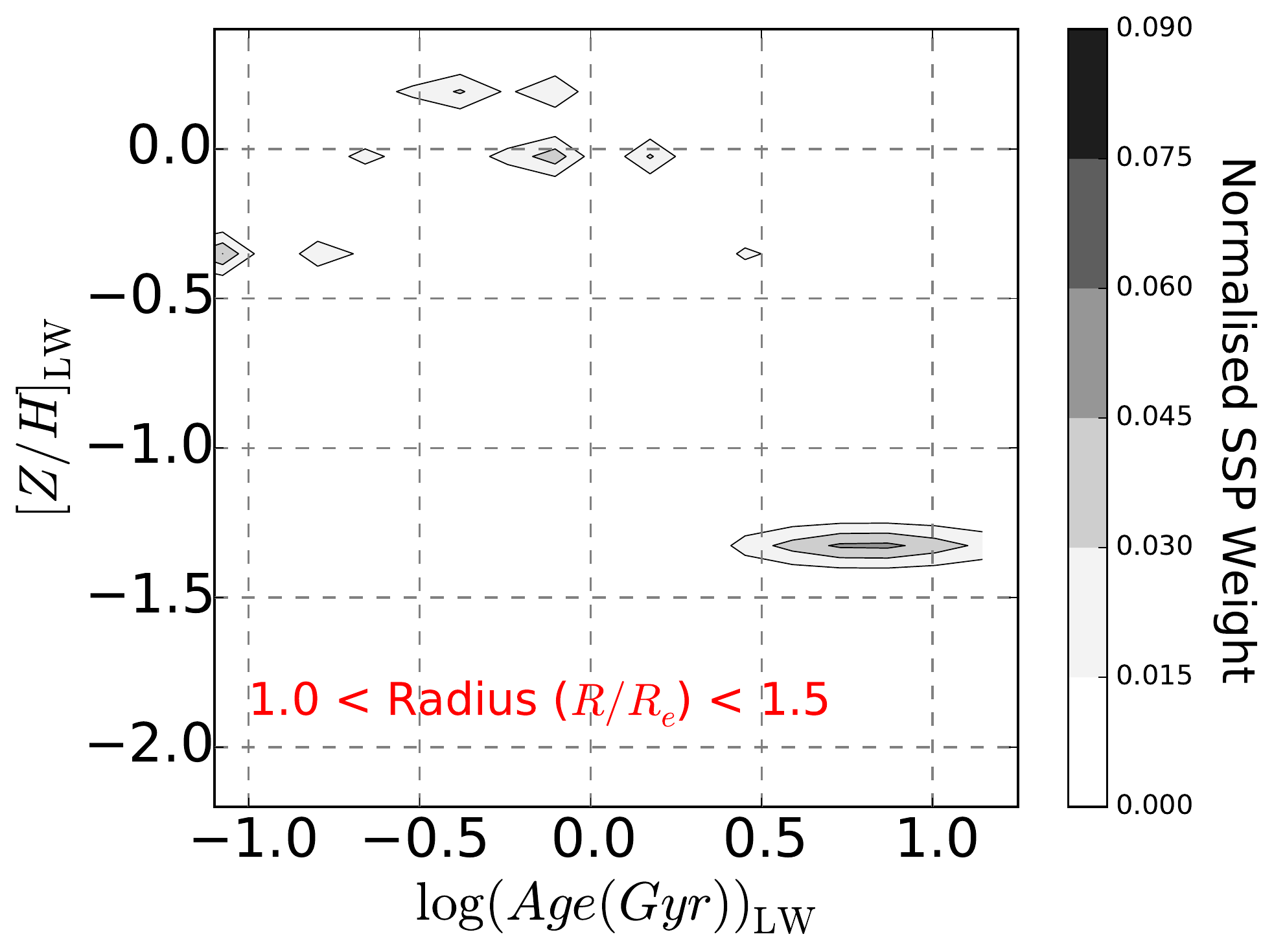}
\caption{Star formation and metal enrichment histories for late-type galaxies as function of galaxy mass and radius. The density scale indicates the relative {\em luminosity}-weights of the stellar populations in the spectral fit in age-metallicity space. The columns are three radial bins with radius increasing from left to right (see labels), rows are the four mass bins with mass increasing from bottom to top (see labels). See Figure~\ref{fig:sfh_lt_mw} for mass-weighted quantities. For more details see Section 4.2.}
\label{fig:sfh_lt_lw}
\end{figure*}
\subsubsection{Metallicity Gradients}
The bottom panels of Figure~\ref{fig:gradients} show the gradients in metallicity for early and late-type galaxies. Early-type galaxies generally show negative gradients in metallicity at all galaxy masses. The light-weighted metallicity gradient around $\sim -0.12\;$dex/$R_{\rm e}$ is slightly shallower but in qualitative agreement with previous literature \citep{saglia2000,mehlert2003,kuntschner2010,rawle2010,greene2012,gonz2015}. A more detailed comparison of this result and the literature can be found in the Discussion. Light and mass-weighted metallicities and their radial dependence are virtually indistinguishable, with an average offset of $\sim 0.05$ dex. This is because there is no well-defined age-metallicity correlation in the (predominantly old) stellar populations of early-type galaxies (see Section~\ref{sec:sfh_et}). This ought to be expected as the stellar populations are generally old in early-type galaxies with no significant recent star formation \citep[e.g.][]{thomas2010}. Our study shows that this is indeed the case within at least 1.5 effective radii. Although visually the light and mass-weighted metallicities appear indistinguishable, there is a mild difference between the gradients obtained as a function of mass. Figure~\ref{fig:gradients} shows that the mass-weighted metallicity gradient becomes more negative as mass increases, ranging from flat gradients in the lowest mass bins ($\sim 0.03$ dex/$R_{\rm e}$) to negative gradients in the most massive ($\sim -0.1$ dex/$R_{\rm e}$). A trend not seen in the light-weighted metallicity.\\
\\
Late-type galaxies metallicity measurements are very different to early-types in this regard. The luminosity-weighted metallicities are systematically larger than mass-weighted by about $\sim 0.35\;$dex. This must be the direct consequence of ongoing chemical enrichment producing young metal-rich stellar populations. We find that this process is largely independent of galactic radius, with the only exception being the highest mass bin with $\log (M/M_{\odot})>11.054$, where the discrepancy between luminosity-weighted and mass-weighted metallicities becomes very small ($\sim 0.09$ dex); quite similar to early-type galaxies. This may be because the highest mass bin is populated by the most bulge-dominated late-type galaxies or due to small number statistics (see Table~\ref{table:sample_classification}) . Both the light-weighted and mass-weighted metallicity gradients are significantly negative at all galaxy masses (except the lowest mass bin at $\log(M/M_{\odot})<9.935$). The negative slope in light-weighted metallicity becomes steeper with increasing galaxy mass, ranging from $\sim -0.1\;$dex/$R_{e}$ in the second lowest mass bin to $\sim -0.3\;$dex/$R_{e}$ for the most massive late-types. The gradients and the trend of steepening slope with galaxy mass is even more pronounced for mass-weighted metallicities, leading to a metallicity gradient of $\sim -0.48\;$dex/$R_{e}$ at the highest masses. These metallicity gradients in the stellar populations of galaxy discs are in good agreement with the radial gradients generally found in the stellar abundances of the Milky Way \citep{carollo2007,hayden2015}, as well as in the gas metallicities of the Milky Way and other disc galaxies \citep{vilchez1996, ho2015, letizia2015}. These results show that the radial dependence of chemical enrichment processes and the effect of gas inflow/metal transport are far more pronounced in late-type galaxies than they are in early-types.

\subsubsection{Dust Gradients}
Figure~\ref{fig:dust_radius} shows the median radial profiles of $E(B-V)$ for both early (left panel) and late-types galaxies (right panel), split by galaxy mass. Early-type galaxies generally have a very small amount of dust and exhibit shallow, relatively flat radial profiles. From the left panel of Figure~\ref{fig:dust_radius}, it is also clear to see that the central part of the galaxy ($R < 1R_{\rm e}$) has a marginally steeper radial profile that the outermost regions ($R > 1R_{\rm e}$) due to the higher dust values found in the most central part. By performing a linear fit to the median radial profiles, we find that the gradient values (in order of ascending mass) are $-0.015 \pm 0.03$, $-0.036 \pm 0.05$, $-0.031 \pm 0.04$ and $0.007 \pm 0.02$ mag/$R_{e}$, respectively. These values are close to flat and there appears to be no obvious trend of radial gradient and galaxy mass.
\\
\\ 
Late-type galaxies generally contain more dust at the centre than early-types, and have radial profiles which trend with stellar mass. Performing a linear fit to the median radial profiles, we find that the gradient values (in order of ascending mass) are $0.00 \pm 0.01$, $-0.03 \pm 0.03$, $-0.06 \pm 0.04$ and $-0.09 \pm 0.05$ mag/$R_{e}$, respectively. The reason for this steepening of gradient as a function of mass is largely driven by the central $E(B-V)$ values. For the most massive late-type galaxies ($\log(M/M_{\odot}) > 11.054$), central $E(B-V)$ values are $\sim 0.18$ mag and this value declines with decreasing mass, eventually reaching a value of $\sim 0.03$ mag in the lowest mass bin ($\log(M/M_{\odot}) < 9.935$). At radii larger than $1R_{e}$ however, the $E(B-V)$ values tend to converge around a similar value of $\sim 0.03$ mag with some scatter. This result suggests that the gravitational potential well in late-type galaxies is key to retaining dust in the central region and that this accumulation of dust drives the radial gradients. In addition to this, the radial profile in the central part of the galaxy ($R < 1R_{\rm e}$) is steeper than in the outermost regions. However, this does not hold true for the lowest mass bin.
\\
\\
The dust gradient results presented here are in good agreement with what was found by the CALIFA survey \citep{gonz2015}. The CALIFA survey found that for galaxies in the mass range $9.1 < \log(M/M_{\odot}) < 10.6$, which is dominated by late-type galaxies, the most massive galaxies had more dust in their central regions and steeper radial gradients. The gradient values range from $\sim 0$ mag/$R_{e}$ for the least massive galaxies, to $\sim -0.1$ mag/$R_{e}$ in the most massive, agreeing quantitatively with what we found for late-type galaxies\footnote{In order to compare our results for $E(B-V)$ gradients to the ones obtained in \citet{gonz2015}, we transform their $A_{\mathrm{V}}$ values into $E(B-V)$ using $R_{\mathrm{V}}$, which is defined as  $A_{\mathrm{V}}/E(B-V)$.}. For early-type galaxies, central $E(B-V)$ values are lower than what is found in late-type galaxies and the dust profiles differ between the inner and outer parts of the galaxy, with radial profiles within $1R_{\rm e}$ being slightly negative and flat at $R > 1R_{\rm e}$. Once again, agreeing with what we present here. In addition to this, the more massive early-type galaxies have less dust in their central region and flatter radial gradients. This trend is fairly weak in our work, due to the large degree of scatter between the mass bins but the most massive early-types (black line) do appear to have less dust in their central regions.

\subsection{Resolved Star Formation Histories}
\label{sec:sfh}
As discussed in the previous Section and shown in Figure~\ref{fig:gradients}, interesting differences exist between light and mass-weighted stellar population properties. In this Section, we present the resolved star formation histories in the age-metallicity plane to shed more light on this variation and to tighten constraints on formation scenarios. To this end, we split the sample into the three radial bins ($R/R_{\rm e}<0.5$, $0.5<R/R_{\rm e}<1.0$, and $1.0<R/R_{\rm e}<1.5$), the same four mass bins as in Figure~\ref{fig:gradients}, as well as early and late-type. The resulting star formation and metal enrichment histories are presented in Figures~\ref{fig:sfh_et_mw} to \ref{fig:sfh_lt_lw}. Mass-weighted and luminosity-weighted quantities for early-type galaxies are shown in Figures~\ref{fig:sfh_et_mw} to \ref{fig:sfh_et_lw}, while Figures~\ref{fig:sfh_lt_mw} to \ref{fig:sfh_lt_lw} present the same for late-types.

\subsubsection{Early-Type Galaxies}
\label{sec:sfh_et}
From Figure~\ref{fig:sfh_et_mw} it can be seen that early-type galaxies are dominated by old stellar populations with some spread in metallicity at all radii. With decreasing galaxy mass (top to bottom in Figure~\ref{fig:sfh_et_mw}), the component of old, metal-poor stellar populations gains weight in mass. At the same time a minor young, metal-rich component also rises with decreasing galaxy mass. Both these trends are in line with the mass-metallicity (MZR) and mass-age relationships of early-type galaxies \citep[e.g.]{2005MNRAS.362...41G, panter2008, thomas2010, gonz2014, gonz2015} and these results will be disseminated further in a forthcoming paper (Goddard et al. in prep). Both these effects are even stronger in light-weighted quantities as shown in Figure~\ref{fig:sfh_et_lw}.\\
\\
Figure~\ref{fig:sfh_et_mw} further reveals the radial dependence: in massive early-types ($\log (M/M_{\odot})>10.552$) the outermost radial bins are dominated by old stellar populations ($t\sim 10\;$Gyr), while a component of intermediate-age ($t\ga 3\;$Gyr), metal-rich populations is present in the centre. This leads to the positive age gradient presented in Figure~\ref{fig:gradients} suggesting "outside-in" progression of star formation. As galaxy centres are more metal-rich than the outskirts, this "outside-in" formation is more likely caused by continuing (or additional) star formation from enriched material \citep{bedregal2011}, rather than by recent star formation activity due to late-time accretion. Also in agreement with Figure~\ref{fig:gradients}, these positive age gradients appear flatter at lower masses, where intermediate-age, metal-rich components are also present at large radii. Some very small contribution of such intermediate-age populations can be seen also in the two massive mass bins in the light-weighted quantities (Figure~\ref{fig:sfh_et_lw}), which explains why luminosity-weighted age gradients are found to be flat. This result suggests that some very small residual star formation is found at large radii, which however does not contribute to the overall mass budget.\\
\\
The positive age gradients implying "outside-in" progression of star formation agree with previous results in the literature (see Introduction, Table~\ref{table:gradient_values} and Figure~\ref{fig:literature_gradients}), and also fit well into recent work by \citet{johnston2012,johnston2014} who find that the bulges of lenticular galaxies in Fornax contain consistently younger and more metal-rich stellar populations than their surrounding discs. This conclusion has also been recently reproduced in an analysis of MaNGA data \citep{johnston2016}, and it will be interesting in future to compare directly with the present sample. \\
\\
Mass-weighted metallicity gradients are shown to be relatively flat in Figure~\ref{fig:gradients}, which is consistent with the fact that the proportion of the metal-rich population does not seem to vary significantly with radius in Figure~\ref{fig:sfh_et_mw}. In light-weighted space, however, the young metal-rich component in the centre as well as an old metal-poor component at large radii becomes more evident (Figure~\ref{fig:sfh_et_lw}), which is in line with the generally negative light-weighted metallicity gradients in Figure~\ref{fig:gradients}.

\subsubsection{Late-Type Galaxies}
The resolved star formation histories of late-type galaxies are distinctly different from the ones of early-types in mass-weighted space. The presence of young, metal-rich populations at all masses and all radial bins also for the mass-weighted quantities is striking (Figure~\ref{fig:sfh_lt_mw}). And in fact a well-defined age-metallicity relation is detected with old, metal-poor and young, metal-rich populations, similar to what is observed in the Milky Way \citep[e.g.][]{edvardsson1993}. This metal enrichment pattern is even more pronounced in light-weighted space (Figure~\ref{fig:sfh_lt_lw}). To our knowledge, this is the first time that a full spectral fitting code has resolved the chemical enrichment history for integrated populations in large galaxy samples.\\
\\
The mass-weighted age gradient is generally flat, because the presence of young populations is equally pronounced at all radii (Figure~\ref{fig:sfh_lt_mw}). However, some slight excess of star formation is detectable in the luminosity-weighted quantities (Figure~\ref{fig:sfh_lt_mw}), which leads to the negative age gradients in late-type galaxies discussed above and shown in Figure~\ref{fig:gradients}. The metal-rich component is particularly evident in the inner radial bins, which explains the strong metallicity gradients observed for late-type galaxies both in luminosity and mass-weighted space (Figure~\ref{fig:gradients}). These gradients become stronger with increasing mass, because the central metal-rich component is more evident in the higher mass bins as can be seen from Figures~\ref{fig:sfh_lt_mw} and~\ref{fig:sfh_lt_lw}.

\subsubsection{Summary}
To conclude this analysis, we find evidence for "outside-in" progression of star formation in early-type galaxies, and only very mild gradients in metallicity. Late-type galaxies exhibit strong metallicity gradients and mild evidence for "inside-out" formation with some small excess of recent star formation activity at large radii. The stellar populations of late-type galaxies follow a well defined age-metallicity relationship spanning several Gyr in age at all radii, while early-type galaxies are dominated by old populations at all radii with significant scatter in metallicity.

\subsection{Correlations with Galaxy Mass}
\begin{figure*}
\includegraphics[width=0.49\textwidth]{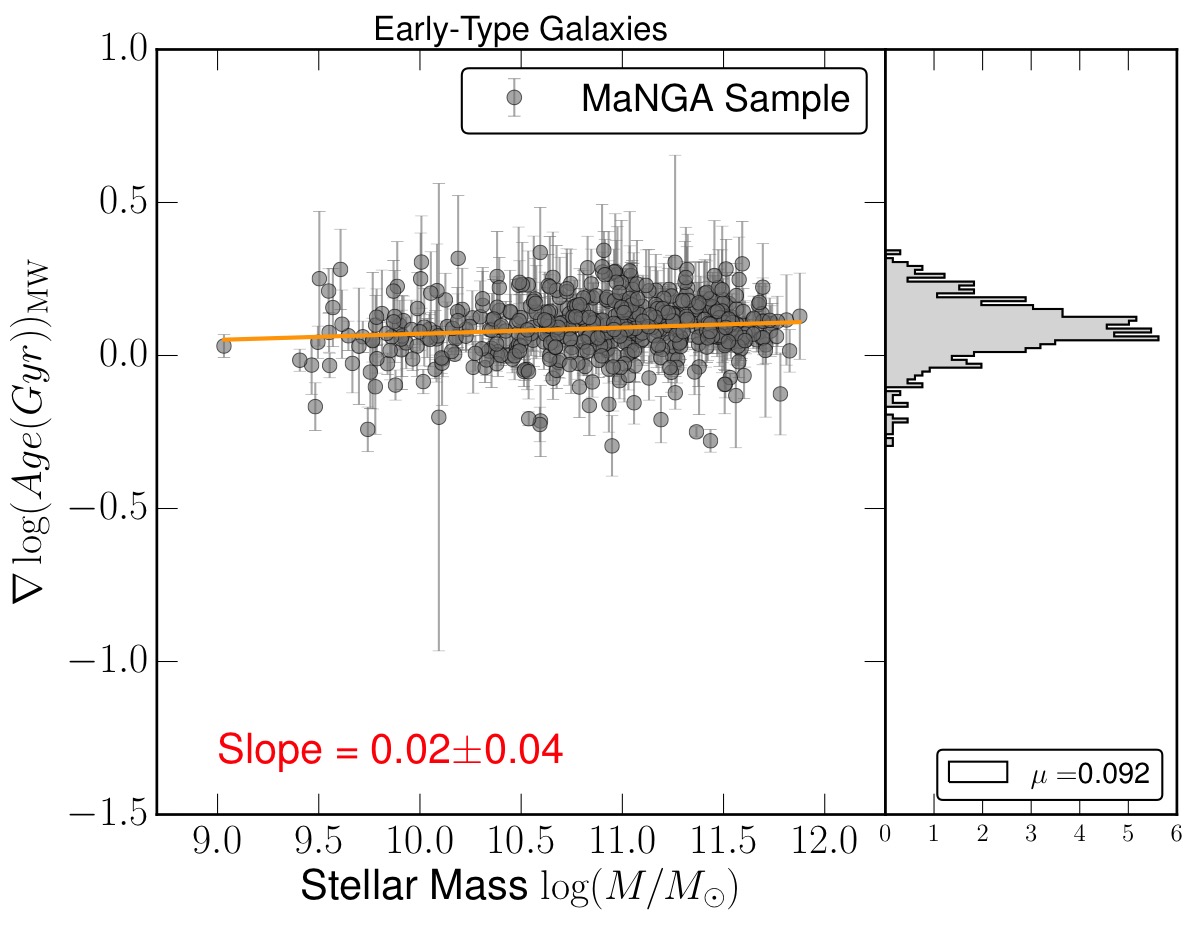}
\includegraphics[width=0.49\textwidth]{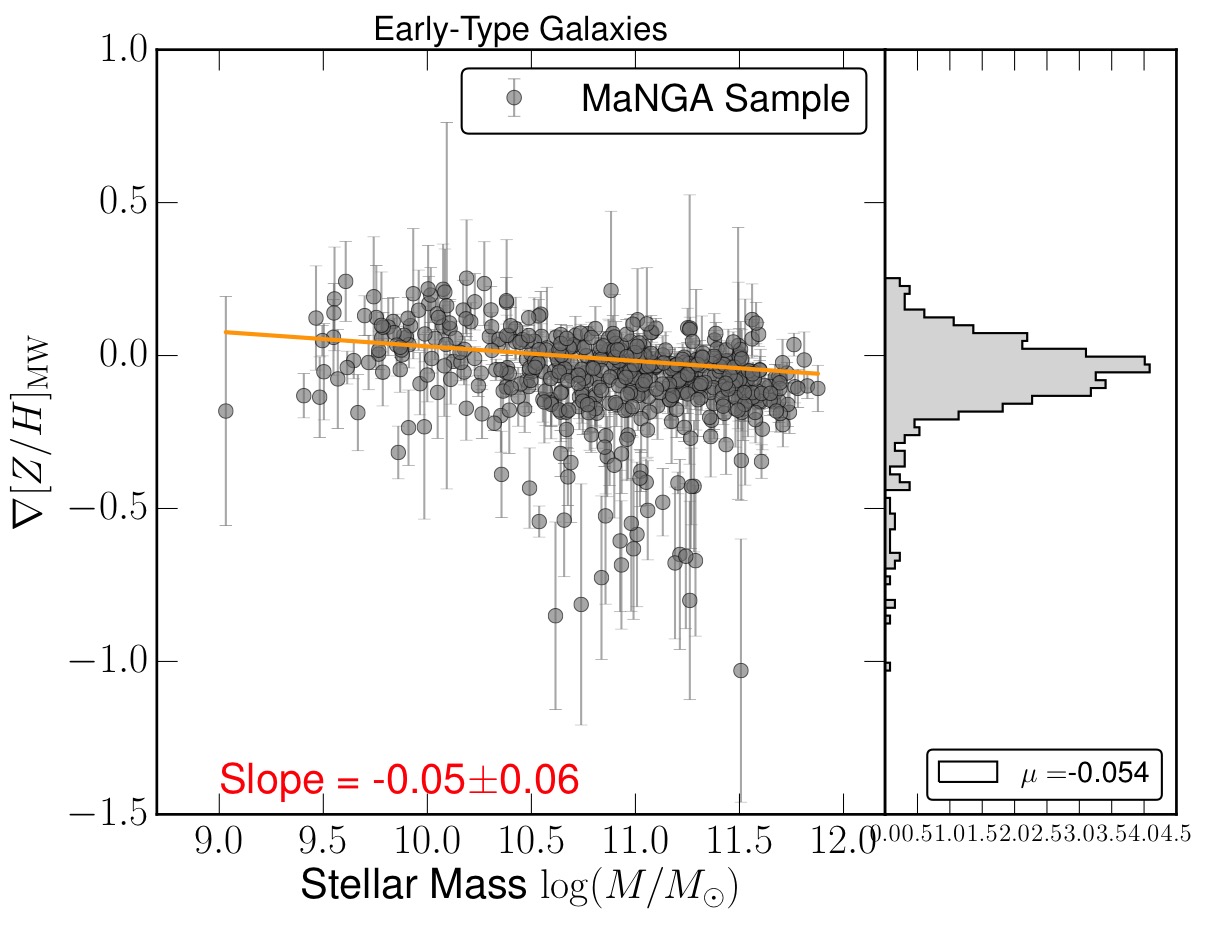}
\includegraphics[width=0.49\textwidth]{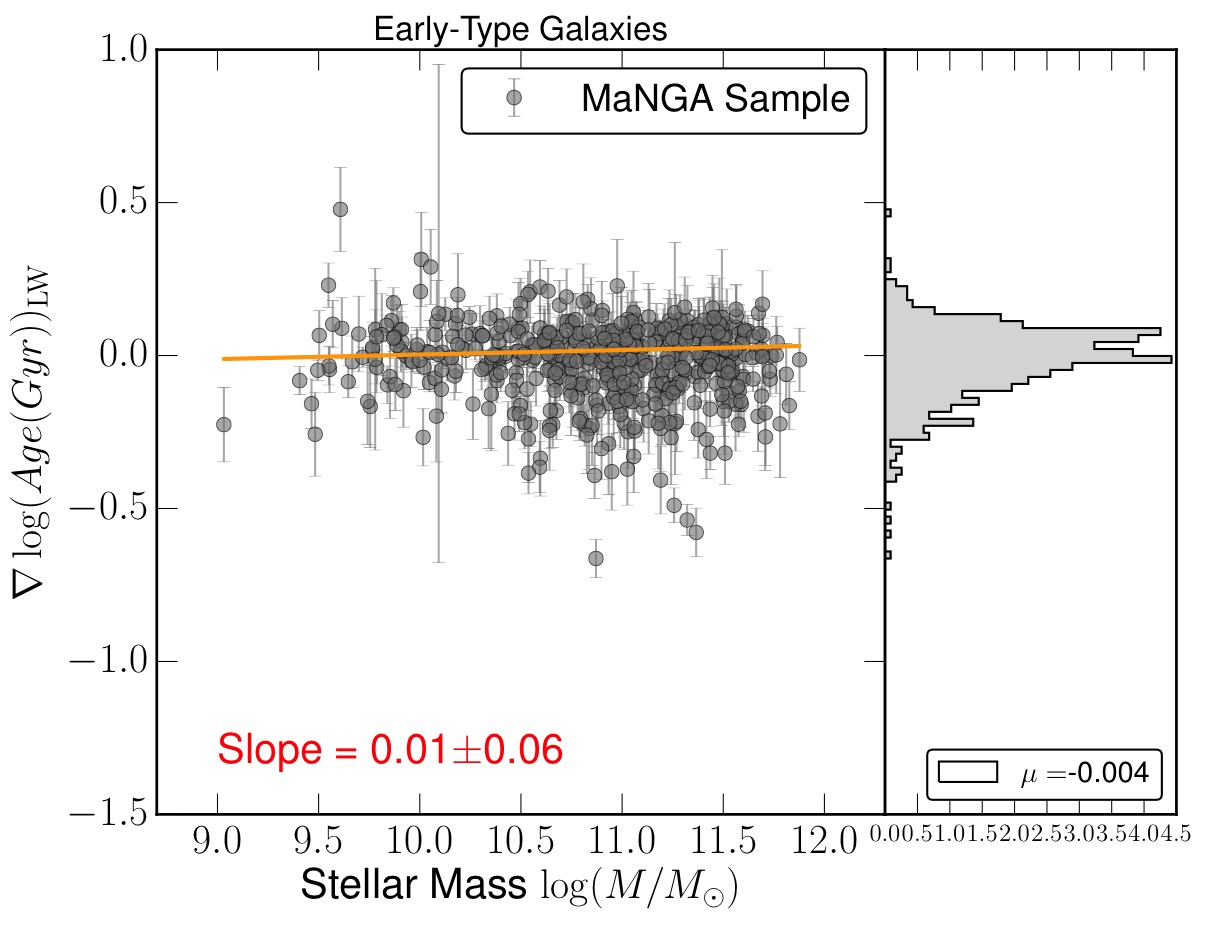}
\includegraphics[width=0.49\textwidth]{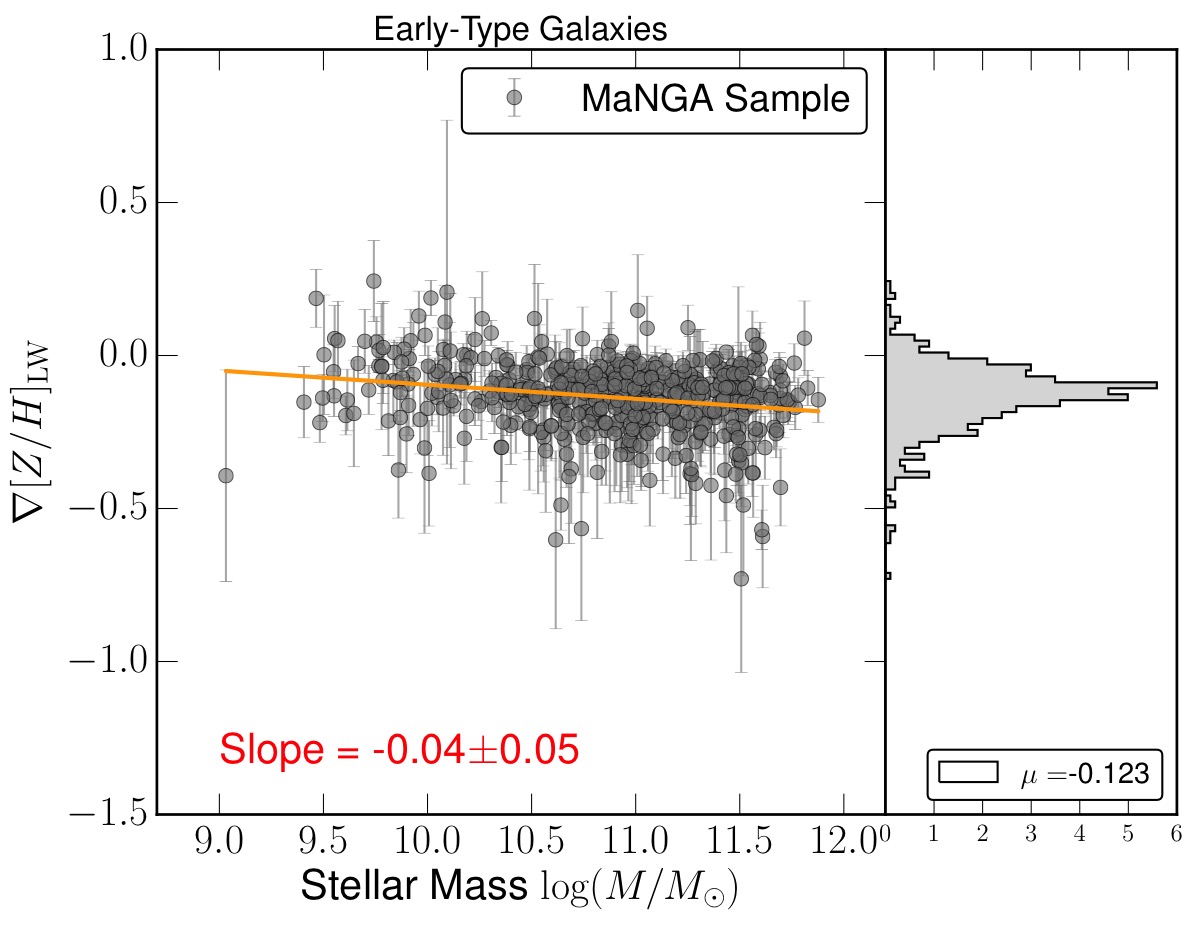}
\begin{centering}
\includegraphics[width=0.49\textwidth]{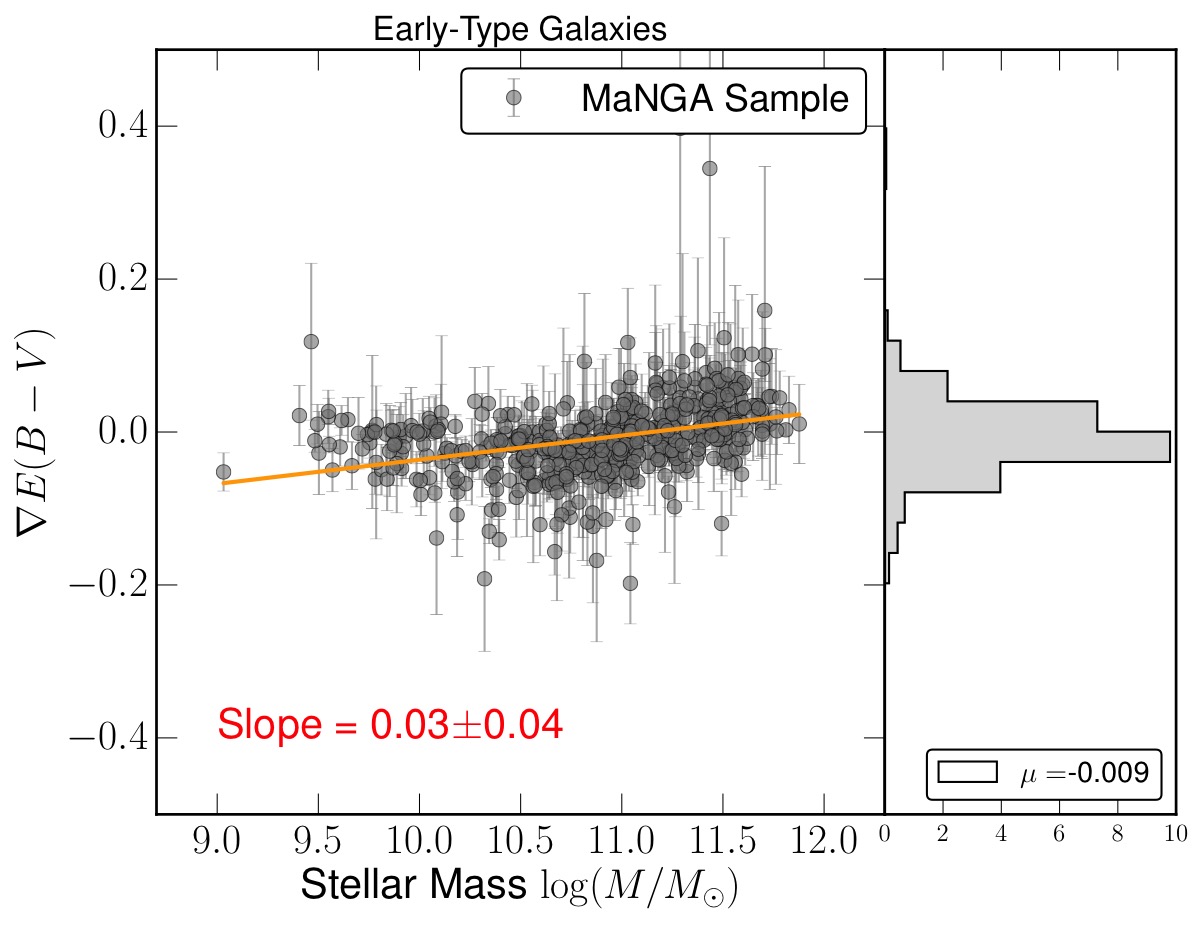}
\end{centering}
\caption{Figure showing the mass and light-weighted stellar population gradients in age (left-hand panels), metallicity (right-hand panels) and $E(B-V)$ (bottom centre) for early-types as a function of galaxy mass. The orange line is a linear fit, with the slope and its error in the legend of each panel. The right-hand sub-panels show the distribution of the gradients. The median value $\mu$ for each distribution is also quoted in the legend.}
\label{fig:early_age_grad}
\end{figure*}

\begin{figure*}
\includegraphics[width=0.49\textwidth]{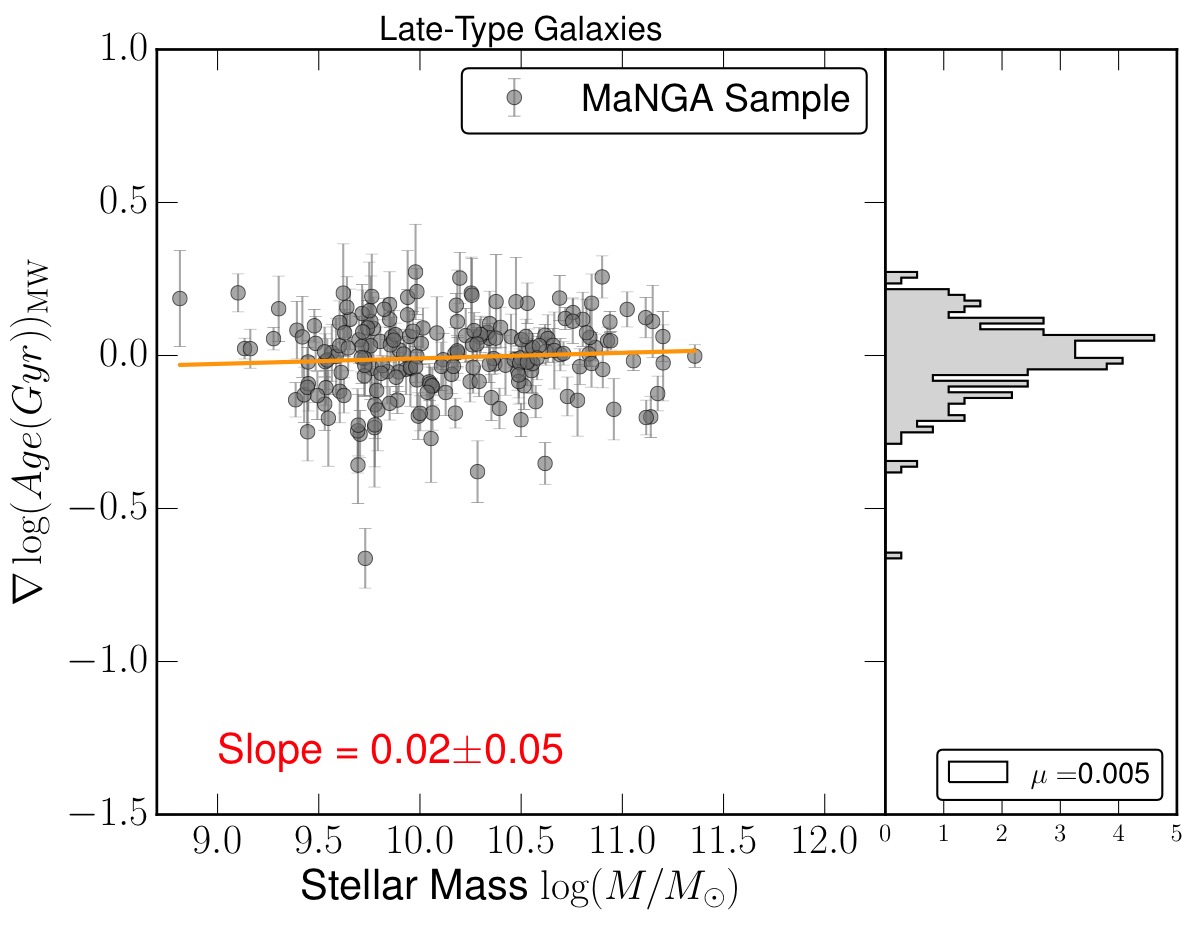}
\includegraphics[width=0.49\textwidth]{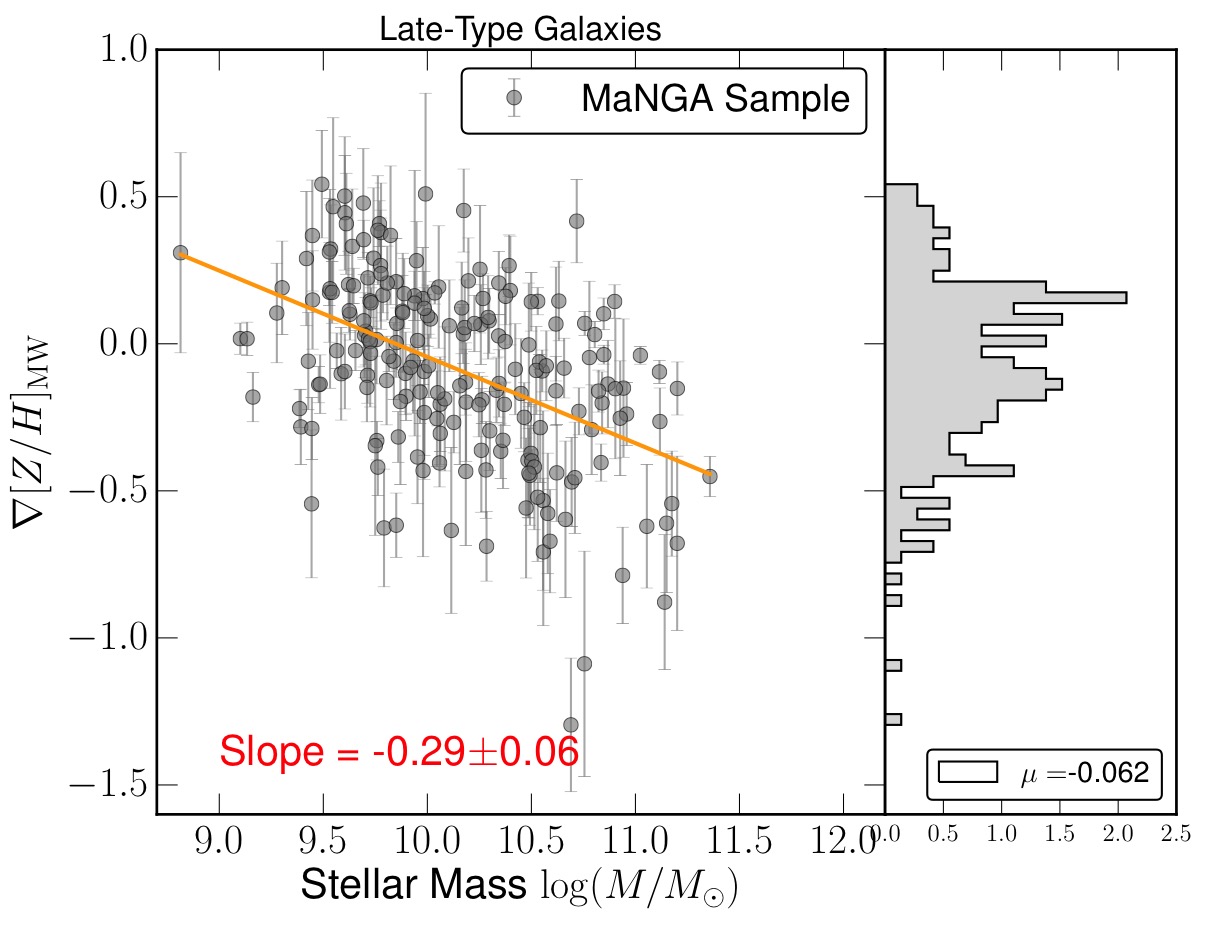}
\includegraphics[width=0.49\textwidth]{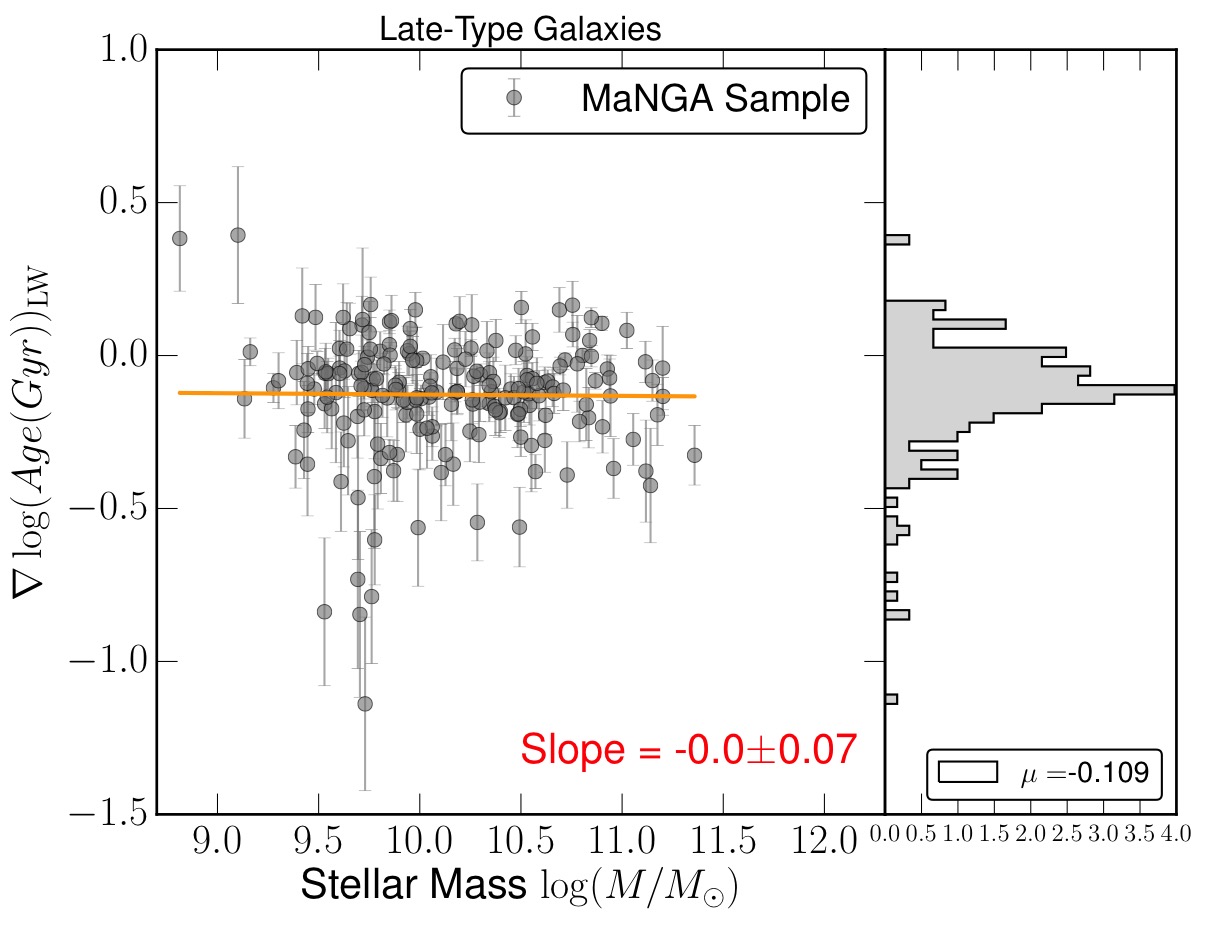}
\includegraphics[width=0.49\textwidth]{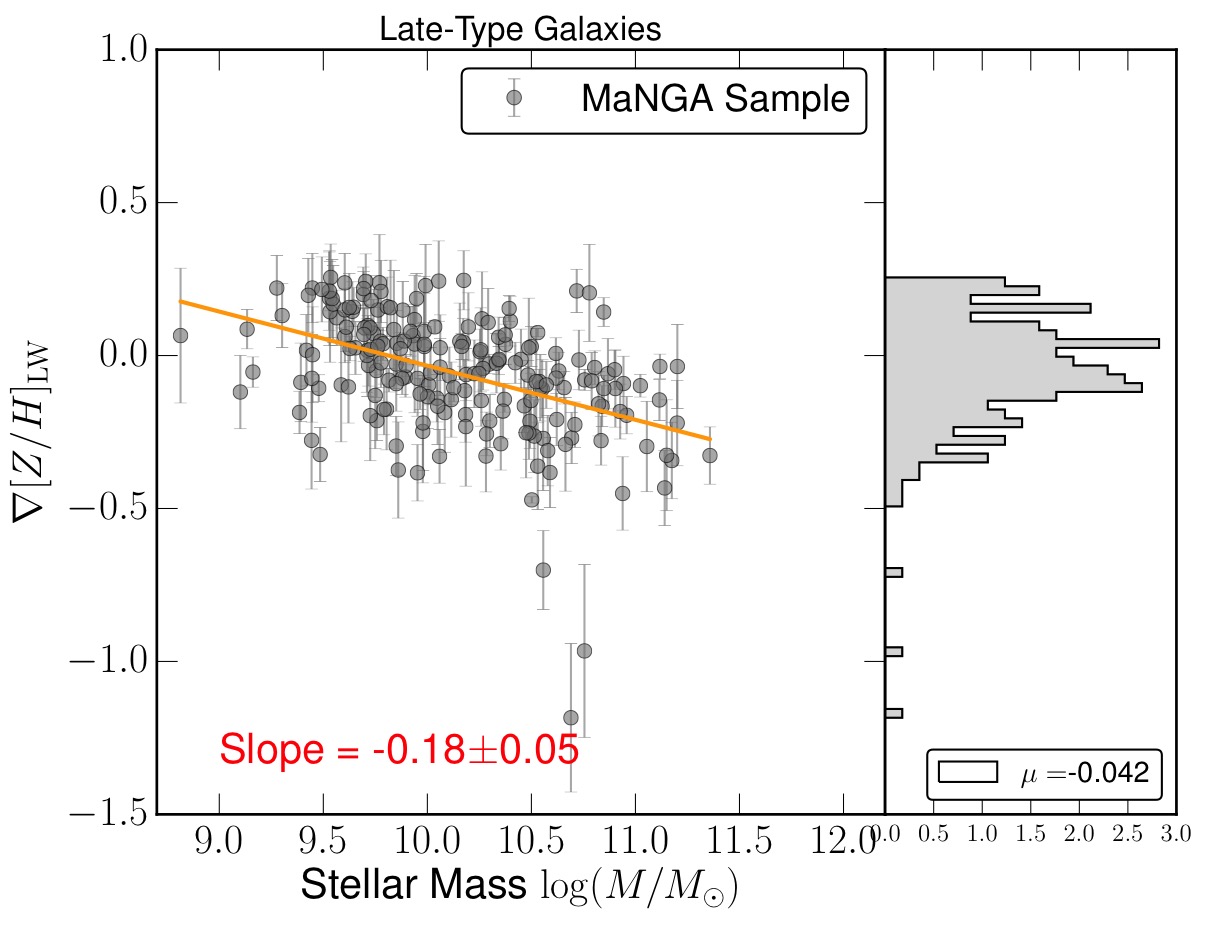}
\begin{centering}
\includegraphics[width=0.49\textwidth]{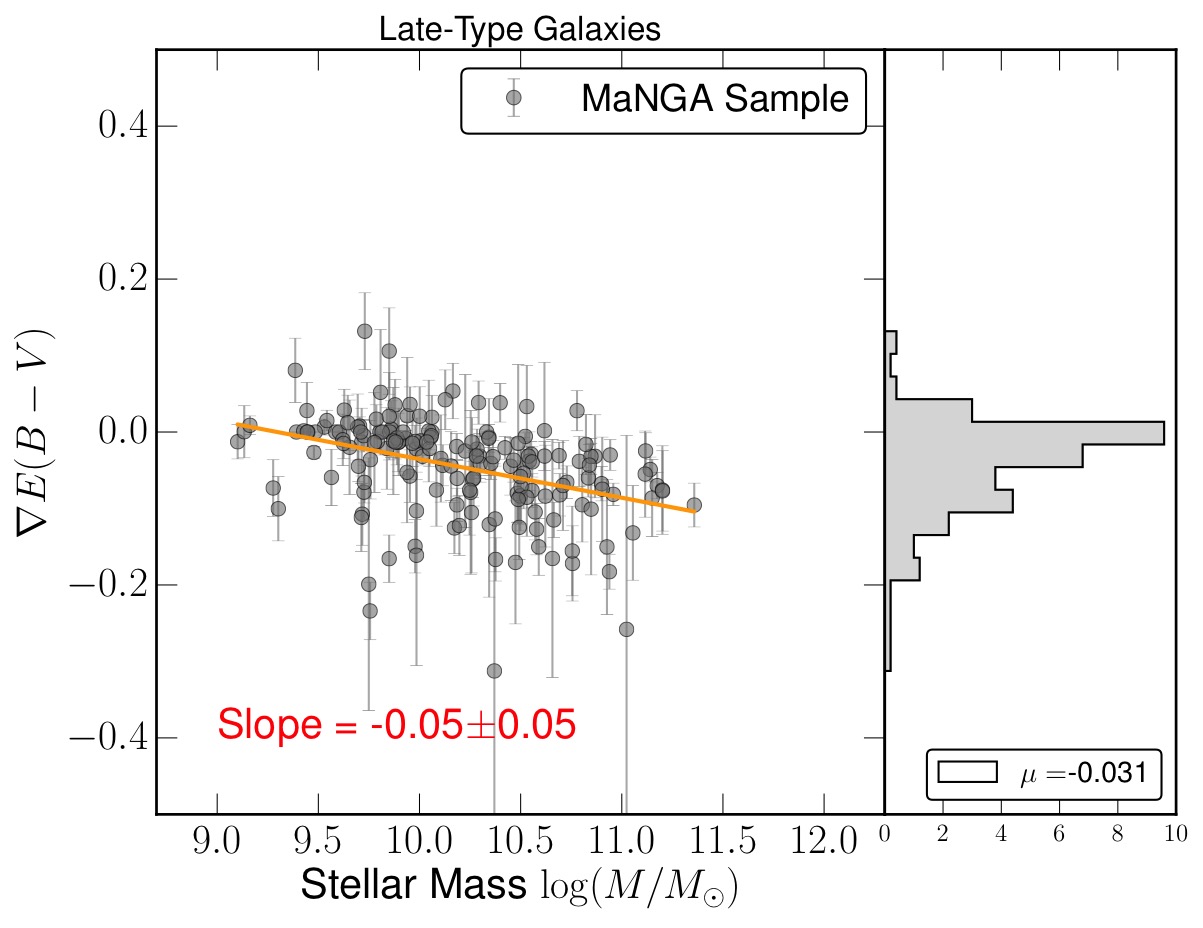}
\end{centering}
\caption{Figure showing the mass and light-weighted stellar population gradients in age (left-hand panels), metallicity (right-hand panels) and $E(B-V)$ (bottom centre) for late-types as a function of galaxy mass. The orange line is a linear fit, with the slope and its error in the legend of each panel. The right-hand sub-panels show the distribution of the gradients. The median value $\mu$ for each distribution is also quoted in the legend.}
\label{fig:late_age_grad}
\end{figure*}

Figures~\ref{fig:early_age_grad} and \ref{fig:late_age_grad} show the derived light and mass-weighted age, metallicity and dust gradients, as a function of stellar mass $\log(M/M_{\odot})$ and their overall distributions, for early and late-type galaxies, respectively. By looking at the overall distribution of gradients in the middle left and top left hand panels of Figure~\ref{fig:early_age_grad}, light-weighted age gradients tend to be flat for early-type galaxies, but slightly positive in mass-weight $\sim0.1$ dex/$R_{e}$. In the middle left and top left hand panels of Figure~\ref{fig:late_age_grad}, late-types galaxies age gradients are slightly negative $\sim-0.1$ dex/$R_{e}$ in light weight, but flat in mass-weight (see also Figure~\ref{fig:gradients}). The bottom central panel in both Figures shows the dust gradients, which are generally flat for early-type galaxies ($\sim0.0$ dex/$R_{e}$) and slightly negative for late-type galaxies ($\sim-0.03$ dex/$R_{e}$). The right hand panels show that metallicity tends to be constantly negative for both morphologies. For early-types, they range from $\sim 0$ to $\sim -0.2\; {\rm dex}/R_{\rm e}$, while for late-types they range from $\sim 0$ to $\sim -0.5\; {\rm dex}/R_{\rm e}$. 
\\
\\
Much of this scatter comes from a well-defined dependence on galaxy mass however, and we can now quantitatively describe the relation between stellar population gradients and stellar mass and assess whether galaxy mass is a significant driver of stellar population gradients. The linear fits and the resulting slopes with their errors are shown in each panel of Figures~\ref{fig:early_age_grad} and \ref{fig:late_age_grad}. It can be seen that luminosity and mass-weighted age gradients generally do not correlate with galaxy mass, both for early and late-type galaxies. The slopes of the gradient-mass relationships are $0.01\pm 0.06$ and $-0.01\pm 0.04$ in light and mass-weight for early-types, and $0.00\pm 0.07$ and $0.02\pm 0.05$ in light and mass-weight for late-types respectively. This is also true for dust gradients in early-type galaxies, where the slope of gradient-mass relationship is $0.03\pm 0.04$.\\ 
\\
The picture is different for metallicity gradients that become steeper with increasing galaxy mass. The gradient-mass relationships are 
\begin{eqnarray}
\nabla [Z/H]_{\rm LW}^{\rm ET} = 0.37 -0.04(\pm 0.05)\times \log(M/M_{\odot}) \\
\nabla [Z/H]_{\rm MW}^{\rm ET} = 0.51 -0.05(\pm 0.06)\times \log(M/M_{\odot}) \\
\nabla [Z/H]_{\rm LW}^{\rm LT} = 1.74 -0.18(\pm 0.05)\times \log(M/M_{\odot}) \\
\nabla [Z/H]_{\rm MW}^{\rm LT} = 2.89 -0.29(\pm 0.06)\times \log(M/M_{\odot}) 
\end{eqnarray}\\
These relationships are quite strong for late-types, and weaker, but still significant, for early-type galaxies. Hence the negative metallicity gradients of galaxies become steeper with increasing galaxy mass but are nearly flat or even positive at masses below $\sim 10^{10}\; M_{\odot}$ for both galaxy types. Given the much weaker correlation of metallicity gradient with galaxy mass for early-type galaxies, we infer that the steepening of the metallicity gradient with galaxy mass in late-type galaxies is mostly driven by gradients of the disc. Such clear dependencies of the metallicity gradient with galaxy mass, in particular for late-type galaxies, had not previously been reported in the literature. In addition to what is presented for metallicity gradients, we also find a weak dependence of the dust gradient in late-type galaxies with mass: the negative dust gradient steepens with increasing galaxy mass.

\section{Discussion}

\begin{figure}
\includegraphics[width=0.49\textwidth]{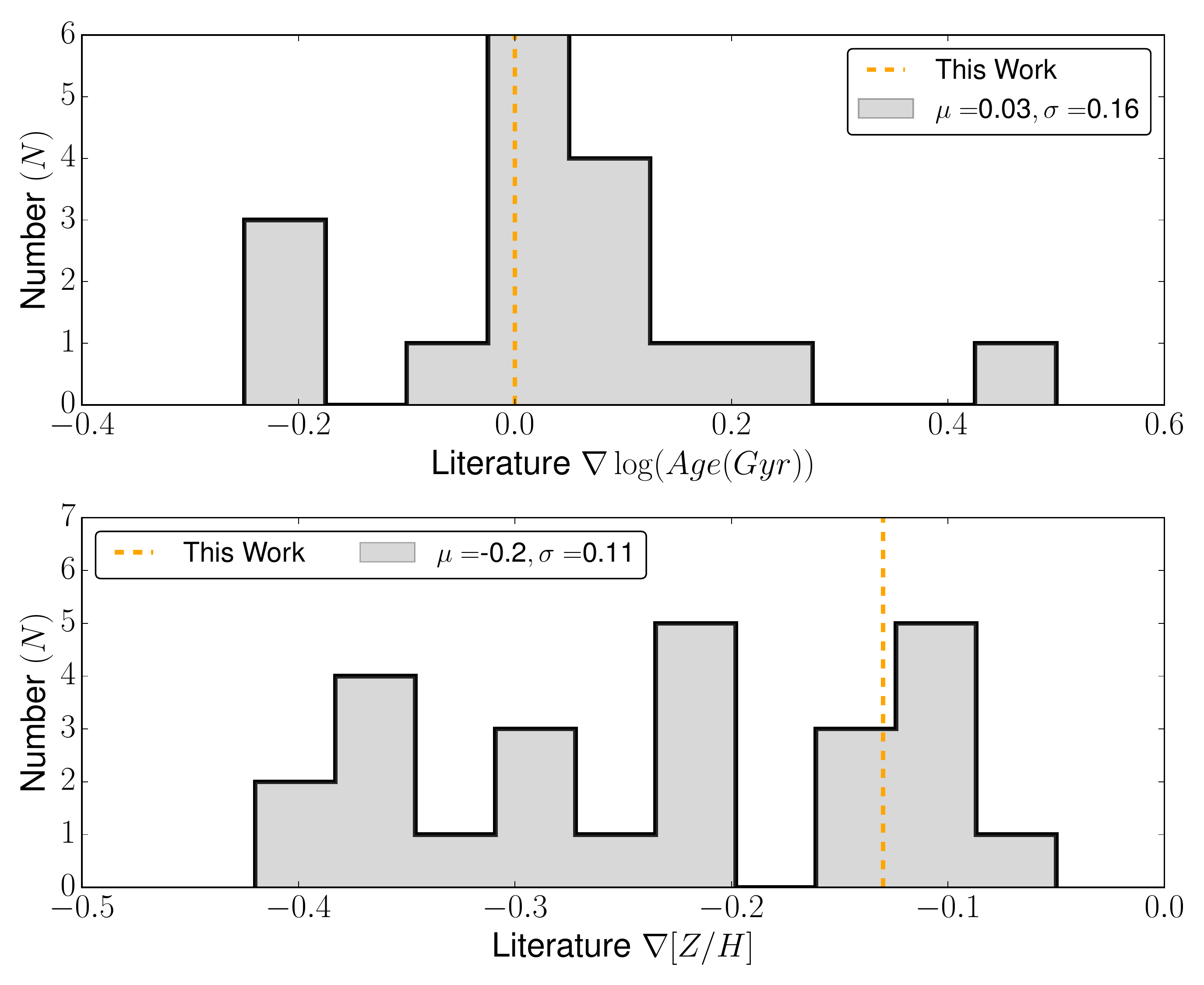}
\caption{Histograms showing the distribution of age (top panel) and metallicity (bottom panel) gradients for early-type galaxies obtained in previous literature studies. All values have been taken from Table~\ref{table:gradient_values}. The orange dashed line represents the light-weighted stellar population gradients obtained in this work.}
\label{fig:literature_gradients}
\end{figure}

An overview of the work on stellar population gradients over the past decade is provided in the Introduction. Here we set our key findings in context with the literature and provide a quantitative comparison for some of the key measurements summarised in Table~\ref{table:gradient_values}, listing gradient measurements in early- and late-type galaxies from 21 and 6 different studies, respectively. The list combines a large variety of different observational approaches, sample sizes and radial coverage. The majority are based on long-slit spectroscopy, some on photometry, and a few on spectroscopic IFU observations. The radial coverage varies between $1$ and $2\ R_{\rm e}$, and most studies present light-weighted stellar population properties. The distributions of the gradients derived in these studies for early-type galaxies are visualised in Figure~\ref{fig:literature_gradients}. We choose not to plot the distributions for late-type galaxies as there are only a few studies covering this area.\\
\\
As discussed in the Introduction, most studies in the literature so far have found that early-type galaxies have a negative metallicity gradient and either a flat or slightly positive age gradient. As can be seen from Table~\ref{table:gradient_values} and Figure~\ref{fig:literature_gradients}, metallicity gradients range from $-0.1$ to $-0.4$ with a median of $-0.2\; {\rm dex}/R_{\rm e}$. The light-weighted metallicity gradient found in the present study ($-0.12\pm 0.05\;$dex/$R_{\rm e}$) is at the shallower side, but well within this distribution. Most notably, the other IFU studies of Table~\ref{table:gradient_values} agree well with our measurement  metallicity gradients (in ${\rm dex}/R_{\rm e}$) of $-0.1$ \citep{gonz2015}, $-0.13$ \citep{rawle2010}, and $-0.09$ \citep{zheng2016}, with \citet{rawle2008} and \citet{kuntschner2010} finding slightly steeper values ($-0.2$ and $-0.28$, respectively). As far as the age gradients of early-type galaxies are concerned, again the measurement reported in the present study is well consistent with the literature. All age gradients in the literature, including other IFU studies with the exception of \citet{gonz2015}, are positive and range from $0$ to $0.45$ with a median of $0.03\; {\rm dex}/R_{\rm e}$. The light-weighted value found in the present study ($0.004\pm 0.06$ dex/$R_{\rm e}$) is in good agreement with this.\\
\\
As discussed in the previous section, the combination of flat to positive age gradients with negative metallicity gradients of early-type galaxies points to an "outside-in" formation scenario in which star formation ceases slightly earlier in the outermost regions with younger, more metal rich stellar populations in the centre. It should be emphasised, however, that the stellar populations are generally old at all radii. Also, our current analysis does not attempt to discriminate between star formation quenching and rejuvenation. Both, inward progression of quenching and/or the rejuvenation of galaxy centres through late, residual star formation will lead to positive age gradients. Hence, the outside-in formation promoted here refers to the radial progression of star formation in galaxies, and not to galaxy assembly. In other words, the conclusions drawn here are not in conflict with the size evolution of galaxies and the fact that galaxies appear to have been more compact in the past \citep[e.g.,][]{2005ApJ...626..680D,2006ApJ...650...18T,2007MNRAS.374..614L,2008A&A...482...21C,2010ApJ...709.1018V,2013ApJ...771...85V,2014ApJ...789...92B}.\\
\\  
The metallicity gradients measured here are much flatter than what is predicted by simple monolithic collapse models alone \citep{1974MNRAS.166..585L,2001ApJ...558..598K, 2004MNRAS.347..740K, 2008A&A...484..679P, pipino2010}, hence the influence of galaxy merging must contribute to the radial distribution of stellar populations in early-type galaxies, which is known to produce flatter gradients \citep{1999ApJ...513..108B,2004MNRAS.347..740K,2009A&A...501L...9D}. It turns out that the age and metallicity gradients found in modern cosmological hydrodynamical simulations of galaxy formation \citep{tortora2011,hirschmann2015,2016arXiv161000014C,2014MNRAS.444.1518V, 2014MNRAS.445..175G} agree well with the observational results reported here. Interestingly, also the dependence on galaxy mass with negative metallicity gradients steepening with increasing galaxy mass is seen in hydrodynamical simulations \citep{tortora2011}.\\
\\
Different from early-type galaxies, late-type galaxies are found to have negative age gradients in light weight ($\nabla \log(Age(Gyr))_{\mathrm{LW}} =-0.11 \pm 0.08$ dex/$R_{e}$), which implies that the central stellar population is older and younger populations reside in the outer parts of a galaxy. Again, this agrees well with other measurements in the literature, where most age gradients are negative (see Table~\ref{table:gradient_values}). Likewise, the negative metallicity gradients found here agree well with other studies in the literature. These gradients will be partly driven by bulge-disc transitions within the galaxies, and true gradients in discs. Bulge and disc need to be disentangled spectroscopically to properly differentiate between these two possibilities \citep[see][]{johnston2016}. Either way, the negative age gradients imply inside-out propagation of star formation in bulge-disc systems as discussed in the previous section. This agrees with the conclusions drawn by \citet{2013ApJ...764L...1P} and \citet{2016MNRAS.463.2799I} who investigate the mass assembly history of galaxies of mixed morphological type.\\
\\
In a companion paper by \citet{zheng2016}, stellar population gradients are derived from MaNGA data following different techniques in sample selection and stellar population analysis, most notably using a different combination of full spectral fitting code and stellar population models (STARLIGHT+BC03 STELIB) than the present work. In Section~\ref{sec:comparison}, we discuss the impact of using different stellar population models and fitting code on the derived stellar population properties and gradients. We show that there is an offset in the age gradients derived for early-types in our work using FIREFLY+M11 MILES and those derived in STARLIGHT+BC03 STELIB of $\mu_{\mathrm{X}-\mathrm{Y}} =0.14$ dex/$R_{\rm e}$ (light-weight) and $\mu_{\mathrm{X}-\mathrm{Y}} =0.15$ dex/$R_{\rm e}$ (mass-weight). There is no significant offset however in the gradients derived for late-type galaxies, instead, where both studies find flat mass-weighted age gradients and negative mass-weighted metallicity gradients. It is clear that the relatively large range in age gradients found in the literature (see Figure~\ref{fig:literature_gradients}) can be attributed to such modelling differences. These systematic uncertainties severely affect current studies, and progress on this end is urgently needed. However, understanding why different fitting codes and stellar population models produce discrepancies, which are of the order on the signal being measured, is challenging. This is due to a combination of factors that contribute to the measurements, such as the method used to calculate the best fit (such as Monte Carlo Markov Chain (MCMC) or $\chi^2$), the adopted stellar library, and the stellar tracks used in the stellar population models. A detailed investigation into this requires sedulous care, and goes beyond the scope of this paper, but is currently being planned within the MaNGA collaboration.

\begin{table*}
\caption{Table providing the results of previous literature studies on stellar population gradients. Early-type galaxies (ET) are classified as both elliptical (E) and lenticular (S0) galaxies. For late-type galaxies, \citet{gonz2015} break down their sample across the Hubble sequence (Sa, Sb, Sbc, Sc, Sd) and we therefore include these results here. Light and mass-weighted gradients are identified in the final column with ${\mathrm{LW}}$ and ${\mathrm{MW}}$ respectively. Additionally, some of the studies below measure galaxy properties at different radial extents, for the use of the reader we describe the nomenclature in Appendix~\ref{sec:appendix}.}
\centering
\begin{tabular}{ccccccccc}
\hline\hline
Author & Galaxy & Sample Size & Observation & Radial &$\nabla$ Age & $\nabla [Z/H]$ & Units & LW/MW\\ 
&  Type & Size & Type & Range &  &  & & \\ [0.5ex]
\hline
This work & ET & 505 & IFU & $1.5R_{\rm e}$ & $0.004 \pm 0.06$ &$-0.12 \pm 0.05$ & dex/$R_{e}$& LW\\[1ex]
This work &  ET & 505 & IFU & $1.5R_{\rm e}$ & $0.09 \pm 0.05$ & $-0.05 \pm 0.07$ &dex/$R_{e}$ & MW \\[1ex]
\citet{baes2007} & ET & 5 & Long-slit & $0.5-3R_{\rm e}$&$-$ &$\sim -0.4$ & dex per radius dex & LW\\[1ex]
\citet{bedregal2011} & S0 & 9 & Long-slit & $1.5R_{\rm e}$ & $\sim 0.2$ &$\sim -0.38$ & dex per radius dex & LW\\[1ex]
\citet{coccato2010} & ET & 1 & Long-slit & $1R_{\rm e}$& $0.5 \pm 1.05$ &$-0.35\pm0.02$ & dex/$R_{e}$ & LW\\[1ex]
\citet{davies1993} & ET &  13 & Long-slit & $1-1.3R_{\rm e}$ & $-$ &  $-0.2 \pm 0.10$ & dex/$R_{e}$ &LW\\[1ex]  
\citet{depropris2005} & ET & 22  & Photometry &$ < 1 \log(R)$ & $-$ & $\sim -0.3$ & dex per radius dex &LW \\[1ex]
\citet{gonz2015} &  E & 41 & IFU & $2$ HLR & $\sim -0.25$ & $\sim -0.1$ & dex/HLR &LW/MW\\[1ex]
\citet{gonz2015} &  S0 & 32 & IFU & $2$ HLR & $\sim -0.23$ & $\sim -0.1$ & dex/HLR &LW/MW\\[1ex]
\citet{gonz2015} &  $<$ET$>$ & 73 & IFU & $2$ HLR & $\sim -0.24$ & $\sim -0.1$ & dex/HLR &LW/MW\\[1ex]
\citet{hirschmann2015} & ET & 10 & Simulation & $2-6R_{\rm e}$ & $\sim0.04$ &$\sim-0.35$ & dex & LW\\[1ex]
\citet{kobayashi1999} & ET & 80 & Long-slit & $ < 2R_{\rm e}$ & $-$ & $-0.3 \pm 0.12$ &dex per radius dex &LW\\[1ex]
\citet{koleva2011} &  ET & 40 & Long-slit & $ < 2R_{\rm e}$ & $\sim 0.1$ & $\sim -0.2$ &dex per radius dex &LW\\[1ex]
\cite{kuntschner2010}  &  ET & 48 & IFU & $ < 1R_{\rm e}$ & $0.02 \pm 0.13$ & $-0.28 \pm 0.12$& dex/$R_{e}$ &LW\\[1ex]
\citet{labarbera2012}  & ET &  674 & Photometric & $ < 8R_{\rm e}$ & $\sim 0.1$ &$\sim -0.35$ & dex/$R_{e}$ &LW\\[1ex]
\citet{mehlert2003} & ET & 35 & Long-slit & $1R_{\rm e}$ & $\sim 0$& $\sim -0.16$ & dex per radius dex &LW\\[1ex]
\citet{rawle2008}  & ET & 12 & IFU & $1R_{\rm e}$& $0.08 \pm 0.08$& $-0.20 \pm 0.05$ & dex &LW\\[1ex]
\citet{rawle2010}  & ET & 19 & IFU  & $1R_{\rm e}$ & $-0.02 \pm 0.06$ & $-0.13 \pm 0.04$ & dex$^{-1}$ &LW\\[1ex]
\citet{reda2007}  & ET & 12 & Long-slit & $1R_{\rm e}$ & $0.04 \pm 0.08$ & $-0.25 \pm 0.05$ & dex &LW\\[1ex]
\citet{roig2015} &  ET & 153,614 & Photometry & $1R_{\rm e}$ & $-$ & $\sim -0.2$ & dex/$R_{e}$ &LW\\[1ex]
\citet{sanchez2007} & ET & 11  &Long-slit & $ < 2R_{\rm e}$ & $0.16 \pm 0.05$ & $-0.33 \pm 0.07$ & dex &LW\\[1ex]
\citet{spolaor2008} & ET & 2 & Long-slit & $ < 1R_{\rm e}$ & $ -0.01 \pm 0.04$ & $-0.42 \pm 0.06$& dex &LW\\[1ex]
\citet{spolaor2009} &  ET & 51 & Long-slit & $1R_{\rm e}$ & $-$ & $\sim -0.16$& dex/$R_{e}$ &LW\\[1ex]
\citet{spolaor2010} & ET & 14 & Long-slit & $1-3R_{\rm e}$& $0.03 \pm 0.17$ & $\sim -0.22$& dex/$R_{e}$ &LW\\[1ex]
\citet{zheng2016} & ET & 463 & IFU & $0.5-1.5R_{\rm e}$& $-0.05 \pm 0.01$ & $-0.09 \pm 0.01$ & dex/$R_{e}$ &MW\\[3ex]
\hline
This work & LT & 216 & IFU  & $1.5R_{\rm e}$ & $-0.11 \pm 0.08$& $-0.007 \pm 0.1$ & dex/$R_{e}$ &LW\\[1ex]
This work &  LT & 216 & IFU & $1.5R_{\rm e}$ & $0.07 \pm 0.07$ & $-0.102 \pm 0.1$ & dex/$R_{e}$ &MW\\[1ex]
\citet{gonz2015}  & Sa & 51 & IFU & $1$HLR & $\sim 0.0$ & $\sim 0.1$ & dex/HLR & LW/MW\\[1ex]
\citet{gonz2015}  & Sb & 53 & IFU & $1$HLR & $\sim -0.3$ & $\sim -0.1$& dex/HLR & LW/MW\\[1ex]
\citet{gonz2015}  & Sbc & 58 & IFU & $1$HLR & $\sim -0.5$ & $\sim -0.15$ & dex/HLR & LW/MW\\[1ex]
\citet{gonz2015}  & Sc & 50 & IFU & $1$HLR & $\sim -0.4$ & $\sim 0.07$ & dex/HLR & LW/MW\\[1ex]
\citet{gonz2015}  & Sd & 15 & IFU & $1$HLR & $\sim -0.25$& $\sim 0.07$ & dex/HLR & LW/MW\\[1ex]
\citet{gonz2015}  & $<$LT$>$& 227 & IFU & $1$HLR & $\sim -0.29$& $\sim -0.02$ & dex/HLR & LW/MW\\[1ex]
\citet{jablonka2007} & LT & 32 & Long-slit &$ < 2R_{\rm e}$ & $0.06 \pm 0.2$ & $-0.16 \pm 0.3$& dex &LW\\[1ex]
\citet{morelli2012}  & LT & 8 & Long-slit &$1.5R_{\rm e}$  & $0.5 \pm 0.9$ & $-0.05 \pm 0.15$& dex &LW\\[1ex]  
\cite{morelli2015}  & LT & 10 & Long-slit & $r_{\mathrm{d95}}-r_{\mathrm{Last}}$&$\sim -0.05 $ & $\sim -0.2 $ & dex &LW\\[1ex]
\citet{sanchez2014} & LT & 62 &  IFU & $1.5R_{\rm e}-r_{\mathrm{Disc}}$ & $-0.04 \pm 0.01$ & $-0.03 \pm 0.006$& dex/$R_{e}$ &LW\\[1ex] 
\citet{sanchez2014} & LT & 62 & IFU & $1.5R_{\rm e}-r_{\mathrm{Disc}}$ & $0.0 \pm 0.006$& $-0.09 \pm 0.008$& dex/$R_{e}$ &MW\\ [1ex]
\citet{zheng2016} & LT & 422 & IFU & $0.5-1.5R_{\rm e}$& $-0.08 \pm 0.02$ & $-0.14 \pm 0.02$ & dex/$R_{e}$ &MW\\[2ex]
\hline
\end{tabular}
\label{table:gradient_values}
\end{table*}

\section{Conclusions}
Mapping Nearby Galaxies at Apache Point Observatory (MaNGA) is a 6-year SDSS-IV survey that is obtaining spatially resolved spectroscopy for a sample of 10,000 nearby galaxies. In this paper, we study the internal gradients of stellar population properties, such as age and metallicity within $1.5\;R_{\rm e}$, for a representative sample of 721 galaxies taken from the first year of MaNGA observations (MPL4, equivalent to DR13) with galaxy masses ranging from $10^{9}\;M_{\odot}$ to $10^{11.5}\;M_{\odot}$. The IFU data was processed by the MaNGA Date Reduction Pipeline and the MaNGA Data Analysis Pipeline to produce data cubes in which cells have been merged through Voronoi binning to a minimum $S/N=5$. We split our galaxy sample into 505 early and 216 late-type galaxies based upon Galaxy Zoo classifications. \\
\\
We then apply our full spectral fitting code FIREFLY on these spectra to derive the stellar population parameters age, metallicity, and dust, as well as full star formation and metal enrichment histories. Very importantly we consider light and mass-weighted average stellar population properties separately, which is key in understanding the true physical formation histories. We use the stellar population models of \cite{maraston2011} (M11), which utilise the MILES stellar library \citep{miles2006} and assume a Kroupa stellar initial mass function (IMF, \cite{kroupa2011}). The median error on age and metallicity at our minimum threshold of $S/N=5$ is 0.28 dex and 0.23 dex, respectively, with a range of scatter up to 0.5 dex. Higher $S/N$ and precision in the determination of stellar population properties are obtained by combining Voronoi cells in radial bins. The resulting typical errors in age and metallicity at $\sim 1.5\;R_{\rm e}$ are 0.06 dex and 0.07 dex, respectively, and even lower at smaller radii. The median errors in age and metallicity gradient are 0.05 dex and 0.07 dex, respectively. To test the dependence of our stellar population measurements both on the spectral fitting technique and the underlying stellar population model, we compare to the results obtained with the spectral fitting code STARLIGHT \citep{starlight2005} using both \citet{maraston2011} and \citet{bruzual2003} STELIB models for a subset of 30 galaxies. We detect significant systematic offsets and large scatter in this comparison. More detailed investigations are needed that go beyond the scope of this paper, but are being considered by the larger MaNGA collaboration. We further look into the effect of beam smearing by studying possible correlations of stellar population gradient on the radial extent to which it is measured using the galaxies from the Secondary MaNGA sample. We conclude that the effect of beam smearing is negligible in our work. \\
\\
In our analysis of stellar population parameters as a function of radius we find that early-type galaxies generally exhibit shallow light-weighted age gradients in agreement with the literature. However, the mass-weighted median age does show some radial dependence with positive gradients ($\sim 0.1\; {\rm dex}/R_{\rm e}$) at all galaxy masses pointing to an "outside-in" progression of star formation. Late-type galaxies, instead, have negative light-weighted age gradients ($\sim -0.1\; {\rm dex}/R_{\rm e}$) in agreement with the literature and observations. We show that mass-weighted age gradients, instead, are flat, which shows that any excess of star formation in the outskirts of galaxy discs compared to the centre must be small and does not contribute significantly to the overall mass budget. We generally detect negative metallicity gradients for both early and late-types at all masses, but these are significantly steeper in late-type compared to early-type galaxies. Very interestingly we find that in late-type galaxies the luminosity-weighted metallicities are systematically larger than the mass-weighted ones by about $\sim 0.35\;$dex, which is the direct consequence of ongoing chemical enrichment producing young metal-rich stellar populations. Our findings show that the radial dependence of chemical enrichment processes and the effect of gas inflow and metal transport are far more pronounced in discs than they are in spheroids. The latter appear to be significantly affected by merging processes rearranging the radial distribution of stellar populations.\\
\\
We further present resolved star formation and metal enrichment histories as function of galaxy type, mass and radius. These show even more clearly that the outermost regions of early-type galaxies are dominated by old stellar populations ($t\sim 10\;$Gyr), while a component of intermediate-age ($t\ga 3\;$Gyr), metal-rich populations are present in the centre suggesting outside-in formation. In late-type galaxies, instead, we detect young, metal-rich populations at all masses and all radii. A well-defined age-metallicity relation is detected with old, metal-poor and young, metal-rich populations, similar to what is observed in the Milky Way. There is mild evidence for inside-out formation with some small excess of recent star formation activity at large radii.\\
\\
We investigate the dependence of stellar population gradients on galaxy mass, and find that age gradients generally do not correlate with galaxy mass, both for early and late-type galaxies. This is different for metallicity gradients. The metallicity gradients of late-type galaxies clearly depend on galaxy mass, with the negative metallicity gradients of disc galaxies becoming steeper with increasing galaxy mass.  They range from $\sim 0$ to $\sim -0.5\; {\rm dex}/R_{\rm e}$ in late-types depending on galaxy mass. The correlation with mass is stronger for late-type galaxies with a slope of $d(\nabla [Z/H])/d(\log M)\sim -0.2\pm 0.05\;$. Early-type galaxies, instead, have metallicity gradients ranging from $\sim 0$ to $\sim -0.2\; {\rm dex}/R_{\rm e}$, and the correlation with mass is weaker with a slope of $d(\nabla [Z/H])/d(\log M)\sim -0.05\pm 0.05\;$. 
\\
\\
It is interesting that more massive early-types galaxies have steeper gradients, which seems counter-intuitive if the evolution of more massive systems are more affected by major mergers. The steeper metallicity gradients must be the result, instead, from the deeper potential wells of massive galaxies affecting the buildup of the metallicity gradient in an outside-in scenario driven by the physics more akin to a monolithic collapse scenario \citep{pipino2010}. Hence this result shows that the merger history plays a relatively small role in shaping metallicity gradients of galaxies. This behaviour appears to be reproduced at least in some cosmological models of galaxy formation. Hydrodynamical simulations of galaxy formation predict age gradients in early-type galaxies to be generally flat and independent of galaxy mass, and negative metallicities to steepen with increasing galaxy mass \citep{tortora2011}. These trends agree well with the findings of this paper. A more comprehensive and direct comparison between MaNGA observations and predictions from galaxy formation simulations will be very valuable in future.

\section*{Acknowledgements}
The authors would like to thank Alfonso Aragon-Salamanca and Matthew Withers for fruitful discussions. DG is supported by an STFC PhD studentship. MAB acknowledges NSF AST-1517006. AW acknowledges support from a Leverhulme Early Career Fellowship. DB is supported by grant RSCF-14-22-00041, RR thanks CNPq and Fapergs for financial support. Numerical computations were done on the Sciama High Performance Compute (HPC) cluster which is supported by the Institute of Cosmology of Gravitation, SEPNet and the University of Portsmouth. Funding for the Sloan Digital Sky Survey IV has been provided by the Alfred P. Sloan Foundation, the U.S. Department of Energy Office of Science, and the Participating Institutions. SDSS- IV acknowledges support and resources from the Center for High-Performance Computing at the University of Utah. The SDSS web site is www.sdss.org. SDSS-IV is managed by the Astrophysical Research Consortium for the Participating Institutions of the SDSS Collaboration including the Brazilian Participation Group, the Carnegie Institution for Science, Carnegie Mellon University, the Chilean Participation Group, the French Participation Group, Harvard-Smithsonian Center for Astrophysics, Instituto de Astrofísica de Canarias, The Johns Hopkins University, Kavli Institute for the Physics and Mathematics of the Universe (IPMU) / University of Tokyo, Lawrence Berkeley National Laboratory, Leibniz Institut für Astrophysik Potsdam (AIP), Max-Planck-Institut für Astronomie (MPIA Heidelberg), Max-Planck-Institut für Astrophysik (MPA Garching), Max-Planck-Institut für Extraterrestrische Physik (MPE), National Astronomical Observatory of China, New Mexico State University, New York University, University of Notre Dame, Observatório Nacional / MCTI, The Ohio State University, Pennsylvania State University, Shanghai Astronomical Observatory, United Kingdom Participation Group, Universidad Nacional Autónoma de México, University of Arizona, University of Colorado Boulder, University of Oxford, University of Portsmouth, University of Utah, University of Virginia, University of Washington, University of Wisconsin, Vanderbilt University, and Yale University. \\
\\
{\it All data taken as part of SDSS-IV is scheduled to be released to the community in fully reduced form at regular intervals through dedicated data releases. The first MaNGA data release was part of the SDSS data release 13 (release date 31 July 2016).}
\\
\\
$^{1}$Institute of Cosmology and Gravitation, University of Portsmouth, Burnaby Road, Portsmouth, UK, PO1 3FX.\\
$^{2}$Instituto de Física, Universidade Federal do Rio Grande do Sul, Campus do Vale, Porto Alegre, Brasil.\\
$^{3}$Laboratório Interinstitucional de e- Astronomia, Rua General José Cristino, 77 Vasco da Gama, Rio de Janeiro, Brasil.\\
$^{4}$National Astronomical Observatories, Chinese Academy of Sciences, A20 Datun Road, Chaoyang District, Beijing 100012, China.\\
$^{5}$Unidad de Astronomía, Fac. Cs. Básicas, U. de Antofagasta, Avda. U. de Antofagasta 02800, Antofagasta, Chile.\\
$^{6}$University of Wisconsin-Madison, Department of Astronomy, 475 N. Charter Street, Madison, WI 53706-1582, USA. \\
$^{7}$Kavli Institute for the Physics and Mathematics of the Universe (WPI), The University of Tokyo Institutes for Advanced Study, Kashiwa, Chiba 277-8583, Japan. \\
$^{8}$McDonald Observatory, Department of Astronomy, University of Texas at Austin, 1 University Station, Austin, TX 78712- 0259, USA.\\
$^{9}$Space Telescope Science Institute, 3700 San Martin Drive, Baltimore, MD 21218, USA.\\
$^{10}$Department of Physics and Astronomy, University of Kentucky, 505 Rose St., Lexington, KY 40506-0057, USA.\\
$^{11}$Department of Physical Sciences, The Open University, Milton Keynes, UK.\\
$^{12}$School of Physics and Astronomy, University of St. Andrews, North Haugh, St. Andrews, KY16 9SS, UK.\\
$^{13}$Apache Point Observatory, P.O. Box 59, Sunspot, NM 88349, USA.\\
$^{14}$Department of Physics and Astronomy, University of Utah, 115 S. 1400 E., Salt Lake City, UT 84112, USA.\\
$^{15}$Instituto de Astrof{\'i}sica, Pontificia Universidad Católica de Chile, Av. Vicuna Mackenna 4860, 782-0436 Macul, Santiago, Chile.\\
$^{16}$University of Cambridge, Cavendish Astrophysics, Cambridge, CB3 0HE, UK.\\
$^{17}$University of Cambridge, Kavli Institute for Cosmology, Cambridge, CB3 0HE, UK.\\
$^{18}$School of Physics and Astronomy, University of Nottingham, University Park, Nottingham NG7 2RD, UK.\\
$^{19}$Unidad de Astronomía, Universidad de Antofagasta, Avenida Angamos 601, Antofagasta 1270300, Chile.\\
$^{20}$Departamento de F{\'i}sica, Facultad de Ciencias, Universidad de La Serena, Cisternas 1200, La Serena, Chile. \\
$^{21}$Department of Astronomy and Astrophysics, The Pennsylvania State University, University Park, PA, 16802, USA. \\
$^{22}$Institute for Gravitation and the Cosmos, The Pennsylvania State University, University Park, PA 16802, USA. \\

\appendix
\section{Radial Gradient Nomenclature}
\label{sec:appendix}
\begin{itemize}
\item{In \citet{gonz2015}, HLR is the half-light radius.}
\item{In \citet{morelli2015}, the analysis is focused on the disc-dominated region between $r_{\mathrm{d95}}$, which is the radius where the disc contributes more than $95\%$ of the galaxy surface brightness, and $r_{\mathrm{Last}}$, which is the farthest radius where the signal-to-noise ratio is sufficient to measure the properties of the stellar populations.}
\item{In \citet{sanchez2014}, $r_{\mathrm{Disc}}$ corresponds to the radius at which the light starts being dominated by the disc.}
\end{itemize}

\bsp	
\label{lastpage}
\end{document}